\documentclass[12pt]{article}
\usepackage{amsfonts}
\usepackage[dvips]{graphicx}
\usepackage{latexsym,amsmath,amssymb,graphics,stmaryrd}
\usepackage{cite}
\usepackage{array}
\usepackage{subfigure}
\usepackage{multirow}
\usepackage{epsfig}
\usepackage[dvips]{graphicx}
\usepackage[dvips]{color}
\usepackage{pstricks}
\usepackage{pst-node}
\usepackage{pst-plot}
\usepackage{dsfont}
\usepackage{amsthm}
\usepackage{graphics}
\usepackage{subfigure}
\usepackage{fancyhdr}
\usepackage[dvips]{color}

\usepackage{feynmp}

\fancypagestyle{plain}{
\fancyhf{}
\newcommand{\fpage}{\iffloatpage{}{\thepage}}
\fancyfoot[C]{\fpage}

}

\newcommand{\col}{~,}
\newcommand{\pnt}{~.}
\newcommand{\AdS}{\text{AdS}}

\newcommand{\YM}{\text{YM}}

\newcommand{\comm}[2]{\left[#1\smash[b]{\mathbin{,}}#2\right]}
\newcommand{\acomm}[2]{\left\{#1\smash[b]{\mathbin{,}}#2\right\}}

\newcommand{\de}{\operatorname{d}\!}

\newcommand{\e}{\operatorname{e}}

\newcommand{\pfour}[4]{{}\{#1,#2,#3,#4\}{}}
\newcommand{\pthree}[3]{{}\{#1,#2,#3\}{}}
\newcommand{\ptwo}[2]{{}\{#1,#2\}{}}
\newcommand{\pone}[1]{{}\{#1\}{}}
\newcommand{\pid}{{}\{ \}{}}

\newlength{\neglength}
\newlength{\diameter}

\newcommand{\svertex}[2]{%
\fmfiequ{#1}{point length(#2)/2 of #2}
}
\newcommand{\dvertex}[3]{%
\fmfiequ{#1}{point length(#3)/3 of #3}
\fmfiequ{#2}{point 2length(#3)/3 of #3}
}
\newcommand{\vvertex}[3]{%
\fmfipath{px}
\fmfiequ{px}{(0,ypart(#2))..(100,ypart(#2))}
\fmfiequ{#1}{point xpart(#3 intersectiontimes px) of #3}
}

\newcommand{\plainwrap}[4]{%
\fmfipath{pi[]}
\fmfiset{pi1}{vloc(__#1) ..controls (-0.175w,ypart(vloc(__#1))) and (-0.175w,-0.15w) .. (xpart(vloc(__#2)),-0.15w)}
\fmfiset{pi2}{(xpart(vloc(__#2)),-0.15w) ..(xpart(vloc(__#3)),-0.15w)}

\fmfiset{pi3}{(xpart(vloc(__#3)),-0.15w) ..controls (1.175w,-0.15w) and (1.175w,ypart(vloc(__#4))) .. vloc(__#4)}
\fmfi{plain}{pi1 ..pi2 ..pi3}
}
\newcommand{\wigglywrap}[4]{%
\fmfipath{pi[]}
\fmfiset{pi1}{#1 ..controls (-0.175w,ypart(#1)) and (-0.175w,-0.15w) .. (xpart(vloc(__#2)),-0.15w)}
\fmfiset{pi2}{(xpart(vloc(__#2)),-0.15w) ..(xpart(vloc(__#3)),-0.15w)}

\fmfiset{pi3}{(xpart(vloc(__#3)),-0.15w) ..controls (1.175w,-0.15w) and (1.175w,ypart(#4)) .. #4}
\fmfi{wiggly}{pi1 ..pi2 ..pi3}
}

\newcommand{\Woneplain}{%
\fmftop{v1}
\fmfbottom{v5}
\fmfforce{(0.125w,h)}{v1}
\fmfforce{(0.125w,0)}{v5}
\fmffixed{(0.25w,0)}{v1,v2}
\fmffixed{(0.25w,0)}{v2,v3}
\fmffixed{(0.25w,0)}{v3,v4}
\fmffixed{(0.25w,0)}{v5,v6}
\fmffixed{(0.25w,0)}{v6,v7}
\fmffixed{(0.25w,0)}{v7,v8}
\fmf{plain,tension=0.5,right=0.25}{v2,vc1}
\fmf{plain,tension=0.5,left=0.25}{v3,vc1}
  \fmf{plain}{vc1,vc2}
\fmf{plain,tension=0.5,left=0.125}{vc3,vc2}
\fmf{plain,tension=0.5,left=0.25}{v6,vc2}
\fmf{plain,tension=0.5,right=0.25}{v7,vc2}
\fmf{plain}{v1,v5}
\fmf{plain}{v4,v8}
\fmf{plain,tension=0.5,right=0,width=1mm}{v5,v8}
\fmfposition
\fmfipath{p[]}
\fmfiset{p1}{vpath(__v1,__v5)}
\fmfiset{p2}{vpath(__v2,__vc1)}
\fmfiset{p3}{vpath(__v3,__vc1)}
\fmfiset{p4}{vpath(__vc1,__vc2)}
\fmfiset{p5}{vpath(__v6,__vc2)}
\fmfiset{p6}{vpath(__v7,__vc2)}
\fmfiset{p7}{vpath(__v4,__v8)}
}

\newcommand{\WoneplainB}{%
\fmftop{v1}
\fmfbottom{v5}
\fmfforce{(0.125w,h)}{v1}
\fmfforce{(0.125w,0)}{v5}
\fmffixed{(0.25w,0)}{v1,v2}
\fmffixed{(0.25w,0)}{v2,v3}
\fmffixed{(0.25w,0)}{v3,v4}
\fmffixed{(0.25w,0)}{v5,v6}
\fmffixed{(0.25w,0)}{v6,v7}
\fmffixed{(0.25w,0)}{v7,v8}
\fmf{plain,tension=0.5,right=0.25}{v2,vc1}
\fmf{plain,tension=0.5,left=0.25}{v3,vc1}
  \fmf{plain}{vc1,vc2}
\fmf{plain,tension=0.5,left=0.125}{vc3,vc2}
\fmf{plain,tension=0.5,left=0.25}{v6,vc2}
\fmf{plain,tension=0.5,right=0.25}{v7,vc2}
\fmfposition
\fmfipath{p[]}
\fmfiset{p1}{vpath(__v1,__v5)}
\fmfiset{p2}{vpath(__v2,__vc1)}
\fmfiset{p3}{vpath(__v3,__vc1)}
\fmfiset{p4}{vpath(__vc1,__vc2)}
\fmfiset{p5}{vpath(__v6,__vc2)}
\fmfiset{p6}{vpath(__v7,__vc2)}
\fmfiset{p7}{vpath(__v4,__v8)}
}

\newcommand{\Wtwoplain}{%
\fmftop{v1}
\fmfbottom{v5}
\fmfforce{(0.125w,h)}{v1}
\fmfforce{(0.125w,0)}{v5}
\fmffixed{(0.25w,0)}{v1,v2}
\fmffixed{(0.25w,0)}{v2,v3}
\fmffixed{(0.25w,0)}{v3,v4}
\fmffixed{(0.25w,0)}{v5,v6}
\fmffixed{(0.25w,0)}{v6,v7}
\fmffixed{(0.25w,0)}{v7,v8}
\fmf{plain,tension=0.5,right=0.25}{v1,vc1}
\fmf{plain,tension=0.5,left=0.25}{v2,vc1}
\fmf{plain,tension=0.5,right=0.25}{v3,vc2}
\fmf{plain,tension=0.5,left=0.25}{v4,vc2}
  \fmf{plain}{vc1,vc3}
  \fmf{plain}{vc2,vc4}
\fmf{plain,tension=0.5,left=0.25}{v5,vc3}
\fmf{plain,tension=0.5,right=0.25}{v6,vc3}
\fmf{plain,tension=0.5,left=0.25}{v7,vc4}
\fmf{plain,tension=0.5,right=0.25}{v8,vc4}
\fmf{plain,tension=0.5,right=0,width=1mm}{v5,v8}
\fmfposition
\fmfipath{p[]}
\fmfiset{p1}{vpath(__v1,__vc1)}
\fmfiset{p2}{vpath(__v2,__vc1)}
\fmfiset{p3}{vpath(__vc1,__vc3)}
\fmfiset{p4}{vpath(__v5,__vc3)}
\fmfiset{p5}{vpath(__v6,__vc3)}
\fmfiset{p6}{vpath(__v3,__vc2)}
\fmfiset{p7}{vpath(__v4,__vc2)}
\fmfiset{p8}{vpath(__vc2,__vc4)}
\fmfiset{p9}{vpath(__v7,__vc4)}
\fmfiset{p10}{vpath(__v8,__vc4)}
}

\newcommand{\Wthreeplain}{%
\fmftop{v1}
\fmfbottom{v5}
\fmfforce{(0.125w,h)}{v1}
\fmfforce{(0.125w,0)}{v5}
\fmffixed{(0.25w,0)}{v1,v2}
\fmffixed{(0.25w,0)}{v2,v3}
\fmffixed{(0.25w,0)}{v3,v4}
\fmffixed{(0.25w,0)}{v5,v6}
\fmffixed{(0.25w,0)}{v6,v7}
\fmffixed{(0.25w,0)}{v7,v8}
\fmffixed{(0,whatever)}{vc1,vc2}
\fmffixed{(0,whatever)}{vc3,vc4}
\fmf{plain,tension=0.5,right=0.25}{v2,vc1}
\fmf{plain,tension=0.5,left=0.25}{v3,vc1}
\fmf{plain,right=0.25}{v1,vc3}
\fmf{plain,tension=0.5,left=0.25}{v5,vc4}
\fmf{plain,tension=0.5,right=0.25}{v6,vc4}
\fmf{plain,right=0.25}{v7,vc2}
\fmf{plain}{v8,v4}
  \fmf{plain,tension=1}{vc1,vc2}
  \fmf{plain,tension=0.5}{vc2,vc3}
  \fmf{plain,tension=1}{vc3,vc4}
\fmf{plain,tension=0.5,right=0,width=1mm}{v5,v8}
\fmfposition
\fmfipath{p[]}
\fmfiset{p1}{vpath(__v1,__vc3)}
\fmfiset{p2}{vpath(__vc3,__vc4)}
\fmfiset{p3}{vpath(__v5,__vc4)}
\fmfiset{p4}{vpath(__v3,__vc1)}
\fmfiset{p5}{vpath(__vc1,__vc2)}
\fmfiset{p6}{vpath(__v7,__vc2)}
\fmfiset{p7}{vpath(__v4,__v8)}
}

\newcommand{\Wfourplain}{%
\fmftop{v1}
\fmfbottom{v5}
\fmfforce{(0.125w,h)}{v1}
\fmfforce{(0.125w,0)}{v5}
\fmffixed{(0.25w,0)}{v1,v2}
\fmffixed{(0.25w,0)}{v2,v3}
\fmffixed{(0.25w,0)}{v3,v4}
\fmffixed{(0.25w,0)}{v5,v6}
\fmffixed{(0.25w,0)}{v6,v7}
\fmffixed{(0.25w,0)}{v7,v8}
\fmffixed{(0,whatever)}{vc1,vc2}
\fmffixed{(0,whatever)}{vc3,vc4}
\fmffixed{(0,whatever)}{vc5,vc6}
\fmf{plain,tension=0.25,right=0.25}{v1,vc1}
\fmf{plain,tension=0.25,left=0.25}{v2,vc1}
\fmf{plain,left=0.25}{v5,vc2}
\fmf{plain,tension=1,left=0.25}{v3,vc3}
\fmf{plain,tension=1,left=0.25}{v4,vc5}
\fmf{plain,left=0.25}{v6,vc4}
\fmf{plain,tension=0.25,left=0.25}{v7,vc6}
\fmf{plain,tension=0.25,right=0.25}{v8,vc6}
  \fmf{plain,tension=0.5}{vc1,vc2}
  \fmf{plain,tension=0.5}{vc2,vc3}
  \fmf{plain,tension=0.5}{vc3,vc4}
  \fmf{plain,tension=0.5}{vc4,vc5}
  \fmf{plain,tension=0.5}{vc5,vc6}
\fmf{plain,tension=0.5,right=0,width=1mm}{v5,v8}
\fmfposition
\fmfipath{p[]}
\fmfiset{p1}{vpath(__v1,__vc1)}
\fmfiset{p2}{vpath(__vc1,__vc2)}
\fmfiset{p3}{vpath(__v5,__vc2)}
\fmfiset{p4}{vpath(__v4,__vc5)}
\fmfiset{p5}{vpath(__vc5,__vc6)}
\fmfiset{p6}{vpath(__v8,__vc6)}
}

\newcommand{\Wfiveplain}{%
\fmftop{v1}
\fmfbottom{v5}
\fmfforce{(0.125w,h)}{v1}
\fmfforce{(0.125w,0)}{v5}
\fmffixed{(0.25w,0)}{v1,v2}
\fmffixed{(0.25w,0)}{v2,v3}
\fmffixed{(0.25w,0)}{v3,v4}
\fmffixed{(0.25w,0)}{v5,v6}
\fmffixed{(0.25w,0)}{v6,v7}
\fmffixed{(0.25w,0)}{v7,v8}
\fmffixed{(0,whatever)}{vc1,vc3}
\fmffixed{(0,whatever)}{vc2,vc4}
\fmffixed{(0,whatever)}{vc5,vc6}
\fmf{plain,tension=1,right=0.25}{v1,vc1}
\fmf{plain,tension=1,left=0.25}{v2,vc1}
\fmf{plain,tension=1,right=0.25}{v3,vc2}
\fmf{plain,tension=1,left=0.25}{v4,vc2}
\fmf{plain,tension=1,left=0.125}{v5,vc3}
\fmf{plain,tension=0.25,left=0.25}{v6,vc6}
\fmf{plain,tension=0.25,right=0.25}{v7,vc6}
\fmf{plain,tension=1,right=0.125}{v8,vc4}
  \fmf{plain,tension=4}{vc1,vc3}
  \fmf{plain,tension=4}{vc2,vc4}
  \fmf{plain,tension=0.5}{vc3,vc5}
  \fmf{plain,tension=0.5}{vc4,vc5}
  \fmf{plain,tension=1}{vc5,vc6}
\fmf{plain,tension=0.5,right=0,width=1mm}{v5,v8}
\fmfposition
\fmfipath{p[]}
\fmfiset{p1}{vpath(__v1,__vc1)}
\fmfiset{p2}{vpath(__vc1,__vc3)}
\fmfiset{p3}{vpath(__v5,__vc3)}
\fmfiset{p4}{vpath(__v4,__vc2)}
\fmfiset{p5}{vpath(__vc2,__vc4)}
\fmfiset{p6}{vpath(__v8,__vc4)}
}

\newcommand{\Wsixplain}{%

\fmftop{v1}
\fmfbottom{v5}
\fmfforce{(0.125w,h)}{v1}
\fmfforce{(0.125w,0)}{v5}
\fmffixed{(0.25w,0)}{v1,v2}
\fmffixed{(0.25w,0)}{v2,v3}
\fmffixed{(0.25w,0)}{v3,v4}
\fmffixed{(0.25w,0)}{v5,v6}
\fmffixed{(0.25w,0)}{v6,v7}
\fmffixed{(0.25w,0)}{v7,v8}
\fmffixed{(0,whatever)}{vc1,vc3}
\fmffixed{(0,whatever)}{vc2,vc4}
\fmffixed{(0,whatever)}{vc5,vc6}
\fmf{plain,tension=1,left=0.25}{v5,vc1}
\fmf{plain,tension=1,right=0.25}{v6,vc1}
\fmf{plain,tension=1,left=0.25}{v7,vc2}
\fmf{plain,tension=1,right=0.25}{v8,vc2}
\fmf{plain,tension=1,right=0.125}{v1,vc3}
\fmf{plain,tension=0.25,right=0.25}{v2,vc6}
\fmf{plain,tension=0.25,left=0.25}{v3,vc6}
\fmf{plain,tension=1,left=0.125}{v4,vc4}
  \fmf{plain,tension=4}{vc1,vc3}
  \fmf{plain,tension=4}{vc2,vc4}
  \fmf{plain,tension=0.5}{vc3,vc5}
  \fmf{plain,tension=0.5}{vc4,vc5}
  \fmf{plain,tension=1}{vc5,vc6}
\fmf{plain,tension=0.5,left=0,width=1mm}{v5,v8}
\fmfposition
\fmfipath{p[]}
\fmfiset{p1}{vpath(__v1,__vc3)}
\fmfiset{p2}{vpath(__vc3,__vc1)}
\fmfiset{p3}{vpath(__v5,__vc1)}
\fmfiset{p4}{vpath(__v4,__vc4)}
\fmfiset{p5}{vpath(__vc4,__vc2)}
\fmfiset{p6}{vpath(__v8,__vc2)}
}

\newcommand{\Wsevenplain}{

\fmftop{v1}
\fmfbottom{v5}
\fmfforce{(0w,h)}{v1}
\fmfforce{(0w,0)}{v5}
\fmffixed{(0.2w,0)}{v1,v2}
\fmffixed{(0.2w,0)}{v2,v3}
\fmffixed{(0.2w,0)}{v3,v4}
\fmffixed{(0.2w,0)}{v4,v9}
\fmffixed{(0.2w,0)}{v5,v6}
\fmffixed{(0.2w,0)}{v6,v7}
\fmffixed{(0.2w,0)}{v7,v8}
\fmffixed{(0.2w,0)}{v8,v10}
\fmffixed{(0,whatever)}{vc1,vc2}
\fmffixed{(0,whatever)}{vc3,vc4}
\fmffixed{(whatever,0)}{vc1,vc6}
\fmffixed{(whatever,0)}{vc2,vc4}
\fmf{plain,tension=0.5,right=0.25}{v2,vc1}
\fmf{plain,tension=0.5,left=0.25}{v3,vc1}
\fmf{plain,right=0.25}{v1,vc3}
\fmf{plain,tension=0.5,left=0.25}{v5,vc4}
\fmf{plain,tension=0.5,right=0.25}{v6,vc4}
\fmf{plain,right=0.25}{v7,vc2}
\fmf{plain,tension=0.5,left=0.25}{v8,vc5}
\fmf{plain,tension=0.5,right=0.25}{v10,vc5}
\fmf{plain,tension=0.5,right=0.25}{v4,vc6}
\fmf{plain,tension=0.5,left=0.25}{v9,vc6}
\fmf{plain}{vc5,vc6}
\fmf{plain,tension=1}{vc1,vc2}
\fmf{plain,tension=0.5}{vc2,vc3}
\fmf{plain,tension=1}{vc3,vc4}
\fmf{plain,tension=0.5,right=0,width=1mm}{v5,v10}
\fmfposition
\fmfipath{p[]}
\fmfiset{p1}{vpath(__v1,__vc3)}
\fmfiset{p2}{vpath(__vc3,__vc4)}
\fmfiset{p3}{vpath(__v5,__vc4)}
\fmfiset{p4}{vpath(__v3,__vc1)}
\fmfiset{p5}{vpath(__vc1,__vc2)}
\fmfiset{p6}{vpath(__v7,__vc2)}
\fmfiset{p7}{vpath(__v4,__v8)}
\fmfiset{p8}{vpath(__v4,__vc6)}
\fmfiset{p9}{vpath(__vc6,__vc5)}
\fmfiset{p10}{vpath(__v8,__vc5)}
}

\newcommand{\Weightplain}{

\fmftop{v1}
\fmfbottom{v5}
\fmfforce{(0w,h)}{v1}
\fmfforce{(0w,0)}{v5}
\fmffixed{(0.2w,0)}{v1,v2}
\fmffixed{(0.2w,0)}{v2,v3}
\fmffixed{(0.2w,0)}{v3,v4}
\fmffixed{(0.2w,0)}{v4,v9}
\fmffixed{(0.2w,0)}{v5,v6}
\fmffixed{(0.2w,0)}{v6,v7}
\fmffixed{(0.2w,0)}{v7,v8}
\fmffixed{(0.2w,0)}{v8,v10}
\fmffixed{(0,whatever)}{vc1,vc2}
\fmffixed{(0,whatever)}{vc3,vc4}
\fmffixed{(whatever,0)}{vc1,vc6}
\fmffixed{(whatever,0)}{vc2,vc4}
\fmf{plain,tension=0.5,right=0.25}{v4,vc1}
\fmf{plain,tension=0.5,left=0.25}{v9,vc1}
\fmf{plain,right=0.25}{v3,vc3}
\fmf{plain,tension=0.5,left=0.25}{v7,vc4}
\fmf{plain,tension=0.5,right=0.25}{v8,vc4}
\fmf{plain,right=0.25}{v10,vc2}
\fmf{plain,tension=0.5,left=0.25}{v5,vc5}
\fmf{plain,tension=0.5,right=0.25}{v6,vc5}
\fmf{plain,tension=0.5,right=0.25}{v1,vc6}
\fmf{plain,tension=0.5,left=0.25}{v2,vc6}
\fmf{plain}{vc5,vc6}
\fmf{plain,tension=1}{vc1,vc2}
\fmf{plain,tension=0.5}{vc2,vc3}
\fmf{plain,tension=1}{vc3,vc4}
\fmf{plain,tension=0.5,right=0,width=1mm}{v5,v10}
\fmfposition
\fmfipath{p[]}
\fmfiset{p1}{vpath(__v3,__vc3)}
\fmfiset{p2}{vpath(__vc3,__vc4)}
\fmfiset{p3}{vpath(__v7,__vc4)}
\fmfiset{p4}{vpath(__v9,__vc1)}
\fmfiset{p5}{vpath(__vc1,__vc2)}
\fmfiset{p6}{vpath(__v10,__vc2)}
\fmfiset{p7}{vpath(__v1,__v8)}
\fmfiset{p8}{vpath(__v2,__vc6)}
\fmfiset{p9}{vpath(__vc6,__vc5)}
\fmfiset{p10}{vpath(__v6,__vc5)}
}

\newcommand{\Wnineplain}{

\fmftop{v1}
\fmfbottom{v5}
\fmfforce{(0w,h)}{v1}
\fmfforce{(0w,0)}{v5}
\fmffixed{(0.2w,0)}{v1,v2}
\fmffixed{(0.2w,0)}{v2,v3}
\fmffixed{(0.2w,0)}{v3,v4}
\fmffixed{(0.2w,0)}{v4,v9}
\fmffixed{(0.2w,0)}{v5,v6}
\fmffixed{(0.2w,0)}{v6,v7}
\fmffixed{(0.2w,0)}{v7,v8}
\fmffixed{(0.2w,0)}{v8,v10}
\fmffixed{(0,whatever)}{vc1,vc2}
\fmffixed{(0,whatever)}{vc3,vc4}
\fmf{plain,tension=0.5,right=0.25}{v1,vc1}
\fmf{plain,tension=0.5,left=0.25}{v2,vc1}
\fmf{plain,tension=0.5,left=0.25}{v5,vc2}
\fmf{plain,tension=0.5,right=0.25}{v6,vc2}
\fmf{plain,tension=0.5,left=0.25}{v8,vc4}
\fmf{plain,tension=0.5,right=0.25}{v10,vc4}
\fmf{plain,tension=0.5,right=0.25}{v4,vc3}
\fmf{plain,tension=0.5,left=0.25}{v9,vc3}
\fmf{plain,tension=1}{vc1,vc2}
\fmf{plain,tension=1}{vc3,vc4}
\fmf{plain,tension=1}{v3,v7}
\fmf{plain,tension=0.5,right=0,width=1mm}{v5,v10}
\fmfposition
\fmfipath{p[]}
\fmfiset{p1}{vpath(__v2,__vc1)}
\fmfiset{p2}{vpath(__vc1,__vc2)}
\fmfiset{p3}{vpath(__v6,__vc2)}
\fmfiset{p4}{vpath(__v3,__v7)}
\fmfiset{p5}{vpath(__v4,__vc3)}
\fmfiset{p6}{vpath(__vc3,__vc4)}
\fmfiset{p7}{vpath(__vc4,__v8)}
}

\DeclareMathOperator{\tr}{tr}

\DeclareMathOperator{\perm}{P}

\numberwithin{equation}{section}
\newlength{\eqoff}
\newlength{\eqofftwo}

\newlength{\unit}
\psset{xunit=\unit,yunit=\unit,runit=\unit}
\newlength{\linew}
\psset{linewidth=\linew}
\begin{document}
\begin{fmffile}{fullgraphs}

\fmfcmd{
wiggly_len := 2mm;
vardef wiggly expr p_arg =
 save wpp,len;
 numeric wpp,alen;
 wpp = ceiling (pixlen (p_arg, 10) / wiggly_len);
 len=length p_arg;
 for k=0 upto wpp - 1:
  point arctime k/(wpp-1)*arclength(p_arg) of p_arg of p_arg
    {direction arctime k/(wpp-1)*arclength(p_arg) of p_arg of p_arg rotated wiggly_slope} ..
  point  arctime (k+.5)/(wpp-1)*arclength(p_arg) of p_arg of p_arg
 {direction arctime (k+.5)/(wpp-1)*arclength(p_arg) of p_arg of p_arg rotated - wiggly_slope} ..
 endfor
 if cycle p_arg: cycle else: point infinity of p_arg fi
enddef;
}
\fmfcmd{%
marksize=2mm;
def draw_mark(expr p,a) =
  begingroup
    save t,tip,dma,dmb; pair tip,dma,dmb;
    t=arctime a of p;
    tip =marksize*unitvector direction t of p;
    dma =marksize*unitvector direction t of p rotated -45;
    dmb =marksize*unitvector direction t of p rotated 45;
    linejoin:=beveled;
    draw (-.5dma.. .5tip-- -.5dmb) shifted point t of p;
  endgroup
enddef;
style_def derplain expr p =
    save amid;
    amid=.5*arclength p;
    draw_mark(p, amid);
    draw p;
enddef;
def draw_point(expr p,a) =
  begingroup
    save t,tip,dma,dmb,dmo; pair tip,dma,dmb,dmo;
    t=arctime a of p;
    tip =marksize*unitvector direction t of p;
    dma =marksize*unitvector direction t of p rotated -45;
    dmb =marksize*unitvector direction t of p rotated 45;
    dmo =marksize*unitvector direction t of p rotated 90;
    linejoin:=beveled;
    draw (-.5dma.. .5tip-- -.5dmb) shifted point t of p withcolor 0white;
    draw (-.5dmo.. .5dmo) shifted point t of p;
  endgroup
enddef;
style_def derplainpt expr p =
    save amid;
    amid=.5*arclength p;
    draw_point(p, amid);
    draw p;
enddef;
style_def dblderplain expr p =
    save amidm;
    save amidp;
    amidm=.5*arclength p-0.75mm;
    amidp=.5*arclength p+0.75mm;
    draw_mark(p, amidm);
    draw_point(p, amidp);
    draw p;
enddef;
}

\begin{titlepage}
\begin{flushright}
IFUM-919-FT\\
\end{flushright}
\vspace{5ex}
\Large
\begin {center}
{\bf
Anomalous dimension with wrapping at four
loops in $\mathcal{N}=4$ SYM}
\end {center}

\renewcommand{\thefootnote}{\fnsymbol{footnote}}

\large
\vspace{1cm}
\centerline{F.\ Fiamberti ${}^{a,b}$, A.\ Santambrogio ${}^b$,
C.\ Sieg ${}^b$, D.\ Zanon ${}^{a,b}$
\footnote[1]{\noindent \tt
francesco.fiamberti@mi.infn.it \\
\hspace*{6.3mm}alberto.santambrogio@mi.infn.it \\
\hspace*{6.3mm}csieg@mi.infn.it \\
\hspace*{6.3mm}daniela.zanon@mi.infn.it}}
\vspace{4ex}
\normalsize
\begin{center}
\emph{$^a$  Dipartimento di Fisica, Universit\`a degli Studi di Milano, \\
Via Celoria 16, 20133 Milano, Italy}\\
\vspace{0.2cm}
\emph{$^b$ INFN--Sezione di Milano,\\
Via Celoria 16, 20133 Milano, Italy}
\end{center}
\vspace{0.5cm}
\rm
\abstract
\normalsize
In this paper we give all the details of the calculation that we presented in
our previous paper arXiv:0712.3522, concerning the four-loop anomalous
dimension of the Konishi descendant $\tr(\phi Z\phi Z-\phi \phi Z Z)$
in the $SU(2)$ sector of the ${\cal{N}}=4$ planar SYM theory.
We explicitly consider all the wrapping diagrams that we compute using an ${\cal{N}}=1$ superspace approach and
Gegenbauer polynomial $x$-space techniques.

\vfill
\end{titlepage}

\section{Introduction}

\label{sec:intro}

Recently we have presented \cite{Fiamberti:2007rj} the calculation of the
anomalous dimension of the composite operator
$\tr(\phi Z\phi Z-\phi \phi Z Z)$ at four loops in the ${\cal N}=4$ planar 
supersymmetric Yang-Mills theory, where $\phi$ and $Z$ are two complex
scalar fields of the theory.

The importance of this calculation resides in the fact that this is the
simplest case in which the so-called {\it wrapping} effects are present:
at a given perturbative order $K$ in
\begin{equation}\label{lambdadef}
\lambda=\frac{g^2N}{16\pi^2}\col
\end{equation}
where $g=g_\YM$ is the Yang-Mills coupling constant, 
the range of the interactions between
adjacent fields grows with the perturbative order as $K+1$. For an
operator with $L$ fields we should expect new effects when the range
exceeds $L$, that is at order $g^{2L}$. It turns out that $L=4$ is the simplest
case in which these effects can be seen (operators with $L=2$ and $L=3$
are automatically BPS).

Understanding these wrapping effects is a crucial step towards a comparison
between the spectrum of the anomalous dimensions of gauge invariant
operators of planar $\mathcal{N}=4$ SYM  and the spectrum of
strings on $\AdS_5\times\text{S}^5$, as predicted by the AdS/CFT conjecture
\cite{maldacena:1998re}.

Indeed, in the case of {\it long} operators, the comparison between the
spectra of the two theories is now at hand, thanks to the recent
progress in understanding their integrability properties. On the gauge
theory side it was crucial the realization \cite{Minahan:2002ve,
Beisert:2003tq} that the planar one-loop dilatation operator of
$\mathcal{N}=4$ SYM maps
into the Hamiltonian of an integrable spin chain.
This spin chain picture was extended to higher orders in perturbation
theory \cite{Beisert:2003ys,Staudacher:2004tk} and thanks to the
integrability properties of the system the computation of the anomalous
dimensions was then reduced to finding solutions of the associated Bethe
equations \cite{Beisert:2004hm}.
This program cannot be applied to {\it short} operators since
the Bethe ansatz is only asymptotic and wrapping effects appear at order
$g^{2L}$: the interaction is no longer localized in some limited region
along the state and asymptotic states cannot  be defined.

Clearly from the perspective of integrable models including wrapping effects 
is very important and several papers have addressed this issue. In \cite{Sieg:2005kd} the properties of wrapping interactions have been analyzed in terms of Feynman diagrams. In \cite{Fischbacher:2004iu} wrapping interactions in 
the BMN matrix model have been discussed. 
In \cite{Ambjorn:2005wa,Janik:2007wt} it was proposed to study wrapping
interactions via the thermodynamic Bethe ansatz, while in \cite{Rej:2005qt} it was assumed that wrapping effects might be
described by the Hubbard model.
In \cite{Kotikov:2007cy} the  form of wrapping contributions at four
loops was conjectured by using restrictions from the BFKL equation
\cite{Lipatov:1976zz,Kuraev:1977fs,Balitsky:1978ic}. In \cite{Penedones:2008rv} wrapping effects in some simple spin chain models were studied. On the string theory side, finite size contributions have been analyzed in several recent papers \cite{Arutyunov:2006gs,SchaferNameki:2006ey,SchaferNameki:2006gk,Gromov:2008ie,Minahan:2008re,Heller:2008at,Ramadanovic:2008qd,Hatsuda:2008gd}.

In order to test these proposals it is crucial to perform an explicit field
theoretical computation of an anomalous dimension at an order in which wrapping contributions become relevant.
As mentioned above the simplest such a case  is the
four-loop anomalous dimension of the Konishi descendant
$\tr(\phi Z\phi Z-\phi \phi Z Z)$ in the $SU(2)$ sector of the theory.

In this paper we describe in detail the computation we have
presented in \cite{Fiamberti:2007rj}, with the aim of supplying the
interested reader with all the details needed to reproduce the calculation.

The complete four-loop Feynman graph calculation is very complicated.
However one can  take advantage of the approach based on the asymptotic
dilatation operator and avoid the explicit computation of some of the most difficult classes of graphs.
Indeed the contribution from all non-wrapping graphs, corresponding
to interactions of range from one to four, is included in the four-loop 
asymptotic dilatation operator together with the interactions of range five. 

Then one is left with two operations to be done: the subtraction of the 
range-five contributions from the dilatation operator, and the computation of 
the wrapping graphs. These two operations are drastically simplified by means
of $\mathcal{N}=1$ superspace techniques. The integrals
that we produce after $D$-algebra manipulations on the supergraphs are
computed by the Gegenbauer polynomial
$x$-space technique (GPXT)
\cite{Chetyrkin:1980pr} and by the method of  uniqueness 
\cite{Kazakov:1983ns}.

The anomalous dimension of a composite operator $\mathcal{O}$ is
extracted from the $\frac{1}{\epsilon}$ pole of the graphs contributing
to its renormalization in a standard way. In the special case in which
the operator is multiplicatively renormalized the anomalous dimension
is given by
\begin{equation}
\label{andim}
\gamma(\mathcal{O})=\lim_{\varepsilon\rightarrow0}\left[\varepsilon
g\frac{\mathrm{d} }{\mathrm{d} g}
\log\mathcal{Z}_{\mathcal{O}}(g,\varepsilon)\right]
\col
\end{equation}
where
\begin{equation}
\mathcal{O}_\text{ren}=\mathcal{Z}_{\mathcal{O}}\,\mathcal{O}_\text{bare}
\pnt
\end{equation}
In the general case of operator mixing, the last equation
should be rewritten in matrix form and the anomalous dimensions
are given by  the eigenvalues of the matrix appearing on the right hand side of (\ref{andim}).
In our case we have two mixing operators but it turns out that the combination
$\tr(\phi Z\phi Z-\phi \phi Z Z)$ is an exact eigenvector of the renormalization
matrix to all orders, so it is multiplicatively renormalized.

The most remarkable outcome of our calculation is that at four-loop order
wrapping interactions give rise to contributions proportional to $\zeta(5)$
increasing the level of transcendentality of the anomalous dimensions.
Our result rules out the previous conjectures on how to treat wrapping
contributions proposed in \cite{Rej:2005qt,Beisert:2006ez,Kotikov:2007cy}.

Another computation of the same anomalous dimension appeared later on
in \cite{Keeler:2008ce}.
The results presented there do not match our findings, but the authors did not attempt
a  comparison with the computations we had reported.
The result of \cite{Keeler:2008ce} contains a trivial error
in the final step, however even after correcting it, the result is
different from ours. Finding the origin of the discrepancy is not an easy
task. The authors of \cite{Keeler:2008ce} used exactly the same strategy as ours but their calculation of the wrapping diagrams is performed using a component field approach. Thus the number of diagrams they have to compute is huge, supersymmetry is not maintained explicitly, renormalization is subtle and partial comparison with our classes of diagrams is almost impossible. Needless to say that keeping under control such a large number of contributions is very difficult. On the contrary using superfields the calculation is much simpler since several cancellations are automatic, and the number of relevant diagrams is significantly reduced.  In any event
we hope that all the detailed explanations
we are going to present in this paper will help to clarify the situation.

Our paper is organized as follows. In Section \ref{sec:N1superspace} we recall the basic ingredients
of the $\mathcal{N}=1$ superspace approach, and we discuss an important result
on the displacement of spinorial derivatives on a supergraph that allows
us to drastically reduce the number of potential contributions.
In Section \ref{sec:r5sub} we describe how to subtract the range-five terms from the
asymptotic Hamiltonian and we build the effective Hamiltonian containing all
the contributions from range one up to range four.
Thanks to a cancellation described in Appendix \ref{app:non-maximal}, 
the range-five subtraction procedure can be performed without computing any 
Feynman
diagram. However in order to provide a further check of our procedure we
have computed all range-five graphs explicitly. This is reported in Appendix \ref{app:range-five}.
Subsection \ref{sec:wrapping} contains the calculation of the wrapping supergraphs. The full list
of them is reported in Appendix \ref{app:wrapping},  with a description of all
the cancellations occurring among them and with all the integrals produced
after completion of the $D$-algebra. The GPXT method used to compute the momentum integrals 
is summarized
in Subsection \ref{sec:GPXT} and Appendix \ref{app:GPXT}.
Using the results obtained so far in Subsection \ref{sec:wrapres} we are able to give an expression for the four-loop wrapping dilatation operator.
In Section \ref{sec:anomalousdim} we collect our results and we give the final answer for
the Konishi anomalous dimension at four loops. 

\section{${\cal N}=1$ superspace approach}

\label{sec:N1superspace}

We are going to describe the calculation of the anomalous dimension of the
composite operator $\tr(\phi Z\phi Z-\phi \phi Z Z)$ at four loops in the planar limit.
In order to streamline the calculation we decided to work using a
${\cal N}=1$ superspace formalism. In this section we review the most relevant features
of this approach needed to perform our calculation.

The action of $\mathcal{N}=4$ SYM can be written using $\mathcal{N}=1$
superspace in terms of one real vector superfield $V$ and three
chiral superfields $\phi^i$ (we follow the notations and conventions of
\cite{Gates:1983nr})

\begin{equation}\label{action}
\begin{aligned}
S &= \int\de^4 x\de^4 \theta \, \tr \left(\e^{-gV} \bar \phi_i \e^{gV}
\phi^i\right) + \frac 1{2g^2} \int \de^4 x \de^2 \theta
\,\tr \left(W^\alpha W_\alpha\right)\\
&\phantom{{}={}}
+i \frac{g}{3!}  \int\de^4 x\de^2 \theta \,\epsilon^{ijk}\,\tr \left(\phi_i
\left[\phi_j , \phi_k\right]\right) + \text{h.c.}\col
\end{aligned}
\end{equation}
where $W_\alpha = i\bar D^2 \left(\e^{-gV} D_\alpha\,\e^{gV}\right)$, and
$V=V^aT^a$, $\phi^i=\phi_i^aT^a$, {\small $i=1,2,3$}, $T^a$ being
matrices satisfying the $SU(N)$ algebra
\begin{equation}
[T_a, T_b] = i f_{abc} T_c
\end{equation}
and being normalized as
\begin{equation}
\tr(T_a T_b) =\delta_{ab}
\pnt
\end{equation}
We will usually denote the three chiral superfields $\phi^i$ as
$(\phi,\psi,Z)$.

We want to study the renormalization of the composite chiral operators
\begin{equation}
\label{ops}
\mathcal{O}_1=\tr(\phi Z\phi Z)\col\qquad\mathcal{O}_2=\tr(\phi\phi ZZ)\col
\end{equation}
which mix under renormalization. To this end, the first step is the
construction
of all the supergraphs containing one operator insertion.
The Feynman rules can be derived from the action (\ref{action}).
In momentum space we have the superfield propagators
\begin{equation}
\langle V^a V^b\rangle= - \frac{\delta^{ab}}{p^2}\col\qquad\qquad
\langle\phi^a_i \bar\phi^b_j\rangle=\delta_{ij} \frac{\delta^{ab}}{p^2}
\col
\label{propagators}
\end{equation}
while
the vertices are read-off directly from the interaction terms in (\ref{action}),
with additional $\bar D^2$, $D^2$ factors for chiral, antichiral lines
respectively.
The ones that we need are the following
\begin{equation}\label{vertices}
\begin{aligned}
V_1&=i g f_{abc}\delta^{ij} \bar\phi^a_i V^b \phi^c_j\col \qquad\qquad
&V_2&=\frac{g^2}{2}  \delta^{ij} f_{adm} f_{bcm} V^aV^b \bar\phi^c_i
\phi^d_j\col\\
V_3&= - \frac{g}{3!} \epsilon^{ijk} f_{abc} \phi_i^a \phi_j^b \phi_k^c
\col\qquad\qquad
&\bar{V}_3&= - \frac{g}{3!} \epsilon^{ijk} f_{abc} \bar\phi_i^a \bar\phi_j^b \bar\phi_k^c
\pnt
\end{aligned}
\end{equation}
Vertices containing three or more vector superfields $V$ do not enter
our calculation.

The general idea is to consider a given supergraph and to perform the $D$-algebra in order to reduce it to a standard graph. In doing so one usually produces a large number
of contributions, but most of them are not relevant for the computation of the anomalous dimension of the
composite operator since they give rise to finite integrals. Thus it is very 
useful
to know and to exploit $D$-algebra manipulations that can be made on the supergraph and that are
legal only up to finite terms. In the next subsection we are going to describe a
property of the supergraphs that was very important in computing our diagrams.

\subsection{A useful proof}

\label{subsec:powercount}

We consider diagrams with one operator insertion, built using only chiral, antichiral,
single-vector and double-vector vertices. We are interested in isolating those contributions which give rise to divergent integrals, i.e. the ones which will contribute to the anomalous dimension. Here
we want to prove that in order to produce divergent contributions we have to perform and complete the $D$-algebra  in such a way that no spinor derivative is moved out of the
diagram onto the external lines, except for derivatives on scalar propagators that do not belong to any loop from the start.
Let us define the following quantities:

\begin{itemize}

\item $V_C=$ number of chiral vertices
\item $V_A=$ number of antichiral vertices
\item $V_V^{(1)}=$ number of single-vector vertices
\item $V_V^{(2)}=$ number of double-vector vertices
\item $p_C=$ number of scalar propagators belonging to at least one loop
\item $p_V=$ number of vector propagators
\item $p=$ total number of propagators belonging to at least one loop
\item $E=$ number of scalar propagators not belonging to any loop. 
\item $N_\ell=$ number of loops
\item $N_D$, $N_{\bar{D}}$: number of spinor derivatives
\end{itemize}
In all our diagrams we have  $V_A=V_C$ and $N_D=N_{\bar{D}}$,
and $E$ is equal to the number of
chiral vertices not belonging to any loop.\\
Moreover the number of outgoing external lines equals the number of fields in the composite operator. From each chiral or antichiral vertex 
three scalar lines start.
From each single-vector vertex, two scalar and one vector propagators start. From each double-vector vertex, two scalar and two vector lines start. \\
Given these relations, we can write
\begin{equation}
p_C=\frac{1}{2}\left[3(V_C+V_A)+2(V_V^{(1)}+V_V^{(2)})\right]-E
\col\qquad
p_V=\frac{1}{2}\left[V_V^{(1)}+2V_V^{(2)}\right]
\pnt
\end{equation}
At each chiral vertex, two out of the three scalar lines have a $\bar{D}^2$.\\
At each antichiral vertex, two out of the three scalar lines have a $D^2$.\\
At each single-vector or double-vector vertex, one of the scalar lines has a $\bar{D}^2$, while the other one has a $D^2$. So we have
\begin{equation}
N_D=4V_C+2(V_V^{(1)}+V_V^{(2)})
\col\qquad
N_{\bar{D}}=4V_A+2(V_V^{(1)}+V_V^{(2)})
\pnt
\end{equation}
Then we can write
\begin{equation}
p=p_C+p_V=\frac{1}{2}\left[V_C+V_A+N_D+V_V^{(1)}+2V_V^{(2)}\right]-E
\pnt
\end{equation}
In order to obtain a logarithmic divergent diagram, the $D$-algebra must produce a number of standard derivatives equal to
\begin{equation}
2p-4N_\ell
\end{equation}
We need a $D$ (and a $\bar{D}$) to create each standard derivative. Furthermore, we need $2N_\ell$ extra $D$'s (and $\bar{D}$'s) to complete the superspace $\de^4\theta$ integration for the $N_\ell$ loops. So after performing the $D$-algebra, a momentum integral with a surface divergence can be found only if at least
\begin{equation}
2p-2N_\ell
\end{equation}
out of all the $D$'s are not brought outside the diagram on the external lines.\\
A $D^2$ can always be moved onto each scalar propagator which is not part of a loop: either it is already there at the beginning of the $D$-algebra, or it can be moved there through an integration by parts at the internal vertex at which the external propagator is linked. So the effective number of $D$'s available in the loops is
\begin{equation}
N_D-2E
\end{equation}
During the $D$-algebra manipulations we will be allowed to move some $D$'s  outside the diagram only if their number at the start exceeds the number we need in order to produce a divergent integral. So we should have
\begin{equation}
N_D-2E>2p-2N_\ell=\left[V_C+V_A+N_D+V_V^{(1)}+2V_V^{(2)}\right]-2E-2N_\ell
\col
\end{equation}
that is
\begin{equation}
N_\ell>\frac{1}{2}\left[V_C+V_A+V_V^{(1)}+2V_V^{(2)}\right]
\pnt
\end{equation}
This condition is never satisfied by the diagrams in the class we are studying, as can be seen using Euler's formula for connected graphs
\begin{equation}
V-P+F=2
\col
\end{equation}
where $V$ denotes the number of vertices, $P$ is the number of propagators and $F$ is the number of 
faces. \\
In our case (remember that the composite operator behaves as an additional vertex) we have
\begin{equation}
V=V_A+V_C+V_V^{(1)}+V_V^{(2)}+1\col\qquad\qquad 
P=p+E\col\qquad\qquad 
F=N_\ell+1
\col
\end{equation}
which leads to
\begin{equation}
N_\ell=\frac{1}{2}\left[V_C+V_A+V_V^{(1)}+2V_V^{(2)}\right]
\pnt
\end{equation}
We conclude that in order to produce a divergent contribution no spinor derivatives can be brought outside the diagram.

All our $D$-algebra manipulations  were performed making use of this property: only the contributions in which all spinor derivatives stay in the loops were kept. Clearly this rule allowed to discard a huge number of irrelevant contributions.

\section{Subtraction of range-five diagrams from the asymptotic Hamiltonian}

\label{sec:r5sub}

In this section we are going to describe how to compute the contributions
from the diagrams with range from
one to four with no wrapping interactions. In order to avoid their explicit, very complicated
Feynman graph computation  
 we take advantage of the
asymptotic four-loop dilatation operator $D_4$, whose expression was given in
the $SU(2)$ sector by \cite{Beisert:2007hz}:

\begin{equation}\label{D4}
\begin{aligned}
D_4&={}-{}(560+4\beta)\pid\\
&\phantom{{}={}}+(1072+12\beta+8\epsilon_{3a})\pone1\\
&\phantom{{}={}}-(84+6\beta+4\epsilon_{3a})\ptwo13
-4\ptwo14-(302+4\beta+8\epsilon_{3a})(\ptwo12+\ptwo21)\\
&\phantom{{}={}}
+(4\beta+4\epsilon_{3a}+2i\epsilon_{3c}-4i\epsilon_{3d})\pthree132
+(4\beta+4\epsilon_{3a}-2i\epsilon_{3c}+4i\epsilon_{3d})\pthree213\\
&\phantom{{}={}}+(4-2i\epsilon_{3c})(\pthree124+\pthree143)
+(4+2i\epsilon_{3c})(\pthree134+\pthree214)\\
&\phantom{{}={}}+(96+4\epsilon_{3a})(\pthree123+\pthree321)\\
&\phantom{{}={}}-(12+2\beta+4\epsilon_{3a})\pfour2132
+(18+4\epsilon_{3a})(\pfour1324+\pfour2143)\\
&\phantom{{}={}}-(8+2\epsilon_{3a}+2i\epsilon_{3b})(\pfour1243+\pfour1432)\\
&\phantom{{}={}}
-(8+2\epsilon_{3a}-2i\epsilon_{3b})(\pfour2134+\pfour3214)\\
&\phantom{{}={}}-10(\pfour1234+\pfour4321)
\col
\end{aligned}
\end{equation}
where $\epsilon_{3a}$, $\epsilon_{3b}$, $\epsilon_{3c}$ and $\epsilon_{3d}$
parameterize the invariance of the Hamiltonian under similarity transformations. In a perturbative approach their values
depend on the 
choice of the renormalization scheme.
The parameter $\beta=4\zeta(3)$ comes from the dressing phase \cite{Arutyunov:2004vx,Hernandez:2006tk,Beisert:2006ib,Beisert:2006ez,Beisert:2007hz}.

The permutation structures appearing in (\ref{D4}) are defined as
\begin{equation}\label{permstrucdef}
\pthree{a_1}{\dots}{a_n}=\sum_{r=0}^{L-1}\perm_{a_1+r\;a_1+r+1}\cdots
\perm_{a_n+r\;a_n+r+1}
\col
\end{equation}
where $\perm_{a\;a+1}$, with the cyclic identification 
$\perm_{a\;a+1}\simeq\perm_{a+L\;a+L+1}$,
permutes the flavours of the $a$th and $(a+1)$th site
when acting on a cyclic state of length $L$.
The number of nearest neighbours, interacting in the flavour space, is
extracted from the list of integers $a_1,\dots,a_n$
in the permutation structure as
\begin{equation}\label{nneighbourint}
\kappa=2+\max_{a_1\dots a_n}-\min_{a_1\dots a_n}\pnt
\end{equation}
It is limited from above by the order $K$ in the perturbation expansion
in powers $\lambda^K$ as $\kappa\le K+1$.
The asymptotic regime in which $D_4$ can be trustfully applied is
determined by the relation $L>K$.
Some rules valid in the asymptotic case for the manipulation of
these structures can be found in
\cite{Beisert:2005wv}.

The expression \eqref{D4} contains contributions which
describe the permutations among $\kappa=5$ neighbouring legs.
Therefore it  is correct only if
applied to a state in the asymptotic sense,
i.e.\ the number of sites in the state
has to be five or more.
We can correct it for the application
on a length four state: the
contributions from all Feynman diagrams which describe the interactions of
five neighbouring legs have to be replaced by the contributions from all
four-loop wrapping interactions. 

Before doing this, it is advantageous to change the basis from
the permutation structures \eqref{permstrucdef}
to a set of functions which are directly related to the chiral structures of
the Feynman supergraphs.
At four loops we are lead to define the following
chiral functions
\begin{equation}\label{chistruc}
\begin{aligned}
\chi(a,b,c,d)&=\pid-4\pone1
+\ptwo ab+\ptwo ac+\ptwo ad+\ptwo bc+\ptwo bd+\ptwo cd\\
&\phantom{{}={}}
-\pthree abc-\pthree abd-\pthree acd-\pthree bcd
+\pfour abcd\col\\
\chi(a,b,c)&=-\pid+3\pone1
-\ptwo ab-\ptwo ac-\ptwo bc+\pthree abc\col\\
\chi(a,b)&=\pid-2\pone1+\ptwo ab\col\\
\chi(1)&=-\pid+\pone1\col\\
\chi() &=\pid
\pnt
\end{aligned}
\normalsize
\end{equation}

\begin{figure}[h]
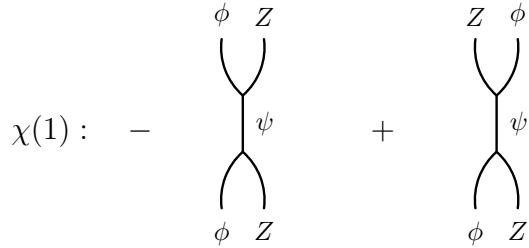

\unitlength=0.75mm
\settoheight{\eqoff}{$\times$}%
\setlength{\eqoff}{0.5\eqoff}%
\addtolength{\eqoff}{-12.5\unitlength}%
\settoheight{\eqofftwo}{$\times$}%
\setlength{\eqofftwo}{0.5\eqofftwo}%
\addtolength{\eqofftwo}{-7.5\unitlength}%
\begin{equation*}
\begin{aligned}
\chi(1):\quad-&
\raisebox{1.5\eqoff}{%
\fmfframe(1,1)(3,4){%
\begin{fmfchar*}(30,30)
\WoneplainB
\fmfipair{w[]}
\svertex{w1}{p4}
\fmfiv{l=\footnotesize{$\phi$},l.a=-90,l.d=5}{vloc(__v6)}
\fmfiv{l=\footnotesize{$Z$},l.a=-90,l.d=5}{vloc(__v7)}
\fmfiv{l=\footnotesize{$\phi$},l.a=90,l.d=5}{vloc(__v2)}
\fmfiv{l=\footnotesize{$Z$},l.a=90,l.d=5}{vloc(__v3)}
\fmfiv{l=\footnotesize{$\psi$},l.a=0,l.d=5}{w1}
\end{fmfchar*}}}
&+&
\raisebox{1.5\eqoff}{%
\fmfframe(3,1)(1,4){%
\begin{fmfchar*}(30,30)
\WoneplainB
\fmfipair{w[]}
\svertex{w1}{p4}
\fmfiv{l=\footnotesize{$\phi$},l.a=-90,l.d=5}{vloc(__v6)}
\fmfiv{l=\footnotesize{$Z$},l.a=-90,l.d=5}{vloc(__v7)}
\fmfiv{l=\footnotesize{$Z$},l.a=90,l.d=5}{vloc(__v2)}
\fmfiv{l=\footnotesize{$\phi$},l.a=90,l.d=5}{vloc(__v3)}
\fmfiv{l=\footnotesize{$\psi$},l.a=0,l.d=5}{w1}
\end{fmfchar*}}}
\end{aligned}
\end{equation*}
\caption{Building block}
\label{buildingblock}
\end{figure}

The relevance of these functions stems from the fact that any Feynman
supergraph produces a permutation structure given by one of them. Moreover
we call them {\it chiral} because their form depends only on the structure
of the chiral interactions of the diagram, while they are completely
insensitive to the presence of vector lines.
\\
The form of these functions can be easily derived: $\chi()$ is
just the identity, corresponding to a Feynman supergraph without chiral
vertices. The first non-trivial function, $\chi(1)$ corresponds to the
diagram given in Figure \ref{buildingblock}.

\vspace{1cm}

\vskip 1cm

This diagram has two vertices $V_3$ and $\bar V_3$ (see \eqref{vertices})
connected by a $\langle\psi\bar\psi\rangle$ superfield propagator,
which is killed by the $D$-algebra. It is a sort of effective four-vertex,
very similar to the one describing the scalar interaction in a component
approach. This diagram is a ``building block'' for the construction
of more complicated ones.
\\
Indeed it is easy to see that every four-loop supergraph  contributing to the one-point function
of a chiral operator in the $SU(2)$ sector, like our operators \eqref{ops},
are built by gluing together up to four of the building blocks given
in Figure \ref{buildingblock}. The form of the chiral functions in \eqref{chistruc} then
follows consequently.
\\
Let us notice that the number of arguments $n$, with $n=1,\dots,4$ at four
loops, corresponds to the number of building blocks in our underlying
Feynman graphs. 
The fact that the coefficients in each chiral structure sum up to zero
guarantees that the two length $L$ BPS operators
$\tr(Z^L)$ and $\tr(\phi Z^{L-1})$ are protected.

In order to rewrite $D_4$ in $\eqref{D4}$ in terms of the chiral functions,
we have to express the permutation structures \eqref{permstrucdef}
in terms of the chiral functions \eqref{chistruc}.
We have
\begin{equation}\label{invchistruc}
\begin{aligned}
\pfour abcd &= \chi(a,b,c,d) + \chi(a,b,c) + \chi(a,b,d) + \chi(a,c,d)
+ \chi(b,c,d)\\
&\phantom{{}={}}
+ \chi(a,b) + \chi(a,c) + \chi(a,d) + \chi(b,c) + \chi(b,d) + \chi(c,d)\\
&\phantom{{}={}}
+4 \chi(1) + \chi()\col\\
\pthree abc &= \chi(a,b,c) + \chi(a,b) + \chi(a,c) + \chi(b,c) + 3\chi(1)
+\chi()\col\\
\ptwo ab &= \chi(a,b) + 2\chi(1) + \chi()\col\\
\pone1 &= \chi(1) + \chi()\col\\
\pid &= \chi()\pnt
\end{aligned}
\end{equation}
Making  use of these relations we reexpress the four-loop dilatation operator
$\eqref{D4}$
in terms of the chiral functions and 
obtain
\begin{equation}
\begin{aligned}\label{D4chi}
D_4&={}+{}200\chi(1)
-150[\chi(1,2)+\chi(2,1)]
+8(10+\epsilon_{3a})\chi(1,3)
-4\chi(1,4)\\
&\phantom{{}={}}
+60[\chi(1,2,3)+\chi(3,2,1)]\\
&\phantom{{}={}}
+(8+2\beta+4\epsilon_{3a}-4i\epsilon_{3b}+2i\epsilon_{3c}-4i\epsilon_{3d})
\chi(1,3,2)\\
&\phantom{{}={}}
+(8+2\beta+4\epsilon_{3a}+4i\epsilon_{3b}-2i\epsilon_{3c}+4i\epsilon_{3d})
\chi(2,1,3)\\
&\phantom{{}={}}
-(4+4i\epsilon_{3b}+2i\epsilon_{3c})[\chi(1,2,4)+\chi(1,4,3)]\\
&\phantom{{}={}}
-(4-4i\epsilon_{3b}-2i\epsilon_{3c})[\chi(1,3,4)+\chi(2,1,4)]\\
&\phantom{{}={}}-(12+2\beta+4\epsilon_{3a})\chi(2,1,3,2)\\
&\phantom{{}={}}
+(18+4\epsilon_{3a})[\chi(1,3,2,4)+\chi(2,1,4,3)]\\
&\phantom{{}={}}
-(8+2\epsilon_{3a}+2i\epsilon_{3b})[\chi(1,2,4,3)+\chi(1,4,3,2)]\\
&\phantom{{}={}}
-(8+2\epsilon_{3a}-2i\epsilon_{3b})[\chi(2,1,3,4)+\chi(3,2,1,4)]\\
&\phantom{{}={}}
-10[\chi(1,2,3,4)+\chi(4,3,2,1)]
\pnt
\end{aligned}
\end{equation}
This representation is particularly useful since the chiral functions $\chi$
and not the permutation structures are directly related to the underlying
Feynman supergraph.  However there is still an ambiguity: two supergraphs which
differ only by the arrangement of the flavour-neutral vector superfield lines
contribute to the coefficient in front of the same chiral 
function. Therefore the chiral functions 
do not allow in general to extract information about the range of the 
interaction of the underlying Feynman supergraph.

In particular range-five diagrams
in which the first or fifth line is connected to the rest of the diagram
only via vector line(s) appear with the same chiral function as the
corresponding range-four diagram in which the corresponding first or fifth
line is a non-interacting spectator.
We call those range-five diagrams which have a chiral function
which does not capture
the range of their interaction {\it non-maximal} range-five diagrams.
They also have to be subtracted from $D_4$ and in general one should compute explicitly
all these Feynman graphs.
Luckily it turns out that there is no need to do that.
In Appendix \ref{app:non-maximal} we show that the {\it non-maximal} range-five 
interactions
do not contribute to $D_4$. This means that  we can determine the subtraction directly from $D_4$.
This operation becomes particularly simple using the representation of $D_4$ in terms of
the chiral functions $\chi$. We only have to neglect all the terms with
chiral functions $\chi(a_1,\dots,a_n)$ which describe a range-five interaction,
i.e.\ which fulfill $\displaystyle\max_{a_1\dots a_n}-\min_{a_1\dots a_n}=3$.
Hence we have to subtract from \eqref{D4chi}
\begin{equation}
\begin{aligned}\label{r5D}
\delta D_4
&={}-{}10[\chi(1,2,3,4)+\chi(4,3,2,1)]+(18+4\epsilon_{3a})[\chi(1,3,2,4)+\chi(2,1,4,3)]\\
&\phantom{{}={}}
-(8+2\epsilon_{3a}+2i\epsilon_{3b})[\chi(1,2,4,3)+\chi(1,4,3,2)]\\
&\phantom{{}={}}
-(8+2\epsilon_{3a}-2i\epsilon_{3b})[\chi(2,1,3,4)+\chi(3,2,1,4)]\\
&\phantom{{}={}}
-(4+4i\epsilon_{3b}+2i\epsilon_{3c})[\chi(1,2,4)+\chi(1,4,3)]\\
&\phantom{{}={}}
-(4-4i\epsilon_{3b}-2i\epsilon_{3c})[\chi(1,3,4)+\chi(2,1,4)]\\
&\phantom{{}={}}
-4\chi(1,4)
\pnt
\end{aligned}
\end{equation}

We should stress that due to the decomposition in \eqref{chistruc},
the subtraction is not only a simple neglection of all range-five permutation
structures as attempted in \cite{Fischbacher:2004iu} for the BMN matrix
model. One has to modify also the coefficients of the
permutation structures of lower range in \eqref{D4}.

In order to have an independent check of the whole procedure,
we have found the above constructed range-five contributions
\eqref{r5D} also through an explicit Feynman graph computation. We give
in Appendix \ref{app:range-five}
all the details of the calculation. As a byproduct we
have found in this way explicit values for the unphysical
 coefficients $\epsilon_{3a}$, $\epsilon_{3b}$, 
$\epsilon_{3c}$.
In the MS-scheme we have used, they are given by
\begin{equation}\label{epsilons}
\epsilon_{3a}=-4\col\qquad
\epsilon_{3b}=-i\frac{4}{3}\col\qquad
\epsilon_{3c}=i\frac{4}{3}\pnt
\end{equation}

Thanks to the absence of non-maximal range-five interactions
the explicit  Feynman graph computation was not necessary.
It was performed just as a further check  on the range-five part of $D_4$.

We emphasize that a straightforward  modification of the proof  in Appendix 
\ref{app:non-maximal} leads to the cancellation of
this type of contributions also at
higher loop orders. This feature is quite interesting since it allows
to determine the necessary subtractions in the general case of
$D_{K}$  applied to a length $L=K$ operator, with no need of Feynman 
diagram computations.

Finally after subtraction of the range-five contributions the four-loop
dilatation operator becomes
\begin{equation}\label{D4sub}
\begin{aligned}
D_4^\text{sub}\equiv D_4 - \delta D_4 &=
{}+{}200\chi(1)
-150[\chi(1,2)+\chi(2,1)]
+8(10+\epsilon_{3a})\chi(1,3)\\
&\phantom{{}={}}
+60[\chi(1,2,3)+\chi(3,2,1)]\\
&\phantom{{}={}}
+(8+2\beta+4\epsilon_{3a}-4i\epsilon_{3b}+2i\epsilon_{3c}-4i\epsilon_{3d})
\chi(1,3,2)\\
&\phantom{{}={}}
+(8+2\beta+4\epsilon_{3a}+4i\epsilon_{3b}-2i\epsilon_{3c}+4i\epsilon_{3d})
\chi(2,1,3)\\
&\phantom{{}={}}-(12+2\beta+4\epsilon_{3a})\chi(2,1,3,2)
\pnt
\end{aligned}
\end{equation}
We can now apply the subtracted dilatation operator to the states with $L=4$
sites.
In the $SU(2)$ subsector there exist two composite operators of length $L=4$
and with two `impurities'
which mix under renormalization. The corresponding states are given by
\begin{equation}\label{Opbasis}
\mathcal{O}_1=\tr(\phi Z\phi Z)\col\qquad\mathcal{O}_2=\tr(\phi\phi ZZ)\pnt
\end{equation}
Defining a two-dimensional vector
$\vec{\mathcal{O}}=(\mathcal{O}_1,\mathcal{O}_2)^\text{t}$,
the application of the chiral structures to $\vec{\mathcal{O}}$ is captured by
the replacements
\begin{equation}\label{chiM}
\begin{gathered}
{}\chi(1)\to M\col\qquad
\chi(1,2)\to-M\col\qquad
\chi(1,3)\to-2M\col\qquad
\chi(2,1)\to-M\col\qquad\\
\chi(1,2,3)\to M\col\qquad
\chi(3,2,1)\to M\col\qquad
\chi(2,1,3)\to2M\col\qquad
\chi(1,3,2)\to2M\col\\
\chi(2,1,3,2)\to-2M
\end{gathered}
\end{equation}
where the common mixing matrix is given by
\begin{equation}
M=\begin{pmatrix} -4 & 4 \\ 2 & -2 \end{pmatrix}
\pnt
\end{equation}
The eigenvectors and eigenvalues of $M$ are given by
\begin{equation}\label{eigenvecval}
\begin{aligned}
\mathcal{O}'_1&=\mathcal{O}_1+2\mathcal{O}_2
=\frac{1}{2}(3\tr(\phi\acomm{Z}{\phi}Z)
-\tr(\phi\comm{Z}{\phi}Z))\col\\
\mathcal{O}'_2&=\mathcal{O}_1-\mathcal{O}_2=\tr(\phi\comm{Z}{\phi}Z)
\col
\end{aligned}\qquad
M\begin{pmatrix}\mathcal{O}'_1 \\ \mathcal{O}'_2 \end{pmatrix}
=\begin{pmatrix} 0 \\ -6\mathcal{O}'_2 \end{pmatrix}
\pnt
\end{equation}
These are directly the eigenvectors of $D_4^\text{sub}$ in \eqref{D4sub},
since in the $L=4$ operator basis \eqref{Opbasis} it is proportional to $M$.
With $\beta=4\zeta(3)$ and $\epsilon_{3a}$ from \eqref{epsilons}
it assumes the form
\begin{equation}\label{D4submatrix}
D_4^\text{sub}\to
4(121+12\zeta(3))M
\pnt
\end{equation}
In the next section we move our attention to the calculation of
wrapping diagrams.

\section{Contributions from wrapping diagrams}

In this section we are going to face the computation of the wrapping
contributions. This is the part of the analysis that necessarily
requires the explicit computation of Feynman supergraphs. The
first step will be to build all wrapping supergraphs, and for this
we rely on the results of \cite{Sieg:2005kd}, where a
systematic Feynman-diagrammatic analysis of wrapping interactions
was performed. 

In the present paper we
are using $\mathcal{N}=1$ superspace techniques, so we need to
adapt that approach for the construction of all the wrapping
supergraphs contributing to the renormalization of the chiral
operators \eqref{Opbasis}. 

The second step will be to compute the four-loop momentum integrals using the GPXT.

Finally we collect these results in a wrapping four-loop dilatation operator.
%
\subsection{Wrapping supergraphs}

\label{sec:wrapping}

\begin{figure}[h]
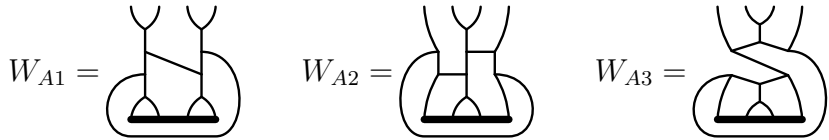

\unitlength=0.75mm
\settoheight{\eqoff}{$\times$}%
\setlength{\eqoff}{0.5\eqoff}%
\addtolength{\eqoff}{-12.5\unitlength}%
\settoheight{\eqofftwo}{$\times$}%
\setlength{\eqofftwo}{0.5\eqofftwo}%
\addtolength{\eqofftwo}{-7.5\unitlength}%
\begin{equation*}
\begin{aligned}
W_{A1}=&
\raisebox{\eqoff}{%
\fmfframe(3,1)(1,4){%
\begin{fmfchar*}(20,20)
\fmftop{v1}
\fmfbottom{v5}
\fmfforce{(0.125w,h)}{v1}
\fmfforce{(0.125w,0)}{v5}
\fmffixed{(0.25w,0)}{v1,v2}
\fmffixed{(0.25w,0)}{v2,v3}
\fmffixed{(0.25w,0)}{v3,v4}
\fmffixed{(0.25w,0)}{v5,v6}
\fmffixed{(0.25w,0)}{v6,v7}
\fmffixed{(0.25w,0)}{v7,v8}
\fmffixed{(0,0.9w)}{v5,vh1}
\fmf{plain,tension=0.5,right=0.25}{v1,vc1}
\fmf{plain,tension=0.5,left=0.25}{v2,vc1}
\fmf{plain,tension=0.5,right=0.25}{v3,vc2}
\fmf{plain,tension=0.5,left=0.25}{v4,vc2}
  \fmf{plain}{vc1,vc3}
    \fmf{plain}{vc3,vc7}
      \fmf{plain}{vc7,vc5}
        \fmf{plain}{vc2,vc8}
      \fmf{plain}{vc8,vc4}
        \fmf{plain}{vc4,vc6}
          \fmf{plain,tension=0}{vc3,vc4}
          \fmf{plain,tension=0.5,left=0.25}{v5,vc5}
          \fmf{plain,tension=0.5,right=0.25}{v6,vc5}
          \fmf{plain,tension=0.5,left=0.25}{v7,vc6}
          \fmf{plain,tension=0.5,right=0.25}{v8,vc6}
          \fmf{plain,tension=0.5,right=0,width=1mm}{v5,v8}
          \fmffreeze
          \fmfposition
          \plainwrap{vc7}{v5}{v8}{vc8}
\end{fmfchar*}}}
\;{\quad}&W_{A2}=&
\raisebox{\eqoff}{%
\fmfframe(3,1)(1,4){%
\begin{fmfchar*}(20,20)
\fmftop{v1}
\fmfforce{(0.125w,h)}{v1}
\fmfforce{(0.125w,0)}{v5}
\fmffixed{(0.25w,0)}{v1,v2}
\fmffixed{(0.25w,0)}{v2,v3}
\fmffixed{(0.25w,0)}{v3,v4}
\fmffixed{(0.25w,0)}{v5,v6}
\fmffixed{(0.25w,0)}{v6,v7}
\fmffixed{(0.25w,0)}{v7,v8}
\fmffixed{(whatever,0)}{vc1,vc3}
\fmffixed{(whatever,0)}{vc5,vc7}
\fmffixed{(whatever,0)}{vc3,vc4}
\fmffixed{(whatever,0)}{vc7,vc8}
\fmf{plain,tension=1,right=0.125}{v1,vc1}
\fmf{plain,tension=0.5,right=0.25}{v2,vc2}
\fmf{plain,tension=0.5,left=0.25}{v3,vc2}
\fmf{plain,tension=1,left=0.125}{v4,vc4}
\fmf{plain,tension=1,left=0.125}{v5,vc5}
\fmf{plain,tension=0.5,left=0.25}{v6,vc6}
\fmf{plain,tension=0.5,right=0.25}{v7,vc6}
\fmf{plain,tension=1,right=0.125}{v8,vc8}
\fmf{plain}{vc1,vc5}
\fmf{plain}{vc4,vc8}
\fmf{plain}{vc2,vc3}
\fmf{plain}{vc6,vc7}
\fmf{plain,tension=3}{vc3,vc7}
\fmf{plain,tension=0.5}{vc3,vc4}
\fmf{plain,tension=0.5}{vc5,vc7}
\fmf{phantom,tension=0.5}{vc7,vc8}
\fmf{phantom,tension=0.5}{vc1,vc3}
\fmf{plain,tension=0.5,right=0,width=1mm}{v5,v8}
\fmffreeze
\fmfposition
\plainwrap{vc1}{v5}{v8}{vc8}
\fmffreeze
\end{fmfchar*}}}
\;{\quad}&W_{A3}=&
\raisebox{\eqoff}{%
\fmfframe(3,1)(1,4){%
\begin{fmfchar*}(20,20)
\fmftop{v1}
\fmfbottom{v5}
\fmfforce{(0.125w,h)}{v1}
\fmfforce{(0.125w,0)}{v5}
\fmffixed{(0.25w,0)}{v1,v2}
\fmffixed{(0.25w,0)}{v2,v3}
\fmffixed{(0.25w,0)}{v3,v4}
\fmffixed{(0.25w,0)}{v5,v6}
\fmffixed{(0.25w,0)}{v6,v7}
\fmffixed{(0.25w,0)}{v7,v8}
\fmffixed{(0,whatever)}{vc1,vc5}
\fmffixed{(0,whatever)}{vc2,vc3}
\fmffixed{(0,whatever)}{vc3,vc6}
\fmffixed{(0,whatever)}{vc6,vc7}
\fmffixed{(0,whatever)}{vc4,vc8}
\fmffixed{(0.5w,0)}{vc1,vc4}
\fmffixed{(0.5w,0)}{vc5,vc8}
\fmf{plain,tension=1,right=0.125}{v1,vc1}
\fmf{plain,tension=0.25,right=0.25}{v2,vc2}
\fmf{plain,tension=0.25,left=0.25}{v3,vc2}
\fmf{plain,tension=1,left=0.125}{v4,vc4}
\fmf{plain,tension=1,left=0.125}{v5,vc5}
\fmf{plain,tension=0.25,left=0.25}{v6,vc6}
\fmf{plain,tension=0.25,right=0.25}{v7,vc6}
\fmf{plain,tension=1,right=0.125}{v8,vc8}
  \fmf{plain,tension=0.5}{vc1,vc3}
    \fmf{plain,tension=0.5}{vc2,vc3}
      \fmf{plain,tension=0.5}{vc3,vc4}
        \fmf{plain,tension=0.5}{vc5,vc7}
      \fmf{plain,tension=0.5}{vc6,vc7}
        \fmf{plain,tension=0.5}{vc7,vc8}
          \fmf{plain,tension=2}{vc1,vc8}
            \fmf{phantom,tension=2}{vc5,vc4}
        \fmffreeze
        \fmfposition
        \plainwrap{vc5}{v5}{v8}{vc4}
        \fmf{plain,tension=1,left=0,width=1mm}{v5,v8}
        \fmffreeze
        \end{fmfchar*}}}
\end{aligned}
\end{equation*}
\caption{Wrapping diagrams with only chiral interactions}
\label{diagrams-chi}
\end{figure}

First of all we have a set of wrapping graphs with only chiral interactions, which are shown in Figure \ref{diagrams-chi}.

\vspace{1cm}

\vspace{1cm}

With the identification of the first and the fifth lines, we can still use the definition~\eqref{permstrucdef} for the permutation operators. Then the chiral structures will be
\begin{equation}\label{chiMr4}
\begin{aligned}
W_{A1}\quad&\rightarrow\quad\chi(2,4,1,3)\quad\rightarrow\quad-4M\\
W_{A2}\quad&\rightarrow\quad\chi(4,1,2,3)\quad\rightarrow\quad-M\\
W_{A3}\quad&\rightarrow\quad\chi(4,3,1,2)\quad\rightarrow\quad-2M
\end{aligned}
\end{equation}
The reflection of $W_{A2}$, which is not symmetric, will have the structure $\chi(1,4,3,2)\rightarrow-M$.\\
Beware that, unlike in the asymptotic case, here the exchange $1\leftrightarrow4$ is not allowed, since lines 1 and 4 are now neighbours. 
Furthermore, there is an ambiguity in the choice of the basis.
Shifts of all arguments by an integer generate four different 
chiral functions for $W_{A1}$ and $W_{A3}$ and two each for $W_{A2}$ and for its reflection. A cyclic rotation of the external legs of $W_{A3}$ generates 
an additional chiral function. The elements in each of the four classes   
are equivalent when applied to a length-four state. From each class we have 
chosen a single representative.

The results after completion of the $D$-algebra for each completely chiral supergraph are given by
\begin{eqnarray}
W_{A1\phantom{0}}&\rightarrow&(g^2 N)^4 (I_4/2)\,\chi(2,4,1,3)\rightarrow-2(g^2 N)^4 I_4 M\col  \\
W_{A2\phantom{0}}&\rightarrow&(g^2 N)^4 I_2\,[\,\chi(1,4,3,2)+\chi(4,1,2,3)\,]\rightarrow-2(g^2 N)^4 I_2 M\col  \\
W_{A3\phantom{0}}&\rightarrow&(g^2 N)^4 I_3\,\chi(4,3,1,2)\rightarrow-2(g^2 N)^4 I_3 M
\col
\end{eqnarray}
where $I_i$ denote the momentum integrals listed in Table \ref{integrals}.\\
Then the total contribution from this class reads
\begin{equation}
\sum W_{A**}\rightarrow-2(g^2 N)^4 (I_4+I_2+I_3)M
\pnt
\end{equation}

Graphs with chiral structure $\chi(1,2,3)$ are shown in Figure \ref{diagrams-123}, while the results of the $D$-algebra for each diagram are summarized schematically in Table \ref{results-123}: here cancellations between pairs of diagrams, as well as graphs which do not produce any divergence are explicitly indicated. Diagrams which do not cancel and therefore must be fully computed are marked with a $\ast$. For these relevant graphs we also give explicitly the value of the relative symmetry factor. Graphs with chiral structure $\chi(3,2,1)$, which is the reflection of $\chi(1,2,3)$, contribute exactly in the same manner. We give a summary of the full results for all the structures in Table \ref{wrapping-results}, where the explicit expressions \eqref{chiM} and \eqref{chiMr4} 
for the chiral structures in the length-four subsector
have been used.\\
So the total contribution from these chiral structures is given by
\begin{equation}
\sum W_{B**}\rightarrow-4(g^2 N)^4 (I_7-I_9)M
\pnt
\end{equation}

We proceed in the same way for the other structures. Diagrams associated to $\chi(1,3,2)$ are listed in Figure \ref{diagrams-132}, while Table \ref{results-132} contains the results from $D$-algebra.
\begin{equation}
\sum W_{C**}\rightarrow-4(g^2 N)^4 I_6 M
\pnt
\end{equation}
There is one more structure which needs only a vector to be completed to four loops, $\chi(2,1,3)$. The corresponding diagrams and the $D$-algebra results can be read from Figure \ref{diagrams-213} and Table \ref{results-213}. The divergent contribution is\footnote{In the first version of the paper there is a symmetry factor of $2$ missing in this expression. We have found this factor after
a revision of our calculation which was suggested by a mismatch between 
the rational part of our previously found Konishi anomalous dimension 
and the one computed in \cite{Bajnok:2008bm}.}
\begin{equation}\label{WD}
\sum W_{D**}\rightarrow-4(g^2 N)^4 I_8 M
\pnt
\end{equation}

Next we turn to chiral structures of shorter range. Two of them require two vectors to be completed to four loops: the first one is $\chi(2,1)$, for which diagrams and $D$-algebra results are given in Figure \ref{diagrams-12} and Table \ref{results-12}. Adding the analogous result from graphs with structure $\chi(1,2)$ we obtain
\begin{equation}
\sum W_{E**}\rightarrow-4(g^2 N)^4 (I_7-I_9)M
\pnt
\end{equation}
The second structure to be completed by two vectors is $\chi(1,3)$ and it is described in Figure \ref{diagrams-13} and Table \ref{results-13}, from which we get
\begin{equation}
\sum W_{F**}\rightarrow-4(g^2 N)^4 (I_{10}+I_{11}-2I_{12})M
\pnt
\end{equation}
The last chiral structure to be considered is $\chi(1)$, which is the only one requiring three vectors. The diagrams for this structure are presented in Figure \ref{diagrams-1}. Using the results summarized in Table \ref{results-1} we obtain
\begin{equation}
\sum W_{G**}\rightarrow2(g^2 N)^4 (I_1-I_2)M
\pnt
\end{equation}

So up to this point we have obtained the total contribution from each relevant chiral structure. Now we  have to substitute the values of the momentum integrals $I_i$. In order to obtain the full wrapping contribution to the dilatation operator on the length-four subsector, we have to take the coefficient of the $\frac{1}{\varepsilon}$ pole of these expressions and multiply them by $-8$ according to eq.~\eqref{andim}.
In the next subsection we describe the computation of these momentum integrals by means of the Gegenbauer polynomial $x$-space technique.

\subsection{Gegenbauer polynomial $x$-space technique}

\label{sec:GPXT}

The integrals with non-trivial numerators which cannot be rephrased in terms
of standard scalar integrals, are calculated with the 
Gegenbauer polynomial $x$-space
technique (GPXT) \cite{Chetyrkin:1980pr}, and are independently
checked with the method of uniqueness \cite{Kazakov:1983ns} and with
the help of {\tt MINCER} \cite{Larin:1998}, a computer program to compute
$3$-loop integrals.
We have furthermore used the GPXT and {\tt MINCER} to reproduce the
known results for the scalar integrals.

The GPXT is based on the fact that in $x$-space integrals all propagators
depend on the difference of only two coordinates and hence can be expanded
in a series involving the Gegenbauer polynomials. The non-trivial
numerators given by scalar products of several momenta become traceless
symmetric products in $x$-space. These products
are easily reexpressed in terms of Gegenbauer polynomials.
Then the Wick rotated integral is solved by introducing spherical coordinates. As a first step 
one performs all angular integrations. Due to the orthogonality
of the Gegenbauer polynomials, in this way the number of independent summations, stemming from the series expansions, is reduced.
Finally, the radial integration is
performed, and the remaining summations are evaluated in order to obtain the
final result.

In particular the GPXT turns out to be very useful in
the following situation:
\begin{enumerate}
\item \label{angcond}
Each angular integration involves maximally two Gegenbauer polynomials.
For this it might be necessary to expand products of Gegenbauer polynomials
with the same argument in a Clebsch-Gordan series.
\item \label{rootvertexcond}
The integrals contain a vertex on which a large number of propagators
end, (in our case provided by the composite operator) which is chosen as the
`root vertex'.
\item \label{derivcond}
By partial integrations the momentum dependence in the numerator can
be rearranged such that it is given by a linear combination of scalar products
of those momenta which run along propagators directly connected to the root
vertex.
\item \label{polescond}
Only the pole structures of the (logarithmically divergent)
integrals are of interest.
\end{enumerate}
Point \ref{angcond} is essential for solving the integral. It guarantees
that the Racah coefficients never appear after having performed the
angular integrals.
Points \ref{rootvertexcond} to \ref{polescond} are not essential, but
significantly simplify the problem. They reduce the number of nested (infinite)
sums.

The first three points are guaranteed
if any of the four loops contains at least a single propagator which
is connected directly to the root vertex and if the non-trivial numerators
do not generate additional vertices which cannot be reduced by a
Clebsch-Gordan series expansion.
Apart from eventually appearing problems with the non-trivial numerators,
the requirement holds for all
diagrams which become tree level or one loop if the root vertex is removed.
In our four-loop calculation at critical order, this is
guaranteed for all wrapping diagrams and range-five diagrams.
At higher loops, it might only be spoiled by the increasing complexity of the
appearing numerators. Therefore, we cannot strictly prove the applicability of
GPXT. However, the found numerators even at higher loop order suggest that it
might hold. In any case, GPXT will
in principle allow for an analytic determination of the pole parts of whole
classes of integrals at arbitrary loop order. A practical restriction is only
given by the increasing number of domains in the radial integrations.

The fact that we are  interested only in the pole structure allows
for a dramatical simplification. We can neglect
the momentum-dependent exponential factor in the $x$-space
integrals, i.e.\ we can set the external momentum of the propagator-type
integrals to zero. We then have to introduce a cutoff $R$ as upper bound of
the integration
domain. This  regularizes the appearing IR divergences in the integral.
Since the poles of the logarithmically divergent
integrals are independent of the momentum, this procedure does not alter the
poles of the integral from which the subdivergences have been subtracted. 
To ensure this, the IR cutoff has also to
be used in the computation of the subdivergences.

GPXT has been proposed in \cite{Chetyrkin:1980pr}, see also 
\cite{Celmaster:1980ji,Kazakov:1983ns} for examples and
some important relations. We will not review the method here, but
only stress an important fact which we could not find in the literature.
For integrals with more than one derivative on
a propagator line, special care has to be taken. 
The corresponding computations are briefly summarized in 
Appendix \ref{app:GPXT}.
We will review the method and the refinements in a separate publication
\cite{usGPXT}.

\subsection{Wrapping dilatation operator}
\label{sec:wrapres}
Having worked out all the integrals presented in Table \ref{integrals}, 
we can now write down the 
wrapping contribution to the four-loop dilatation operator in terms 
of the chiral
functions $\chi$ as follows
\begin{equation}
\begin{aligned}
\label{D4w}
D_4^\text{w}
&=-{}8\Big(2\zeta(3)\chi(1)-(3\zeta(3)-5\zeta(5))[\chi(1,2)+\chi(2,1)]
-(1+3\zeta(3)-5\zeta(5))\chi(1,3)\\
&\phantom{{}={}-8}
+(3\zeta(3)-5\zeta(5))[\chi(1,2,3)+\chi(3,2,1)]
+\frac{7}{6}\chi(1,3,2)
+\frac{11}{6}\chi(2,1,3)\\
&\phantom{{}={}-8}
-\frac{1}{2}(1-\zeta(3))\chi(2,4,1,3)
+\Big(\frac{5}{4}-\zeta(3)\Big)[\chi(1,4,3,2)+\chi(4,1,2,3)]\\
&\phantom{{}={}-8}-\Big(\frac{1}{2}-\zeta(3)\Big)\chi(4,1,3,2)
\Big)
\pnt
\end{aligned}
\end{equation}
Applying it to our operator basis \eqref{Opbasis} we then have
\begin{equation}\label{D4wrapping}
D_4^\text{w}\to-8\Big(\frac{17}{2}+18\zeta(3)-30\zeta(5)\Big)M
\pnt
\end{equation}

\section{The Konishi anomalous dimension at four loops}

\label{sec:anomalousdim}

We collect the results from the previous three sections.
They can be encoded in a dilatation operator at four loops which
includes wrapping. It can be obtained by adding together the range-four 
subtracted dilatation operator \eqref{D4sub} and the wrapping
contribution \eqref{D4w}. We find
\begin{equation}
\begin{aligned}
D_4^\text{sub}+D_4^\text{w}
&=(200-16\zeta(3))\chi(1)-(150-24\zeta(3)+40\zeta(5))[\chi(1,2)+\chi(2,1)]\\
&\phantom{{}={}}
+(88+8\epsilon_{3a}+24\zeta(3)-40\zeta(5))\chi(1,3)\\
&\phantom{{}={}}+(60-24\zeta(3)+40\zeta(5))[\chi(1,2,3)+\chi(3,2,1)]\\
&\phantom{{}={}}-\Big(\frac{4}{3}-2\beta-4\epsilon_{3a}+4i\epsilon_{3b}-2i\epsilon_{3c}+4i\epsilon_{3d}\Big)\chi(1,3,2)\\
&\phantom{{}={}}-\Big(\frac{20}{3}-2\beta-4\epsilon_{3a}-4i\epsilon_{3b}+2i\epsilon_{3c}-4i\epsilon_{3d}\Big)\chi(2,1,3)\\
&\phantom{{}={}}+4(1-\zeta(3))\chi(2,4,1,3)
-(10-8\zeta(3))[\chi(1,4,3,2)+\chi(4,1,2,3)]\\
&\phantom{{}={}}-(12+2\beta+4\epsilon_{3a})\chi(2,1,3,2)
+(4-8\zeta(3))\chi(4,1,3,2)
\pnt
\end{aligned}
\end{equation}
On the basis \eqref{Opbasis} it gives
\begin{equation}\label{D4matrix}
D_4^\text{sub}+D_4^\text{w}
\to\big(416-96\zeta(3)+240\zeta(5)\big)M \pnt
\end{equation}
The non-vanishing eigenvalue is hence found to be
\begin{equation}
\gamma_4=-2496+576\zeta(3)-1440\zeta(5) \pnt
\end{equation}
Restoring the dependence on the coupling constant \eqref{lambdadef}, and 
including also the contributions at lower order \cite{Beisert:2006ez}, the
final result for the range-four Konishi-descendant up to four loops
reads\footnote{Note that due to a sign change of the $\chi(1)$ contribution
in \eqref{D4w}, 
the coefficient of the $\zeta(3)$ term differs from the one in the first 
version of \cite{Fiamberti:2007rj}.

Moreover, as remarked in the footnote to \eqref{WD}, the rational part 
has been corrected w.r.t to the one in the first version of the present paper. 
It agrees with the one found in \cite{Bajnok:2008bm}.}
\begin{equation}
\label{finalgamma}
\gamma=4+12\lambda-48\lambda^2+336\lambda^3
+\lambda^4(-2496+576\zeta(3)-1440\zeta(5))
\pnt
\end{equation}

In the literature there also exist several conjectures on the value of this anomalous dimension \cite{Rej:2005qt,Kotikov:2007cy,Beisert:2006ez} and moreover, as mentioned in the introduction, an explicit component field theory computation \cite{Keeler:2008ce} has been presented following our superspace calculation. None of these findings shows an agreement. Therefore it is crucial to have an explicit result that can be trusted.

The aim of this paper has been to present all the steps of our calculation. The superspace approach that we have used allowed drastic simplifications: several cancellations were automatic and the number of graphs that one really had to compute was reasonably small. With all the ingredients given in this paper an interested reader might want to reproduce the whole calculation.

\section*{Acknowledgements}

\noindent This work has been supported in
part by INFN  and the European Commission RTN
program MRTN--CT--2004--005104.

\newpage

\appendix

\newenvironment{mytable}
{
\refstepcounter{table}
\begin{center}
Table \thetable:}
{
\end{center}
}

\newenvironment{myfigure}
{
\refstepcounter{figure}
\begin{center}
Figure \thefigure:}
{
\end{center}
}

\renewcommand{\thefigure}{A.\arabic{figure}}
\setcounter{figure}{0}
\renewcommand{\thetable}{A.\arabic{table}}
\setcounter{table}{0}

\section{Cancellation of non-maximal range-five supergraphs}
\label{app:non-maximal}

\begin{figure}[p]
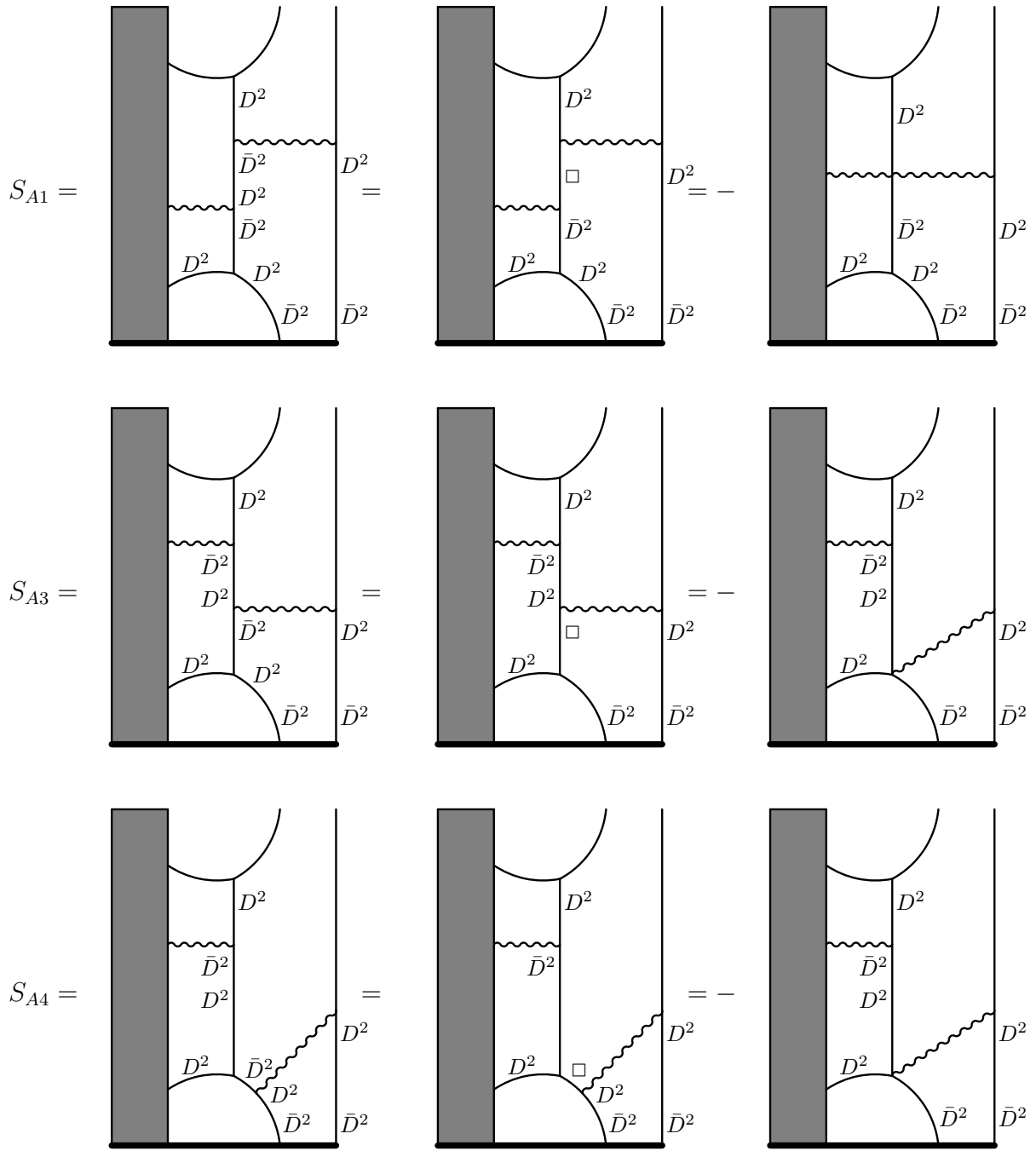

\unitlength=1.75mm
\settoheight{\eqoff}{$\times$}%
\setlength{\eqoff}{0.5\eqoff}%
\addtolength{\eqoff}{-12.5\unitlength}%
\settoheight{\eqofftwo}{$\times$}%
\setlength{\eqofftwo}{0.5\eqofftwo}%
\addtolength{\eqofftwo}{-7.5\unitlength}%
\begin{equation*}
\begin{aligned}
S_{A1}=
&
\begin{minipage}{4.3cm}
\raisebox{\eqoff}{%
\fmfframe(3,1)(1,4){%
\begin{fmfchar*}(20,30)
\fmftop{v1}
\fmfbottom{v5}
\fmfforce{(0w,h)}{v1}
\fmfforce{(0w,0)}{v5}
\fmffixed{(0.25w,0)}{v1,v2}
\fmffixed{(0.5w,0)}{v2,v3}
\fmffixed{(0.25w,0)}{v3,v4}
\fmffixed{(0.25w,0)}{v5,v6}
\fmffixed{(0.5w,0)}{v6,v7}
\fmffixed{(0.25w,0)}{v7,v8}
\fmf{phantom,tension=0.35,right=0.25}{v2,vc1}
\fmf{phantom,tension=0.35,left=0.25}{v6,vc2}
\fmf{plain,tension=0.5,left=0.25}{v3,vc1}
\fmf{plain,tension=0.5,right=0.25}{v7,vc2}
\fmf{plain,tension=0.3}{vc1,vc2}
\fmf{plain,tension=0.5,left=0.125}{vc3,vc2}
\fmf{plain}{v1,v5}
\fmf{plain}{v4,v8}
\fmf{plain,tension=0.5,right=0,width=1mm}{v5,v8}
\fmf{plain}{v1,v2}
\fmf{plain}{v6,v2}
\fmf{phantom}{v2,vc4}
\fmf{phantom,tension=0.25}{vc4,vc5}
\fmf{phantom}{vc5,v6}
\fmffreeze
\fmf{plain,left=0.2}{vc1,vc4}
\fmf{plain,right=0.2}{vc2,vc5}
\fmfposition
\fmfipath{p[]}
\fmfiset{p1}{vpath(__v1,__v5)}
\fmfiset{p2}{vpath(__v2,__vc1)}
\fmfiset{p3}{vpath(__v3,__vc1)}
\fmfiset{p4}{vpath(__vc1,__vc2)}
\fmfiset{p5}{vpath(__v6,__vc2)}
\fmfiset{p6}{vpath(__v7,__vc2)}
\fmfiset{p7}{vpath(__v4,__v8)}
\fmfiset{p8}{vpath(__v2,__v6)}
\fmfiset{p9}{vpath(__vc1,__vc4)}
\fmfiset{p10}{vpath(__vc2,__vc5)}
\fmfipair{w[]}
\fmfipair{wc[]}
\fmfiequ{w1}{point length(p8)/4 of p8}
\fmfiequ{w2}{(xpart(w3),ypart(wc2))}
\fmfiequ{w3}{point 3length(p8)/4 of p8}
\fmfiequ{w4}{point length(p3) of p3}
\fmfiequ{w5}{point length(p4) of p4}
\fmfiequ{w6}{point length(p6)/5 of p6}
\fmfiequ{w7}{point 4length(p6)/5 of p6}
\fmfiequ{w8}{point length(p10)/3 of p10}
\fmfiequ{w9}{(xpart(w10),ypart(wc1))}
\svertex{w10}{p7}
\fmfiequ{w11}{(xpart(w10),ypart(w6))}
\fmfiequ{w12}{(xpart(wc1),ypart(wc1)-(ypart(wc1)-ypart(wc2))/2)}
\fmfiequ{w13}{wc2}
\fmfiequ{w14}{(xpart(w10),ypart(w12))}
\fmfiequ{w15}{(xpart(wc1),ypart(vloc(__vc1))-(ypart(vloc(__vc1))-ypart(wc1))/2)}
\fmfiequ{wc1}{point length(p4)/3 of p4}
\fmfiequ{wc2}{point 2length(p4)/3 of p4}
\fmfiequ{wc3}{point 5length(p4)/6 of p4}
\fmfi{wiggly}{w2..wc2}
\fmfi{wiggly}{w9..wc1}
\fmfcmd{path ph; ph:=vloc(__v1)--vloc(__v5)--vloc(__v6)--vloc(__v2)--cycle;fill ph withcolor 0.5white;}
\fmfiv{l=\footnotesize{$D^2$},l.a=28,l.d=2}{w7}
\fmfiv{l=\footnotesize{$D^2$},l.a=152,l.d=0.25}{w8}
\fmfiv{l=\footnotesize{$\bar{D}^2$},l.a=28,l.d=2}{w6}
\fmfiv{l=\footnotesize{$\bar{D}^2$},l.a=28,l.d=2}{w11}
\fmfiv{l=\footnotesize{$\bar{D}^2$},l.a=28,l.d=2}{w12}
\fmfiv{l=\footnotesize{$D^2$},l.a=28,l.d=2}{w13}
\fmfiv{l=\footnotesize{$D^2$},l.a=28,l.d=2}{w14}
\fmfiv{l=\footnotesize{$\bar{D}^2$},l.a=28,l.d=2}{wc3}
\fmfiv{l=\footnotesize{$D^2$},l.a=28,l.d=2}{w15}
\end{fmfchar*}}}
\end{minipage}
=
&
\begin{minipage}{4.3cm}
\raisebox{\eqoff}{%
\fmfframe(3,1)(1,4){%
\begin{fmfchar*}(20,30)
\fmftop{v1}
\fmfbottom{v5}
\fmfforce{(0w,h)}{v1}
\fmfforce{(0w,0)}{v5}
\fmffixed{(0.25w,0)}{v1,v2}
\fmffixed{(0.5w,0)}{v2,v3}
\fmffixed{(0.25w,0)}{v3,v4}
\fmffixed{(0.25w,0)}{v5,v6}
\fmffixed{(0.5w,0)}{v6,v7}
\fmffixed{(0.25w,0)}{v7,v8}
\fmf{phantom,tension=0.35,right=0.25}{v2,vc1}
\fmf{phantom,tension=0.35,left=0.25}{v6,vc2}
\fmf{plain,tension=0.5,left=0.25}{v3,vc1}
\fmf{plain,tension=0.5,right=0.25}{v7,vc2}
\fmf{plain,tension=0.3}{vc1,vc2}
\fmf{plain,tension=0.5,left=0.125}{vc3,vc2}
\fmf{plain}{v1,v5}
\fmf{plain}{v4,v8}
\fmf{plain,tension=0.5,right=0,width=1mm}{v5,v8}
\fmf{plain}{v1,v2}
\fmf{plain}{v6,v2}
\fmf{phantom}{v2,vc4}
\fmf{phantom,tension=0.25}{vc4,vc5}
\fmf{phantom}{vc5,v6}
\fmffreeze
\fmf{plain,left=0.2}{vc1,vc4}
\fmf{plain,right=0.2}{vc2,vc5}
\fmfposition
\fmfipath{p[]}
\fmfiset{p1}{vpath(__v1,__v5)}
\fmfiset{p2}{vpath(__v2,__vc1)}
\fmfiset{p3}{vpath(__v3,__vc1)}
\fmfiset{p4}{vpath(__vc1,__vc2)}
\fmfiset{p5}{vpath(__v6,__vc2)}
\fmfiset{p6}{vpath(__v7,__vc2)}
\fmfiset{p7}{vpath(__v4,__v8)}
\fmfiset{p8}{vpath(__v2,__v6)}
\fmfiset{p9}{vpath(__vc1,__vc4)}
\fmfiset{p10}{vpath(__vc2,__vc5)}
\fmfipair{w[]}
\fmfipair{wc[]}
\fmfiequ{w1}{point length(p8)/4 of p8}
\fmfiequ{w2}{(xpart(w3),ypart(wc2))}
\fmfiequ{w3}{point 3length(p8)/4 of p8}
\fmfiequ{w4}{point length(p3) of p3}
\fmfiequ{w5}{point length(p4) of p4}
\fmfiequ{w6}{point length(p6)/5 of p6}
\fmfiequ{w7}{point 4length(p6)/5 of p6}
\fmfiequ{w8}{point length(p10)/3 of p10}
\fmfiequ{w9}{(xpart(w10),ypart(wc1))}
\svertex{w10}{p7}
\fmfiequ{w11}{(xpart(w10),ypart(w6))}
\fmfiequ{w12}{(xpart(wc1),ypart(wc1)-2(ypart(wc1)-ypart(wc2))/3)}
\fmfiequ{w13}{wc2}
\fmfiequ{w14}{(xpart(w10),ypart(w12))}
\fmfiequ{w15}{(xpart(wc1),ypart(vloc(__vc1))-(ypart(vloc(__vc1))-ypart(wc1))/2)}
\fmfiequ{wc1}{point length(p4)/3 of p4}
\fmfiequ{wc2}{point 2length(p4)/3 of p4}
\fmfiequ{wc3}{point 5length(p4)/6 of p4}
\fmfi{wiggly}{w2..wc2}
\fmfi{wiggly}{w9..wc1}
\fmfcmd{path ph; ph:=vloc(__v1)--vloc(__v5)--vloc(__v6)--vloc(__v2)--cycle;fill ph withcolor 0.5white;}
\fmfiv{l=\footnotesize{$D^2$},l.a=28,l.d=2}{w7}
\fmfiv{l=\footnotesize{$D^2$},l.a=152,l.d=0.25}{w8}
\fmfiv{l=\footnotesize{$\bar{D}^2$},l.a=28,l.d=2}{w6}
\fmfiv{l=\footnotesize{$\bar{D}^2$},l.a=28,l.d=2}{w11}
\fmfiv{l=\footnotesize{$\Box$},l.a=28,l.d=2}{w12}
\fmfiv{l=\footnotesize{$D^2$},l.a=28,l.d=2}{w14}
\fmfiv{l=\footnotesize{$\bar{D}^2$},l.a=28,l.d=2}{wc3}
\fmfiv{l=\footnotesize{$D^2$},l.a=28,l.d=2}{w15}
\end{fmfchar*}}}
\end{minipage}
=-
&
\begin{minipage}{4.3cm}
\raisebox{\eqoff}{%
\fmfframe(3,1)(1,4){%
\begin{fmfchar*}(20,30)
\fmftop{v1}
\fmfbottom{v5}
\fmfforce{(0w,h)}{v1}
\fmfforce{(0w,0)}{v5}
\fmffixed{(0.25w,0)}{v1,v2}
\fmffixed{(0.5w,0)}{v2,v3}
\fmffixed{(0.25w,0)}{v3,v4}
\fmffixed{(0.25w,0)}{v5,v6}
\fmffixed{(0.5w,0)}{v6,v7}
\fmffixed{(0.25w,0)}{v7,v8}
\fmf{phantom,tension=0.35,right=0.25}{v2,vc1}
\fmf{phantom,tension=0.35,left=0.25}{v6,vc2}
\fmf{plain,tension=0.5,left=0.25}{v3,vc1}
\fmf{plain,tension=0.5,right=0.25}{v7,vc2}
\fmf{plain,tension=0.3}{vc1,vc2}
\fmf{plain,tension=0.5,left=0.125}{vc3,vc2}
\fmf{plain}{v1,v5}
\fmf{plain}{v4,v8}
\fmf{plain,tension=0.5,right=0,width=1mm}{v5,v8}
\fmf{plain}{v1,v2}
\fmf{plain}{v6,v2}
\fmf{phantom}{v2,vc4}
\fmf{phantom,tension=0.25}{vc4,vc5}
\fmf{phantom}{vc5,v6}
\fmffreeze
\fmf{plain,left=0.2}{vc1,vc4}
\fmf{plain,right=0.2}{vc2,vc5}
\fmfposition
\fmfipath{p[]}
\fmfiset{p1}{vpath(__v1,__v5)}
\fmfiset{p2}{vpath(__v2,__vc1)}
\fmfiset{p3}{vpath(__v3,__vc1)}
\fmfiset{p4}{vpath(__vc1,__vc2)}
\fmfiset{p5}{vpath(__v6,__vc2)}
\fmfiset{p6}{vpath(__v7,__vc2)}
\fmfiset{p7}{vpath(__v4,__v8)}
\fmfiset{p8}{vpath(__v2,__v6)}
\fmfiset{p9}{vpath(__vc1,__vc4)}
\fmfiset{p10}{vpath(__vc2,__vc5)}
\fmfipair{w[]}
\fmfipair{wc[]}
\fmfiequ{w1}{point length(p8)/4 of p8}
\fmfiequ{w2}{(xpart(w3),ypart(wc2))}
\fmfiequ{w3}{point 3length(p8)/4 of p8}
\fmfiequ{w4}{point length(p3) of p3}
\fmfiequ{w5}{point length(p4) of p4}
\fmfiequ{w6}{point length(p6)/5 of p6}
\fmfiequ{w7}{point 4length(p6)/5 of p6}
\fmfiequ{w8}{point length(p10)/3 of p10}
\fmfiequ{w9}{(xpart(w10),ypart(wc1))}
\svertex{w10}{p7}
\fmfiequ{w11}{(xpart(w10),ypart(w6))}
\fmfiequ{w12}{(xpart(wc1),ypart(wc1)-2(ypart(wc1)-ypart(wc2))/3)}
\fmfiequ{w13}{wc2}
\fmfiequ{w14}{(xpart(w10),ypart(wc3))}
\fmfiequ{w15}{(xpart(wc1),ypart(vloc(__vc1))-(ypart(vloc(__vc1))-ypart(wc1))/2)}
\fmfiequ{wc1}{point length(p4)/2 of p4}
\fmfiequ{wc2}{point length(p4)/2 of p4}
\fmfiequ{wc3}{point 5length(p4)/6 of p4}
\fmfi{wiggly}{w2..wc2}
\fmfi{wiggly}{w9..wc1}
\fmfcmd{path ph; ph:=vloc(__v1)--vloc(__v5)--vloc(__v6)--vloc(__v2)--cycle;fill ph withcolor 0.5white;}
\fmfiv{l=\footnotesize{$D^2$},l.a=28,l.d=2}{w7}
\fmfiv{l=\footnotesize{$D^2$},l.a=152,l.d=0.25}{w8}
\fmfiv{l=\footnotesize{$\bar{D}^2$},l.a=28,l.d=2}{w6}
\fmfiv{l=\footnotesize{$\bar{D}^2$},l.a=28,l.d=2}{w11}
\fmfiv{l=\footnotesize{$D^2$},l.a=28,l.d=2}{w14}
\fmfiv{l=\footnotesize{$\bar{D}^2$},l.a=28,l.d=2}{wc3}
\fmfiv{l=\footnotesize{$D^2$},l.a=28,l.d=2}{w15}
\end{fmfchar*}}}
\end{minipage}
\\
S_{A3}=&
\begin{minipage}{4.3cm}
\raisebox{\eqoff}{%
\fmfframe(3,1)(1,4){%
\begin{fmfchar*}(20,30)
\fmftop{v1}
\fmfbottom{v5}
\fmfforce{(0w,h)}{v1}
\fmfforce{(0w,0)}{v5}
\fmffixed{(0.25w,0)}{v1,v2}
\fmffixed{(0.5w,0)}{v2,v3}
\fmffixed{(0.25w,0)}{v3,v4}
\fmffixed{(0.25w,0)}{v5,v6}
\fmffixed{(0.5w,0)}{v6,v7}
\fmffixed{(0.25w,0)}{v7,v8}
\fmf{phantom,tension=0.35,right=0.25}{v2,vc1}
\fmf{phantom,tension=0.35,left=0.25}{v6,vc2}
\fmf{plain,tension=0.5,left=0.25}{v3,vc1}
\fmf{plain,tension=0.5,right=0.25}{v7,vc2}
\fmf{plain,tension=0.3}{vc1,vc2}
\fmf{plain,tension=0.5,left=0.125}{vc3,vc2}
\fmf{plain}{v1,v5}
\fmf{plain}{v4,v8}
\fmf{plain,tension=0.5,right=0,width=1mm}{v5,v8}
\fmf{plain}{v1,v2}
\fmf{plain}{v6,v2}
\fmf{phantom}{v2,vc4}
\fmf{phantom,tension=0.25}{vc4,vc5}
\fmf{phantom}{vc5,v6}
\fmffreeze
\fmf{plain,left=0.2}{vc1,vc4}
\fmf{plain,right=0.2}{vc2,vc5}
\fmfposition
\fmfipath{p[]}
\fmfiset{p1}{vpath(__v1,__v5)}
\fmfiset{p2}{vpath(__v2,__vc1)}
\fmfiset{p3}{vpath(__v3,__vc1)}
\fmfiset{p4}{vpath(__vc1,__vc2)}
\fmfiset{p5}{vpath(__v6,__vc2)}
\fmfiset{p6}{vpath(__v7,__vc2)}
\fmfiset{p7}{vpath(__v4,__v8)}
\fmfiset{p8}{vpath(__v2,__v6)}
\fmfiset{p9}{vpath(__vc1,__vc4)}
\fmfiset{p10}{vpath(__vc2,__vc5)}
\fmfipair{w[]}
\fmfipair{wc[]}
\fmfiequ{wc1}{point length(p4)/3 of p4}
\fmfiequ{wc2}{point 2length(p4)/3 of p4}
\fmfiequ{wc3}{point 5length(p4)/6 of p4}
\fmfiequ{w1}{point length(p8)/4 of p8}
\fmfiequ{w2}{(xpart(w3),ypart(wc1))}
\fmfiequ{w3}{point 3length(p8)/4 of p8}
\fmfiequ{w4}{point length(p3) of p3}
\fmfiequ{w5}{point length(p4) of p4}
\fmfiequ{w6}{point length(p6)/5 of p6}
\fmfiequ{w7}{point 4length(p6)/5 of p6}
\fmfiequ{w8}{point length(p10)/3 of p10}
\fmfiequ{w9}{(xpart(w10),ypart(wc2))}
\svertex{w10}{p7}
\fmfiequ{w11}{(xpart(w10),ypart(w6))}
\fmfiequ{w12}{(xpart(wc1),ypart(wc1)-(ypart(wc1)-ypart(wc2))/2)}
\fmfiequ{w13}{wc2}
\fmfiequ{w14}{(xpart(w10),ypart(wc3))}
\fmfiequ{w15}{(xpart(wc1),ypart(vloc(__vc1))-(ypart(vloc(__vc1))-ypart(wc1))/2)}
\fmfi{wiggly}{w2..wc1}
\fmfi{wiggly}{w9..wc2}
\fmfcmd{path ph; ph:=vloc(__v1)--vloc(__v5)--vloc(__v6)--vloc(__v2)--cycle;fill ph withcolor 0.5white;}
\fmfiv{l=\footnotesize{$D^2$},l.a=28,l.d=2}{w7}
\fmfiv{l=\footnotesize{$D^2$},l.a=152,l.d=0.25}{w8}
\fmfiv{l=\footnotesize{$\bar{D}^2$},l.a=28,l.d=2}{w6}
\fmfiv{l=\footnotesize{$\bar{D}^2$},l.a=28,l.d=2}{w11}
\fmfiv{l=\footnotesize{$\bar{D}^2$},l.a=152,l.d=2}{w12}
\fmfiv{l=\footnotesize{$D^2$},l.a=152,l.d=2}{w13}
\fmfiv{l=\footnotesize{$D^2$},l.a=28,l.d=2}{w14}
\fmfiv{l=\footnotesize{$\bar{D}^2$},l.a=28,l.d=2}{wc3}
\fmfiv{l=\footnotesize{$D^2$},l.a=28,l.d=2}{w15}
\end{fmfchar*}}}
\end{minipage}
=&
\begin{minipage}{4.3cm}
\raisebox{\eqoff}{%
\fmfframe(3,1)(1,4){%
\begin{fmfchar*}(20,30)
\fmftop{v1}
\fmfbottom{v5}
\fmfforce{(0w,h)}{v1}
\fmfforce{(0w,0)}{v5}
\fmffixed{(0.25w,0)}{v1,v2}
\fmffixed{(0.5w,0)}{v2,v3}
\fmffixed{(0.25w,0)}{v3,v4}
\fmffixed{(0.25w,0)}{v5,v6}
\fmffixed{(0.5w,0)}{v6,v7}
\fmffixed{(0.25w,0)}{v7,v8}
\fmf{phantom,tension=0.35,right=0.25}{v2,vc1}
\fmf{phantom,tension=0.35,left=0.25}{v6,vc2}
\fmf{plain,tension=0.5,left=0.25}{v3,vc1}
\fmf{plain,tension=0.5,right=0.25}{v7,vc2}
\fmf{plain,tension=0.3}{vc1,vc2}
\fmf{plain,tension=0.5,left=0.125}{vc3,vc2}
\fmf{plain}{v1,v5}
\fmf{plain}{v4,v8}
\fmf{plain,tension=0.5,right=0,width=1mm}{v5,v8}
\fmf{plain}{v1,v2}
\fmf{plain}{v6,v2}
\fmf{phantom}{v2,vc4}
\fmf{phantom,tension=0.25}{vc4,vc5}
\fmf{phantom}{vc5,v6}
\fmffreeze
\fmf{plain,left=0.2}{vc1,vc4}
\fmf{plain,right=0.2}{vc2,vc5}
\fmfposition
\fmfipath{p[]}
\fmfiset{p1}{vpath(__v1,__v5)}
\fmfiset{p2}{vpath(__v2,__vc1)}
\fmfiset{p3}{vpath(__v3,__vc1)}
\fmfiset{p4}{vpath(__vc1,__vc2)}
\fmfiset{p5}{vpath(__v6,__vc2)}
\fmfiset{p6}{vpath(__v7,__vc2)}
\fmfiset{p7}{vpath(__v4,__v8)}
\fmfiset{p8}{vpath(__v2,__v6)}
\fmfiset{p9}{vpath(__vc1,__vc4)}
\fmfiset{p10}{vpath(__vc2,__vc5)}
\fmfipair{w[]}
\fmfipair{wc[]}
\fmfiequ{wc1}{point length(p4)/3 of p4}
\fmfiequ{wc2}{point 2length(p4)/3 of p4}
\fmfiequ{wc3}{point 5length(p4)/6 of p4}
\fmfiequ{w1}{point length(p8)/4 of p8}
\fmfiequ{w2}{(xpart(w3),ypart(wc1))}
\fmfiequ{w3}{point 3length(p8)/4 of p8}
\fmfiequ{w4}{point length(p3) of p3}
\fmfiequ{w5}{point length(p4) of p4}
\fmfiequ{w6}{point length(p6)/5 of p6}
\fmfiequ{w7}{point 4length(p6)/5 of p6}
\fmfiequ{w8}{point length(p10)/3 of p10}
\fmfiequ{w9}{(xpart(w10),ypart(wc2))}
\svertex{w10}{p7}
\fmfiequ{w11}{(xpart(w10),ypart(w6))}
\fmfiequ{w12}{(xpart(wc1),ypart(wc1)-(ypart(wc1)-ypart(wc2))/2)}
\fmfiequ{w13}{wc2}
\fmfiequ{w14}{(xpart(w10),ypart(wc3))}
\fmfiequ{w15}{(xpart(wc1),ypart(vloc(__vc1))-(ypart(vloc(__vc1))-ypart(wc1))/2)}
\fmfi{wiggly}{w2..wc1}
\fmfi{wiggly}{w9..wc2}
\fmfcmd{path ph; ph:=vloc(__v1)--vloc(__v5)--vloc(__v6)--vloc(__v2)--cycle;fill ph withcolor 0.5white;}
\fmfiv{l=\footnotesize{$D^2$},l.a=152,l.d=0.25}{w8}
\fmfiv{l=\footnotesize{$\bar{D}^2$},l.a=28,l.d=2}{w6}
\fmfiv{l=\footnotesize{$\bar{D}^2$},l.a=28,l.d=2}{w11}
\fmfiv{l=\footnotesize{$\bar{D}^2$},l.a=152,l.d=2}{w12}
\fmfiv{l=\footnotesize{$D^2$},l.a=152,l.d=2}{w13}
\fmfiv{l=\footnotesize{$D^2$},l.a=28,l.d=2}{w14}
\fmfiv{l=\footnotesize{$\Box$},l.a=28,l.d=2}{wc3}
\fmfiv{l=\footnotesize{$D^2$},l.a=28,l.d=2}{w15}
\end{fmfchar*}}}
\end{minipage}
=-&
\begin{minipage}{4.3cm}
\raisebox{\eqoff}{%
\fmfframe(3,1)(1,4){%
\begin{fmfchar*}(20,30)
\fmftop{v1}
\fmfbottom{v5}
\fmfforce{(0w,h)}{v1}
\fmfforce{(0w,0)}{v5}
\fmffixed{(0.25w,0)}{v1,v2}
\fmffixed{(0.5w,0)}{v2,v3}
\fmffixed{(0.25w,0)}{v3,v4}
\fmffixed{(0.25w,0)}{v5,v6}
\fmffixed{(0.5w,0)}{v6,v7}
\fmffixed{(0.25w,0)}{v7,v8}
\fmf{phantom,tension=0.35,right=0.25}{v2,vc1}
\fmf{phantom,tension=0.35,left=0.25}{v6,vc2}
\fmf{plain,tension=0.5,left=0.25}{v3,vc1}
\fmf{plain,tension=0.5,right=0.25}{v7,vc2}
\fmf{plain,tension=0.3}{vc1,vc2}
\fmf{plain,tension=0.5,left=0.125}{vc3,vc2}
\fmf{plain}{v1,v5}
\fmf{plain}{v4,v8}
\fmf{plain,tension=0.5,right=0,width=1mm}{v5,v8}
\fmf{plain}{v1,v2}
\fmf{plain}{v6,v2}
\fmf{phantom}{v2,vc4}
\fmf{phantom,tension=0.25}{vc4,vc5}
\fmf{phantom}{vc5,v6}
\fmffreeze
\fmf{plain,left=0.2}{vc1,vc4}
\fmf{plain,right=0.2}{vc2,vc5}
\fmfposition
\fmfipath{p[]}
\fmfiset{p1}{vpath(__v1,__v5)}
\fmfiset{p2}{vpath(__v2,__vc1)}
\fmfiset{p3}{vpath(__v3,__vc1)}
\fmfiset{p4}{vpath(__vc1,__vc2)}
\fmfiset{p5}{vpath(__v6,__vc2)}
\fmfiset{p6}{vpath(__v7,__vc2)}
\fmfiset{p7}{vpath(__v4,__v8)}
\fmfiset{p8}{vpath(__v2,__v6)}
\fmfiset{p9}{vpath(__vc1,__vc4)}
\fmfiset{p10}{vpath(__vc2,__vc5)}
\fmfipair{w[]}
\fmfipair{wc[]}
\fmfiequ{wc1}{point length(p4)/3 of p4}
\fmfiequ{wc2}{point 2length(p4)/3 of p4}
\fmfiequ{wc3}{point 5length(p4)/6 of p4}
\fmfiequ{w1}{point length(p8)/4 of p8}
\fmfiequ{w2}{(xpart(w3),ypart(wc1))}
\fmfiequ{w3}{point 3length(p8)/4 of p8}
\fmfiequ{w4}{point length(p3) of p3}
\fmfiequ{w5}{point length(p4) of p4}
\fmfiequ{w6}{point length(p6)/5 of p6}
\fmfiequ{w7}{point 4length(p6)/5 of p6}
\fmfiequ{w8}{point length(p10)/3 of p10}
\fmfiequ{w9}{(xpart(w10),ypart(wc2))}
\svertex{w10}{p7}
\fmfiequ{w11}{(xpart(w10),ypart(w6))}
\fmfiequ{w12}{(xpart(wc1),ypart(wc1)-(ypart(wc1)-ypart(wc2))/2)}
\fmfiequ{w13}{wc2}
\fmfiequ{w14}{(xpart(w10),ypart(wc3))}
\fmfiequ{w15}{(xpart(wc1),ypart(vloc(__vc1))-(ypart(vloc(__vc1))-ypart(wc1))/2)}
\fmfi{wiggly}{w2..wc1}
\fmfi{wiggly}{w9..vloc(__vc2)}
\fmfcmd{path ph; ph:=vloc(__v1)--vloc(__v5)--vloc(__v6)--vloc(__v2)--cycle;fill ph withcolor 0.5white;}
\fmfiv{l=\footnotesize{$D^2$},l.a=152,l.d=0.25}{w8}
\fmfiv{l=\footnotesize{$\bar{D}^2$},l.a=28,l.d=2}{w6}
\fmfiv{l=\footnotesize{$\bar{D}^2$},l.a=28,l.d=2}{w11}
\fmfiv{l=\footnotesize{$\bar{D}^2$},l.a=152,l.d=2}{w12}
\fmfiv{l=\footnotesize{$D^2$},l.a=152,l.d=2}{w13}
\fmfiv{l=\footnotesize{$D^2$},l.a=28,l.d=2}{w14}
\fmfiv{l=\footnotesize{$D^2$},l.a=28,l.d=2}{w15}
\end{fmfchar*}}}
\end{minipage}
\\
S_{A4}=&
\begin{minipage}{4.3cm}
\raisebox{\eqoff}{%
\fmfframe(3,1)(1,4){%
\begin{fmfchar*}(20,30)
\fmftop{v1}
\fmfbottom{v5}
\fmfforce{(0w,h)}{v1}
\fmfforce{(0w,0)}{v5}
\fmffixed{(0.25w,0)}{v1,v2}
\fmffixed{(0.5w,0)}{v2,v3}
\fmffixed{(0.25w,0)}{v3,v4}
\fmffixed{(0.25w,0)}{v5,v6}
\fmffixed{(0.5w,0)}{v6,v7}
\fmffixed{(0.25w,0)}{v7,v8}
\fmf{phantom,tension=0.35,right=0.25}{v2,vc1}
\fmf{phantom,tension=0.35,left=0.25}{v6,vc2}
\fmf{plain,tension=0.5,left=0.25}{v3,vc1}
\fmf{plain,tension=0.5,right=0.25}{v7,vc2}
\fmf{plain,tension=0.3}{vc1,vc2}
\fmf{plain,tension=0.5,left=0.125}{vc3,vc2}
\fmf{plain}{v1,v5}
\fmf{plain}{v4,v8}
\fmf{plain,tension=0.5,right=0,width=1mm}{v5,v8}
\fmf{plain}{v1,v2}
\fmf{plain}{v6,v2}
\fmf{phantom}{v2,vc4}
\fmf{phantom,tension=0.25}{vc4,vc5}
\fmf{phantom}{vc5,v6}
\fmffreeze
\fmf{plain,left=0.2}{vc1,vc4}
\fmf{plain,right=0.2}{vc2,vc5}
\fmfposition
\fmfipath{p[]}
\fmfiset{p1}{vpath(__v1,__v5)}
\fmfiset{p2}{vpath(__v2,__vc1)}
\fmfiset{p3}{vpath(__v3,__vc1)}
\fmfiset{p4}{vpath(__vc1,__vc2)}
\fmfiset{p5}{vpath(__v6,__vc2)}
\fmfiset{p6}{vpath(__v7,__vc2)}
\fmfiset{p7}{vpath(__v4,__v8)}
\fmfiset{p8}{vpath(__v2,__v6)}
\fmfiset{p9}{vpath(__vc1,__vc4)}
\fmfiset{p10}{vpath(__vc2,__vc5)}
\fmfipair{w[]}
\fmfipair{wc[]}
\fmfiequ{wc1}{point length(p4)/3 of p4}
\fmfiequ{wc2}{point 2length(p4)/3 of p4}
\fmfiequ{wc3}{point 5length(p4)/6 of p4}
\fmfiequ{w1}{point length(p8)/4 of p8}
\fmfiequ{w2}{(xpart(w3),ypart(wc1))}
\fmfiequ{w3}{point 3length(p8)/4 of p8}
\fmfiequ{w4}{point length(p3) of p3}
\fmfiequ{w5}{point length(p4) of p4}
\fmfiequ{w6}{point 2length(p6)/3 of p6}
\fmfiequ{w7}{point 10length(p6)/11 of p6}
\fmfiequ{w8}{point length(p10)/3 of p10}
\fmfiequ{w9}{(xpart(w10),ypart(wc2))}
\svertex{w10}{p7}
\fmfiequ{w11}{(xpart(w10),ypart(w16))}
\fmfiequ{w12}{(xpart(wc1),ypart(wc1)-(ypart(wc1)-ypart(wc2))/2)}
\fmfiequ{w13}{wc2}
\fmfiequ{w14}{(xpart(w10),ypart(wc3))}
\fmfiequ{w15}{(xpart(wc1),ypart(vloc(__vc1))-(ypart(vloc(__vc1))-ypart(wc1))/2)}
\fmfiequ{w16}{point length(p6)/8 of p6}
\fmfiequ{w17}{point 4length(p6)/8 of p6}
\fmfi{wiggly}{w2..wc1}
\fmfi{wiggly}{w9..w6}
\fmfcmd{path ph; ph:=vloc(__v1)--vloc(__v5)--vloc(__v6)--vloc(__v2)--cycle;fill ph withcolor 0.5white;}
\fmfiv{l=\footnotesize{$\bar{D}^2$},l.a=28,l.d=2}{w7}
\fmfiv{l=\footnotesize{$D^2$},l.a=152,l.d=0.25}{w8}
\fmfiv{l=\footnotesize{$\bar{D}^2$},l.a=28,l.d=2}{w11}
\fmfiv{l=\footnotesize{$\bar{D}^2$},l.a=152,l.d=2}{w12}
\fmfiv{l=\footnotesize{$D^2$},l.a=152,l.d=2}{w13}
\fmfiv{l=\footnotesize{$D^2$},l.a=28,l.d=2}{w14}
\fmfiv{l=\footnotesize{$D^2$},l.a=28,l.d=2}{w15}
\fmfiv{l=\footnotesize{$D^2$},l.a=28,l.d=2}{w17}
\fmfiv{l=\footnotesize{$\bar{D}^2$},l.a=28,l.d=2}{w16}
\end{fmfchar*}}}
\end{minipage}
=&
\begin{minipage}{4.3cm}
\raisebox{\eqoff}{%
\fmfframe(3,1)(1,4){%
\begin{fmfchar*}(20,30)
\fmftop{v1}
\fmfbottom{v5}
\fmfforce{(0w,h)}{v1}
\fmfforce{(0w,0)}{v5}
\fmffixed{(0.25w,0)}{v1,v2}
\fmffixed{(0.5w,0)}{v2,v3}
\fmffixed{(0.25w,0)}{v3,v4}
\fmffixed{(0.25w,0)}{v5,v6}
\fmffixed{(0.5w,0)}{v6,v7}
\fmffixed{(0.25w,0)}{v7,v8}
\fmf{phantom,tension=0.35,right=0.25}{v2,vc1}
\fmf{phantom,tension=0.35,left=0.25}{v6,vc2}
\fmf{plain,tension=0.5,left=0.25}{v3,vc1}
\fmf{plain,tension=0.5,right=0.25}{v7,vc2}
\fmf{plain,tension=0.3}{vc1,vc2}
\fmf{plain,tension=0.5,left=0.125}{vc3,vc2}
\fmf{plain}{v1,v5}
\fmf{plain}{v4,v8}
\fmf{plain,tension=0.5,right=0,width=1mm}{v5,v8}
\fmf{plain}{v1,v2}
\fmf{plain}{v6,v2}
\fmf{phantom}{v2,vc4}
\fmf{phantom,tension=0.25}{vc4,vc5}
\fmf{phantom}{vc5,v6}
\fmffreeze
\fmf{plain,left=0.2}{vc1,vc4}
\fmf{plain,right=0.2}{vc2,vc5}
\fmfposition
\fmfipath{p[]}
\fmfiset{p1}{vpath(__v1,__v5)}
\fmfiset{p2}{vpath(__v2,__vc1)}
\fmfiset{p3}{vpath(__v3,__vc1)}
\fmfiset{p4}{vpath(__vc1,__vc2)}
\fmfiset{p5}{vpath(__v6,__vc2)}
\fmfiset{p6}{vpath(__v7,__vc2)}
\fmfiset{p7}{vpath(__v4,__v8)}
\fmfiset{p8}{vpath(__v2,__v6)}
\fmfiset{p9}{vpath(__vc1,__vc4)}
\fmfiset{p10}{vpath(__vc2,__vc5)}
\fmfipair{w[]}
\fmfipair{wc[]}
\fmfiequ{wc1}{point length(p4)/3 of p4}
\fmfiequ{wc2}{point 2length(p4)/3 of p4}
\fmfiequ{wc3}{point 5length(p4)/6 of p4}
\fmfiequ{w1}{point length(p8)/4 of p8}
\fmfiequ{w2}{(xpart(w3),ypart(wc1))}
\fmfiequ{w3}{point 3length(p8)/4 of p8}
\fmfiequ{w4}{point length(p3) of p3}
\fmfiequ{w5}{point length(p4) of p4}
\fmfiequ{w6}{point 2length(p6)/3 of p6}
\fmfiequ{w7}{point 10length(p6)/11 of p6}
\fmfiequ{w8}{point length(p10)/3 of p10}
\fmfiequ{w9}{(xpart(w10),ypart(wc2))}
\svertex{w10}{p7}
\fmfiequ{w11}{(xpart(w10),ypart(w16))}
\fmfiequ{w12}{(xpart(wc1),ypart(wc1)-(ypart(wc1)-ypart(wc2))/2)}
\fmfiequ{w13}{wc2}
\fmfiequ{w14}{(xpart(w10),ypart(wc3))}
\fmfiequ{w15}{(xpart(wc1),ypart(vloc(__vc1))-(ypart(vloc(__vc1))-ypart(wc1))/2)}
\fmfiequ{w16}{point length(p6)/8 of p6}
\fmfiequ{w17}{point 4length(p6)/8 of p6}
\fmfi{wiggly}{w2..wc1}
\fmfi{wiggly}{w9..w6}
\fmfcmd{path ph; ph:=vloc(__v1)--vloc(__v5)--vloc(__v6)--vloc(__v2)--cycle;fill ph withcolor 0.5white;}
\fmfiv{l=\footnotesize{$\Box$},l.a=28,l.d=2}{w7}
\fmfiv{l=\footnotesize{$D^2$},l.a=152,l.d=0.25}{w8}
\fmfiv{l=\footnotesize{$\bar{D}^2$},l.a=28,l.d=2}{w11}
\fmfiv{l=\footnotesize{$\bar{D}^2$},l.a=152,l.d=2}{w12}
\fmfiv{l=\footnotesize{$D^2$},l.a=28,l.d=2}{w14}
\fmfiv{l=\footnotesize{$D^2$},l.a=28,l.d=2}{w15}
\fmfiv{l=\footnotesize{$D^2$},l.a=28,l.d=2}{w17}
\fmfiv{l=\footnotesize{$\bar{D}^2$},l.a=28,l.d=2}{w16}
\end{fmfchar*}}}
\end{minipage}
=-&
\begin{minipage}{4.3cm}
\raisebox{\eqoff}{%
\fmfframe(3,1)(1,4){%
\begin{fmfchar*}(20,30)
\fmftop{v1}
\fmfbottom{v5}
\fmfforce{(0w,h)}{v1}
\fmfforce{(0w,0)}{v5}
\fmffixed{(0.25w,0)}{v1,v2}
\fmffixed{(0.5w,0)}{v2,v3}
\fmffixed{(0.25w,0)}{v3,v4}
\fmffixed{(0.25w,0)}{v5,v6}
\fmffixed{(0.5w,0)}{v6,v7}
\fmffixed{(0.25w,0)}{v7,v8}
\fmf{phantom,tension=0.35,right=0.25}{v2,vc1}
\fmf{phantom,tension=0.35,left=0.25}{v6,vc2}
\fmf{plain,tension=0.5,left=0.25}{v3,vc1}
\fmf{plain,tension=0.5,right=0.25}{v7,vc2}
\fmf{plain,tension=0.3}{vc1,vc2}
\fmf{plain,tension=0.5,left=0.125}{vc3,vc2}
\fmf{plain}{v1,v5}
\fmf{plain}{v4,v8}
\fmf{plain,tension=0.5,right=0,width=1mm}{v5,v8}
\fmf{plain}{v1,v2}
\fmf{plain}{v6,v2}
\fmf{phantom}{v2,vc4}
\fmf{phantom,tension=0.25}{vc4,vc5}
\fmf{phantom}{vc5,v6}
\fmffreeze
\fmf{plain,left=0.2}{vc1,vc4}
\fmf{plain,right=0.2}{vc2,vc5}
\fmfposition
\fmfipath{p[]}
\fmfiset{p1}{vpath(__v1,__v5)}
\fmfiset{p2}{vpath(__v2,__vc1)}
\fmfiset{p3}{vpath(__v3,__vc1)}
\fmfiset{p4}{vpath(__vc1,__vc2)}
\fmfiset{p5}{vpath(__v6,__vc2)}
\fmfiset{p6}{vpath(__v7,__vc2)}
\fmfiset{p7}{vpath(__v4,__v8)}
\fmfiset{p8}{vpath(__v2,__v6)}
\fmfiset{p9}{vpath(__vc1,__vc4)}
\fmfiset{p10}{vpath(__vc2,__vc5)}
\fmfipair{w[]}
\fmfipair{wc[]}
\fmfiequ{wc1}{point length(p4)/3 of p4}
\fmfiequ{wc2}{point 2length(p4)/3 of p4}
\fmfiequ{wc3}{point 5length(p4)/6 of p4}
\fmfiequ{w1}{point length(p8)/4 of p8}
\fmfiequ{w2}{(xpart(w3),ypart(wc1))}
\fmfiequ{w3}{point 3length(p8)/4 of p8}
\fmfiequ{w4}{point length(p3) of p3}
\fmfiequ{w5}{point length(p4) of p4}
\fmfiequ{w6}{point length(p6)/5 of p6}
\fmfiequ{w7}{point 4length(p6)/5 of p6}
\fmfiequ{w8}{point length(p10)/3 of p10}
\fmfiequ{w9}{(xpart(w10),ypart(wc2))}
\svertex{w10}{p7}
\fmfiequ{w11}{(xpart(w10),ypart(w6))}
\fmfiequ{w12}{(xpart(wc1),ypart(wc1)-(ypart(wc1)-ypart(wc2))/2)}
\fmfiequ{w13}{wc2}
\fmfiequ{w14}{(xpart(w10),ypart(wc3))}
\fmfiequ{w15}{(xpart(wc1),ypart(vloc(__vc1))-(ypart(vloc(__vc1))-ypart(wc1))/2)}
\fmfi{wiggly}{w2..wc1}
\fmfi{wiggly}{w9..vloc(__vc2)}
\fmfcmd{path ph; ph:=vloc(__v1)--vloc(__v5)--vloc(__v6)--vloc(__v2)--cycle;fill ph withcolor 0.5white;}
\fmfiv{l=\footnotesize{$D^2$},l.a=152,l.d=0.25}{w8}
\fmfiv{l=\footnotesize{$\bar{D}^2$},l.a=28,l.d=2}{w6}
\fmfiv{l=\footnotesize{$\bar{D}^2$},l.a=28,l.d=2}{w11}
\fmfiv{l=\footnotesize{$\bar{D}^2$},l.a=152,l.d=2}{w12}
\fmfiv{l=\footnotesize{$D^2$},l.a=152,l.d=2}{w13}
\fmfiv{l=\footnotesize{$D^2$},l.a=28,l.d=2}{w14}
\fmfiv{l=\footnotesize{$D^2$},l.a=28,l.d=2}{w15}
\end{fmfchar*}}}
\end{minipage}
\end{aligned}
\end{equation*}
\caption{$D$-algebra steps showing explicit cancellations for class A}
\label{diagrams-A-dem}
\end{figure}

In this appendix we use the result of subsection \ref{subsec:powercount} to show that the divergent parts of range-five diagrams where one line interacts with the rest of the graph only through a vector line sum up to zero.\\
Let us consider a generic three-loop, range-four diagram. We want to extend it to range five and four loops by connecting it to a free line from the operator, using a vector propagator. There are three possible structures for the part of the three-loop diagram which directly interacts with the new line. All the possible diagrams coming from each class are shown in figures~\ref{diagrams-A}, \ref{diagrams-B} and \ref{diagrams-C}.
\begin{itemize}
\item Class A (Figure \ref{diagrams-A})
:\\
As can be seen from the first line of Figure \ref{diagrams-A-dem}, performing part of the $D$-algebra for diagram $S_{A1}$ is enough to obtain the same structure of diagram $S_{A2}$, with a different sign due to the $\Box=-p^2$ cancelling one propagator. Since the two diagrams have the same color factor, their divergent parts sum up to zero.\\
Now consider diagrams $S_{A3}$ and $S_{A4}$: they will produce the same result after $D$-algebra, since in a few steps they can be reduced to the same structure (see lines 2 and 3 of Figure \ref{diagrams-A-dem}). However, the two diagrams have opposite color factors and so their divergent parts cancel.\\
Diagram $S_{A5}$ is finite.

\item Class B (Figure \ref{diagrams-B}):\\
For diagrams $S_{B1}$ and $S_{B2}$ the discussion is similar to the one for $S_{A3}$ and $S_{A4}$: a partial $D$-algebra for both diagrams shows that they will produce the same result (see lines 1 and 2 of Figure \ref{diagrams-BC-dem}), and the opposite color factors make the total divergent part vanish.\\
Diagram $S_{B3}$ is finite.

\item Class C (Figure \ref{diagrams-C}):\\
For diagrams $S_{C1}$ and $S_{C2}$ we have the same situation as for $S_{A1}$ and $S_{A2}$: diagram $S_{C1}$ can be reduced to $S_{C2}$ through the cancellation of a propagator with a $\Box$, which gives a minus sign (line 3 of Figure \ref{diagrams-BC-dem}). This difference in sign is responsible for the cancellation of divergent parts, since the two color factors are equal.\\
Diagram $S_{C3}$ is finite.

\end{itemize}
\begin{figure}[h]
\unitlength=1.75mm
\settoheight{\eqoff}{$\times$}%
\setlength{\eqoff}{0.5\eqoff}%
\addtolength{\eqoff}{-12.5\unitlength}%
\settoheight{\eqofftwo}{$\times$}%
\setlength{\eqofftwo}{0.5\eqofftwo}%
\addtolength{\eqofftwo}{-7.5\unitlength}%
\centering
\subfigure[$S_{A1}$]{
\raisebox{\eqoff}{%
\fmfframe(3,1)(1,4){%
\begin{fmfchar*}(20,30)
\fmftop{v1}
\fmfbottom{v5}
\fmfforce{(0w,h)}{v1}
\fmfforce{(0w,0)}{v5}
\fmffixed{(0.25w,0)}{v1,v2}
\fmffixed{(0.5w,0)}{v2,v3}
\fmffixed{(0.25w,0)}{v3,v4}
\fmffixed{(0.25w,0)}{v5,v6}
\fmffixed{(0.5w,0)}{v6,v7}
\fmffixed{(0.25w,0)}{v7,v8}
\fmf{phantom,tension=0.35,right=0.25}{v2,vc1}
\fmf{phantom,tension=0.35,left=0.25}{v6,vc2}
\fmf{plain,tension=0.5,left=0.25}{v3,vc1}
\fmf{plain,tension=0.5,right=0.25}{v7,vc2}
\fmf{plain,tension=0.3}{vc1,vc2}
\fmf{plain,tension=0.5,left=0.125}{vc3,vc2}
\fmf{plain}{v1,v5}
\fmf{plain}{v4,v8}
\fmf{plain,tension=0.5,right=0,width=1mm}{v5,v8}
\fmf{plain}{v1,v2}
\fmf{plain}{v6,v2}
\fmf{phantom}{v2,vc4}
\fmf{phantom,tension=0.25}{vc4,vc5}
\fmf{phantom}{vc5,v6}
\fmffreeze
\fmf{plain,left=0.2}{vc1,vc4}
\fmf{plain,right=0.2}{vc2,vc5}
\fmfposition
\fmfipath{p[]}
\fmfiset{p1}{vpath(__v1,__v5)}
\fmfiset{p2}{vpath(__v2,__vc1)}
\fmfiset{p3}{vpath(__v3,__vc1)}
\fmfiset{p4}{vpath(__vc1,__vc2)}
\fmfiset{p5}{vpath(__v6,__vc2)}
\fmfiset{p6}{vpath(__v7,__vc2)}
\fmfiset{p7}{vpath(__v4,__v8)}
\fmfiset{p8}{vpath(__v2,__v6)}
\fmfiset{p9}{vpath(__vc1,__vc4)}
\fmfiset{p10}{vpath(__vc2,__vc5)}
\fmfipair{w[]}
\fmfipair{wc[]}
\fmfiequ{w1}{point length(p8)/4 of p8}
\fmfiequ{w2}{(xpart(w3),ypart(wc2))}
\fmfiequ{w3}{point 3length(p8)/4 of p8}
\fmfiequ{w4}{point length(p3) of p3}
\fmfiequ{w5}{point length(p4) of p4}
\fmfiequ{w6}{point length(p6)/5 of p6}
\fmfiequ{w7}{point 4length(p6)/5 of p6}
\fmfiequ{w8}{point length(p10)/3 of p10}
\fmfiequ{w9}{(xpart(w10),ypart(wc1))}
\svertex{w10}{p7}
\fmfiequ{w11}{(xpart(w10),ypart(w6))}
\fmfiequ{w12}{(xpart(wc1),ypart(wc1)-(ypart(wc1)-ypart(wc2))/2)}
\fmfiequ{w13}{wc2}
\fmfiequ{w14}{(xpart(w10),ypart(w12))}
\fmfiequ{w15}{(xpart(wc1),ypart(vloc(__vc1))-(ypart(vloc(__vc1))-ypart(wc1))/2)}
\fmfiequ{wc1}{point length(p4)/3 of p4}
\fmfiequ{wc2}{point 2length(p4)/3 of p4}
\fmfiequ{wc3}{point 5length(p4)/6 of p4}
\fmfi{wiggly}{w2..wc2}
\fmfi{wiggly}{w9..wc1}
\fmfcmd{path ph; ph:=vloc(__v1)--vloc(__v5)--vloc(__v6)--vloc(__v2)--cycle;fill ph withcolor 0.5white;}
\fmfiv{l=\footnotesize{$D^2$},l.a=28,l.d=2}{w7}
\fmfiv{l=\footnotesize{$D^2$},l.a=152,l.d=0.25}{w8}
\fmfiv{l=\footnotesize{$\bar{D}^2$},l.a=28,l.d=2}{w6}
\fmfiv{l=\footnotesize{$\bar{D}^2$},l.a=28,l.d=2}{w11}
\fmfiv{l=\footnotesize{$\bar{D}^2$},l.a=28,l.d=2}{w12}
\fmfiv{l=\footnotesize{$D^2$},l.a=28,l.d=2}{w13}
\fmfiv{l=\footnotesize{$D^2$},l.a=28,l.d=2}{w14}
\fmfiv{l=\footnotesize{$\bar{D}^2$},l.a=28,l.d=2}{wc3}
\fmfiv{l=\footnotesize{$D^2$},l.a=28,l.d=2}{w15}
\end{fmfchar*}}}}
\subfigure[$S_{A2}$]{
\raisebox{\eqoff}{%
\fmfframe(3,1)(1,4){%
\begin{fmfchar*}(20,30)
\fmftop{v1}
\fmfbottom{v5}
\fmfforce{(0w,h)}{v1}
\fmfforce{(0w,0)}{v5}
\fmffixed{(0.25w,0)}{v1,v2}
\fmffixed{(0.5w,0)}{v2,v3}
\fmffixed{(0.25w,0)}{v3,v4}
\fmffixed{(0.25w,0)}{v5,v6}
\fmffixed{(0.5w,0)}{v6,v7}
\fmffixed{(0.25w,0)}{v7,v8}
\fmf{phantom,tension=0.35,right=0.25}{v2,vc1}
\fmf{phantom,tension=0.35,left=0.25}{v6,vc2}
\fmf{plain,tension=0.5,left=0.25}{v3,vc1}
\fmf{plain,tension=0.5,right=0.25}{v7,vc2}
\fmf{plain,tension=0.3}{vc1,vc2}
\fmf{plain,tension=0.5,left=0.125}{vc3,vc2}
\fmf{plain}{v1,v5}
\fmf{plain}{v4,v8}
\fmf{plain,tension=0.5,right=0,width=1mm}{v5,v8}
\fmf{plain}{v1,v2}
\fmf{plain}{v6,v2}
\fmf{phantom}{v2,vc4}
\fmf{phantom,tension=0.25}{vc4,vc5}
\fmf{phantom}{vc5,v6}
\fmffreeze
\fmf{plain,left=0.2}{vc1,vc4}
\fmf{plain,right=0.2}{vc2,vc5}
\fmfposition
\fmfipath{p[]}
\fmfiset{p1}{vpath(__v1,__v5)}
\fmfiset{p2}{vpath(__v2,__vc1)}
\fmfiset{p3}{vpath(__v3,__vc1)}
\fmfiset{p4}{vpath(__vc1,__vc2)}
\fmfiset{p5}{vpath(__v6,__vc2)}
\fmfiset{p6}{vpath(__v7,__vc2)}
\fmfiset{p7}{vpath(__v4,__v8)}
\fmfiset{p8}{vpath(__v2,__v6)}
\fmfiset{p9}{vpath(__vc1,__vc4)}
\fmfiset{p10}{vpath(__vc2,__vc5)}
\fmfipair{w[]}
\fmfipair{wc[]}
\fmfiequ{w1}{point length(p8)/4 of p8}
\fmfiequ{w2}{(xpart(w3),ypart(wc2))}
\fmfiequ{w3}{point 3length(p8)/4 of p8}
\fmfiequ{w4}{point length(p3) of p3}
\fmfiequ{w5}{point length(p4) of p4}
\fmfiequ{w6}{point length(p6)/5 of p6}
\fmfiequ{w7}{point 4length(p6)/5 of p6}
\fmfiequ{w8}{point length(p10)/3 of p10}
\fmfiequ{w9}{(xpart(w10),ypart(wc1))}
\svertex{w10}{p7}
\fmfiequ{w11}{(xpart(w10),ypart(w6))}
\fmfiequ{w12}{(xpart(wc1),ypart(wc1)-2(ypart(wc1)-ypart(wc2))/3)}
\fmfiequ{w13}{wc2}
\fmfiequ{w14}{(xpart(w10),ypart(wc3))}
\fmfiequ{w15}{(xpart(wc1),ypart(vloc(__vc1))-(ypart(vloc(__vc1))-ypart(wc1))/2)}
\fmfiequ{wc1}{point length(p4)/2 of p4}
\fmfiequ{wc2}{point length(p4)/2 of p4}
\fmfiequ{wc3}{point 5length(p4)/6 of p4}
\fmfi{wiggly}{w2..wc2}
\fmfi{wiggly}{w9..wc1}
\fmfcmd{path ph; ph:=vloc(__v1)--vloc(__v5)--vloc(__v6)--vloc(__v2)--cycle;fill ph withcolor 0.5white;}
\fmfiv{l=\footnotesize{$D^2$},l.a=28,l.d=2}{w7}
\fmfiv{l=\footnotesize{$D^2$},l.a=152,l.d=0.25}{w8}
\fmfiv{l=\footnotesize{$\bar{D}^2$},l.a=28,l.d=2}{w6}
\fmfiv{l=\footnotesize{$\bar{D}^2$},l.a=28,l.d=2}{w11}
\fmfiv{l=\footnotesize{$D^2$},l.a=28,l.d=2}{w14}
\fmfiv{l=\footnotesize{$\bar{D}^2$},l.a=28,l.d=2}{wc3}
\fmfiv{l=\footnotesize{$D^2$},l.a=28,l.d=2}{w15}
\end{fmfchar*}}}}
\subfigure[$S_{A3}$]{
\raisebox{\eqoff}{%
\fmfframe(3,1)(1,4){%
\begin{fmfchar*}(20,30)
\fmftop{v1}
\fmfbottom{v5}
\fmfforce{(0w,h)}{v1}
\fmfforce{(0w,0)}{v5}
\fmffixed{(0.25w,0)}{v1,v2}
\fmffixed{(0.5w,0)}{v2,v3}
\fmffixed{(0.25w,0)}{v3,v4}
\fmffixed{(0.25w,0)}{v5,v6}
\fmffixed{(0.5w,0)}{v6,v7}
\fmffixed{(0.25w,0)}{v7,v8}
\fmf{phantom,tension=0.35,right=0.25}{v2,vc1}
\fmf{phantom,tension=0.35,left=0.25}{v6,vc2}
\fmf{plain,tension=0.5,left=0.25}{v3,vc1}
\fmf{plain,tension=0.5,right=0.25}{v7,vc2}
\fmf{plain,tension=0.3}{vc1,vc2}
\fmf{plain,tension=0.5,left=0.125}{vc3,vc2}
\fmf{plain}{v1,v5}
\fmf{plain}{v4,v8}
\fmf{plain,tension=0.5,right=0,width=1mm}{v5,v8}
\fmf{plain}{v1,v2}
\fmf{plain}{v6,v2}
\fmf{phantom}{v2,vc4}
\fmf{phantom,tension=0.25}{vc4,vc5}
\fmf{phantom}{vc5,v6}
\fmffreeze
\fmf{plain,left=0.2}{vc1,vc4}
\fmf{plain,right=0.2}{vc2,vc5}
\fmfposition
\fmfipath{p[]}
\fmfiset{p1}{vpath(__v1,__v5)}
\fmfiset{p2}{vpath(__v2,__vc1)}
\fmfiset{p3}{vpath(__v3,__vc1)}
\fmfiset{p4}{vpath(__vc1,__vc2)}
\fmfiset{p5}{vpath(__v6,__vc2)}
\fmfiset{p6}{vpath(__v7,__vc2)}
\fmfiset{p7}{vpath(__v4,__v8)}
\fmfiset{p8}{vpath(__v2,__v6)}
\fmfiset{p9}{vpath(__vc1,__vc4)}
\fmfiset{p10}{vpath(__vc2,__vc5)}
\fmfipair{w[]}
\fmfipair{wc[]}
\fmfiequ{wc1}{point length(p4)/3 of p4}
\fmfiequ{wc2}{point 2length(p4)/3 of p4}
\fmfiequ{wc3}{point 5length(p4)/6 of p4}
\fmfiequ{w1}{point length(p8)/4 of p8}
\fmfiequ{w2}{(xpart(w3),ypart(wc1))}
\fmfiequ{w3}{point 3length(p8)/4 of p8}
\fmfiequ{w4}{point length(p3) of p3}
\fmfiequ{w5}{point length(p4) of p4}
\fmfiequ{w6}{point length(p6)/5 of p6}
\fmfiequ{w7}{point 4length(p6)/5 of p6}
\fmfiequ{w8}{point length(p10)/3 of p10}
\fmfiequ{w9}{(xpart(w10),ypart(wc2))}
\svertex{w10}{p7}
\fmfiequ{w11}{(xpart(w10),ypart(w6))}
\fmfiequ{w12}{(xpart(wc1),ypart(wc1)-(ypart(wc1)-ypart(wc2))/2)}
\fmfiequ{w13}{wc2}
\fmfiequ{w14}{(xpart(w10),ypart(wc3))}
\fmfiequ{w15}{(xpart(wc1),ypart(vloc(__vc1))-(ypart(vloc(__vc1))-ypart(wc1))/2)}
\fmfi{wiggly}{w2..wc1}
\fmfi{wiggly}{w9..wc2}
\fmfcmd{path ph; ph:=vloc(__v1)--vloc(__v5)--vloc(__v6)--vloc(__v2)--cycle;fill ph withcolor 0.5white;}
\fmfiv{l=\footnotesize{$D^2$},l.a=28,l.d=2}{w7}
\fmfiv{l=\footnotesize{$D^2$},l.a=152,l.d=0.25}{w8}
\fmfiv{l=\footnotesize{$\bar{D}^2$},l.a=28,l.d=2}{w6}
\fmfiv{l=\footnotesize{$\bar{D}^2$},l.a=28,l.d=2}{w11}
\fmfiv{l=\footnotesize{$\bar{D}^2$},l.a=152,l.d=2}{w12}
\fmfiv{l=\footnotesize{$D^2$},l.a=152,l.d=2}{w13}
\fmfiv{l=\footnotesize{$D^2$},l.a=28,l.d=2}{w14}
\fmfiv{l=\footnotesize{$\bar{D}^2$},l.a=28,l.d=2}{wc3}
\fmfiv{l=\footnotesize{$D^2$},l.a=28,l.d=2}{w15}
\end{fmfchar*}}}}
\\\subfigure[$S_{A4}$]{
\raisebox{\eqoff}{%
\fmfframe(3,1)(1,4){%
\begin{fmfchar*}(20,30)
\fmftop{v1}
\fmfbottom{v5}
\fmfforce{(0w,h)}{v1}
\fmfforce{(0w,0)}{v5}
\fmffixed{(0.25w,0)}{v1,v2}
\fmffixed{(0.5w,0)}{v2,v3}
\fmffixed{(0.25w,0)}{v3,v4}
\fmffixed{(0.25w,0)}{v5,v6}
\fmffixed{(0.5w,0)}{v6,v7}
\fmffixed{(0.25w,0)}{v7,v8}
\fmf{phantom,tension=0.35,right=0.25}{v2,vc1}
\fmf{phantom,tension=0.35,left=0.25}{v6,vc2}
\fmf{plain,tension=0.5,left=0.25}{v3,vc1}
\fmf{plain,tension=0.5,right=0.25}{v7,vc2}
\fmf{plain,tension=0.3}{vc1,vc2}
\fmf{plain,tension=0.5,left=0.125}{vc3,vc2}
\fmf{plain}{v1,v5}
\fmf{plain}{v4,v8}
\fmf{plain,tension=0.5,right=0,width=1mm}{v5,v8}
\fmf{plain}{v1,v2}
\fmf{plain}{v6,v2}
\fmf{phantom}{v2,vc4}
\fmf{phantom,tension=0.25}{vc4,vc5}
\fmf{phantom}{vc5,v6}
\fmffreeze
\fmf{plain,left=0.2}{vc1,vc4}
\fmf{plain,right=0.2}{vc2,vc5}
\fmfposition
\fmfipath{p[]}
\fmfiset{p1}{vpath(__v1,__v5)}
\fmfiset{p2}{vpath(__v2,__vc1)}
\fmfiset{p3}{vpath(__v3,__vc1)}
\fmfiset{p4}{vpath(__vc1,__vc2)}
\fmfiset{p5}{vpath(__v6,__vc2)}
\fmfiset{p6}{vpath(__v7,__vc2)}
\fmfiset{p7}{vpath(__v4,__v8)}
\fmfiset{p8}{vpath(__v2,__v6)}
\fmfiset{p9}{vpath(__vc1,__vc4)}
\fmfiset{p10}{vpath(__vc2,__vc5)}
\fmfipair{w[]}
\fmfipair{wc[]}
\fmfiequ{wc1}{point length(p4)/3 of p4}
\fmfiequ{wc2}{point 2length(p4)/3 of p4}
\fmfiequ{wc3}{point 5length(p4)/6 of p4}
\fmfiequ{w1}{point length(p8)/4 of p8}
\fmfiequ{w2}{(xpart(w3),ypart(wc1))}
\fmfiequ{w3}{point 3length(p8)/4 of p8}
\fmfiequ{w4}{point length(p3) of p3}
\fmfiequ{w5}{point length(p4) of p4}
\fmfiequ{w6}{point 2length(p6)/3 of p6}
\fmfiequ{w7}{point 10length(p6)/11 of p6}
\fmfiequ{w8}{point length(p10)/3 of p10}
\fmfiequ{w9}{(xpart(w10),ypart(wc2))}
\svertex{w10}{p7}
\fmfiequ{w11}{(xpart(w10),ypart(w16))}
\fmfiequ{w12}{(xpart(wc1),ypart(wc1)-(ypart(wc1)-ypart(wc2))/2)}
\fmfiequ{w13}{wc2}
\fmfiequ{w14}{(xpart(w10),ypart(wc3))}
\fmfiequ{w15}{(xpart(wc1),ypart(vloc(__vc1))-(ypart(vloc(__vc1))-ypart(wc1))/2)}
\fmfiequ{w16}{point length(p6)/8 of p6}
\fmfiequ{w17}{point 4length(p6)/8 of p6}
\fmfi{wiggly}{w2..wc1}
\fmfi{wiggly}{w9..w6}
\fmfcmd{path ph; ph:=vloc(__v1)--vloc(__v5)--vloc(__v6)--vloc(__v2)--cycle;fill ph withcolor 0.5white;}
\fmfiv{l=\footnotesize{$\bar{D}^2$},l.a=28,l.d=2}{w7}
\fmfiv{l=\footnotesize{$D^2$},l.a=152,l.d=0.25}{w8}
\fmfiv{l=\footnotesize{$\bar{D}^2$},l.a=28,l.d=2}{w11}
\fmfiv{l=\footnotesize{$\bar{D}^2$},l.a=152,l.d=2}{w12}
\fmfiv{l=\footnotesize{$D^2$},l.a=152,l.d=2}{w13}
\fmfiv{l=\footnotesize{$D^2$},l.a=28,l.d=2}{w14}
\fmfiv{l=\footnotesize{$D^2$},l.a=28,l.d=2}{w15}
\fmfiv{l=\footnotesize{$D^2$},l.a=28,l.d=2}{w17}
\fmfiv{l=\footnotesize{$\bar{D}^2$},l.a=28,l.d=2}{w16}
\end{fmfchar*}}}}
\subfigure[$S_{A5}$]{
\raisebox{\eqoff}{%
\fmfframe(3,1)(1,4){%
\begin{fmfchar*}(20,30)
\fmftop{v1}
\fmfbottom{v5}
\fmfforce{(0w,h)}{v1}
\fmfforce{(0w,0)}{v5}
\fmffixed{(0.25w,0)}{v1,v2}
\fmffixed{(0.5w,0)}{v2,v3}
\fmffixed{(0.25w,0)}{v3,v4}
\fmffixed{(0.25w,0)}{v5,v6}
\fmffixed{(0.5w,0)}{v6,v7}
\fmffixed{(0.25w,0)}{v7,v8}
\fmf{phantom,tension=0.35,right=0.25}{v2,vc1}
\fmf{phantom,tension=0.35,left=0.25}{v6,vc2}
\fmf{plain,tension=0.5,left=0.25}{v3,vc1}
\fmf{plain,tension=0.5,right=0.25}{v7,vc2}
\fmf{plain,tension=0.3}{vc1,vc2}
\fmf{plain,tension=0.5,left=0.125}{vc3,vc2}
\fmf{plain}{v1,v5}
\fmf{plain}{v4,v8}
\fmf{plain,tension=0.5,right=0,width=1mm}{v5,v8}
\fmf{plain}{v1,v2}
\fmf{plain}{v6,v2}
\fmf{phantom}{v2,vc4}
\fmf{phantom,tension=0.25}{vc4,vc5}
\fmf{phantom}{vc5,v6}
\fmffreeze
\fmf{plain,left=0.2}{vc1,vc4}
\fmf{plain,right=0.2}{vc2,vc5}
\fmfposition
\fmfipath{p[]}
\fmfiset{p1}{vpath(__v1,__v5)}
\fmfiset{p2}{vpath(__v2,__vc1)}
\fmfiset{p3}{vpath(__v3,__vc1)}
\fmfiset{p4}{vpath(__vc1,__vc2)}
\fmfiset{p5}{vpath(__v6,__vc2)}
\fmfiset{p6}{vpath(__v7,__vc2)}
\fmfiset{p7}{vpath(__v4,__v8)}
\fmfiset{p8}{vpath(__v2,__v6)}
\fmfiset{p9}{vpath(__vc1,__vc4)}
\fmfiset{p10}{vpath(__vc2,__vc5)}
\fmfipair{w[]}
\fmfipair{wc[]}
\fmfiequ{wc1}{point length(p4)/2 of p4}
\fmfiequ{wc2}{point 2length(p4)/3 of p4}
\fmfiequ{wc3}{point 5length(p4)/6 of p4}
\fmfiequ{w1}{point length(p8)/4 of p8}
\fmfiequ{w2}{(xpart(w3),ypart(wc1))}
\fmfiequ{w3}{point 3length(p8)/4 of p8}
\fmfiequ{w4}{point length(p3) of p3}
\fmfiequ{w5}{point length(p4) of p4}
\fmfiequ{w6}{point length(p6)/5 of p6}
\fmfiequ{w7}{point 4length(p6)/5 of p6}
\fmfiequ{w8}{point length(p10)/3 of p10}
\fmfiequ{w9}{(xpart(w10),ypart(w16))}
\svertex{w10}{p7}
\fmfiequ{w11}{(xpart(w10),ypart(w6))}
\fmfiequ{w12}{(xpart(wc1),ypart(wc1)-(ypart(wc1)-ypart(vloc(__vc2)))/2)}
\fmfiequ{w13}{wc2}
\fmfiequ{w14}{(xpart(w10),ypart(w15))}
\fmfiequ{w15}{(xpart(wc1),ypart(vloc(__vc1))-(ypart(vloc(__vc1))-ypart(wc1))/2)}
\fmfiequ{w16}{point length(p3)/2 of p3}
\fmfiequ{w17}{point 2length(p3)/3 of p3}
\fmfi{wiggly}{w2..wc1}
\fmfi{wiggly}{w9..w16}
\fmfcmd{path ph; ph:=vloc(__v1)--vloc(__v5)--vloc(__v6)--vloc(__v2)--cycle;fill ph withcolor 0.5white;}
\fmfiv{l=\footnotesize{$D^2$},l.a=28,l.d=2}{w7}
\fmfiv{l=\footnotesize{$D^2$},l.a=152,l.d=0.25}{w8}
\fmfiv{l=\footnotesize{$\bar{D}^2$},l.a=28,l.d=2}{w6}
\fmfiv{l=\footnotesize{$\bar{D}^2$},l.a=28,l.d=2}{w11}
\fmfiv{l=\footnotesize{$\bar{D}^2$},l.a=28,l.d=2}{w12}
\fmfiv{l=\footnotesize{$D^2$},l.a=28,l.d=2}{wc1}
\fmfiv{l=\footnotesize{$D^2$},l.a=28,l.d=2}{w14}
\fmfiv{l=\footnotesize{$\bar{D}^2$},l.a=28,l.d=2}{w15}
\fmfiv{l=\footnotesize{$D^2$},l.a=-28,l.d=2}{w17}
\end{fmfchar*}}}}
\caption{Four-loop non-maximal diagrams of class A}
\label{diagrams-A}
\end{figure}

\begin{figure}[p]
\centering
\setcounter{subfigure}{0}
\unitlength=1.75mm
\settoheight{\eqoff}{$\times$}%
\setlength{\eqoff}{0.5\eqoff}%
\addtolength{\eqoff}{-12.5\unitlength}%
\settoheight{\eqofftwo}{$\times$}%
\setlength{\eqofftwo}{0.5\eqofftwo}%
\addtolength{\eqofftwo}{-7.5\unitlength}%
\begin{equation*}
\begin{aligned}
\subfigure[$S_{B1}$]{
\raisebox{\eqoff}{%
\fmfframe(3,1)(1,4){%
\begin{fmfchar*}(20,30)
\fmftop{v1}
\fmfbottom{v5}
\fmfforce{(0w,h)}{v1}
\fmfforce{(0w,0)}{v5}
\fmffixed{(0.25w,0)}{v1,v2}
\fmffixed{(0.5w,0)}{v2,v3}
\fmffixed{(0.25w,0)}{v3,v4}
\fmffixed{(0.25w,0)}{v5,v6}
\fmffixed{(0.5w,0)}{v6,v7}
\fmffixed{(0.25w,0)}{v7,v8}
\fmf{phantom,tension=0.35,right=0.25}{v2,vc1}
\fmf{phantom,tension=0.35,left=0.25}{v6,vc2}
\fmf{plain,tension=0.5,left=0.25}{v3,vc1}
\fmf{plain,tension=0.5,right=0.25}{v7,vc2}
\fmf{plain,tension=0.3}{vc1,vc2}
\fmf{plain,tension=0.5,left=0.125}{vc3,vc2}
\fmf{plain}{v1,v5}
\fmf{plain}{v4,v8}
\fmf{plain,tension=0.5,right=0,width=1mm}{v5,v8}
\fmf{plain}{v1,v2}
\fmf{plain}{v6,v2}
\fmf{phantom}{v2,vc4}
\fmf{phantom,tension=0.25}{vc4,vc5}
\fmf{phantom}{vc5,v6}
\fmffreeze
\fmf{plain,left=0.2}{vc1,vc4}
\fmf{plain,right=0.2}{vc2,vc5}
\fmfposition
\fmfipath{p[]}
\fmfiset{p1}{vpath(__v1,__v5)}
\fmfiset{p2}{vpath(__v2,__vc1)}
\fmfiset{p3}{vpath(__v3,__vc1)}
\fmfiset{p4}{vpath(__vc1,__vc2)}
\fmfiset{p5}{vpath(__v6,__vc2)}
\fmfiset{p6}{vpath(__v7,__vc2)}
\fmfiset{p7}{vpath(__v4,__v8)}
\fmfiset{p8}{vpath(__v2,__v6)}
\fmfiset{p9}{vpath(__vc1,__vc4)}
\fmfiset{p10}{vpath(__vc2,__vc5)}
\fmfipair{w[]}
\fmfipair{wc[]}
\fmfiequ{w1}{point length(p8)/4 of p8}
\fmfiequ{w2}{(xpart(w3),ypart(wc2))}
\fmfiequ{w3}{point 3length(p8)/4 of p8}
\fmfiequ{w4}{point length(p3) of p3}
\fmfiequ{w5}{point length(p4) of p4}
\fmfiequ{w6}{point length(p6)/5 of p6}
\fmfiequ{w7}{point 4length(p6)/5 of p6}
\fmfiequ{w8}{point length(p10)/3 of p10}
\fmfiequ{w9}{(xpart(w10),ypart(wc1))}
\svertex{w10}{p7}
\fmfiequ{w11}{(xpart(w10),ypart(w6))}
\fmfiequ{w12}{(xpart(wc1),ypart(wc1)-2(ypart(wc1)-ypart(wc2))/3)}
\fmfiequ{w13}{wc2}
\fmfiequ{w14}{(xpart(w10),ypart(wc3))}
\fmfiequ{w15}{(xpart(wc1),ypart(vloc(__vc1))-(ypart(vloc(__vc1))-ypart(wc1))/2)}
\fmfiequ{wc1}{point length(p4)/2 of p4}
\fmfiequ{wc2}{point length(p4)/2 of p4}
\fmfiequ{wc3}{point 2length(p4)/3 of p4}
\fmfi{wiggly}{w9..wc1}
\fmfcmd{path ph; ph:=vloc(__v1)--vloc(__v5)--vloc(__v6)--vloc(__v2)--cycle;fill ph withcolor 0.5white;}
\fmfiv{l=\footnotesize{$D^2$},l.a=28,l.d=2}{w7}
\fmfiv{l=\footnotesize{$D^2$},l.a=152,l.d=0.25}{w8}
\fmfiv{l=\footnotesize{$\bar{D}^2$},l.a=28,l.d=2}{w6}
\fmfiv{l=\footnotesize{$\bar{D}^2$},l.a=28,l.d=2}{w11}
\fmfiv{l=\footnotesize{$D^2$},l.a=28,l.d=2}{w14}
\fmfiv{l=\footnotesize{$\bar{D}^2$},l.a=28,l.d=2}{wc3}
\fmfiv{l=\footnotesize{$D^2$},l.a=28,l.d=2}{w15}
\end{fmfchar*}}}}
\qquad&
\subfigure[$S_{B2}$]{
\raisebox{\eqoff}{%
\fmfframe(3,1)(1,4){%
\begin{fmfchar*}(20,30)
\fmftop{v1}
\fmfbottom{v5}
\fmfforce{(0w,h)}{v1}
\fmfforce{(0w,0)}{v5}
\fmffixed{(0.25w,0)}{v1,v2}
\fmffixed{(0.5w,0)}{v2,v3}
\fmffixed{(0.25w,0)}{v3,v4}
\fmffixed{(0.25w,0)}{v5,v6}
\fmffixed{(0.5w,0)}{v6,v7}
\fmffixed{(0.25w,0)}{v7,v8}
\fmf{phantom,tension=0.35,right=0.25}{v2,vc1}
\fmf{phantom,tension=0.35,left=0.25}{v6,vc2}
\fmf{plain,tension=0.5,left=0.25}{v3,vc1}
\fmf{plain,tension=0.5,right=0.25}{v7,vc2}
\fmf{plain,tension=0.3}{vc1,vc2}
\fmf{plain,tension=0.5,left=0.125}{vc3,vc2}
\fmf{plain}{v1,v5}
\fmf{plain}{v4,v8}
\fmf{plain,tension=0.5,right=0,width=1mm}{v5,v8}
\fmf{plain}{v1,v2}
\fmf{plain}{v6,v2}
\fmf{phantom}{v2,vc4}
\fmf{phantom,tension=0.25}{vc4,vc5}
\fmf{phantom}{vc5,v6}
\fmffreeze
\fmf{plain,left=0.2}{vc1,vc4}
\fmf{plain,right=0.2}{vc2,vc5}
\fmfposition
\fmfipath{p[]}
\fmfiset{p1}{vpath(__v1,__v5)}
\fmfiset{p2}{vpath(__v2,__vc1)}
\fmfiset{p3}{vpath(__v3,__vc1)}
\fmfiset{p4}{vpath(__vc1,__vc2)}
\fmfiset{p5}{vpath(__v6,__vc2)}
\fmfiset{p6}{vpath(__v7,__vc2)}
\fmfiset{p7}{vpath(__v4,__v8)}
\fmfiset{p8}{vpath(__v2,__v6)}
\fmfiset{p9}{vpath(__vc1,__vc4)}
\fmfiset{p10}{vpath(__vc2,__vc5)}
\fmfipair{w[]}
\fmfipair{wc[]}
\fmfiequ{wc1}{point length(p4)/3 of p4}
\fmfiequ{wc2}{point 2length(p4)/3 of p4}
\fmfiequ{wc3}{point 5length(p4)/6 of p4}
\fmfiequ{w1}{point length(p8)/4 of p8}
\fmfiequ{w2}{(xpart(w3),ypart(wc1))}
\fmfiequ{w3}{point 3length(p8)/4 of p8}
\fmfiequ{w4}{point length(p3) of p3}
\fmfiequ{w5}{point length(p4) of p4}
\fmfiequ{w6}{point 2length(p6)/3 of p6}
\fmfiequ{w7}{point 10length(p6)/11 of p6}
\fmfiequ{w8}{point length(p10)/3 of p10}
\fmfiequ{w9}{(xpart(w10),ypart(wc2))}
\svertex{w10}{p7}
\fmfiequ{w11}{(xpart(w10),ypart(w16))}
\fmfiequ{w12}{(xpart(wc1),ypart(wc1)-(ypart(wc1)-ypart(wc2))/2)}
\fmfiequ{w13}{wc2}
\fmfiequ{w14}{(xpart(w10),ypart(wc3))}
\fmfiequ{w15}{(xpart(wc1),ypart(vloc(__vc1))-(ypart(vloc(__vc1))-ypart(wc1))/2)}
\fmfiequ{w16}{point length(p6)/8 of p6}
\fmfiequ{w17}{point 4length(p6)/8 of p6}
\fmfi{wiggly}{w9..w6}
\fmfcmd{path ph; ph:=vloc(__v1)--vloc(__v5)--vloc(__v6)--vloc(__v2)--cycle;fill ph withcolor 0.5white;}
\fmfiv{l=\footnotesize{$\bar{D}^2$},l.a=28,l.d=2}{w7}
\fmfiv{l=\footnotesize{$D^2$},l.a=152,l.d=0.25}{w8}
\fmfiv{l=\footnotesize{$\bar{D}^2$},l.a=28,l.d=2}{w11}
\fmfiv{l=\footnotesize{$D^2$},l.a=152,l.d=2}{w13}
\fmfiv{l=\footnotesize{$D^2$},l.a=28,l.d=2}{w14}
\fmfiv{l=\footnotesize{$D^2$},l.a=28,l.d=2}{w17}
\fmfiv{l=\footnotesize{$\bar{D}^2$},l.a=28,l.d=2}{w16}
\end{fmfchar*}}}}
\qquad&
\subfigure[$S_{B3}$]{
\raisebox{\eqoff}{%
\fmfframe(3,1)(1,4){%
\begin{fmfchar*}(20,30)
\fmftop{v1}
\fmfbottom{v5}
\fmfforce{(0w,h)}{v1}
\fmfforce{(0w,0)}{v5}
\fmffixed{(0.25w,0)}{v1,v2}
\fmffixed{(0.5w,0)}{v2,v3}
\fmffixed{(0.25w,0)}{v3,v4}
\fmffixed{(0.25w,0)}{v5,v6}
\fmffixed{(0.5w,0)}{v6,v7}
\fmffixed{(0.25w,0)}{v7,v8}
\fmf{phantom,tension=0.35,right=0.25}{v2,vc1}
\fmf{phantom,tension=0.35,left=0.25}{v6,vc2}
\fmf{plain,tension=0.5,left=0.25}{v3,vc1}
\fmf{plain,tension=0.5,right=0.25}{v7,vc2}
\fmf{plain,tension=0.3}{vc1,vc2}
\fmf{plain,tension=0.5,left=0.125}{vc3,vc2}
\fmf{plain}{v1,v5}
\fmf{plain}{v4,v8}
\fmf{plain,tension=0.5,right=0,width=1mm}{v5,v8}
\fmf{plain}{v1,v2}
\fmf{plain}{v6,v2}
\fmf{phantom}{v2,vc4}
\fmf{phantom,tension=0.25}{vc4,vc5}
\fmf{phantom}{vc5,v6}
\fmffreeze
\fmf{plain,left=0.2}{vc1,vc4}
\fmf{plain,right=0.2}{vc2,vc5}
\fmfposition
\fmfipath{p[]}
\fmfiset{p1}{vpath(__v1,__v5)}
\fmfiset{p2}{vpath(__v2,__vc1)}
\fmfiset{p3}{vpath(__v3,__vc1)}
\fmfiset{p4}{vpath(__vc1,__vc2)}
\fmfiset{p5}{vpath(__v6,__vc2)}
\fmfiset{p6}{vpath(__v7,__vc2)}
\fmfiset{p7}{vpath(__v4,__v8)}
\fmfiset{p8}{vpath(__v2,__v6)}
\fmfiset{p9}{vpath(__vc1,__vc4)}
\fmfiset{p10}{vpath(__vc2,__vc5)}
\fmfipair{w[]}
\fmfipair{wc[]}
\fmfiequ{wc1}{point length(p4)/2 of p4}
\fmfiequ{wc2}{point 2length(p4)/3 of p4}
\fmfiequ{wc3}{point 5length(p4)/6 of p4}
\fmfiequ{w1}{point length(p8)/4 of p8}
\fmfiequ{w2}{(xpart(w3),ypart(wc1))}
\fmfiequ{w3}{point 3length(p8)/4 of p8}
\fmfiequ{w4}{point length(p3) of p3}
\fmfiequ{w5}{point length(p4) of p4}
\fmfiequ{w6}{point length(p6)/5 of p6}
\fmfiequ{w7}{point 4length(p6)/5 of p6}
\fmfiequ{w8}{point length(p10)/3 of p10}
\fmfiequ{w9}{(xpart(w10),ypart(w16))}
\svertex{w10}{p7}
\fmfiequ{w11}{(xpart(w10),ypart(w6))}
\fmfiequ{w12}{(xpart(wc1),ypart(wc1)-(ypart(wc1)-ypart(vloc(__vc2)))/2)}
\fmfiequ{w13}{wc2}
\fmfiequ{w14}{(xpart(w10),ypart(w15))}
\fmfiequ{w15}{(xpart(wc1),ypart(vloc(__vc1))-(ypart(vloc(__vc1))-ypart(wc1))/2)}
\fmfiequ{w16}{point length(p3)/2 of p3}
\fmfiequ{w17}{point 2length(p3)/3 of p3}
\fmfi{wiggly}{w9..w16}
\fmfcmd{path ph; ph:=vloc(__v1)--vloc(__v5)--vloc(__v6)--vloc(__v2)--cycle;fill ph withcolor 0.5white;}
\fmfiv{l=\footnotesize{$D^2$},l.a=28,l.d=2}{w7}
\fmfiv{l=\footnotesize{$D^2$},l.a=152,l.d=0.25}{w8}
\fmfiv{l=\footnotesize{$\bar{D}^2$},l.a=28,l.d=2}{w6}
\fmfiv{l=\footnotesize{$\bar{D}^2$},l.a=28,l.d=2}{w11}
\fmfiv{l=\footnotesize{$D^2$},l.a=28,l.d=2}{w14}
\fmfiv{l=\footnotesize{$\bar{D}^2$},l.a=28,l.d=2}{w15}
\fmfiv{l=\footnotesize{$D^2$},l.a=-28,l.d=2}{w17}
\end{fmfchar*}}}}
\end{aligned}
\end{equation*}
\caption{Four-loop non-maximal diagrams of class B}
\label{diagrams-B}
\end{figure}

\begin{figure}[p]
\centering
\setcounter{subfigure}{0}
\unitlength=1.75mm
\settoheight{\eqoff}{$\times$}%
\setlength{\eqoff}{0.5\eqoff}%
\addtolength{\eqoff}{-12.5\unitlength}%
\settoheight{\eqofftwo}{$\times$}%
\setlength{\eqofftwo}{0.5\eqofftwo}%
\addtolength{\eqofftwo}{-7.5\unitlength}%
\begin{equation*}
\begin{aligned}
\subfigure[$S_{C1}$]{
\raisebox{\eqoff}{%
\fmfframe(3,1)(1,4){%
\begin{fmfchar*}(20,30)
\fmftop{v1}
\fmfbottom{v5}
\fmfforce{(0w,h)}{v1}
\fmfforce{(0w,0)}{v5}
\fmffixed{(0.25w,0)}{v1,v2}
\fmffixed{(0.35w,0)}{v2,v2b}
\fmffixed{(0.2w,0)}{v2b,v3}
\fmffixed{(0.2w,0)}{v3,v4}
\fmffixed{(0.25w,0)}{v5,v6}
\fmffixed{(0.35w,0)}{v6,v6b}
\fmffixed{(0.2w,0)}{v6b,v7}
\fmffixed{(0.2w,0)}{v7,v8}
\fmf{phantom,tension=0.35,right=0.25}{v2,vc1}
\fmf{phantom,tension=0.35,left=0.25}{v6,vc2}
\fmf{phantom,tension=0.5,left=0.25}{v3,vc1}
\fmf{phantom,tension=0.5,right=0.25}{v7,vc2}
\fmf{phantom,tension=0.3}{vc1,vc2}
\fmf{phantom,tension=0.5,left=0.125}{vc3,vc2}
\fmf{plain}{v1,v5}
\fmf{plain}{v4,v8}
\fmf{plain,tension=0.5,right=0,width=1mm}{v5,v8}
\fmf{plain}{v1,v2}
\fmf{plain}{v6,v2}
\fmf{plain}{v2b,v6b}
\fmf{phantom}{v2,vc4}
\fmf{phantom,tension=0.25}{vc4,vc5}
\fmf{phantom}{vc5,v6}
\fmffreeze
\fmf{phantom,left=0.2}{vc1,vc4}
\fmf{phantom,right=0.2}{vc2,vc5}
\fmfposition
\fmfipath{p[]}
\fmfiset{p1}{vpath(__v1,__v5)}
\fmfiset{p2}{vpath(__v2,__vc1)}
\fmfiset{p3}{vpath(__v3,__vc1)}
\fmfiset{p4}{vpath(__vc1,__vc2)}
\fmfiset{p5}{vpath(__v6,__vc2)}
\fmfiset{p6}{vpath(__v7,__vc2)}
\fmfiset{p7}{vpath(__v4,__v8)}
\fmfiset{p8}{vpath(__v2,__v6)}
\fmfiset{p9}{vpath(__vc1,__vc4)}
\fmfiset{p10}{vpath(__vc2,__vc5)}
\fmfiset{p11}{vpath(__v2b,__v6b)}
\fmfiset{p12}{vpath(__v4,__v8)}
\fmfipair{w[]}
\fmfipair{wc[]}
\fmfiequ{wc1}{point length(p11)/3 of p11}
\fmfiequ{wc2}{point 2length(p11)/3 of p11}
\fmfiequ{w1}{point length(p8)/3 of p8}
\fmfiequ{w2}{point 2length(p8)/3 of p8}
\fmfiequ{w3}{point length(p12)/3 of p12}
\fmfiequ{w4}{point 2length(p12)/3 of p12}
\fmfiequ{w5}{point 13length(p12)/14 of p12}
\fmfiequ{w6}{point 11length(p12)/14 of p12}
\vvertex{w7}{w5}{p11}
\vvertex{w8}{w6}{p11}
\fmfiequ{w9}{point 9length(p12)/15 of p11}
\fmfiequ{w10}{point 7length(p12)/15 of p11}
\fmfi{wiggly}{w4..wc2}
\fmfi{wiggly}{w2..wc1}
\fmfcmd{path ph; ph:=vloc(__v1)--vloc(__v5)--vloc(__v6)--vloc(__v2)--cycle;fill ph withcolor 0.5white;}
\fmfiv{l=\footnotesize{$\bar{D}^2$},l.a=28,l.d=2}{w5}
\fmfiv{l=\footnotesize{$D^2$},l.a=28,l.d=2}{w6}
\fmfiv{l=\footnotesize{$\bar{D}^2$},l.a=28,l.d=2}{w7}
\fmfiv{l=\footnotesize{$D^2$},l.a=28,l.d=2}{w8}
\fmfiv{l=\footnotesize{$\bar{D}^2$},l.a=28,l.d=2}{w9}
\fmfiv{l=\footnotesize{$D^2$},l.a=28,l.d=2}{w10}
\end{fmfchar*}}}}
\qquad&
\subfigure[$S_{C2}$]{
\raisebox{\eqoff}{%
\fmfframe(3,1)(1,4){%
\begin{fmfchar*}(20,30)
\fmftop{v1}
\fmfbottom{v5}
\fmfforce{(0w,h)}{v1}
\fmfforce{(0w,0)}{v5}
\fmffixed{(0.25w,0)}{v1,v2}
\fmffixed{(0.35w,0)}{v2,v2b}
\fmffixed{(0.2w,0)}{v2b,v3}
\fmffixed{(0.2w,0)}{v3,v4}
\fmffixed{(0.25w,0)}{v5,v6}
\fmffixed{(0.35w,0)}{v6,v6b}
\fmffixed{(0.2w,0)}{v6b,v7}
\fmffixed{(0.2w,0)}{v7,v8}
\fmf{phantom,tension=0.35,right=0.25}{v2,vc1}
\fmf{phantom,tension=0.35,left=0.25}{v6,vc2}
\fmf{phantom,tension=0.5,left=0.25}{v3,vc1}
\fmf{phantom,tension=0.5,right=0.25}{v7,vc2}
\fmf{phantom,tension=0.3}{vc1,vc2}
\fmf{phantom,tension=0.5,left=0.125}{vc3,vc2}
\fmf{plain}{v1,v5}
\fmf{plain}{v4,v8}
\fmf{plain,tension=0.5,right=0,width=1mm}{v5,v8}
\fmf{plain}{v1,v2}
\fmf{plain}{v6,v2}
\fmf{plain}{v2b,v6b}
\fmf{phantom}{v2,vc4}
\fmf{phantom,tension=0.25}{vc4,vc5}
\fmf{phantom}{vc5,v6}
\fmffreeze
\fmf{phantom,left=0.2}{vc1,vc4}
\fmf{phantom,right=0.2}{vc2,vc5}
\fmfposition
\fmfipath{p[]}
\fmfiset{p1}{vpath(__v1,__v5)}
\fmfiset{p2}{vpath(__v2,__vc1)}
\fmfiset{p3}{vpath(__v3,__vc1)}
\fmfiset{p4}{vpath(__vc1,__vc2)}
\fmfiset{p5}{vpath(__v6,__vc2)}
\fmfiset{p6}{vpath(__v7,__vc2)}
\fmfiset{p7}{vpath(__v4,__v8)}
\fmfiset{p8}{vpath(__v2,__v6)}
\fmfiset{p9}{vpath(__vc1,__vc4)}
\fmfiset{p10}{vpath(__vc2,__vc5)}
\fmfiset{p11}{vpath(__v2b,__v6b)}
\fmfiset{p12}{vpath(__v4,__v8)}
\fmfipair{w[]}
\fmfipair{wc[]}
\fmfiequ{wc1}{point length(p11)/3 of p11}
\fmfiequ{wc2}{point 2length(p11)/3 of p11}
\fmfiequ{w1}{point length(p8)/3 of p8}
\fmfiequ{w2}{point 2length(p8)/3 of p8}
\fmfiequ{w3}{point length(p12)/3 of p12}
\fmfiequ{w4}{point 2length(p12)/3 of p12}
\fmfiequ{w5}{point 13length(p12)/14 of p12}
\fmfiequ{w6}{point 11length(p12)/14 of p12}
\vvertex{w7}{w5}{p11}
\vvertex{w8}{w6}{p11}
\fmfiequ{w9}{point 9length(p12)/15 of p11}
\fmfiequ{w10}{point 7length(p12)/15 of p11}
\fmfi{wiggly}{w4..wc2}
\fmfi{wiggly}{w1..wc2}
\fmfcmd{path ph; ph:=vloc(__v1)--vloc(__v5)--vloc(__v6)--vloc(__v2)--cycle;fill ph withcolor 0.5white;}
\fmfiv{l=\footnotesize{$\bar{D}^2$},l.a=28,l.d=2}{w5}
\fmfiv{l=\footnotesize{$D^2$},l.a=28,l.d=2}{w6}
\fmfiv{l=\footnotesize{$\bar{D}^2$},l.a=28,l.d=2}{w7}
\fmfiv{l=\footnotesize{$D^2$},l.a=28,l.d=2}{w8}
\end{fmfchar*}}}}
\subfigure[$S_{C3}$]{
\raisebox{\eqoff}{%
\fmfframe(3,1)(1,4){%
\begin{fmfchar*}(20,30)
\fmftop{v1}
\fmfbottom{v5}
\fmfforce{(0w,h)}{v1}
\fmfforce{(0w,0)}{v5}
\fmffixed{(0.25w,0)}{v1,v2}
\fmffixed{(0.35w,0)}{v2,v2b}
\fmffixed{(0.2w,0)}{v2b,v3}
\fmffixed{(0.2w,0)}{v3,v4}
\fmffixed{(0.25w,0)}{v5,v6}
\fmffixed{(0.35w,0)}{v6,v6b}
\fmffixed{(0.2w,0)}{v6b,v7}
\fmffixed{(0.2w,0)}{v7,v8}
\fmf{phantom,tension=0.35,right=0.25}{v2,vc1}
\fmf{phantom,tension=0.35,left=0.25}{v6,vc2}
\fmf{phantom,tension=0.5,left=0.25}{v3,vc1}
\fmf{phantom,tension=0.5,right=0.25}{v7,vc2}
\fmf{phantom,tension=0.3}{vc1,vc2}
\fmf{phantom,tension=0.5,left=0.125}{vc3,vc2}
\fmf{plain}{v1,v5}
\fmf{plain}{v4,v8}
\fmf{plain,tension=0.5,right=0,width=1mm}{v5,v8}
\fmf{plain}{v1,v2}
\fmf{plain}{v6,v2}
\fmf{plain}{v2b,v6b}
\fmf{phantom}{v2,vc4}
\fmf{phantom,tension=0.25}{vc4,vc5}
\fmf{phantom}{vc5,v6}
\fmffreeze
\fmf{phantom,left=0.2}{vc1,vc4}
\fmf{phantom,right=0.2}{vc2,vc5}
\fmfposition
\fmfipath{p[]}
\fmfiset{p1}{vpath(__v1,__v5)}
\fmfiset{p2}{vpath(__v2,__vc1)}
\fmfiset{p3}{vpath(__v3,__vc1)}
\fmfiset{p4}{vpath(__vc1,__vc2)}
\fmfiset{p5}{vpath(__v6,__vc2)}
\fmfiset{p6}{vpath(__v7,__vc2)}
\fmfiset{p7}{vpath(__v4,__v8)}
\fmfiset{p8}{vpath(__v2,__v6)}
\fmfiset{p9}{vpath(__vc1,__vc4)}
\fmfiset{p10}{vpath(__vc2,__vc5)}
\fmfiset{p11}{vpath(__v2b,__v6b)}
\fmfiset{p12}{vpath(__v4,__v8)}
\fmfipair{w[]}
\fmfipair{wc[]}
\fmfiequ{wc1}{point length(p11)/3 of p11}
\fmfiequ{wc2}{point 2length(p11)/3 of p11}
\fmfiequ{w1}{point length(p8)/3 of p8}
\fmfiequ{w2}{point 2length(p8)/3 of p8}
\fmfiequ{w3}{point length(p12)/3 of p12}
\fmfiequ{w4}{point 2length(p12)/3 of p12}
\fmfiequ{w5}{point 13length(p12)/14 of p12}
\fmfiequ{w6}{point 11length(p12)/14 of p12}
\vvertex{w7}{w5}{p11}
\vvertex{w8}{w6}{p11}
\fmfiequ{w9}{point 9length(p12)/15 of p11}
\fmfiequ{w10}{point 7length(p12)/15 of p11}
\fmfiequ{w11}{(xpart(w5),ypart(w10))}
\fmfi{wiggly}{w3..wc1}
\fmfi{wiggly}{w1..wc2}
\fmfcmd{path ph; ph:=vloc(__v1)--vloc(__v5)--vloc(__v6)--vloc(__v2)--cycle;fill ph withcolor 0.5white;}
\fmfiv{l=\footnotesize{$\bar{D}^2$},l.a=28,l.d=2}{w5}
\fmfiv{l=\footnotesize{$D^2$},l.a=28,l.d=2}{w11}
\fmfiv{l=\footnotesize{$\bar{D}^2$},l.a=28,l.d=2}{w7}
\fmfiv{l=\footnotesize{$D^2$},l.a=28,l.d=2}{w8}
\fmfiv{l=\footnotesize{$\bar{D}^2$},l.a=28,l.d=2}{w9}
\fmfiv{l=\footnotesize{$D^2$},l.a=28,l.d=2}{w10}
\end{fmfchar*}}}}
\end{aligned}
\end{equation*}
\caption{Four-loop non-maximal diagrams of class C}
\label{diagrams-C}
\end{figure}

\begin{figure}[p]
\unitlength=1.75mm
\settoheight{\eqoff}{$\times$}%
\setlength{\eqoff}{0.5\eqoff}%
\addtolength{\eqoff}{-12.5\unitlength}%
\settoheight{\eqofftwo}{$\times$}%
\setlength{\eqofftwo}{0.5\eqofftwo}%
\addtolength{\eqofftwo}{-7.5\unitlength}%
\begin{equation*}
\begin{aligned}
S_{B1}=&
\begin{minipage}{4.3cm}
\raisebox{\eqoff}{%
\fmfframe(3,1)(1,4){%
\begin{fmfchar*}(20,30)
\fmftop{v1}
\fmfbottom{v5}
\fmfforce{(0w,h)}{v1}
\fmfforce{(0w,0)}{v5}
\fmffixed{(0.25w,0)}{v1,v2}
\fmffixed{(0.5w,0)}{v2,v3}
\fmffixed{(0.25w,0)}{v3,v4}
\fmffixed{(0.25w,0)}{v5,v6}
\fmffixed{(0.5w,0)}{v6,v7}
\fmffixed{(0.25w,0)}{v7,v8}
\fmf{phantom,tension=0.35,right=0.25}{v2,vc1}
\fmf{phantom,tension=0.35,left=0.25}{v6,vc2}
\fmf{plain,tension=0.5,left=0.25}{v3,vc1}
\fmf{plain,tension=0.5,right=0.25}{v7,vc2}
\fmf{plain,tension=0.3}{vc1,vc2}
\fmf{plain,tension=0.5,left=0.125}{vc3,vc2}
\fmf{plain}{v1,v5}
\fmf{plain}{v4,v8}
\fmf{plain,tension=0.5,right=0,width=1mm}{v5,v8}
\fmf{plain}{v1,v2}
\fmf{plain}{v6,v2}
\fmf{phantom}{v2,vc4}
\fmf{phantom,tension=0.25}{vc4,vc5}
\fmf{phantom}{vc5,v6}
\fmffreeze
\fmf{plain,left=0.2}{vc1,vc4}
\fmf{plain,right=0.2}{vc2,vc5}
\fmfposition
\fmfipath{p[]}
\fmfiset{p1}{vpath(__v1,__v5)}
\fmfiset{p2}{vpath(__v2,__vc1)}
\fmfiset{p3}{vpath(__v3,__vc1)}
\fmfiset{p4}{vpath(__vc1,__vc2)}
\fmfiset{p5}{vpath(__v6,__vc2)}
\fmfiset{p6}{vpath(__v7,__vc2)}
\fmfiset{p7}{vpath(__v4,__v8)}
\fmfiset{p8}{vpath(__v2,__v6)}
\fmfiset{p9}{vpath(__vc1,__vc4)}
\fmfiset{p10}{vpath(__vc2,__vc5)}
\fmfipair{w[]}
\fmfipair{wc[]}
\fmfiequ{w1}{point length(p8)/4 of p8}
\fmfiequ{w2}{(xpart(w3),ypart(wc2))}
\fmfiequ{w3}{point 3length(p8)/4 of p8}
\fmfiequ{w4}{point length(p3) of p3}
\fmfiequ{w5}{point length(p4) of p4}
\fmfiequ{w6}{point length(p6)/5 of p6}
\fmfiequ{w7}{point 4length(p6)/5 of p6}
\fmfiequ{w8}{point length(p10)/3 of p10}
\fmfiequ{w9}{(xpart(w10),ypart(wc1))}
\svertex{w10}{p7}
\fmfiequ{w11}{(xpart(w10),ypart(w6))}
\fmfiequ{w12}{(xpart(wc1),ypart(wc1)-2(ypart(wc1)-ypart(wc2))/3)}
\fmfiequ{w13}{wc2}
\fmfiequ{w14}{(xpart(w10),ypart(wc3)-10)}
\fmfiequ{w15}{(xpart(wc1),ypart(vloc(__vc1))-(ypart(vloc(__vc1))-ypart(wc1))/2)}
\fmfiequ{wc1}{point length(p4)/2 of p4}
\fmfiequ{wc2}{point length(p4)/2 of p4}
\fmfiequ{wc3}{point 2length(p4)/3 of p4}
\fmfi{wiggly}{w9..wc1}
\fmfcmd{path ph; ph:=vloc(__v1)--vloc(__v5)--vloc(__v6)--vloc(__v2)--cycle;fill ph withcolor 0.5white;}
\fmfiv{l=\footnotesize{$D^2$},l.a=28,l.d=2}{w7}
\fmfiv{l=\footnotesize{$D^2$},l.a=152,l.d=0.25}{w8}
\fmfiv{l=\footnotesize{$\bar{D}^2$},l.a=28,l.d=2}{w6}
\fmfiv{l=\footnotesize{$\bar{D}^2$},l.a=28,l.d=2}{w11}
\fmfiv{l=\footnotesize{$D^2$},l.a=28,l.d=2}{w14}
\fmfiv{l=\footnotesize{$\bar{D}^2$},l.a=28,l.d=2}{wc3}
\fmfiv{l=\footnotesize{$D^2$},l.a=28,l.d=2}{w15}
\end{fmfchar*}}}
\end{minipage}
=&
\begin{minipage}{4.3cm}
\raisebox{\eqoff}{%
\fmfframe(3,1)(1,4){%
\begin{fmfchar*}(20,30)
\fmftop{v1}
\fmfbottom{v5}
\fmfforce{(0w,h)}{v1}
\fmfforce{(0w,0)}{v5}
\fmffixed{(0.25w,0)}{v1,v2}
\fmffixed{(0.5w,0)}{v2,v3}
\fmffixed{(0.25w,0)}{v3,v4}
\fmffixed{(0.25w,0)}{v5,v6}
\fmffixed{(0.5w,0)}{v6,v7}
\fmffixed{(0.25w,0)}{v7,v8}
\fmf{phantom,tension=0.35,right=0.25}{v2,vc1}
\fmf{phantom,tension=0.35,left=0.25}{v6,vc2}
\fmf{plain,tension=0.5,left=0.25}{v3,vc1}
\fmf{plain,tension=0.5,right=0.25}{v7,vc2}
\fmf{plain,tension=0.3}{vc1,vc2}
\fmf{plain,tension=0.5,left=0.125}{vc3,vc2}
\fmf{plain}{v1,v5}
\fmf{plain}{v4,v8}
\fmf{plain,tension=0.5,right=0,width=1mm}{v5,v8}
\fmf{plain}{v1,v2}
\fmf{plain}{v6,v2}
\fmf{phantom}{v2,vc4}
\fmf{phantom,tension=0.25}{vc4,vc5}
\fmf{phantom}{vc5,v6}
\fmffreeze
\fmf{plain,left=0.2}{vc1,vc4}
\fmf{plain,right=0.2}{vc2,vc5}
\fmfposition
\fmfipath{p[]}
\fmfiset{p1}{vpath(__v1,__v5)}
\fmfiset{p2}{vpath(__v2,__vc1)}
\fmfiset{p3}{vpath(__v3,__vc1)}
\fmfiset{p4}{vpath(__vc1,__vc2)}
\fmfiset{p5}{vpath(__v6,__vc2)}
\fmfiset{p6}{vpath(__v7,__vc2)}
\fmfiset{p7}{vpath(__v4,__v8)}
\fmfiset{p8}{vpath(__v2,__v6)}
\fmfiset{p9}{vpath(__vc1,__vc4)}
\fmfiset{p10}{vpath(__vc2,__vc5)}
\fmfipair{w[]}
\fmfipair{wc[]}
\fmfiequ{w1}{point length(p8)/4 of p8}
\fmfiequ{w2}{(xpart(w3),ypart(wc2))}
\fmfiequ{w3}{point 3length(p8)/4 of p8}
\fmfiequ{w4}{point length(p3) of p3}
\fmfiequ{w5}{point length(p4) of p4}
\fmfiequ{w6}{point length(p6)/5 of p6}
\fmfiequ{w7}{point 4length(p6)/5 of p6}
\fmfiequ{w8}{point length(p10)/3 of p10}
\fmfiequ{w9}{(xpart(w10),ypart(wc1))}
\svertex{w10}{p7}
\fmfiequ{w11}{(xpart(w10),ypart(w6))}
\fmfiequ{w12}{(xpart(wc1),ypart(wc1)-2(ypart(wc1)-ypart(wc2))/3)}
\fmfiequ{w13}{wc2}
\fmfiequ{w14}{(xpart(w10),ypart(wc3)-10)}
\fmfiequ{w15}{(xpart(wc1),ypart(vloc(__vc1))-(ypart(vloc(__vc1))-ypart(wc1))/2)}
\fmfiequ{wc1}{point length(p4)/2 of p4}
\fmfiequ{wc2}{point length(p4)/2 of p4}
\fmfiequ{wc3}{point 2length(p4)/3 of p4}
\fmfiequ{wc4}{point 3length(p4)/4 of p4}
\fmfi{wiggly}{w9..wc1}
\fmfcmd{path ph; ph:=vloc(__v1)--vloc(__v5)--vloc(__v6)--vloc(__v2)--cycle;fill ph withcolor 0.5white;}
\fmfiv{l=\footnotesize{$D^2$},l.a=152,l.d=0.25}{w8}
\fmfiv{l=\footnotesize{$\bar{D}^2$},l.a=28,l.d=2}{w6}
\fmfiv{l=\footnotesize{$\bar{D}^2$},l.a=28,l.d=2}{w11}
\fmfiv{l=\footnotesize{$D^2$},l.a=28,l.d=2}{w14}
\fmfiv{l=\footnotesize{$\Box$},l.a=28,l.d=2}{wc4}
\fmfiv{l=\footnotesize{$D^2$},l.a=28,l.d=2}{w15}
\end{fmfchar*}}}
\end{minipage}
=-&
\begin{minipage}{4.3cm}
\raisebox{\eqoff}{%
\fmfframe(3,1)(1,4){%
\begin{fmfchar*}(20,30)
\fmftop{v1}
\fmfbottom{v5}
\fmfforce{(0w,h)}{v1}
\fmfforce{(0w,0)}{v5}
\fmffixed{(0.25w,0)}{v1,v2}
\fmffixed{(0.5w,0)}{v2,v3}
\fmffixed{(0.25w,0)}{v3,v4}
\fmffixed{(0.25w,0)}{v5,v6}
\fmffixed{(0.5w,0)}{v6,v7}
\fmffixed{(0.25w,0)}{v7,v8}
\fmf{phantom,tension=0.35,right=0.25}{v2,vc1}
\fmf{phantom,tension=0.35,left=0.25}{v6,vc2}
\fmf{plain,tension=0.5,left=0.25}{v3,vc1}
\fmf{plain,tension=0.5,right=0.25}{v7,vc2}
\fmf{plain,tension=0.3}{vc1,vc2}
\fmf{plain,tension=0.5,left=0.125}{vc3,vc2}
\fmf{plain}{v1,v5}
\fmf{plain}{v4,v8}
\fmf{plain,tension=0.5,right=0,width=1mm}{v5,v8}
\fmf{plain}{v1,v2}
\fmf{plain}{v6,v2}
\fmf{phantom}{v2,vc4}
\fmf{phantom,tension=0.25}{vc4,vc5}
\fmf{phantom}{vc5,v6}
\fmffreeze
\fmf{plain,left=0.2}{vc1,vc4}
\fmf{plain,right=0.2}{vc2,vc5}
\fmfposition
\fmfipath{p[]}
\fmfiset{p1}{vpath(__v1,__v5)}
\fmfiset{p2}{vpath(__v2,__vc1)}
\fmfiset{p3}{vpath(__v3,__vc1)}
\fmfiset{p4}{vpath(__vc1,__vc2)}
\fmfiset{p5}{vpath(__v6,__vc2)}
\fmfiset{p6}{vpath(__v7,__vc2)}
\fmfiset{p7}{vpath(__v4,__v8)}
\fmfiset{p8}{vpath(__v2,__v6)}
\fmfiset{p9}{vpath(__vc1,__vc4)}
\fmfiset{p10}{vpath(__vc2,__vc5)}
\fmfipair{w[]}
\fmfipair{wc[]}
\fmfiequ{w1}{point length(p8)/4 of p8}
\fmfiequ{w2}{(xpart(w3),ypart(wc2))}
\fmfiequ{w3}{point 3length(p8)/4 of p8}
\fmfiequ{w4}{point length(p3) of p3}
\fmfiequ{w5}{point length(p4) of p4}
\fmfiequ{w6}{point length(p6)/5 of p6}
\fmfiequ{w7}{point 4length(p6)/5 of p6}
\fmfiequ{w8}{point length(p10)/3 of p10}
\fmfiequ{w9}{(xpart(w10),ypart(wc1))}
\svertex{w10}{p7}
\fmfiequ{w11}{(xpart(w10),ypart(w6))}
\fmfiequ{w12}{(xpart(wc1),ypart(wc1)-2(ypart(wc1)-ypart(wc2))/3)}
\fmfiequ{w13}{wc2}
\fmfiequ{w14}{(xpart(w10),ypart(wc3))}
\fmfiequ{w15}{(xpart(wc1),ypart(vloc(__vc1))-(ypart(vloc(__vc1))-ypart(wc1))/2)}
\fmfiequ{wc1}{point length(p4)/2 of p4}
\fmfiequ{wc2}{point length(p4)/2 of p4}
\fmfiequ{wc3}{point 2length(p4)/3 of p4}
\fmfiequ{wc4}{point 3length(p4)/4 of p4}
\fmfi{wiggly}{w9..vloc(__vc2)}
\fmfcmd{path ph; ph:=vloc(__v1)--vloc(__v5)--vloc(__v6)--vloc(__v2)--cycle;fill ph withcolor 0.5white;}
\fmfiv{l=\footnotesize{$D^2$},l.a=152,l.d=0.25}{w8}
\fmfiv{l=\footnotesize{$\bar{D}^2$},l.a=28,l.d=2}{w6}
\fmfiv{l=\footnotesize{$\bar{D}^2$},l.a=28,l.d=2}{w11}
\fmfiv{l=\footnotesize{$D^2$},l.a=28,l.d=2}{w14}
\fmfiv{l=\footnotesize{$D^2$},l.a=28,l.d=2}{w15}
\end{fmfchar*}}}
\end{minipage}
\\
S_{B2}=&
\begin{minipage}{4.3cm}
\raisebox{\eqoff}{%
\fmfframe(3,1)(1,4){%
\begin{fmfchar*}(20,30)
\fmftop{v1}
\fmfbottom{v5}
\fmfforce{(0w,h)}{v1}
\fmfforce{(0w,0)}{v5}
\fmffixed{(0.25w,0)}{v1,v2}
\fmffixed{(0.5w,0)}{v2,v3}
\fmffixed{(0.25w,0)}{v3,v4}
\fmffixed{(0.25w,0)}{v5,v6}
\fmffixed{(0.5w,0)}{v6,v7}
\fmffixed{(0.25w,0)}{v7,v8}
\fmf{phantom,tension=0.35,right=0.25}{v2,vc1}
\fmf{phantom,tension=0.35,left=0.25}{v6,vc2}
\fmf{plain,tension=0.5,left=0.25}{v3,vc1}
\fmf{plain,tension=0.5,right=0.25}{v7,vc2}
\fmf{plain,tension=0.3}{vc1,vc2}
\fmf{plain,tension=0.5,left=0.125}{vc3,vc2}
\fmf{plain}{v1,v5}
\fmf{plain}{v4,v8}
\fmf{plain,tension=0.5,right=0,width=1mm}{v5,v8}
\fmf{plain}{v1,v2}
\fmf{plain}{v6,v2}
\fmf{phantom}{v2,vc4}
\fmf{phantom,tension=0.25}{vc4,vc5}
\fmf{phantom}{vc5,v6}
\fmffreeze
\fmf{plain,left=0.2}{vc1,vc4}
\fmf{plain,right=0.2}{vc2,vc5}
\fmfposition
\fmfipath{p[]}
\fmfiset{p1}{vpath(__v1,__v5)}
\fmfiset{p2}{vpath(__v2,__vc1)}
\fmfiset{p3}{vpath(__v3,__vc1)}
\fmfiset{p4}{vpath(__vc1,__vc2)}
\fmfiset{p5}{vpath(__v6,__vc2)}
\fmfiset{p6}{vpath(__v7,__vc2)}
\fmfiset{p7}{vpath(__v4,__v8)}
\fmfiset{p8}{vpath(__v2,__v6)}
\fmfiset{p9}{vpath(__vc1,__vc4)}
\fmfiset{p10}{vpath(__vc2,__vc5)}
\fmfipair{w[]}
\fmfipair{wc[]}
\fmfiequ{wc1}{point length(p4)/3 of p4}
\fmfiequ{wc2}{point 2length(p4)/3 of p4}
\fmfiequ{wc3}{point 5length(p4)/6 of p4}
\fmfiequ{w1}{point length(p8)/4 of p8}
\fmfiequ{w2}{(xpart(w3),ypart(wc1))}
\fmfiequ{w3}{point 3length(p8)/4 of p8}
\fmfiequ{w4}{point length(p3) of p3}
\fmfiequ{w5}{point length(p4) of p4}
\fmfiequ{w6}{point 2length(p6)/3 of p6}
\fmfiequ{w7}{point 10length(p6)/11 of p6}
\fmfiequ{w8}{point length(p10)/3 of p10}
\fmfiequ{w9}{(xpart(w10),ypart(wc2))}
\svertex{w10}{p7}
\fmfiequ{w11}{(xpart(w10),ypart(w16))}
\fmfiequ{w12}{(xpart(wc1),ypart(wc1)-(ypart(wc1)-ypart(wc2))/2)}
\fmfiequ{w13}{wc2}
\fmfiequ{w14}{(xpart(w10),ypart(wc3))}
\fmfiequ{w15}{(xpart(wc1),ypart(vloc(__vc1))-(ypart(vloc(__vc1))-ypart(wc1))/2)}
\fmfiequ{w16}{point length(p6)/8 of p6}
\fmfiequ{w17}{point 4length(p6)/8 of p6}
\fmfi{wiggly}{w9..w6}
\fmfcmd{path ph; ph:=vloc(__v1)--vloc(__v5)--vloc(__v6)--vloc(__v2)--cycle;fill ph withcolor 0.5white;}
\fmfiv{l=\footnotesize{$\bar{D}^2$},l.a=28,l.d=2}{w7}
\fmfiv{l=\footnotesize{$D^2$},l.a=152,l.d=0.25}{w8}
\fmfiv{l=\footnotesize{$\bar{D}^2$},l.a=28,l.d=2}{w11}
\fmfiv{l=\footnotesize{$D^2$},l.a=152,l.d=2}{w13}
\fmfiv{l=\footnotesize{$D^2$},l.a=28,l.d=2}{w14}
\fmfiv{l=\footnotesize{$D^2$},l.a=28,l.d=2}{w17}
\fmfiv{l=\footnotesize{$\bar{D}^2$},l.a=28,l.d=2}{w16}
\end{fmfchar*}}}
\end{minipage}
=&
\begin{minipage}{4.3cm}
\raisebox{\eqoff}{%
\fmfframe(3,1)(1,4){%
\begin{fmfchar*}(20,30)
\fmftop{v1}
\fmfbottom{v5}
\fmfforce{(0w,h)}{v1}
\fmfforce{(0w,0)}{v5}
\fmffixed{(0.25w,0)}{v1,v2}
\fmffixed{(0.5w,0)}{v2,v3}
\fmffixed{(0.25w,0)}{v3,v4}
\fmffixed{(0.25w,0)}{v5,v6}
\fmffixed{(0.5w,0)}{v6,v7}
\fmffixed{(0.25w,0)}{v7,v8}
\fmf{phantom,tension=0.35,right=0.25}{v2,vc1}
\fmf{phantom,tension=0.35,left=0.25}{v6,vc2}
\fmf{plain,tension=0.5,left=0.25}{v3,vc1}
\fmf{plain,tension=0.5,right=0.25}{v7,vc2}
\fmf{plain,tension=0.3}{vc1,vc2}
\fmf{plain,tension=0.5,left=0.125}{vc3,vc2}
\fmf{plain}{v1,v5}
\fmf{plain}{v4,v8}
\fmf{plain,tension=0.5,right=0,width=1mm}{v5,v8}
\fmf{plain}{v1,v2}
\fmf{plain}{v6,v2}
\fmf{phantom}{v2,vc4}
\fmf{phantom,tension=0.25}{vc4,vc5}
\fmf{phantom}{vc5,v6}
\fmffreeze
\fmf{plain,left=0.2}{vc1,vc4}
\fmf{plain,right=0.2}{vc2,vc5}
\fmfposition
\fmfipath{p[]}
\fmfiset{p1}{vpath(__v1,__v5)}
\fmfiset{p2}{vpath(__v2,__vc1)}
\fmfiset{p3}{vpath(__v3,__vc1)}
\fmfiset{p4}{vpath(__vc1,__vc2)}
\fmfiset{p5}{vpath(__v6,__vc2)}
\fmfiset{p6}{vpath(__v7,__vc2)}
\fmfiset{p7}{vpath(__v4,__v8)}
\fmfiset{p8}{vpath(__v2,__v6)}
\fmfiset{p9}{vpath(__vc1,__vc4)}
\fmfiset{p10}{vpath(__vc2,__vc5)}
\fmfipair{w[]}
\fmfipair{wc[]}
\fmfiequ{wc1}{point length(p4)/3 of p4}
\fmfiequ{wc2}{point 2length(p4)/3 of p4}
\fmfiequ{wc3}{point 5length(p4)/6 of p4}
\fmfiequ{w1}{point length(p8)/4 of p8}
\fmfiequ{w2}{(xpart(w3),ypart(wc1))}
\fmfiequ{w3}{point 3length(p8)/4 of p8}
\fmfiequ{w4}{point length(p3) of p3}
\fmfiequ{w5}{point length(p4) of p4}
\fmfiequ{w6}{point 2length(p6)/3 of p6}
\fmfiequ{w7}{point 10length(p6)/11 of p6}
\fmfiequ{w8}{point length(p10)/3 of p10}
\fmfiequ{w9}{(xpart(w10),ypart(wc2))}
\svertex{w10}{p7}
\fmfiequ{w11}{(xpart(w10),ypart(w16))}
\fmfiequ{w12}{(xpart(wc1),ypart(wc1)-(ypart(wc1)-ypart(wc2))/2)}
\fmfiequ{w13}{wc2}
\fmfiequ{w14}{(xpart(w10),ypart(wc3))}
\fmfiequ{w15}{(xpart(wc1),ypart(vloc(__vc1))-(ypart(vloc(__vc1))-ypart(wc1))/2)}
\fmfiequ{w16}{point length(p6)/8 of p6}
\fmfiequ{w17}{point 4length(p6)/8 of p6}
\fmfi{wiggly}{w9..w6}
\fmfcmd{path ph; ph:=vloc(__v1)--vloc(__v5)--vloc(__v6)--vloc(__v2)--cycle;fill ph withcolor 0.5white;}
\fmfiv{l=\footnotesize{$\Box$},l.a=28,l.d=2}{w7}
\fmfiv{l=\footnotesize{$D^2$},l.a=152,l.d=0.25}{w8}
\fmfiv{l=\footnotesize{$\bar{D}^2$},l.a=28,l.d=2}{w11}
\fmfiv{l=\footnotesize{$D^2$},l.a=28,l.d=2}{w14}
\fmfiv{l=\footnotesize{$D^2$},l.a=28,l.d=2}{w17}
\fmfiv{l=\footnotesize{$\bar{D}^2$},l.a=28,l.d=2}{w16}
\end{fmfchar*}}}
\end{minipage}
=-&
\begin{minipage}{4.3cm}
\raisebox{\eqoff}{%
\fmfframe(3,1)(1,4){%
\begin{fmfchar*}(20,30)
\fmftop{v1}
\fmfbottom{v5}
\fmfforce{(0w,h)}{v1}
\fmfforce{(0w,0)}{v5}
\fmffixed{(0.25w,0)}{v1,v2}
\fmffixed{(0.5w,0)}{v2,v3}
\fmffixed{(0.25w,0)}{v3,v4}
\fmffixed{(0.25w,0)}{v5,v6}
\fmffixed{(0.5w,0)}{v6,v7}
\fmffixed{(0.25w,0)}{v7,v8}
\fmf{phantom,tension=0.35,right=0.25}{v2,vc1}
\fmf{phantom,tension=0.35,left=0.25}{v6,vc2}
\fmf{plain,tension=0.5,left=0.25}{v3,vc1}
\fmf{plain,tension=0.5,right=0.25}{v7,vc2}
\fmf{plain,tension=0.3}{vc1,vc2}
\fmf{plain,tension=0.5,left=0.125}{vc3,vc2}
\fmf{plain}{v1,v5}
\fmf{plain}{v4,v8}
\fmf{plain,tension=0.5,right=0,width=1mm}{v5,v8}
\fmf{plain}{v1,v2}
\fmf{plain}{v6,v2}
\fmf{phantom}{v2,vc4}
\fmf{phantom,tension=0.25}{vc4,vc5}
\fmf{phantom}{vc5,v6}
\fmffreeze
\fmf{plain,left=0.2}{vc1,vc4}
\fmf{plain,right=0.2}{vc2,vc5}
\fmfposition
\fmfipath{p[]}
\fmfiset{p1}{vpath(__v1,__v5)}
\fmfiset{p2}{vpath(__v2,__vc1)}
\fmfiset{p3}{vpath(__v3,__vc1)}
\fmfiset{p4}{vpath(__vc1,__vc2)}
\fmfiset{p5}{vpath(__v6,__vc2)}
\fmfiset{p6}{vpath(__v7,__vc2)}
\fmfiset{p7}{vpath(__v4,__v8)}
\fmfiset{p8}{vpath(__v2,__v6)}
\fmfiset{p9}{vpath(__vc1,__vc4)}
\fmfiset{p10}{vpath(__vc2,__vc5)}
\fmfipair{w[]}
\fmfipair{wc[]}
\fmfiequ{wc1}{point length(p4)/3 of p4}
\fmfiequ{wc2}{point 2length(p4)/3 of p4}
\fmfiequ{wc3}{point 5length(p4)/6 of p4}
\fmfiequ{w1}{point length(p8)/4 of p8}
\fmfiequ{w2}{(xpart(w3),ypart(wc1))}
\fmfiequ{w3}{point 3length(p8)/4 of p8}
\fmfiequ{w4}{point length(p3) of p3}
\fmfiequ{w5}{point length(p4) of p4}
\fmfiequ{w6}{point 2length(p6)/3 of p6}
\fmfiequ{w7}{point 10length(p6)/11 of p6}
\fmfiequ{w8}{point length(p10)/3 of p10}
\fmfiequ{w9}{(xpart(w10),ypart(wc2))}
\svertex{w10}{p7}
\fmfiequ{w11}{(xpart(w10),ypart(w16))}
\fmfiequ{w12}{(xpart(wc1),ypart(wc1)-(ypart(wc1)-ypart(wc2))/2)}
\fmfiequ{w13}{wc2}
\fmfiequ{w14}{(xpart(w10),ypart(wc3))}
\fmfiequ{w15}{(xpart(wc1),ypart(vloc(__vc1))-(ypart(vloc(__vc1))-ypart(wc1))/2)}
\fmfiequ{w16}{point length(p6)/8 of p6}
\fmfiequ{w17}{point 4length(p6)/8 of p6}
\fmfi{wiggly}{w9..vloc(__vc2)}
\fmfcmd{path ph; ph:=vloc(__v1)--vloc(__v5)--vloc(__v6)--vloc(__v2)--cycle;fill ph withcolor 0.5white;}
\fmfiv{l=\footnotesize{$D^2$},l.a=152,l.d=0.25}{w8}
\fmfiv{l=\footnotesize{$\bar{D}^2$},l.a=28,l.d=2}{w11}
\fmfiv{l=\footnotesize{$D^2$},l.a=28,l.d=2}{w14}
\fmfiv{l=\footnotesize{$D^2$},l.a=28,l.d=2}{w17}
\fmfiv{l=\footnotesize{$\bar{D}^2$},l.a=28,l.d=2}{w16}
\end{fmfchar*}}}
\end{minipage}
\\
S_{C1}=&
\begin{minipage}{4.3cm}
\raisebox{\eqoff}{%
\fmfframe(3,1)(1,4){%
\begin{fmfchar*}(20,30)
\fmftop{v1}
\fmfbottom{v5}
\fmfforce{(0w,h)}{v1}
\fmfforce{(0w,0)}{v5}
\fmffixed{(0.25w,0)}{v1,v2}
\fmffixed{(0.35w,0)}{v2,v2b}
\fmffixed{(0.2w,0)}{v2b,v3}
\fmffixed{(0.2w,0)}{v3,v4}
\fmffixed{(0.25w,0)}{v5,v6}
\fmffixed{(0.35w,0)}{v6,v6b}
\fmffixed{(0.2w,0)}{v6b,v7}
\fmffixed{(0.2w,0)}{v7,v8}
\fmf{phantom,tension=0.35,right=0.25}{v2,vc1}
\fmf{phantom,tension=0.35,left=0.25}{v6,vc2}
\fmf{phantom,tension=0.5,left=0.25}{v3,vc1}
\fmf{phantom,tension=0.5,right=0.25}{v7,vc2}
\fmf{phantom,tension=0.3}{vc1,vc2}
\fmf{phantom,tension=0.5,left=0.125}{vc3,vc2}
\fmf{plain}{v1,v5}
\fmf{plain}{v4,v8}
\fmf{plain,tension=0.5,right=0,width=1mm}{v5,v8}
\fmf{plain}{v1,v2}
\fmf{plain}{v6,v2}
\fmf{plain}{v2b,v6b}
\fmf{phantom}{v2,vc4}
\fmf{phantom,tension=0.25}{vc4,vc5}
\fmf{phantom}{vc5,v6}
\fmffreeze
\fmf{phantom,left=0.2}{vc1,vc4}
\fmf{phantom,right=0.2}{vc2,vc5}
\fmfposition
\fmfipath{p[]}
\fmfiset{p1}{vpath(__v1,__v5)}
\fmfiset{p2}{vpath(__v2,__vc1)}
\fmfiset{p3}{vpath(__v3,__vc1)}
\fmfiset{p4}{vpath(__vc1,__vc2)}
\fmfiset{p5}{vpath(__v6,__vc2)}
\fmfiset{p6}{vpath(__v7,__vc2)}
\fmfiset{p7}{vpath(__v4,__v8)}
\fmfiset{p8}{vpath(__v2,__v6)}
\fmfiset{p9}{vpath(__vc1,__vc4)}
\fmfiset{p10}{vpath(__vc2,__vc5)}
\fmfiset{p11}{vpath(__v2b,__v6b)}
\fmfiset{p12}{vpath(__v4,__v8)}
\fmfipair{w[]}
\fmfipair{wc[]}
\fmfiequ{wc1}{point length(p11)/3 of p11}
\fmfiequ{wc2}{point 2length(p11)/3 of p11}
\fmfiequ{w1}{point length(p8)/3 of p8}
\fmfiequ{w2}{point 2length(p8)/3 of p8}
\fmfiequ{w3}{point length(p12)/3 of p12}
\fmfiequ{w4}{point 2length(p12)/3 of p12}
\fmfiequ{w5}{point 13length(p12)/14 of p12}
\fmfiequ{w6}{point 11length(p12)/14 of p12}
\vvertex{w7}{w5}{p11}
\vvertex{w8}{w6}{p11}
\fmfiequ{w9}{point 9length(p12)/15 of p11}
\fmfiequ{w10}{point 7length(p12)/15 of p11}
\fmfi{wiggly}{w4..wc2}
\fmfi{wiggly}{w2..wc1}
\fmfcmd{path ph; ph:=vloc(__v1)--vloc(__v5)--vloc(__v6)--vloc(__v2)--cycle;fill ph withcolor 0.5white;}
\fmfiv{l=\footnotesize{$\bar{D}^2$},l.a=28,l.d=2}{w5}
\fmfiv{l=\footnotesize{$D^2$},l.a=28,l.d=2}{w6}
\fmfiv{l=\footnotesize{$\bar{D}^2$},l.a=28,l.d=2}{w7}
\fmfiv{l=\footnotesize{$D^2$},l.a=28,l.d=2}{w8}
\fmfiv{l=\footnotesize{$\bar{D}^2$},l.a=28,l.d=2}{w9}
\fmfiv{l=\footnotesize{$D^2$},l.a=28,l.d=2}{w10}
\end{fmfchar*}}}
\end{minipage}
=&
\begin{minipage}{4.3cm}
\raisebox{\eqoff}{%
\fmfframe(3,1)(1,4){%
\begin{fmfchar*}(20,30)
\fmftop{v1}
\fmfbottom{v5}
\fmfforce{(0w,h)}{v1}
\fmfforce{(0w,0)}{v5}
\fmffixed{(0.25w,0)}{v1,v2}
\fmffixed{(0.35w,0)}{v2,v2b}
\fmffixed{(0.2w,0)}{v2b,v3}
\fmffixed{(0.2w,0)}{v3,v4}
\fmffixed{(0.25w,0)}{v5,v6}
\fmffixed{(0.35w,0)}{v6,v6b}
\fmffixed{(0.2w,0)}{v6b,v7}
\fmffixed{(0.2w,0)}{v7,v8}
\fmf{phantom,tension=0.35,right=0.25}{v2,vc1}
\fmf{phantom,tension=0.35,left=0.25}{v6,vc2}
\fmf{phantom,tension=0.5,left=0.25}{v3,vc1}
\fmf{phantom,tension=0.5,right=0.25}{v7,vc2}
\fmf{phantom,tension=0.3}{vc1,vc2}
\fmf{phantom,tension=0.5,left=0.125}{vc3,vc2}
\fmf{plain}{v1,v5}
\fmf{plain}{v4,v8}
\fmf{plain,tension=0.5,right=0,width=1mm}{v5,v8}
\fmf{plain}{v1,v2}
\fmf{plain}{v6,v2}
\fmf{plain}{v2b,v6b}
\fmf{phantom}{v2,vc4}
\fmf{phantom,tension=0.25}{vc4,vc5}
\fmf{phantom}{vc5,v6}
\fmffreeze
\fmf{phantom,left=0.2}{vc1,vc4}
\fmf{phantom,right=0.2}{vc2,vc5}
\fmfposition
\fmfipath{p[]}
\fmfiset{p1}{vpath(__v1,__v5)}
\fmfiset{p2}{vpath(__v2,__vc1)}
\fmfiset{p3}{vpath(__v3,__vc1)}
\fmfiset{p4}{vpath(__vc1,__vc2)}
\fmfiset{p5}{vpath(__v6,__vc2)}
\fmfiset{p6}{vpath(__v7,__vc2)}
\fmfiset{p7}{vpath(__v4,__v8)}
\fmfiset{p8}{vpath(__v2,__v6)}
\fmfiset{p9}{vpath(__vc1,__vc4)}
\fmfiset{p10}{vpath(__vc2,__vc5)}
\fmfiset{p11}{vpath(__v2b,__v6b)}
\fmfiset{p12}{vpath(__v4,__v8)}
\fmfipair{w[]}
\fmfipair{wc[]}
\fmfiequ{wc1}{point length(p11)/3 of p11}
\fmfiequ{wc2}{point 2length(p11)/3 of p11}
\fmfiequ{w1}{point length(p8)/3 of p8}
\fmfiequ{w2}{point 2length(p8)/3 of p8}
\fmfiequ{w3}{point length(p12)/3 of p12}
\fmfiequ{w4}{point 2length(p12)/3 of p12}
\fmfiequ{w5}{point 13length(p12)/14 of p12}
\fmfiequ{w6}{point 11length(p12)/14 of p12}
\vvertex{w7}{w5}{p11}
\vvertex{w8}{w6}{p11}
\fmfiequ{w9}{point 8length(p12)/15 of p11}
\fmfiequ{w10}{point 7length(p12)/15 of p11}
\fmfi{wiggly}{w4..wc2}
\fmfi{wiggly}{w2..wc1}
\fmfcmd{path ph; ph:=vloc(__v1)--vloc(__v5)--vloc(__v6)--vloc(__v2)--cycle;fill ph withcolor 0.5white;}
\fmfiv{l=\footnotesize{$\bar{D}^2$},l.a=28,l.d=2}{w5}
\fmfiv{l=\footnotesize{$D^2$},l.a=28,l.d=2}{w6}
\fmfiv{l=\footnotesize{$\bar{D}^2$},l.a=28,l.d=2}{w7}
\fmfiv{l=\footnotesize{$D^2$},l.a=28,l.d=2}{w8}
\fmfiv{l=\footnotesize{$\Box$},l.a=28,l.d=2}{w9}
\end{fmfchar*}}}
\end{minipage}
=-&
\begin{minipage}{4.3cm}
\raisebox{\eqoff}{%
\fmfframe(3,1)(1,4){%
\begin{fmfchar*}(20,30)
\fmftop{v1}
\fmfbottom{v5}
\fmfforce{(0w,h)}{v1}
\fmfforce{(0w,0)}{v5}
\fmffixed{(0.25w,0)}{v1,v2}
\fmffixed{(0.35w,0)}{v2,v2b}
\fmffixed{(0.2w,0)}{v2b,v3}
\fmffixed{(0.2w,0)}{v3,v4}
\fmffixed{(0.25w,0)}{v5,v6}
\fmffixed{(0.35w,0)}{v6,v6b}
\fmffixed{(0.2w,0)}{v6b,v7}
\fmffixed{(0.2w,0)}{v7,v8}
\fmf{phantom,tension=0.35,right=0.25}{v2,vc1}
\fmf{phantom,tension=0.35,left=0.25}{v6,vc2}
\fmf{phantom,tension=0.5,left=0.25}{v3,vc1}
\fmf{phantom,tension=0.5,right=0.25}{v7,vc2}
\fmf{phantom,tension=0.3}{vc1,vc2}
\fmf{phantom,tension=0.5,left=0.125}{vc3,vc2}
\fmf{plain}{v1,v5}
\fmf{plain}{v4,v8}
\fmf{plain,tension=0.5,right=0,width=1mm}{v5,v8}
\fmf{plain}{v1,v2}
\fmf{plain}{v6,v2}
\fmf{plain}{v2b,v6b}
\fmf{phantom}{v2,vc4}
\fmf{phantom,tension=0.25}{vc4,vc5}
\fmf{phantom}{vc5,v6}
\fmffreeze
\fmf{phantom,left=0.2}{vc1,vc4}
\fmf{phantom,right=0.2}{vc2,vc5}
\fmfposition
\fmfipath{p[]}
\fmfiset{p1}{vpath(__v1,__v5)}
\fmfiset{p2}{vpath(__v2,__vc1)}
\fmfiset{p3}{vpath(__v3,__vc1)}
\fmfiset{p4}{vpath(__vc1,__vc2)}
\fmfiset{p5}{vpath(__v6,__vc2)}
\fmfiset{p6}{vpath(__v7,__vc2)}
\fmfiset{p7}{vpath(__v4,__v8)}
\fmfiset{p8}{vpath(__v2,__v6)}
\fmfiset{p9}{vpath(__vc1,__vc4)}
\fmfiset{p10}{vpath(__vc2,__vc5)}
\fmfiset{p11}{vpath(__v2b,__v6b)}
\fmfiset{p12}{vpath(__v4,__v8)}
\fmfipair{w[]}
\fmfipair{wc[]}
\fmfiequ{wc1}{point length(p11)/3 of p11}
\fmfiequ{wc2}{point 2length(p11)/3 of p11}
\fmfiequ{w1}{point length(p8)/3 of p8}
\fmfiequ{w2}{point 2length(p8)/3 of p8}
\fmfiequ{w3}{point length(p12)/3 of p12}
\fmfiequ{w4}{point 2length(p12)/3 of p12}
\fmfiequ{w5}{point 13length(p12)/14 of p12}
\fmfiequ{w6}{point 11length(p12)/14 of p12}
\vvertex{w7}{w5}{p11}
\vvertex{w8}{w6}{p11}
\fmfiequ{w9}{point 9length(p12)/15 of p11}
\fmfiequ{w10}{point 7length(p12)/15 of p11}
\fmfi{wiggly}{w4..wc2}
\fmfi{wiggly}{w1..wc2}
\fmfcmd{path ph; ph:=vloc(__v1)--vloc(__v5)--vloc(__v6)--vloc(__v2)--cycle;fill ph withcolor 0.5white;}
\fmfiv{l=\footnotesize{$\bar{D}^2$},l.a=28,l.d=2}{w5}
\fmfiv{l=\footnotesize{$D^2$},l.a=28,l.d=2}{w6}
\fmfiv{l=\footnotesize{$\bar{D}^2$},l.a=28,l.d=2}{w7}
\fmfiv{l=\footnotesize{$D^2$},l.a=28,l.d=2}{w8}
\end{fmfchar*}}}
\end{minipage}
\end{aligned}
\end{equation*}
\caption{$D$-algebra steps showing explicit cancellations for classes B and C}
\label{diagrams-BC-dem}
\end{figure}

\clearpage
\newpage

\renewcommand{\thefigure}{B.\arabic{figure}}
\setcounter{figure}{-1}
\renewcommand{\thetable}{B.\arabic{table}}
\setcounter{table}{0}

\section{Range-five diagrams}

\label{app:range-five}
In this appendix we present the results of the computation of range-five diagrams. While this calculation would not be strictly required for the subtraction of range-five interactions from the asymptotic dilatation operator, it has been performed as a check of our procedure. Moreover, it allowed us to compute the values of three out of the four $\epsilon$ coefficients in our scheme.

The completely chiral four-loop diagrams contributing to the asymptotic dilatation operator are shown in Figure \ref{r5diagrams-chiral}. They are all range five, except for the last one which is range four. Each diagram has a different chiral structure, so the corresponding coefficient in the dilatation operator is obtained directly from the value of the diagram.

\begin{figure}[p]
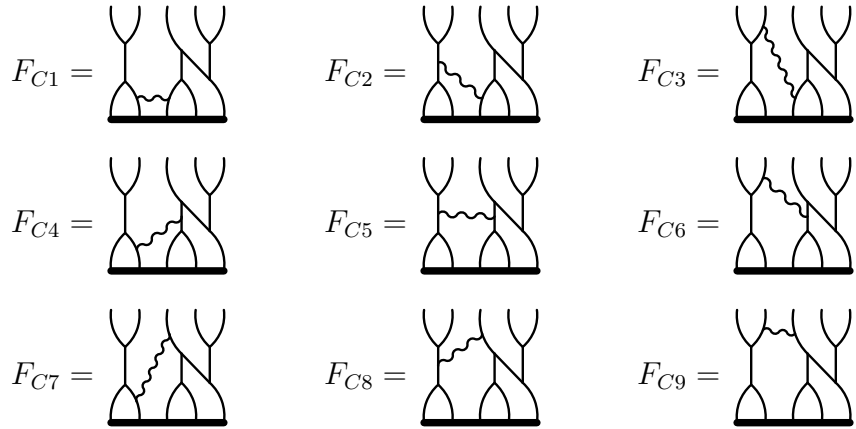

\unitlength=0.75mm
\settoheight{\eqoff}{$\times$}%
\setlength{\eqoff}{0.5\eqoff}%
\addtolength{\eqoff}{-12.5\unitlength}%
\settoheight{\eqofftwo}{$\times$}%
\setlength{\eqofftwo}{0.5\eqofftwo}%
\addtolength{\eqofftwo}{-7.5\unitlength}%
\begin{equation*}
\begin{aligned}
F_{C1}=&
\raisebox{\eqoff}{%
\fmfframe(3,1)(1,4){%
\begin{fmfchar*}(25,20)
\Weightplain
\fmfipair{w[]}
\svertex{w1}{p3}
\svertex{w2}{p10}
\fmfi{wiggly}{w1..w2}
\end{fmfchar*}}}
\;{\quad}&F_{C2}=&
\raisebox{\eqoff}{%
\fmfframe(3,1)(1,4){%
\begin{fmfchar*}(25,20)
\Weightplain
\fmfipair{w[]}
\svertex{w1}{p3}
\svertex{w2}{p9}
\fmfi{wiggly}{w1..w2}
\end{fmfchar*}}}
\;{\quad}&F_{C3}=&
\raisebox{\eqoff}{%
\fmfframe(3,1)(1,4){%
\begin{fmfchar*}(25,20)
\Weightplain
\fmfipair{w[]}
\svertex{w1}{p3}
\svertex{w2}{p8}
\fmfi{wiggly}{w1..w2}
\end{fmfchar*}}}
\\
F_{C4}=&
\raisebox{\eqoff}{%
\fmfframe(3,1)(1,4){%
\begin{fmfchar*}(25,20)
\Weightplain
\fmfipair{w[]}
\svertex{w1}{p2}
\svertex{w2}{p10}
\fmfi{wiggly}{w1..w2}
\end{fmfchar*}}}
\;{\quad}&F_{C5}=&
\raisebox{\eqoff}{%
\fmfframe(3,1)(1,4){%
\begin{fmfchar*}(25,20)
\Weightplain
\fmfipair{w[]}
\svertex{w1}{p2}
\svertex{w2}{p9}
\fmfi{wiggly}{w1..w2}
\end{fmfchar*}}}
\;{\quad}&F_{C6}=&
\raisebox{\eqoff}{%
\fmfframe(3,1)(1,4){%
\begin{fmfchar*}(25,20)
\Weightplain
\fmfipair{w[]}
\svertex{w1}{p2}
\svertex{w2}{p8}
\fmfi{wiggly}{w1..w2}
\end{fmfchar*}}}
\\
F_{C7}=&
\raisebox{\eqoff}{%
\fmfframe(3,1)(1,4){%
\begin{fmfchar*}(25,20)
\Weightplain
\fmfipair{w[]}
\svertex{w1}{p1}
\svertex{w2}{p10}
\fmfi{wiggly}{w1..w2}
\end{fmfchar*}}}
\;{\quad}&F_{C8}=&
\raisebox{\eqoff}{%
\fmfframe(3,1)(1,4){%
\begin{fmfchar*}(25,20)
\Weightplain
\fmfipair{w[]}
\svertex{w1}{p1}
\svertex{w2}{p9}
\fmfi{wiggly}{w1..w2}
\end{fmfchar*}}}
\;{\quad}&F_{C9}=&
\raisebox{\eqoff}{%
\fmfframe(3,1)(1,4){%
\begin{fmfchar*}(25,20)
\Weightplain
\fmfipair{w[]}
\svertex{w1}{p1}
\svertex{w2}{p8}
\fmfi{wiggly}{w1..w2}
\end{fmfchar*}}}
\end{aligned}
\end{equation*}
\caption{Diagrams with structure $\chi(1,4,3)$}
\label{r5diagrams-124}
\end{figure}

The diagrams with structures $\chi(2,1,4)$ (which is the reflection of $\chi(1,3,4)$), $\chi(1,4,3)$ (reflection of $\chi(1,2,4)$) and $\chi(1,4)$ are shown in figures~\ref{r5diagrams-214}, \ref{r5diagrams-124} and~\ref{r5diagrams-14} respectively. The results of $D$-algebra for every diagram in these classes are summarized in tables~\ref{r5results-214}, \ref{r5results-124} and~\ref{r5results-14}: cancellations between pairs of supergraphs, as well as diagrams which do not produce any divergent part, are explicitly indicated. In all the other cases, a $*$ indicates that the diagram gives a non-vanishing contribution which must be computed. For these relevant diagrams, we also give the relative 
symmetry factor.
In Table \ref{results-r5} all these relevant contributions are collected.
The integrals $J_{i}$ are given in Table \ref{integrals-r5}.

Let us define $\mathcal{C}(\chi(\ldots))$ as the coefficient of structure $\chi(\ldots)$ in the dilatation operator~\eqref{D4chi}. According to equation~\eqref{andim}, in order to compute $\mathcal{C}(\chi(\ldots))$ from the sum of the corresponding diagrams, we have to take the coefficient of the $\frac{1}{\varepsilon}$ pole and multiply it by $-8$. We find:
\begin{itemize}

\item $\mathcal{C}(\chi(1,2,3,4))=-10$, which is a check

\item $\mathcal{C}(\chi(3,2,1,4))+\mathcal{C}(\chi(1,4,3,2))=8/3-8/3=0$ \\
This must be equal to $-2(8+2\epsilon_{3a})$ from~\eqref{D4chi}. So we obtain $\epsilon_{3a}=-4$

\item $\mathcal{C}(\chi(3,2,1,4))-\mathcal{C}(\chi(1,4,3,2))=8/3-(-8/3)=16/3$ \\
Since this has to be equal to $4i\epsilon_{3b}$, we get $\epsilon_{3b}=-i4/3$

\item $\mathcal{C}(\chi(2,4,1,3))=2=18+4\epsilon_{3a}$ which is another check.

\item $\mathcal{C}(\chi(2,1,3,2))=4-8\zeta(3)$ \\
This must be equal to $(-12-2\beta-4\epsilon_{3a})$, so $\beta=4\zeta(3)$

\item $\mathcal{C}(\chi(2,1,4))+\mathcal{C}(\chi(1,2,4))=-4/3-20/3=-8$ is a check.

\item $\mathcal{C}(\chi(2,1,4))-\mathcal{C}(\chi(1,2,4))=-4/3-(-20/3)=16/3$ \\
This allows to compute $\epsilon_{3c}$, since from $8i\epsilon_{3b}+4i\epsilon_{3c}=16/3$ we obtain $\epsilon_{3c}=i4/3$

\item $\mathcal{C}(\chi(1,4))=-4$ is a check.\\

\end{itemize}

\renewcommand{\thesubfigure}{}

\begin{figure}[!h]
\unitlength=0.75mm
\settoheight{\eqoff}{$\times$}%
\setlength{\eqoff}{0.5\eqoff}%
\addtolength{\eqoff}{-12.5\unitlength}%
\settoheight{\eqofftwo}{$\times$}%
\setlength{\eqofftwo}{0.5\eqofftwo}%
\addtolength{\eqofftwo}{-7.5\unitlength}%
\begin{equation*}
\begin{aligned}
F_{A1}=&
\subfigure[$\chi(1,2,3,4)\ \ \ $]{
\raisebox{\eqoff}{%
\fmfframe(3,1)(1,4){%
\begin{fmfchar*}(25,20)
\fmftop{v1}
\fmfbottom{v5}
\fmfforce{(0w,h)}{v1}
\fmfforce{(0w,0)}{v5}
\fmffixed{(0.2w,0)}{v1,v2}
\fmffixed{(0.2w,0)}{v2,v3}
\fmffixed{(0.2w,0)}{v3,v4}
\fmffixed{(0.2w,0)}{v4,v9}
\fmffixed{(0.2w,0)}{v5,v6}
\fmffixed{(0.2w,0)}{v6,v7}
\fmffixed{(0.2w,0)}{v7,v8}
\fmffixed{(0.2w,0)}{v8,v10}
\fmffixed{(0,whatever)}{vc1,vc2}
\fmffixed{(0,whatever)}{vc3,vc4}
\fmffixed{(0,whatever)}{vc5,vc6}
\fmffixed{(0,whatever)}{vc7,vc8}
\fmf{plain,tension=0.25,right=0.25}{v1,vc1}
\fmf{plain,tension=0.25,left=0.25}{v2,vc1}
\fmf{plain,left=0.25}{v5,vc2}
\fmf{plain,tension=1,left=0.25}{v3,vc3}
\fmf{plain,tension=1,left=0.25}{v4,vc5}
\fmf{plain,tension=1,left=0.25}{v9,vc7}
\fmf{plain,left=0.25}{v7,vc6}
\fmf{plain,tension=0.25,left=0.25}{v8,vc8}
\fmf{plain,tension=0.25,right=0.25}{v10,vc8}
\fmf{plain,left=0.25}{v6,vc4}
\fmf{plain,tension=0.5}{vc1,vc2}
\fmf{plain,tension=0.5}{vc2,vc3}
\fmf{plain,tension=0.5}{vc3,vc4}
\fmf{plain,tension=0.5}{vc4,vc5}
\fmf{plain,tension=0.5}{vc5,vc6}
\fmf{plain,tension=0.5}{vc6,vc7}
\fmf{plain,tension=0.5}{vc7,vc8}
\fmf{plain,tension=0.5,right=0,width=1mm}{v5,v10}
\fmfposition
\end{fmfchar*}}}
}
\;{\quad}&
F_{A2}=&
\subfigure[$\chi(3,2,1,4)\ \ \ $]{
\raisebox{\eqoff}{%
\fmfframe(3,1)(1,4){%
\begin{fmfchar*}(25,20)
\fmftop{v1}
\fmfbottom{v5}
\fmfforce{(0w,h)}{v1}
\fmfforce{(0w,0)}{v5}
\fmffixed{(0.2w,0)}{v1,v2}
\fmffixed{(0.2w,0)}{v2,v3}
\fmffixed{(0.2w,0)}{v3,v4}
\fmffixed{(0.2w,0)}{v4,v9}
\fmffixed{(0.2w,0)}{v5,v6}
\fmffixed{(0.2w,0)}{v6,v7}
\fmffixed{(0.2w,0)}{v7,v8}
\fmffixed{(0.2w,0)}{v8,v10}
\fmffixed{(0,whatever)}{vb1,vc2}
\fmffixed{(0,whatever)}{vc3,vb4}
\fmffixed{(0,whatever)}{vb7,vc8}
\fmf{plain,tension=0.25,left=0.25}{v5,vc2}
\fmf{plain,tension=0.25,right=0.25}{v6,vc2}
\fmf{phantom,tension=0.25,right=0.25}{v1,vb1}
\fmf{phantom,tension=0.25,left=0.25}{v2,vb1}
\fmf{phantom,tension=0.5}{vc2,vb1}
\fmf{plain,tension=0.25,left=0.25}{v8,vc8}
\fmf{plain,tension=0.25,right=0.25}{v10,vc8}
\fmf{phantom,tension=0.25,right=0.25}{v4,vb7}
\fmf{phantom,tension=0.25,left=0.25}{v9,vb7}
\fmf{phantom,tension=0.5}{vb7,vc8}
\fmf{plain,tension=0.25,right=0.25}{v3,vc3}
\fmf{plain,tension=0.25,left=0.25}{v4,vc3}
\fmf{phantom,tension=0.25,left=0.25}{v7,vb4}
\fmf{phantom,tension=0.25,right=0.25}{v8,vb4}
\fmf{phantom,tension=0.2}{vc3,vb4}
\fmffreeze
\fmffixed{(0,whatever)}{vc1,vc2}
\fmffixed{(0,whatever)}{vc3,vc4}
\fmffixed{(0,whatever)}{vc5,vc6}
\fmffixed{(0,whatever)}{vc7,vc8}
\fmffixed{(whatever,0)}{vc7,vc5}
\fmf{plain,tension=0.5,right=0.25}{v1,vc1}
\fmf{plain,tension=0.15,right=0.25}{v2,vc5}
\fmf{plain,tension=0.5,left=0.25}{v9,vc7}
\fmf{plain,tension=0.25}{vc3,vc4}
\fmf{plain,tension=0.5}{vc1,vc2}
\fmf{plain,tension=0.5}{vc6,vc1}
\fmf{plain,tension=0.5}{vc7,vc4}
\fmf{plain,tension=0.5}{vc4,vc5}
\fmf{plain,tension=0.5}{vc5,vc6}
\fmf{plain,tension=0.5,left=0.25}{vc6,v7}
\fmf{plain,tension=1}{vc7,vc8}
\fmf{plain,tension=0.5,right=0,width=1mm}{v5,v10}
\fmfposition
\fmfipath{p[]}
\end{fmfchar*}}}
}
\;{\quad}&
F_{A3}=&
\subfigure[$\chi(1,4,3,2)\ \ \ $]{
\raisebox{\eqoff}{%
\fmfframe(3,1)(1,4){%
\begin{fmfchar*}(25,20)
\fmftop{v1}
\fmfbottom{v5}
\fmfforce{(0w,h)}{v1}
\fmfforce{(0w,0)}{v5}
\fmffixed{(0.2w,0)}{v1,v2}
\fmffixed{(0.2w,0)}{v2,v3}
\fmffixed{(0.2w,0)}{v3,v4}
\fmffixed{(0.2w,0)}{v4,v9}
\fmffixed{(0.2w,0)}{v5,v6}
\fmffixed{(0.2w,0)}{v6,v7}
\fmffixed{(0.2w,0)}{v7,v8}
\fmffixed{(0.2w,0)}{v8,v10}
\fmffixed{(0,whatever)}{vb1,vc1}
\fmffixed{(0,whatever)}{vb2,vc2}
\fmffixed{(0,whatever)}{vb7,vc7}
\fmf{plain,tension=0.25,right=0.25}{v1,vc1}
\fmf{plain,tension=0.25,left=0.25}{v2,vc1}
\fmf{phantom,tension=0.25,right=0.25}{v5,vb1}
\fmf{phantom,tension=0.25,left=0.25}{v6,vb1}
\fmf{phantom,tension=0.5}{vc1,vb1}
\fmf{plain,tension=0.25,left=0.25}{v6,vc2}
\fmf{plain,tension=0.25,right=0.25}{v7,vc2}
\fmf{phantom,tension=0.25,right=0.25}{v2,vb2}
\fmf{phantom,tension=0.25,left=0.25}{v3,vb2}
\fmf{phantom,tension=0.5}{vc2,vb2}
\fmf{plain,tension=0.25,right=0.25}{v4,vc7}
\fmf{plain,tension=0.25,left=0.25}{v9,vc7}
\fmf{phantom,tension=0.25,right=0.25}{v8,vb7}
\fmf{phantom,tension=0.25,left=0.25}{v10,vb7}
\fmf{phantom,tension=0.5}{vc7,vb7}
\fmffreeze
\fmffixed{(0,whatever)}{vc1,vc3}
\fmffixed{(0,whatever)}{vc2,vc4}
\fmffixed{(0,whatever)}{vc5,vc6}
\fmffixed{(0,whatever)}{vc7,vc8}
\fmffixed{(whatever,0)}{vc3,vc5}
\fmf{plain,tension=1}{vc1,vc3}
\fmf{plain,tension=0.25,left=0.25}{v5,vc3}
\fmf{plain,tension=0.5}{vc3,vc4}
\fmf{plain,tension=0.5}{vc2,vc4}
\fmf{plain,tension=0.5}{vc4,vc5}
\fmf{plain,tension=0.25,left=0.25}{vc5,v8}
\fmf{plain,tension=1}{vc5,vc6}
\fmf{plain,tension=1,right=0.25}{v3,vc6}
\fmf{plain,tension=0.5,right=0.25}{v10,vc8}
\fmf{plain,tension=0.5}{vc6,vc8}
\fmf{plain,tension=0.5}{vc7,vc8}
\fmf{plain,tension=0.5,right=0,width=1mm}{v5,v10}
\fmfposition
\fmfipath{p[]}
\end{fmfchar*}}}
}
\\
F_{A4}=&
\subfigure[$\chi(2,4,1,3)\ \ \ $]{
\raisebox{\eqoff}{%
\fmfframe(3,1)(1,4){%
\begin{fmfchar*}(25,20)
\fmftop{v1}
\fmfbottom{v5}
\fmfforce{(0w,h)}{v1}
\fmfforce{(0w,0)}{v5}
\fmffixed{(0.2w,0)}{v1,v2}
\fmffixed{(0.2w,0)}{v2,v3}
\fmffixed{(0.2w,0)}{v3,v4}
\fmffixed{(0.2w,0)}{v4,v9}
\fmffixed{(0.2w,0)}{v5,v6}
\fmffixed{(0.2w,0)}{v6,v7}
\fmffixed{(0.2w,0)}{v7,v8}
\fmffixed{(0.2w,0)}{v8,v10}
\fmffixed{(0,whatever)}{vb2,vc2}
\fmffixed{(0,whatever)}{vb3,vc3}
\fmffixed{(0,whatever)}{vb5,vc5}
\fmffixed{(0,whatever)}{vb7,vc7}
\fmf{plain,tension=0.25,left=0.25}{v5,vc2}
\fmf{plain,tension=0.25,right=0.25}{v6,vc2}
\fmf{phantom,tension=0.25,right=0.25}{v1,vb2}
\fmf{phantom,tension=0.25,left=0.25}{v2,vb2}
\fmf{phantom,tension=0.5}{vc2,vb2}
\fmf{plain,tension=0.25,right=0.25}{v2,vc3}
\fmf{plain,tension=0.25,left=0.25}{v3,vc3}
\fmf{phantom,tension=0.25,right=0.25}{v6,vb3}
\fmf{phantom,tension=0.25,left=0.25}{v7,vb3}
\fmf{phantom,tension=0.5}{vc3,vb3}
\fmf{plain,tension=0.25,left=0.25}{v7,vc5}
\fmf{plain,tension=0.25,right=0.25}{v8,vc5}
\fmf{phantom,tension=0.25,right=0.25}{v3,vb5}
\fmf{phantom,tension=0.25,left=0.25}{v4,vb5}
\fmf{phantom,tension=0.5}{vc5,vb5}
\fmf{plain,tension=0.25,right=0.25}{v4,vc7}
\fmf{plain,tension=0.25,left=0.25}{v9,vc7}
\fmf{phantom,tension=0.25,right=0.25}{v8,vb7}
\fmf{phantom,tension=0.25,left=0.25}{v10,vb7}
\fmf{phantom,tension=0.5}{vc7,vb7}
\fmffreeze
\fmffixed{(0,whatever)}{vc1,vc2}
\fmffixed{(0,whatever)}{vc3,vc4}
\fmffixed{(0,whatever)}{vc5,vc6}
\fmffixed{(0,whatever)}{vc7,vc8}
\fmffixed{(whatever,0)}{vc1,vc6}
\fmffixed{(whatever,0)}{vc4,vc8}
\fmf{plain,tension=0.25,right=0.25}{v1,vc1}
\fmf{plain,tension=0.5}{vc1,vc2}
\fmf{plain,tension=0.5}{vc3,vc4}
\fmf{plain,tension=0.5}{vc1,vc4}
\fmf{plain,tension=0.5}{vc5,vc6}
\fmf{plain,tension=0.5}{vc4,vc6}
\fmf{plain,tension=0.25,right=0.25}{v10,vc8}
\fmf{plain,tension=0.5}{vc7,vc8}
\fmf{plain,tension=0.5}{vc6,vc8}
\fmf{plain,tension=0.5,right=0,width=1mm}{v5,v10}
\fmfposition
\fmfipath{p[]}
\end{fmfchar*}}}
}
\;{\quad}&
F_{A5}=&
\subfigure[$\chi(2,1,3,2)$]{
\raisebox{\eqoff}{%
\fmfframe(3,1)(1,4){%
\begin{fmfchar*}(20,20)
\fmftop{v1}
\fmfbottom{v5}
\fmfforce{(0.125w,h)}{v1}
\fmfforce{(0.125w,0)}{v5}
\fmffixed{(0.25w,0)}{v1,v2}
\fmffixed{(0.25w,0)}{v2,v3}
\fmffixed{(0.25w,0)}{v3,v4}
\fmffixed{(0.25w,0)}{v5,v6}
\fmffixed{(0.25w,0)}{v6,v7}
\fmffixed{(0.25w,0)}{v7,v8}
\fmffixed{(0,whatever)}{vc1,vc5}
\fmffixed{(0,whatever)}{vc2,vc3}
\fmffixed{(0,whatever)}{vc3,vc6}
\fmffixed{(0,whatever)}{vc6,vc7}
\fmffixed{(0,whatever)}{vc4,vc8}
\fmffixed{(0.5w,0)}{vc1,vc4}
\fmffixed{(0.5w,0)}{vc5,vc8}
\fmf{plain,tension=1,right=0.125}{v1,vc1}
\fmf{plain,tension=0.25,right=0.25}{v2,vc2}
\fmf{plain,tension=0.25,left=0.25}{v3,vc2}
\fmf{plain,tension=1,left=0.125}{v4,vc4}
\fmf{plain,tension=1,left=0.125}{v5,vc5}
\fmf{plain,tension=0.25,left=0.25}{v6,vc6}
\fmf{plain,tension=0.25,right=0.25}{v7,vc6}
\fmf{plain,tension=1,right=0.125}{v8,vc8}
  \fmf{plain,tension=0.5}{vc1,vc3}
  \fmf{plain,tension=0.5}{vc2,vc3}
  \fmf{plain,tension=0.5}{vc3,vc4}
  \fmf{plain,tension=0.5}{vc5,vc7}
  \fmf{plain,tension=0.5}{vc6,vc7}
  \fmf{plain,tension=0.5}{vc7,vc8}
  \fmf{plain,tension=2}{vc1,vc5}
  \fmf{plain,tension=2}{vc4,vc8}
  \fmf{phantom,tension=2}{vc5,vc4}
\fmffreeze
\fmfposition
\fmf{plain,tension=1,left=0,width=1mm}{v5,v8}
\fmffreeze
\end{fmfchar*}}}
}
\end{aligned}
\end{equation*}
\caption{Diagrams with only chiral interactions}
\label{r5diagrams-chiral}
%
\vspace{2cm}
\unitlength=0.75mm
\settoheight{\eqoff}{$\times$}%
\setlength{\eqoff}{0.5\eqoff}%
\addtolength{\eqoff}{-12.5\unitlength}%
\settoheight{\eqofftwo}{$\times$}%
\setlength{\eqofftwo}{0.5\eqofftwo}%
\addtolength{\eqofftwo}{-7.5\unitlength}%
\begin{equation*}
\begin{aligned}
F_{B1}=&
\raisebox{\eqoff}{%
\fmfframe(3,1)(1,4){%
\begin{fmfchar*}(25,20)
\Wsevenplain
\fmfipair{w[]}
\svertex{w1}{p10}
\svertex{w2}{p6}
\fmfi{wiggly}{w1..w2}
\end{fmfchar*}}}
\;{\quad}&F_{B2}=&
\raisebox{\eqoff}{%
\fmfframe(3,1)(1,4){%
\begin{fmfchar*}(25,20)
\Wsevenplain
\fmfipair{w[]}
\svertex{w1}{p10}
\svertex{w2}{p5}
\fmfi{wiggly}{w1..w2}
\end{fmfchar*}}}
\;{\quad}&F_{B3}=&
\raisebox{\eqoff}{%
\fmfframe(3,1)(1,4){%
\begin{fmfchar*}(25,20)
\Wsevenplain
\fmfipair{w[]}
\svertex{w1}{p10}
\svertex{w2}{p4}
\fmfi{wiggly}{w1..w2}
\end{fmfchar*}}}
\\
F_{B4}=&
\raisebox{\eqoff}{%
\fmfframe(3,1)(1,4){%
\begin{fmfchar*}(25,20)
\Wsevenplain
\fmfipair{w[]}
\svertex{w1}{p9}
\svertex{w2}{p6}
\fmfi{wiggly}{w1..w2}
\end{fmfchar*}}}
\;{\quad}&F_{B5}=&
\raisebox{\eqoff}{%
\fmfframe(3,1)(1,4){%
\begin{fmfchar*}(25,20)
\Wsevenplain
\fmfipair{w[]}
\svertex{w1}{p9}
\svertex{w2}{p5}
\fmfi{wiggly}{w1..w2}
\end{fmfchar*}}}
\;{\quad}&F_{B6}=&
\raisebox{\eqoff}{%
\fmfframe(3,1)(1,4){%
\begin{fmfchar*}(25,20)
\Wsevenplain
\fmfipair{w[]}
\svertex{w1}{p9}
\svertex{w2}{p4}
\fmfi{wiggly}{w1..w2}
\end{fmfchar*}}}
\\
F_{B7}=&
\raisebox{\eqoff}{%
\fmfframe(3,1)(1,4){%
\begin{fmfchar*}(25,20)
\Wsevenplain
\fmfipair{w[]}
\svertex{w1}{p8}
\svertex{w2}{p6}
\fmfi{wiggly}{w1..w2}
\end{fmfchar*}}}
\;{\quad}&F_{B8}=&
\raisebox{\eqoff}{%
\fmfframe(3,1)(1,4){%
\begin{fmfchar*}(25,20)
\Wsevenplain
\fmfipair{w[]}
\svertex{w1}{p8}
\svertex{w2}{p5}
\fmfi{wiggly}{w1..w2}
\end{fmfchar*}}}
\;{\quad}&F_{B9}=&
\raisebox{\eqoff}{%
\fmfframe(3,1)(1,4){%
\begin{fmfchar*}(25,20)
\Wsevenplain
\fmfipair{w[]}
\svertex{w1}{p8}
\svertex{w2}{p4}
\fmfi{wiggly}{w1..w2}
\end{fmfchar*}}}
\end{aligned}
\end{equation*}
\begin{myfigure}
Diagrams with structure $\chi(2,1,4)$
\label{r5diagrams-214}
\end{myfigure}
\vspace{1cm}
\begin{center}
\fbox{
\small
\begin{tabular}{|m{35pt}m{30pt}|m{6pt}|}
\hline
& & \\[-2ex]
$F_{B1\phantom{0}}\rightarrow$ & $\ast\phantom{W_{A}}$ & 1  \\
$F_{B2\phantom{0}}\rightarrow$ & $\ast\phantom{W_{A}}$ & 1  \\
$F_{B3\phantom{0}}\rightarrow$ & $-F_{B6}$ &  \\
\hline
\end{tabular}
\begin{tabular}{|m{35pt}m{30pt}|m{6pt}|}
\hline
& & \\[-2ex]
$F_{B4\phantom{0}}\rightarrow$ & $\ast\phantom{W_{A}}$ & 1 \\
$F_{B5\phantom{0}}\rightarrow$ & finite &  \\
$F_{B6\phantom{0}}\rightarrow$ & $-F_{B3}$ &  \\
\hline
\end{tabular}
\begin{tabular}{|m{35pt}m{30pt}|m{6pt}|}
\hline
& & \\[-2ex]
$F_{B7\phantom{0}}\rightarrow$ & $-F_{B8}$ &  \\
$F_{B8\phantom{0}}\rightarrow$ & $-F_{B7}$ &  \\
$F_{B9\phantom{0}}\rightarrow$ & finite &  \\
\hline
\end{tabular}}
\end{center}
\begin{mytable}
Results of $D$-algebra for diagrams with structure $\chi(2,1,4)$
\label{r5results-214}
\end{mytable}
\end{figure}
\normalsize

\clearpage

\begin{table}[p]
\begin{center}
\fbox{
\small
\begin{tabular}{|m{35pt}m{30pt}|m{6pt}|}
\hline
& & \\[-2ex]
$F_{C1\phantom{0}}\rightarrow$ & $\ast\phantom{W_{A}}$ & 1 \\
$F_{C2\phantom{0}}\rightarrow$ & $\ast\phantom{W_{A}}$ & 1 \\
$F_{C3\phantom{0}}\rightarrow$ & $-F_{C6}$ & \\
\hline
\end{tabular}
\begin{tabular}{|m{35pt}m{30pt}|m{6pt}|}
\hline
& & \\[-2ex]
$F_{C4\phantom{0}}\rightarrow$ & $\ast\phantom{W_{A}}$ & 1  \\
$F_{C5\phantom{0}}\rightarrow$ & finite &  \\
$F_{C6\phantom{0}}\rightarrow$ & $-F_{C3}$ & \\
\hline
\end{tabular}
\begin{tabular}{|m{35pt}m{35pt}|m{6pt}|}
\hline
& & \\[-2ex]
$F_{C7\phantom{0}}\rightarrow$ & $-F_{C8}$ &  \\
$F_{C8\phantom{0}}\rightarrow$ & $-F_{C7}$ &  \\
$F_{C9\phantom{0}}\rightarrow$ & finite & \\
\hline
\end{tabular}
}
\end{center}
\caption{Results of $D$-algebra for diagrams with structure $\chi(1,4,3)$}
\label{r5results-124}
\end{table}
\normalsize

\clearpage

\begin{figure}[p]
\unitlength=0.75mm
\settoheight{\eqoff}{$\times$}%
\setlength{\eqoff}{0.5\eqoff}%
\addtolength{\eqoff}{-12.5\unitlength}%
\settoheight{\eqofftwo}{$\times$}%
\setlength{\eqofftwo}{0.5\eqofftwo}%
\addtolength{\eqofftwo}{-7.5\unitlength}%
\begin{equation*}
\begin{aligned}
F_{D1\phantom{0}}=&
\raisebox{\eqoff}{%
\fmfframe(3,1)(1,4){%
\begin{fmfchar*}(25,20)
\Wnineplain
\fmfipair{w[]}
\svertex{w1}{p4}
\svertex{w2}{p3}
\svertex{w3}{p7}
\vvertex{w4}{w3}{p4}
\fmfi{wiggly}{w1..w2}
\fmfi{wiggly}{w3..w4}
\end{fmfchar*}}}
\;{\quad}&F_{D2\phantom{0}}=&
\raisebox{\eqoff}{%
\fmfframe(3,1)(1,4){%
\begin{fmfchar*}(25,20)
\Wnineplain
\fmfipair{w[]}
\svertex{w1}{p4}
\svertex{w2}{p3}
\svertex{w3}{p7}
\svertex{w5}{p6}
\vvertex{w4}{w3}{p4}
\fmfi{wiggly}{w1..w2}
\fmfi{wiggly}{w5..w4}
\end{fmfchar*}}}
\;{\quad}&F_{D3\phantom{0}}=&
\raisebox{\eqoff}{%
\fmfframe(3,1)(1,4){%
\begin{fmfchar*}(25,20)
\Wnineplain
\fmfipair{w[]}
\svertex{w1}{p4}
\svertex{w2}{p3}
\svertex{w3}{p7}
\svertex{w5}{p5}
\vvertex{w4}{w3}{p4}
\fmfi{wiggly}{w1..w2}
\fmfi{wiggly}{w5..w4}
\end{fmfchar*}}}
\;{\quad}&F_{D4\phantom{0}}=&
\raisebox{\eqoff}{%
\fmfframe(3,1)(1,4){%
\begin{fmfchar*}(25,20)
\Wnineplain
\fmfipair{w[]}
\svertex{w1}{p4}
\svertex{w2}{p3}
\svertex{w3}{p7}
\svertex{w5}{p6}
\vvertex{w4}{w3}{p4}
\fmfi{wiggly}{w2..w4}
\fmfi{wiggly}{w3..w4}
\end{fmfchar*}}}
\\
F_{D5\phantom{0}}=&
\raisebox{\eqoff}{%
\fmfframe(3,1)(1,4){%
\begin{fmfchar*}(25,20)
\Wnineplain
\fmfipair{w[]}
\svertex{w1}{p4}
\svertex{w2}{p3}
\svertex{w3}{p6}
\fmfi{wiggly}{w1..w2}
\fmfi{wiggly}{w3..w1}
\end{fmfchar*}}}
\;{\quad}&F_{D6\phantom{0}}=&
\raisebox{\eqoff}{%
\fmfframe(3,1)(1,4){%
\begin{fmfchar*}(25,20)
\Wnineplain
\fmfipair{w[]}
\svertex{w1}{p4}
\svertex{w2}{p3}
\svertex{w3}{p5}
\fmfi{wiggly}{w1..w2}
\fmfi{wiggly}{w1..w3}
\end{fmfchar*}}}
\;{\quad}&F_{D7\phantom{0}}=&
\raisebox{\eqoff}{%
\fmfframe(3,1)(1,4){%
\begin{fmfchar*}(25,20)
\Wnineplain
\fmfipair{w[]}
\svertex{w1}{p4}
\svertex{w2}{p3}
\svertex{w3}{p7}
\svertex{w5}{p6}
\vvertex{w4}{w2}{p4}
\fmfi{wiggly}{w1..w5}
\fmfi{wiggly}{w2..w4}
\end{fmfchar*}}}
\;{\quad}&F_{D8\phantom{0}}=&
\raisebox{\eqoff}{%
\fmfframe(3,1)(1,4){%
\begin{fmfchar*}(25,20)
\Wnineplain
\fmfipair{w[]}
\svertex{w1}{p4}
\svertex{w2}{p3}
\svertex{w3}{p7}
\svertex{w5}{p5}
\vvertex{w4}{w2}{p4}
\fmfi{wiggly}{w1..w5}
\fmfi{wiggly}{w2..w4}
\end{fmfchar*}}}
\\
F_{D9\phantom{0}}=&
\raisebox{\eqoff}{%
\fmfframe(3,1)(1,4){%
\begin{fmfchar*}(25,20)
\Wnineplain
\fmfipair{w[]}
\svertex{w1}{p4}
\svertex{w2}{p3}
\svertex{w3}{p2}
\svertex{w5}{p6}
\vvertex{w4}{w2}{p4}
\fmfi{wiggly}{w1..w3}
\fmfi{wiggly}{w5..w4}
\end{fmfchar*}}}
\;{\quad}&F_{D10}=&
\raisebox{\eqoff}{%
\fmfframe(3,1)(1,4){%
\begin{fmfchar*}(25,20)
\Wnineplain
\fmfipair{w[]}
\svertex{w1}{p4}
\svertex{w2}{p3}
\svertex{w3}{p2}
\svertex{w5}{p5}
\vvertex{w4}{w2}{p4}
\fmfi{wiggly}{w1..w3}
\fmfi{wiggly}{w5..w4}
\end{fmfchar*}}}
\;{\quad}&F_{D11}=&
\raisebox{\eqoff}{%
\fmfframe(3,1)(1,4){%
\begin{fmfchar*}(25,20)
\Wnineplain
\fmfipair{w[]}
\svertex{w1}{p4}
\svertex{w2}{p3}
\svertex{w3}{p2}
\svertex{w5}{p6}
\vvertex{w4}{w2}{p4}
\fmfi{wiggly}{w1..w3}
\fmfi{wiggly}{w5..w1}
\end{fmfchar*}}}
\;{\quad}&F_{D12}=&
\raisebox{\eqoff}{%
\fmfframe(3,1)(1,4){%
\begin{fmfchar*}(25,20)
\Wnineplain
\fmfipair{w[]}
\svertex{w1}{p4}
\svertex{w2}{p3}
\svertex{w3}{p2}
\svertex{w5}{p5}
\vvertex{w4}{w2}{p4}
\fmfi{wiggly}{w1..w3}
\fmfi{wiggly}{w5..w1}
\end{fmfchar*}}}
\\
F_{D13}=&
\raisebox{\eqoff}{%
\fmfframe(3,1)(1,4){%
\begin{fmfchar*}(25,20)
\Wnineplain
\fmfipair{w[]}
\svertex{w1}{p4}
\svertex{w2}{p3}
\svertex{w3}{p2}
\svertex{w5}{p5}
\vvertex{w4}{w5}{p4}
\fmfi{wiggly}{w1..w3}
\fmfi{wiggly}{w5..w4}
\end{fmfchar*}}}
\;{\quad}&F_{D14}=&
\raisebox{\eqoff}{%
\fmfframe(3,1)(1,4){%
\begin{fmfchar*}(25,20)
\Wnineplain
\fmfipair{w[]}
\svertex{w1}{p1}
\svertex{w2}{p5}
\vvertex{w3}{w1}{p4}
\fmfi{wiggly}{w1..w3}
\fmfi{wiggly}{w2..w3}
\end{fmfchar*}}}
\end{aligned}
\end{equation*}
\caption{Diagrams with structure $\chi(1,4)$}
\label{r5diagrams-14}
\end{figure}

\begin{table}[p]
\begin{center}
\small
\fbox{
\begin{tabular}{|m{35pt}m{30pt}|m{6pt}|}
\hline
& & \\[-2ex]
$F_{D1\phantom{0}}\rightarrow$ & $-F_{D7}$ &  \\
$F_{D2\phantom{0}}\rightarrow$ & $-F_{D3}$ &  \\
$F_{D3\phantom{0}}\rightarrow$ & $-F_{D2}$ &  \\
$F_{D4\phantom{0}}\rightarrow$ & $\ast\phantom{W_{A}}$ & 1 \\
$F_{D5\phantom{0}}\rightarrow$ & $\ast\phantom{W_{A}}$ & 2 \\
\hline
\end{tabular}
\begin{tabular}{|m{35pt}m{30pt}|m{6pt}|}
\hline
& & \\[-2ex]
$F_{D6\phantom{0}}\rightarrow$ & $-F_{D12}$ &  \\
$F_{D7\phantom{0}}\rightarrow$ & $-F_{D1}$ &  \\
$F_{D8\phantom{0}}\rightarrow$ & finite &  \\
$F_{D9\phantom{0}}\rightarrow$ & $-F_{D10}$ &  \\
$F_{D10}\rightarrow$ & $-F_{D9}$ &  \\
\hline
\end{tabular}
\begin{tabular}{|m{35pt}m{30pt}|m{6pt}|}
\hline
& & \\[-2ex]
$F_{D11}\rightarrow$ & finite & \\
$F_{D12}\rightarrow$ & $-F_{D6}$ &  \\
$F_{D13}\rightarrow$ & finite & \\
$F_{D14}\rightarrow$ & finite &  \\
& &  \\
\hline
\end{tabular}}
\end{center}
\caption{Results of $D$-algebra for diagrams with structure $\chi(1,4)$}
\label{r5results-14}
\end{table}
\normalsize

\begin{table}[p]
\begin{center}
\small
\begin{tabular}{|ll|}
\hline
& \\[-2ex]
$F_{A1\phantom{0}}\rightarrow$ & $(g^2 N)^4 J_{1}\ \chi(1,2,3,4) $  \\
$F_{A2\phantom{0}}\rightarrow$ & $(g^2 N)^4 J_{2}\ \chi(3,2,1,4) $  \\
$F_{A3\phantom{0}}\rightarrow$ & $(g^2 N)^4 J_{3}\ \chi(1,4,3,2)$  \\
$F_{A4\phantom{0}}\rightarrow$ & $(g^2 N)^4 J_{4}\ \chi(2,4,1,3)$  \\
$F_{A5\phantom{0}}\rightarrow$ & $(g^2 N)^4 J_{5}\ \chi(2,1,3,2)$  \\
[1ex]
\hline
\hline
& \\[-2ex]
$F_{B1\phantom{0}}\rightarrow$ & $(g^2 N)^4 (J_1+J_{2}-2J_{6})\ \chi(2,1,4) $  \\
$F_{B2\phantom{0}}\rightarrow$ & $-(g^2 N)^4 J_{1}\ \chi(2,1,4) $  \\
$F_{B4\phantom{0}}\rightarrow$ & $-(g^2 N)^4 J_{2}\ \chi(2,1,4)$  \\
[1ex]
\hline
\multicolumn{2}{|c|}{}\\[-1ex]
\multicolumn{2}{|c|}{
$\sum F_{B**}
\rightarrow -2(g^2 N)^4 J_{6}\ \chi(2,1,4)
$
} \\[1ex]
\hline
\hline
& \\[-2ex]
$F_{C1\phantom{0}}\rightarrow$ & $(g^2 N)^4 (J_1+J_{3}-2J_{7})\ \chi(1,2,4)$  \\
$F_{C2\phantom{0}}\rightarrow$ & $-(g^2 N)^4 J_1\ \chi(1,2,4)$  \\
$F_{C4\phantom{0}}\rightarrow$ & $-(g^2 N)^4 J_{3}\ \chi(1,2,4)$  \\
[1ex]
\hline
\multicolumn{2}{|c|}{}\\[-1ex]
\multicolumn{2}{|c|}{
$\sum F_{C**}
\rightarrow -2(g^2 N)^4 J_{7}\ \chi(1,2,4)
$
} \\[1ex]
\hline
\hline
& \\[-2ex]
$F_{D4\phantom{0}}\rightarrow$ & $-2(g^2 N)^4 (J_1+J_8)\ \chi(1,4)$  \\
$F_{D5\phantom{0}}\rightarrow$ & $2(g^2 N)^4 J_1\ \chi(1,4)$  \\
[1ex]
\hline
\multicolumn{2}{|c|}{}\\[-1ex]
\multicolumn{2}{|c|}{
$\sum F_{D**}
\rightarrow -2(g^2 N)^4 J_8\ \chi(1,4)
$
} \\[1ex]
\hline
\end{tabular}
\end{center}
\caption{Summary of $D$-algebra results.}
\label{results-r5}
\end{table}

\begin{table}[p]
\settoheight{\eqoff}{$\times$}%
\setlength{\eqoff}{0.5\eqoff}%
\addtolength{\eqoff}{-7.5\unitlength}
\begin{equation*}
\begin{aligned}
J_1=
\raisebox{\eqoff}{%
\begin{fmfchar*}(20,15)
\fmfleft{in}
\fmfright{out}
\fmf{plain}{in,v1}
\fmf{plain,left=0.25}{v1,v2}
\fmf{plain,left=0}{v2,v4}
\fmf{plain,left=0.25}{v4,v3}
\fmf{plain,tension=0.5,right=0.25}{v1,v0,v1}
\fmf{plain,right=0.25}{v0,v3}
\fmf{plain}{v0,v2}
\fmf{plain}{v0,v4}
\fmf{plain}{v3,out}
\fmffixed{(0.9w,0)}{v1,v3}
\fmffixed{(0.4w,0)}{v2,v4}
\fmfpoly{phantom}{v4,v2,v0}
\fmffreeze
\end{fmfchar*}}
&=\frac{1}{(4\pi)^8}\Big(
-\frac{1}{24\varepsilon^4}+\frac{1}{4\varepsilon^3}
-\frac{19}{24\varepsilon^2}
+\frac{5}{4\varepsilon}
\Big)
\\
J_2=
\raisebox{\eqoff}{%
\begin{fmfchar*}(20,15)
\fmfleft{in}
\fmfright{out}
\fmf{plain}{in,v1}
\fmf{plain,left=0}{v1,v2}
\fmf{plain,left=0}{v2,v4}
\fmf{plain,left=0}{v3,v4}
\fmf{plain,left=0}{v0,v2}
\fmf{plain,tension=0.35,right=0.25}{v1,v0,v1}
\fmf{plain,tension=0.25,right=0.25}{v3,v0,v3}
\fmf{plain}{v4,out}
\fmffixed{(0.9w,0)}{v1,v4}
\fmffixed{(whatever,0.5h)}{v0,v2}
\fmffreeze
\end{fmfchar*}}
&=\frac{1}{(4\pi)^8}\Big(
-\frac{1}{8\varepsilon^4}+\frac{1}{3\varepsilon^3}
-\frac{5}{24\varepsilon^2}
-\frac{1}{3\varepsilon}
\Big)\\
J_3=
\raisebox{\eqoff}{%
\begin{fmfchar*}(20,15)
\fmfleft{in}
\fmfright{out}
\fmf{plain}{in,v1}
\fmf{plain,left=0}{v1,v2}
\fmf{plain,tension=0.35,left=0}{v2,v3}
\fmf{plain,left=0}{v3,v4}
\fmf{plain,left=0}{v0,v1}
\fmf{plain,tension=0.35,left=0}{v0,v2}
\fmf{plain,left=0}{v0,v4}
\fmf{plain,tension=0.35,right=0.25}{v3,v0,v3}
\fmf{plain}{v4,out}
\fmffixed{(0.9w,0)}{v1,v4}
\fmffixed{(whatever,0.5h)}{v0,v3}
\fmffixed{(whatever,0.4h)}{v0,v2}
\fmffreeze
\end{fmfchar*}}
&=\frac{1}{(4\pi)^8}\Big(
-\frac{1}{8\varepsilon^4}+\frac{1}{2\varepsilon^3}
-\frac{7}{8\varepsilon^2}
+\frac{1}{3\varepsilon}
\Big)\\
J_4=
\raisebox{\eqoff}{%
\begin{fmfchar*}(20,15)
\fmfleft{in}
\fmfright{out}
\fmf{plain}{in,v1}
\fmf{plain,tension=0.5,left=0}{v1,v2}
\fmf{plain,tension=0,left=0}{v2,v4}
\fmf{plain,tension=0.5,left=0}{v1,v3}
\fmf{plain,left=0}{v0,v4}
\fmf{plain,tension=0.35,right=0.5}{v3,v0,v3}
\fmf{plain,tension=0.35,right=0.5}{v2,v0,v2}
\fmf{plain}{v4,out}
\fmffixed{(0.9w,0)}{v1,v4}
\fmffixed{(0,0.5h)}{v2,v3}
\fmffixed{(0,whatever)}{v0,v3}
\fmffreeze
\end{fmfchar*}}
&=\frac{1}{(4\pi)^8}\Big(
-\frac{5}{24\varepsilon^4}+\frac{5}{12\varepsilon^3}
+\frac{1}{24\varepsilon^2}
-\frac{1}{4\varepsilon}
\Big)\\
J_5=\raisebox{\eqoff}{%
\begin{fmfchar*}(20,15)
\fmfleft{in}
\fmfright{out}
\fmf{plain}{in,v1}
\fmf{plain,left=0.25}{v1,v2}
\fmf{plain,left=0.25}{v2,v3}
\fmf{plain,left=0.25}{v3,v4}
\fmf{plain,left=0.25}{v4,v1}
\fmf{plain,tension=0.5,right=0.25}{v1,v0,v1}
\fmf{phantom}{v0,v3}
\fmf{plain}{v2,v0}
\fmf{plain}{v0,v4}
\fmf{plain}{v3,out}
\fmffixed{(0.9w,0)}{v1,v3}
\fmffixed{(0,0.45w)}{v4,v2}
\fmffreeze
\end{fmfchar*}}
&=\frac{1}{(4\pi)^8}\Big(
-\frac{1}{12\varepsilon^4}+\frac{1}{3\varepsilon^3}
-\frac{5}{12\varepsilon^2}
-\frac{1}{\varepsilon}\Big(\frac{1}{2}-\zeta(3)\Big)\Big)
\\
J_6=
\settoheight{\eqoff}{$\times$}%
\setlength{\eqoff}{0.5\eqoff}%
\addtolength{\eqoff}{-7.5\unitlength}%
\raisebox{\eqoff}{%
\begin{fmfchar*}(20,15)
  \fmfleft{in}
  \fmfright{out}
  \fmf{plain}{in,v1}
  \fmf{phantom,tension=2,left=0.25}{v1,v2}
  \fmf{plain,tension=2,left=0.25}{v2,v3}
  \fmf{derplain,left=0.25}{v4,v1}
  \fmf{plain,left=0.25}{v0,v4}
  \fmf{plain,right=0}{v0,v1}
  \fmf{plain,right=0.25}{v0,v5}
  \fmf{plain,right=0.75}{v4,v5}
  \fmf{phantom,right=0}{v3,v0}
  \fmf{derplain,right=0.25}{v5,v3}
  \fmf{plain}{v3,out}
\fmffixed{(0.9w,0)}{v1,v3}
\fmfpoly{phantom}{v2,v4,v5}
\fmffixed{(0.5w,0)}{v4,v5}
\fmf{plain,tension=0.25,right=0.25}{v2,v0}
\fmf{plain,tension=0.25,right=0.25}{v0,v2}
\fmffreeze
\fmfshift{(0,0.15w)}{in,out,v1,v2,v3,v4,v5,v0}
\end{fmfchar*}}
&=\frac{1}{(4\pi)^8}\Big(
\frac{1}{12\varepsilon^2}
-\frac{1}{12\varepsilon}
\Big)\\
J_7=
\raisebox{\eqoff}{%
\begin{fmfchar*}(20,15)
  \fmfleft{in}
  \fmfright{out}
  \fmf{plain}{in,v1}
  \fmf{plain,right=0.25}{v1,v4}
  \fmf{plain,right=0.25}{v4,v0}
  \fmf{plain,right=0.25}{v0,v1}
  \fmf{plain,right=0.25}{v0,v5}
  \fmf{derplain,left=0.25}{v6,v4}
  \fmf{plain,left=0.25}{v5,v6}
  \fmf{plain,left=0.25}{v0,v3}
  \fmf{derplain,right=0.25}{v5,v3}
  \fmf{plain}{v3,out}
\fmffixed{(0.9w,0)}{v1,v3}
\fmfpoly{phantom}{v0,v4,v6,v5}
\fmffixed{(0.45w,0)}{v4,v5}
\fmf{plain}{v6,v0}
\fmffreeze
\fmfshift{(0,0.15w)}{in,out,v1,v2,v3,v4,v5,v6,v0}
\end{fmfchar*}}
&=\frac{1}{(4\pi)^8}\Big(
\frac{1}{4\varepsilon^2}
-\frac{5}{12\varepsilon}
\Big)\\
J_8
=\raisebox{\eqoff}{%
\begin{fmfchar*}(20,15)
  \fmfleft{in}
  \fmfright{out}
  \fmf{plain}{in,v1}
  \fmf{plain,right=0.25}{v1,v4}
  \fmf{plain,right=0.25}{v4,v0}
  \fmf{derplain,right=0.25}{v0,v1}
  \fmf{plain,right=0.25}{v0,v5}
  \fmf{plain,right=0.25}{v4,v6}
  \fmf{plain,left=0.25}{v5,v6}
  \fmf{derplain,left=0.25}{v0,v3}
  \fmf{plain,right=0.25}{v5,v3}
  \fmf{plain}{v3,out}
\fmffixed{(0.9w,0)}{v1,v3}
\fmfpoly{phantom}{v0,v4,v6,v5}
\fmffixed{(0.45w,0)}{v4,v5}
\fmf{plain}{v6,v0}
\fmffreeze
\fmfshift{(0,0.15w)}{in,out,v1,v2,v3,v4,v5,v6,v0}
\end{fmfchar*}}
&=\frac{1}{(4\pi)^8}\Big(
-\frac{1}{4\varepsilon}
\Big)
\end{aligned}
\end{equation*}
\caption{Loop integrals for range-five diagrams. The arrows indicate contracted 
spacetime derivatives.}
\label{integrals-r5}
\end{table}

\clearpage

\renewcommand{\thefigure}{C.\arabic{figure}}
\setcounter{figure}{0}
\renewcommand{\thetable}{C.\arabic{table}}
\setcounter{table}{0}

\newpage
\section{Four-loop wrapping diagrams with vectors}
\label{app:wrapping}

\vspace{3cm}
\begin{figure}[h]
\unitlength=0.75mm
\settoheight{\eqoff}{$\times$}%
\setlength{\eqoff}{0.5\eqoff}%
\addtolength{\eqoff}{-12.5\unitlength}%
\settoheight{\eqofftwo}{$\times$}%
\setlength{\eqofftwo}{0.5\eqofftwo}%
\addtolength{\eqofftwo}{-7.5\unitlength}%
\begin{equation*}
\begin{aligned}
W_{B1\phantom{0}}=&
\raisebox{\eqoff}{%
\fmfframe(3,1)(1,4){%
\begin{fmfchar*}(20,20)
\Wfourplain
\fmfipair{wu[]}
\fmfipair{w[]}
\fmfipair{wd[]}
\svertex{w6}{p6}
\vvertex{w7}{w6}{p3}
\wigglywrap{w7}{v5}{v8}{w6}
\end{fmfchar*}}}
\;{\quad}&W_{B2\phantom{0}}=&
\raisebox{\eqoff}{%
\fmfframe(3,1)(1,4){%
\begin{fmfchar*}(20,20)
\Wfourplain
\fmfipair{wu[]}
\fmfipair{w[]}
\fmfipair{wd[]}
\svertex{w5}{p5}
\svertex{w3}{p3}
\wigglywrap{w3}{v5}{v8}{w5}
\end{fmfchar*}}}
\;{\quad}&W_{B3\phantom{0}}=&
\raisebox{\eqoff}{%
\fmfframe(3,1)(1,4){%
\begin{fmfchar*}(20,20)
\Wfourplain
\fmfipair{wu[]}
\fmfipair{w[]}
\fmfipair{wd[]}
\svertex{w4}{p4}
\svertex{w3}{p3}
\wigglywrap{w3}{v5}{v8}{w4}
\end{fmfchar*}}}
\\
W_{B4\phantom{0}}=&
\raisebox{\eqoff}{%
\fmfframe(3,1)(1,4){%
\begin{fmfchar*}(20,20)
\Wfourplain
\fmfipair{wu[]}
\fmfipair{w[]}
\fmfipair{wd[]}
\svertex{w2}{p2}
\svertex{w6}{p6}
\wigglywrap{w2}{v5}{v8}{w6}
\end{fmfchar*}}}
\;{\quad}&W_{B5\phantom{0}}=&
\raisebox{\eqoff}{%
\fmfframe(3,1)(1,4){%
\begin{fmfchar*}(20,20)
\Wfourplain
\fmfipair{wu[]}
\fmfipair{w[]}
\fmfipair{wd[]}
\svertex{w2}{p2}
\svertex{w5}{p5}
\wigglywrap{w2}{v5}{v8}{w5}
\end{fmfchar*}}}
\;{\quad}&W_{B6\phantom{0}}=&
\raisebox{\eqoff}{%
\fmfframe(3,1)(1,4){%
\begin{fmfchar*}(20,20)
\Wfourplain
\fmfipair{wu[]}
\fmfipair{w[]}
\fmfipair{wd[]}
\svertex{w2}{p2}
\svertex{w4}{p4}
\wigglywrap{w2}{v5}{v8}{w4}
\end{fmfchar*}}}
\\
W_{B7\phantom{0}}=&
\raisebox{\eqoff}{%
\fmfframe(3,1)(1,4){%
\begin{fmfchar*}(20,20)
\Wfourplain
\fmfipair{wu[]}
\fmfipair{w[]}
\fmfipair{wd[]}
\svertex{w1}{p1}
\svertex{w6}{p6}
\wigglywrap{w1}{v5}{v8}{w6}
\end{fmfchar*}}}
\;{\quad}&W_{B8\phantom{0}}=&
\raisebox{\eqoff}{%
\fmfframe(3,1)(1,4){%
\begin{fmfchar*}(20,20)
\Wfourplain
\fmfipair{wu[]}
\fmfipair{w[]}
\fmfipair{wd[]}
\svertex{w1}{p1}
\svertex{w5}{p5}
\wigglywrap{w1}{v5}{v8}{w5}
\end{fmfchar*}}}
\;{\quad}&W_{B9\phantom{0}}=&
\raisebox{\eqoff}{%
\fmfframe(3,1)(1,4){%
\begin{fmfchar*}(20,20)
\Wfourplain
\fmfipair{wu[]}
\fmfipair{w[]}
\fmfipair{wd[]}
\svertex{w1}{p1}
\svertex{w4}{p4}
\wigglywrap{w1}{v5}{v8}{w4}
\end{fmfchar*}}}
\end{aligned}
\end{equation*}
\caption{Wrapping diagrams with chiral structure $\chi(1,2,3)$}
\label{diagrams-123}

\end{figure}
\vspace{2cm}
\begin{table}[h]
\begin{center}
\fbox{
\small
\begin{tabular}{|m{35pt}m{30pt}|m{6pt}|}
\hline
& & \\[-2ex]
$W_{B1\phantom{0}}\rightarrow$ & $\ast\phantom{W_{A}}$ & 1 \\
$W_{B2\phantom{0}}\rightarrow$ & $\ast\phantom{W_{A}}$ & 1 \\
$W_{B3\phantom{0}}\rightarrow$ & $\ast\phantom{W_{A}}$ & 1 \\
\hline
\end{tabular}
\begin{tabular}{|m{35pt}m{35pt}|m{6pt}|}
\hline
& & \\[-2ex]
$W_{B4\phantom{0}}\rightarrow$ & $\ast\phantom{W_{A}}$ & 1 \\
$W_{B5\phantom{0}}\rightarrow$ & $-W_{B6}$ &  \\
$W_{B6\phantom{0}}\rightarrow$ & $-W_{B5}$ &  \\
\hline
\end{tabular}
\begin{tabular}{|m{35pt}m{35pt}|m{6pt}|}
\hline
& & \\[-2ex]
$W_{B7\phantom{0}}\rightarrow$ & finite &  \\
$W_{B8\phantom{0}}\rightarrow$ & finite &  \\
$W_{B9\phantom{0}}\rightarrow$ & finite &  \\
\hline
\end{tabular}}
\end{center}
\caption{Results of $D$-algebra for diagrams with structure $\chi(1,2,3)$}
\label{results-123}
\end{table}
\normalsize

\clearpage
\thispagestyle{empty}
\begin{figure}[!h]
\unitlength=0.75mm
\settoheight{\eqoff}{$\times$}%
\setlength{\eqoff}{0.5\eqoff}%
\addtolength{\eqoff}{-12.5\unitlength}%
\settoheight{\eqofftwo}{$\times$}%
\setlength{\eqofftwo}{0.5\eqofftwo}%
\addtolength{\eqofftwo}{-7.5\unitlength}%
\begin{equation*}
\begin{aligned}
W_{C1\phantom{0}}=&
\raisebox{\eqoff}{%
\fmfframe(3,1)(1,4){%
\begin{fmfchar*}(20,20)
\Wfiveplain
\fmfipair{wu[]}
\fmfipair{w[]}
\fmfipair{wd[]}
\svertex{w3}{p3}
\svertex{w6}{p6}
\wigglywrap{w3}{v5}{v8}{w6}
\end{fmfchar*}}}
\;{\quad}&W_{C2\phantom{0}}=&
\raisebox{\eqoff}{%
\fmfframe(3,1)(1,4){%
\begin{fmfchar*}(20,20)
\Wfiveplain
\fmfipair{wu[]}
\fmfipair{w[]}
\fmfipair{wd[]}
\svertex{w2}{p2}
\svertex{w6}{p6}
\wigglywrap{w2}{v5}{v8}{w6}
\end{fmfchar*}}}
\;{\quad}&W_{C3\phantom{0}}=&
\raisebox{\eqoff}{%
\fmfframe(3,1)(1,4){%
\begin{fmfchar*}(20,20)
\Wfiveplain
\fmfipair{wu[]}
\fmfipair{w[]}
\fmfipair{wd[]}
\svertex{w1}{p1}
\svertex{w6}{p6}
\wigglywrap{w1}{v5}{v8}{w6}
\end{fmfchar*}}}
\\
W_{C4\phantom{0}}=&
\raisebox{\eqoff}{%
\fmfframe(3,1)(1,4){%
\begin{fmfchar*}(20,20)
\Wfiveplain
\fmfipair{wu[]}
\fmfipair{w[]}
\fmfipair{wd[]}
\svertex{w2}{p2}
\svertex{w5}{p5}
\wigglywrap{w2}{v5}{v8}{w5}
\end{fmfchar*}}}
\;{\quad}&W_{C5\phantom{0}}=&
\raisebox{\eqoff}{%
\fmfframe(3,1)(1,4){%
\begin{fmfchar*}(20,20)
\Wfiveplain
\fmfipair{wu[]}
\fmfipair{w[]}
\fmfipair{wd[]}
\svertex{w1}{p1}
\svertex{w5}{p5}
\wigglywrap{w1}{v5}{v8}{w5}
\end{fmfchar*}}}
\;{\quad}&W_{C6\phantom{0}}=&
\raisebox{\eqoff}{%
\fmfframe(3,1)(1,4){%
\begin{fmfchar*}(20,20)
\Wfiveplain
\fmfipair{wu[]}
\fmfipair{w[]}
\fmfipair{wd[]}
\svertex{w1}{p1}
\svertex{w4}{p4}
\wigglywrap{w1}{v5}{v8}{w4}
\end{fmfchar*}}}
\end{aligned}
\end{equation*}
\caption{Wrapping diagrams with chiral structure $\chi(1,3,2)$}
\label{diagrams-132}

\vspace{1cm}
\begin{center}
\small
\fbox{
\begin{tabular}{|m{35pt}m{30pt}|m{6pt}|}
\hline
& & \\[-2ex]
$W_{C1\phantom{0}}\rightarrow$ & $\ast\phantom{W_{A}}$ & 1  \\
$W_{C2\phantom{0}}\rightarrow$ & $\ast\phantom{W_{A}}$ & 2  \\
$W_{C3\phantom{0}}\rightarrow$ & $-W_{C5}$ &   \\
\hline
\end{tabular}
\begin{tabular}{|m{35pt}m{30pt}|m{6pt}|}
\hline
& &  \\[-2ex]
$W_{C4\phantom{0}}\rightarrow$ & finite &  \\
$W_{C5\phantom{0}}\rightarrow$ & $-W_{C3}$ &   \\
$W_{C6\phantom{0}}\rightarrow$ & finite &  \\
\hline
\end{tabular}}
\end{center}
\begin{mytable}
Results of $D$-algebra for diagrams with structure $\chi(1,3,2)$
\label{results-132}
\end{mytable}


\vspace{2cm}
\begin{equation*}
\begin{aligned}
W_{D1\phantom{0}}=&
\raisebox{\eqoff}{%
\fmfframe(3,1)(1,4){%
\begin{fmfchar*}(20,20)
\Wsixplain
\fmfipair{wu[]}
\fmfipair{w[]}
\fmfipair{wd[]}
\svertex{w3}{p3}
\svertex{w6}{p6}
\wigglywrap{w3}{v5}{v8}{w6}
\end{fmfchar*}}}
\;{\quad}&W_{D2\phantom{0}}=&
\raisebox{\eqoff}{%
\fmfframe(3,1)(1,4){%
\begin{fmfchar*}(20,20)
\Wsixplain
\fmfipair{wu[]}
\fmfipair{w[]}
\fmfipair{wd[]}
\svertex{w2}{p2}
\svertex{w6}{p6}
\wigglywrap{w2}{v5}{v8}{w6}
\end{fmfchar*}}}
\;{\quad}&W_{D3\phantom{0}}=&
\raisebox{\eqoff}{%
\fmfframe(3,1)(1,4){%
\begin{fmfchar*}(20,20)
\Wsixplain
\fmfipair{wu[]}
\fmfipair{w[]}
\fmfipair{wd[]}
\svertex{w1}{p1}
\svertex{w6}{p6}
\wigglywrap{w1}{v5}{v8}{w6}
\end{fmfchar*}}}
\\
W_{D4\phantom{0}}=&
\raisebox{\eqoff}{%
\fmfframe(3,1)(1,4){%
\begin{fmfchar*}(20,20)
\Wsixplain
\fmfipair{wu[]}
\fmfipair{w[]}
\fmfipair{wd[]}
\svertex{w2}{p2}
\svertex{w5}{p5}
\wigglywrap{w2}{v5}{v8}{w5}
\end{fmfchar*}}}
\;{\quad}&W_{D5\phantom{0}}=&
\raisebox{\eqoff}{%
\fmfframe(3,1)(1,4){%
\begin{fmfchar*}(20,20)
\Wsixplain
\fmfipair{wu[]}
\fmfipair{w[]}
\fmfipair{wd[]}
\svertex{w1}{p1}
\svertex{w5}{p5}
\wigglywrap{w1}{v5}{v8}{w5}
\end{fmfchar*}}}
\;{\quad}&W_{D6\phantom{0}}=&
\raisebox{\eqoff}{%
\fmfframe(3,1)(1,4){%
\begin{fmfchar*}(20,20)
\Wsixplain
\fmfipair{wu[]}
\fmfipair{w[]}
\fmfipair{wd[]}
\svertex{w1}{p1}
\svertex{w4}{p4}
\wigglywrap{w1}{v5}{v8}{w4}
\end{fmfchar*}}}
\end{aligned}
\end{equation*}
\caption{Wrapping diagrams with chiral structure $\chi(2,1,3)$}
\label{diagrams-213}
\vspace{1cm}
\begin{center}
\fbox{
\begin{tabular}{|m{35pt}m{30pt}|m{6pt}|}
\hline
& & \\[-2ex]
$W_{D1\phantom{0}}\rightarrow$ & $\ast\phantom{W_{A}}$ & 1  \\
$W_{D2\phantom{0}}\rightarrow$ & $\ast\phantom{W_{A}}$ & 2  \\
$W_{D3\phantom{0}}\rightarrow$ & $-W_{D5}$ &  \\
\hline
\end{tabular}
\begin{tabular}{|m{35pt}m{30pt}|m{6pt}|}
\hline
& & \\[-2ex]
$W_{D4\phantom{0}}\rightarrow$ & finite &  \\
$W_{D5\phantom{0}}\rightarrow$ & $-W_{D3}$ &  \\
$W_{D6\phantom{0}}\rightarrow$ & finite &  \\
\hline
\end{tabular}}
\end{center}
\begin{mytable}
Results of $D$-algebra for diagrams with structure $\chi(2,1,3)$
\label{results-213}
\end{mytable}
\end{figure}
\normalsize

\clearpage
\newpage
\begin{figure}[!h]
\unitlength=0.75mm
\settoheight{\eqoff}{$\times$}%
\setlength{\eqoff}{0.5\eqoff}%
\addtolength{\eqoff}{-12.5\unitlength}%
\settoheight{\eqofftwo}{$\times$}%
\setlength{\eqofftwo}{0.5\eqofftwo}%
\addtolength{\eqofftwo}{-7.5\unitlength}%
\begin{equation*}
\begin{aligned}
W_{E1\phantom{0}}=&
\raisebox{\eqoff}{%
\fmfframe(3,1)(1,4){%
\begin{fmfchar*}(20,20)
\Wthreeplain
\fmfipair{wu[]}
\fmfipair{w[]}
\fmfipair{wd[]}
\svertex{w3}{p3}
\svertex{w6}{p6}
\svertex{w5}{p5}
\vvertex{w8}{w6}{p7}
\vvertex{w9}{w5}{p7}
\fmfi{wiggly}{w6..w9}
\wigglywrap{w3}{v5}{v8}{w8}
\end{fmfchar*}}}
\;{\quad}&W_{E2\phantom{0}}=&
\raisebox{\eqoff}{%
\fmfframe(3,1)(1,4){%
\begin{fmfchar*}(20,20)
\Wthreeplain
\fmfipair{wu[]}
\fmfipair{w[]}
\fmfipair{wd[]}
\svertex{w3}{p3}
\svertex{w6}{p6}
\vvertex{w8}{w6}{p7}
\fmfi{wiggly}{w6..w8}
\wigglywrap{w3}{v5}{v8}{w8}
\end{fmfchar*}}}
\;{\quad}&W_{E3\phantom{0}}=&
\raisebox{\eqoff}{%
\fmfframe(3,1)(1,4){%
\begin{fmfchar*}(20,20)
\Wthreeplain
\fmfipair{wu[]}
\fmfipair{w[]}
\fmfipair{wd[]}
\svertex{w3}{p3}
\svertex{w6}{p6}
\svertex{w5}{p5}
\vvertex{w8}{w6}{p7}
\vvertex{w9}{w5}{p7}
\fmfi{wiggly}{w6..w8}
\wigglywrap{w3}{v5}{v8}{w9}
\end{fmfchar*}}}
\;{\quad}&W_{E4\phantom{0}}=&
\raisebox{\eqoff}{%
\fmfframe(3,1)(1,4){%
\begin{fmfchar*}(20,20)
\Wthreeplain
\fmfipair{wu[]}
\fmfipair{w[]}
\fmfipair{wd[]}
\svertex{w2}{p2}
\svertex{w6}{p6}
\svertex{w5}{p5}
\vvertex{w8}{w6}{p7}
\vvertex{w9}{w5}{p7}
\fmfi{wiggly}{w6..w9}
\wigglywrap{w2}{v5}{v8}{w8}
\end{fmfchar*}}}
\\
W_{E5\phantom{0}}=&
\raisebox{\eqoff}{%
\fmfframe(3,1)(1,4){%
\begin{fmfchar*}(20,20)
\Wthreeplain
\fmfipair{wu[]}
\fmfipair{w[]}
\fmfipair{wd[]}
\svertex{w2}{p2}
\svertex{w6}{p6}
\vvertex{w8}{w6}{p7}
\fmfi{wiggly}{w6..w8}
\wigglywrap{w2}{v5}{v8}{w8}
\end{fmfchar*}}}
\;{\quad}&W_{E6\phantom{0}}=&
\raisebox{\eqoff}{%
\fmfframe(3,1)(1,4){%
\begin{fmfchar*}(20,20)
\Wthreeplain
\fmfipair{wu[]}
\fmfipair{w[]}
\fmfipair{wd[]}
\svertex{w2}{p2}
\svertex{w5}{p5}
\svertex{w6}{p6}
\vvertex{w8}{w6}{p7}
\vvertex{w9}{w2}{p7}
\fmfi{wiggly}{w6..w8}
\wigglywrap{w2}{v5}{v8}{w9}
\end{fmfchar*}}}
\;{\quad}&W_{E7\phantom{0}}=&
\raisebox{\eqoff}{%
\fmfframe(3,1)(1,4){%
\begin{fmfchar*}(20,20)
\Wthreeplain
\fmfipair{wu[]}
\fmfipair{w[]}
\fmfipair{wd[]}
\svertex{w1}{p1}
\svertex{w6}{p6}
\svertex{w5}{p5}
\vvertex{w8}{w6}{p7}
\vvertex{w9}{w5}{p7}
\fmfi{wiggly}{w6..w9}
\wigglywrap{w1}{v5}{v8}{w8}
\end{fmfchar*}}}
\;{\quad}&W_{E8\phantom{0}}=&
\raisebox{\eqoff}{%
\fmfframe(3,1)(1,4){%
\begin{fmfchar*}(20,20)
\Wthreeplain
\fmfipair{wu[]}
\fmfipair{w[]}
\fmfipair{wd[]}
\svertex{w1}{p1}
\svertex{w6}{p6}
\svertex{w5}{p5}
\vvertex{w8}{w6}{p7}
\vvertex{w9}{w5}{p7}
\fmfi{wiggly}{w6..w8}
\wigglywrap{w1}{v5}{v8}{w8}
\end{fmfchar*}}}
\\
W_{E9\phantom{0}}=&
\raisebox{\eqoff}{%
\fmfframe(3,1)(1,4){%
\begin{fmfchar*}(20,20)
\Wthreeplain
\fmfipair{wu[]}
\fmfipair{w[]}
\fmfipair{wd[]}
\svertex{w1}{p1}
\svertex{w6}{p6}
\vvertex{w8}{w6}{p7}
\vvertex{w9}{w1}{p7}
\fmfi{wiggly}{w6..w8}
\wigglywrap{w1}{v5}{v8}{w9}
\end{fmfchar*}}}
\;{\quad}&W_{E10}=&
\raisebox{\eqoff}{%
\fmfframe(3,1)(1,4){%
\begin{fmfchar*}(20,20)
\Wthreeplain
\fmfipair{wu[]}
\fmfipair{w[]}
\fmfipair{wd[]}
\svertex{w3}{p3}
\svertex{w5}{p5}
\vvertex{w8}{w3}{p7}
\vvertex{w9}{w5}{p7}
\fmfi{wiggly}{w5..w9}
\wigglywrap{w3}{v5}{v8}{w8}
\end{fmfchar*}}}
\;{\quad}&W_{E11}=&
\raisebox{\eqoff}{%
\fmfframe(3,1)(1,4){%
\begin{fmfchar*}(20,20)
\Wthreeplain
\fmfipair{wu[]}
\fmfipair{w[]}
\fmfipair{wd[]}
\svertex{w3}{p3}
\svertex{w5}{p5}
\vvertex{w8}{w5}{p7}
\fmfi{wiggly}{w5..w8}
\wigglywrap{w3}{v5}{v8}{w8}
\end{fmfchar*}}}
\;{\quad}&W_{E12}=&
\raisebox{\eqoff}{%
\fmfframe(3,1)(1,4){%
\begin{fmfchar*}(20,20)
\Wthreeplain
\fmfipair{wu[]}
\fmfipair{w[]}
\fmfipair{wd[]}
\svertex{w3}{p3}
\svertex{w4}{p4}
\svertex{w5}{p5}
\vvertex{w8}{w4}{p7}
\vvertex{w9}{w5}{p7}
\fmfi{wiggly}{w5..w9}
\wigglywrap{w3}{v5}{v8}{w8}
\end{fmfchar*}}}
\\
W_{E13}=&
\raisebox{\eqoff}{%
\fmfframe(3,1)(1,4){%
\begin{fmfchar*}(20,20)
\Wthreeplain
\fmfipair{wu[]}
\fmfipair{w[]}
\fmfipair{wd[]}
\svertex{w2}{p2}
\svertex{w5}{p5}
\svertex{w6}{p6}
\vvertex{w8}{w6}{p7}
\vvertex{w9}{w5}{p7}
\fmfi{wiggly}{w5..w9}
\wigglywrap{w2}{v5}{v8}{w8}
\end{fmfchar*}}}
\;{\quad}&W_{E14}=&
\raisebox{\eqoff}{%
\fmfframe(3,1)(1,4){%
\begin{fmfchar*}(20,20)
\Wthreeplain
\fmfipair{wu[]}
\fmfipair{w[]}
\fmfipair{wd[]}
\svertex{w2}{p2}
\svertex{w5}{p5}
\vvertex{w8}{w5}{p7}
\fmfi{wiggly}{w5..w8}
\wigglywrap{w2}{v5}{v8}{w8}
\end{fmfchar*}}}
\;{\quad}&W_{E15}=&
\raisebox{\eqoff}{%
\fmfframe(3,1)(1,4){%
\begin{fmfchar*}(20,20)
\Wthreeplain
\fmfipair{wu[]}
\fmfipair{w[]}
\fmfipair{wd[]}
\svertex{w2}{p2}
\svertex{w4}{p4}
\svertex{w5}{p5}
\vvertex{w8}{w5}{p7}
\vvertex{w9}{w4}{p7}
\fmfi{wiggly}{w5..w8}
\wigglywrap{w2}{v5}{v8}{w9}
\end{fmfchar*}}}
\;{\quad}&W_{E16}=&
\raisebox{\eqoff}{%
\fmfframe(3,1)(1,4){%
\begin{fmfchar*}(20,20)
\Wthreeplain
\fmfipair{wu[]}
\fmfipair{w[]}
\fmfipair{wd[]}
\svertex{w1}{p1}
\svertex{w6}{p6}
\svertex{w5}{p5}
\vvertex{w8}{w6}{p7}
\vvertex{w9}{w5}{p7}
\fmfi{wiggly}{w5..w9}
\wigglywrap{w1}{v5}{v8}{w8}
\end{fmfchar*}}}
\\
W_{E17}=&
\raisebox{\eqoff}{%
\fmfframe(3,1)(1,4){%
\begin{fmfchar*}(20,20)
\Wthreeplain
\fmfipair{wu[]}
\fmfipair{w[]}
\fmfipair{wd[]}
\svertex{w1}{p1}
\svertex{w5}{p5}
\vvertex{w8}{w5}{p7}
\fmfi{wiggly}{w5..w8}
\wigglywrap{w1}{v5}{v8}{w8}
\end{fmfchar*}}}
\;{\quad}&W_{E18}=&
\raisebox{\eqoff}{%
\fmfframe(3,1)(1,4){%
\begin{fmfchar*}(20,20)
\Wthreeplain
\fmfipair{wu[]}
\fmfipair{w[]}
\fmfipair{wd[]}
\svertex{w1}{p1}
\svertex{w5}{p5}
\svertex{w4}{p4}
\vvertex{w8}{w5}{p7}
\vvertex{w9}{w4}{p7}
\fmfi{wiggly}{w5..w8}
\wigglywrap{w1}{v5}{v8}{w9}
\end{fmfchar*}}}
\;{\quad}&W_{E19}=&
\raisebox{\eqoff}{%
\fmfframe(3,1)(1,4){%
\begin{fmfchar*}(20,20)
\Wthreeplain
\fmfipair{wu[]}
\fmfipair{w[]}
\fmfipair{wd[]}
\svertex{w3}{p3}
\svertex{w4}{p4}
\vvertex{w8}{w3}{p7}
\vvertex{w9}{w4}{p7}
\fmfi{wiggly}{w4..w9}
\wigglywrap{w3}{v5}{v8}{w8}
\end{fmfchar*}}}
\;{\quad}&W_{E20}=&
\raisebox{\eqoff}{%
\fmfframe(3,1)(1,4){%
\begin{fmfchar*}(20,20)
\Wthreeplain
\fmfipair{wu[]}
\fmfipair{w[]}
\fmfipair{wd[]}
\svertex{w3}{p3}
\svertex{w4}{p4}
\svertex{w7}{p7}
\fmfi{wiggly}{w4..w7}
\wigglywrap{w3}{v5}{v8}{w7}
\end{fmfchar*}}}
\\
W_{E21}=&
\raisebox{\eqoff}{%
\fmfframe(3,1)(1,4){%
\begin{fmfchar*}(20,20)
\Wthreeplain
\fmfipair{wu[]}
\fmfipair{w[]}
\fmfipair{wd[]}
\svertex{w3}{p3}
\svertex{w4}{p4}
\svertex{w7}{p7}
\vvertex{w8}{w4}{p7}
\fmfi{wiggly}{w4..w7}
\wigglywrap{w3}{v5}{v8}{w8}
\end{fmfchar*}}}
\;{\quad}&W_{E22}=&
\raisebox{\eqoff}{%
\fmfframe(3,1)(1,4){%
\begin{fmfchar*}(20,20)
\Wthreeplain
\fmfipair{wu[]}
\fmfipair{w[]}
\fmfipair{wd[]}
\svertex{w2}{p2}
\svertex{w4}{p4}
\vvertex{w8}{w2}{p7}
\vvertex{w9}{w4}{p7}
\fmfi{wiggly}{w4..w9}
\wigglywrap{w2}{v5}{v8}{w8}
\end{fmfchar*}}}
\;{\quad}&W_{E23}=&
\raisebox{\eqoff}{%
\fmfframe(3,1)(1,4){%
\begin{fmfchar*}(20,20)
\Wthreeplain
\fmfipair{wu[]}
\fmfipair{w[]}
\fmfipair{wd[]}
\svertex{w2}{p2}
\svertex{w4}{p4}
\svertex{w7}{p7}
\fmfi{wiggly}{w4..w7}
\wigglywrap{w2}{v5}{v8}{w7}
\end{fmfchar*}}}
\;{\quad}&W_{E24}=&
\raisebox{\eqoff}{%
\fmfframe(3,1)(1,4){%
\begin{fmfchar*}(20,20)
\Wthreeplain
\fmfipair{wu[]}
\fmfipair{w[]}
\fmfipair{wd[]}
\svertex{w2}{p2}
\svertex{w4}{p4}
\svertex{w7}{p7}
\vvertex{w8}{w4}{p7}
\fmfi{wiggly}{w4..w7}
\wigglywrap{w2}{v5}{v8}{w8}
\end{fmfchar*}}}
\\
W_{E25}=&
\raisebox{\eqoff}{%
\fmfframe(3,1)(1,4){%
\begin{fmfchar*}(20,20)
\Wthreeplain
\fmfipair{wu[]}
\fmfipair{w[]}
\fmfipair{wd[]}
\svertex{w1}{p1}
\svertex{w4}{p4}
\svertex{w7}{p7}
\vvertex{w8}{w4}{p7}
\vvertex{w9}{w1}{p7}
\fmfi{wiggly}{w4..w8}
\wigglywrap{w1}{v5}{v8}{w7}
\end{fmfchar*}}}
\;{\quad}&W_{E26}=&
\raisebox{\eqoff}{%
\fmfframe(3,1)(1,4){%
\begin{fmfchar*}(20,20)
\Wthreeplain
\fmfipair{wu[]}
\fmfipair{w[]}
\fmfipair{wd[]}
\svertex{w1}{p1}
\svertex{w4}{p4}
\vvertex{w8}{w4}{p7}
\vvertex{w9}{w4}{p7}
\fmfi{wiggly}{w4..w8}
\wigglywrap{w1}{v5}{v8}{w8}
\end{fmfchar*}}}
\;{\quad}&W_{E27}=&
\raisebox{\eqoff}{%
\fmfframe(3,1)(1,4){%
\begin{fmfchar*}(20,20)
\Wthreeplain
\fmfipair{wu[]}
\fmfipair{w[]}
\fmfipair{wd[]}
\svertex{w1}{p1}
\svertex{w4}{p4}
\svertex{w7}{p7}
\vvertex{w8}{w4}{p7}
\fmfi{wiggly}{w4..w7}
\wigglywrap{w1}{v5}{v8}{w8}
\end{fmfchar*}}}
\end{aligned}
\end{equation*}
\caption{Wrapping diagrams with chiral structure $\chi(2,1)$}
\label{diagrams-12}

\begin{center}
\small
\fbox{
\begin{tabular}{|m{35pt}m{30pt}|m{6pt}|}
\hline
& & \\[-2ex]
$W_{E1\phantom{0}}\rightarrow$ & $-W_{E10}$ &  \\
$W_{E2\phantom{0}}\rightarrow$ & $\ast\phantom{W_{A}}$ & 1  \\
$W_{E3\phantom{0}}\rightarrow$ & $-W_{E6}$ &  \\
$W_{E4\phantom{0}}\rightarrow$ & $-W_{E13}$ &  \\
$W_{E5\phantom{0}}\rightarrow$ & $\ast\phantom{W_{A}}$ & 1  \\
$W_{E6\phantom{0}}\rightarrow$ & $-W_{E3}$ &  \\
$W_{E7\phantom{0}}\rightarrow$ & $-W_{E8}$ &  \\
\hline
\end{tabular}
\begin{tabular}{|m{35pt}m{30pt}|m{6pt}|}
\hline
& & \\[-2ex]
$W_{E8\phantom{0}}\rightarrow$ & $-W_{E7}$ &  \\
$W_{E9\phantom{0}}\rightarrow$ & finite &  \\
$W_{E10}\rightarrow$ & $-W_{E1}$ &  \\
$W_{E11}\rightarrow$ & $\ast\phantom{W_{A}}$ & 1  \\
$W_{E12}\rightarrow$ & $-W_{E15}$ &  \\
$W_{E13}\rightarrow$ & $-W_{E4}$ &  \\
$W_{E14}\rightarrow$ & $\ast\phantom{W_{A}}$ & 1 \\
\hline
\end{tabular}
\begin{tabular}{|m{35pt}m{30pt}|m{6pt}|}
\hline
& & \\[-2ex]
$W_{E15}\rightarrow$ & $-W_{E12}$ &  \\
$W_{E16}\rightarrow$ & $-W_{E17}$ &  \\
$W_{E17}\rightarrow$ & $-W_{E16}$ &  \\
$W_{E18}\rightarrow$ & finite &  \\
$W_{E19}\rightarrow$ & finite &  \\
$W_{E20}\rightarrow$ & finite &  \\
$W_{E21}\rightarrow$ & finite &  \\
\hline
\end{tabular}
\begin{tabular}{|m{35pt}m{30pt}|m{6pt}|}
\hline
& & \\[-2ex]
$W_{E22}\rightarrow$ & finite &  \\
$W_{E23}\rightarrow$ & finite &  \\
$W_{E24}\rightarrow$ & finite &  \\
$W_{E25}\rightarrow$ & finite &  \\
$W_{E26}\rightarrow$ & finite &  \\
$W_{E27}\rightarrow$ & finite &  \\
& &  \\
\hline
\end{tabular}}
\end{center}
\begin{mytable}
Results of $D$-algebra for diagrams with structure $\chi(2,1)$
\label{results-12}
\end{mytable}
\end{figure}
\normalsize

\clearpage

\begin{figure}[!h]
\unitlength=0.75mm
\settoheight{\eqoff}{$\times$}%
\setlength{\eqoff}{0.5\eqoff}%
\addtolength{\eqoff}{-12.5\unitlength}%
\settoheight{\eqofftwo}{$\times$}%
\setlength{\eqofftwo}{0.5\eqofftwo}%
\addtolength{\eqofftwo}{-7.5\unitlength}%
\begin{equation*}
\begin{aligned}
W_{F1\phantom{0}}=&
\raisebox{\eqoff}{%
\fmfframe(3,1)(1,4){%
\begin{fmfchar*}(20,20)
\Wtwoplain
\fmfipair{wu[]}
\fmfipair{w[]}
\fmfipair{wd[]}
\svertex{w5}{p5}
\svertex{w9}{p9}
\svertex{w4}{p4}
\svertex{w10}{p10}
\fmfi{wiggly}{w5..w9}
\wigglywrap{w4}{v5}{v8}{w10}
\end{fmfchar*}}}
\;{\quad}&W_{F2\phantom{0}}=&
\raisebox{\eqoff}{%
\fmfframe(3,1)(1,4){%
\begin{fmfchar*}(20,20)
\Wtwoplain
\fmfipair{wu[]}
\fmfipair{w[]}
\fmfipair{wd[]}
\svertex{w5}{p5}
\svertex{w9}{p9}
\svertex{w3}{p3}
\svertex{w10}{p10}
\fmfi{wiggly}{w5..w9}
\wigglywrap{w3}{v5}{v8}{w10}
\end{fmfchar*}}}
\;{\quad}&W_{F3\phantom{0}}=&
\raisebox{\eqoff}{%
\fmfframe(3,1)(1,4){%
\begin{fmfchar*}(20,20)
\Wtwoplain
\fmfipair{wu[]}
\fmfipair{w[]}
\fmfipair{wd[]}
\svertex{w5}{p5}
\svertex{w9}{p9}
\svertex{w1}{p1}
\svertex{w10}{p10}
\fmfi{wiggly}{w5..w9}
\wigglywrap{w1}{v5}{v8}{w10}
\end{fmfchar*}}}
\;{\quad}&W_{F4\phantom{0}}=&
\raisebox{\eqoff}{%
\fmfframe(3,1)(1,4){%
\begin{fmfchar*}(20,20)
\Wtwoplain
\fmfipair{wu[]}
\fmfipair{w[]}
\fmfipair{wd[]}
\svertex{w5}{p5}
\svertex{w9}{p9}
\svertex{w3}{p3}
\svertex{w8}{p8}
\fmfi{wiggly}{w5..w9}
\wigglywrap{w3}{v5}{v8}{w8}
\end{fmfchar*}}}
\\
W_{F5\phantom{0}}=&
\raisebox{\eqoff}{%
\fmfframe(3,1)(1,4){%
\begin{fmfchar*}(20,20)
\Wtwoplain
\fmfipair{wu[]}
\fmfipair{w[]}
\fmfipair{wd[]}
\svertex{w5}{p5}
\svertex{w9}{p9}
\svertex{w1}{p1}
\svertex{w8}{p8}
\fmfi{wiggly}{w5..w9}
\wigglywrap{w1}{v5}{v8}{w8}
\end{fmfchar*}}}
\;{\quad}&W_{F6\phantom{0}}=&
\raisebox{\eqoff}{%
\fmfframe(3,1)(1,4){%
\begin{fmfchar*}(20,20)
\Wtwoplain
\fmfipair{wu[]}
\fmfipair{w[]}
\fmfipair{wd[]}
\svertex{w5}{p5}
\svertex{w9}{p9}
\svertex{w1}{p1}
\svertex{w7}{p7}
\fmfi{wiggly}{w5..w9}
\wigglywrap{w1}{v5}{v8}{w7}
\end{fmfchar*}}}
\;{\quad}&W_{F7\phantom{0}}=&
\raisebox{\eqoff}{%
\fmfframe(3,1)(1,4){%
\begin{fmfchar*}(20,20)
\Wtwoplain
\fmfipair{wu[]}
\fmfipair{w[]}
\fmfipair{wd[]}
\svertex{w5}{p5}
\svertex{w8}{p8}
\svertex{w3}{p3}
\svertex{w10}{p10}
\fmfi{wiggly}{w5..w8}
\wigglywrap{w3}{v5}{v8}{w10}
\end{fmfchar*}}}
\;{\quad}&W_{F8\phantom{0}}=&
\raisebox{\eqoff}{%
\fmfframe(3,1)(1,4){%
\begin{fmfchar*}(20,20)
\Wtwoplain
\fmfipair{wu[]}
\fmfipair{w[]}
\fmfipair{wd[]}
\svertex{w5}{p5}
\svertex{w8}{p8}
\svertex{w1}{p1}
\svertex{w10}{p10}
\fmfi{wiggly}{w5..w8}
\wigglywrap{w1}{v5}{v8}{w10}
\end{fmfchar*}}}
\\
W_{F9\phantom{0}}=&
\raisebox{\eqoff}{%
\fmfframe(3,1)(1,4){%
\begin{fmfchar*}(20,20)
\Wtwoplain
\fmfipair{wu[]}
\fmfipair{w[]}
\fmfipair{wd[]}
\svertex{w5}{p5}
\svertex{w4}{p4}
\fmfiequ{wu8}{point length(p8)/4 of p8}
\fmfiequ{wd8}{point 3length(p8)/4 of p8}
\fmfi{wiggly}{w5..wu8}
\wigglywrap{w4}{v5}{v8}{wd8}
\end{fmfchar*}}}
\;{\quad}&W_{F10}=&
\raisebox{\eqoff}{%
\fmfframe(3,1)(1,4){%
\begin{fmfchar*}(20,20)
\Wtwoplain
\fmfipair{wu[]}
\fmfipair{w[]}
\fmfipair{wd[]}
\svertex{w5}{p5}
\svertex{w3}{p3}
\fmfiequ{wu8}{point length(p8)/4 of p8}
\fmfiequ{wd8}{point 3length(p8)/4 of p8}
\fmfi{wiggly}{wu8..w5}
\wigglywrap{w3}{v5}{v8}{wd8}
\end{fmfchar*}}}
\;{\quad}&W_{F11}=&
\raisebox{\eqoff}{%
\fmfframe(3,1)(1,4){%
\begin{fmfchar*}(20,20)
\Wtwoplain
\fmfipair{wu[]}
\fmfipair{w[]}
\fmfipair{wd[]}
\svertex{w5}{p5}
\svertex{w1}{p1}
\fmfiequ{wu8}{point length(p8)/4 of p8}
\fmfiequ{wd8}{point 3length(p8)/4 of p8}
\fmfi{wiggly}{w5..wu8}
\wigglywrap{w1}{v5}{v8}{wd8}
\end{fmfchar*}}}
\;{\quad}&W_{F12}=&
\raisebox{\eqoff}{%
\fmfframe(3,1)(1,4){%
\begin{fmfchar*}(20,20)
\Wtwoplain
\fmfipair{wu[]}
\fmfipair{w[]}
\fmfipair{wd[]}
\svertex{w5}{p5}
\svertex{w8}{p8}
\svertex{w4}{p4}
\fmfi{wiggly}{w5..w8}
\wigglywrap{w4}{v5}{v8}{w8}
\end{fmfchar*}}}
\\
W_{F13}=&
\raisebox{\eqoff}{%
\fmfframe(3,1)(1,4){%
\begin{fmfchar*}(20,20)
\Wtwoplain
\fmfipair{wu[]}
\fmfipair{w[]}
\fmfipair{wd[]}
\svertex{w5}{p5}
\svertex{w3}{p3}
\svertex{w8}{p8}
\fmfi{wiggly}{w5..w8}
\wigglywrap{w3}{v5}{v8}{w8}
\end{fmfchar*}}}
\;{\quad}&W_{F14}=&
\raisebox{\eqoff}{%
\fmfframe(3,1)(1,4){%
\begin{fmfchar*}(20,20)
\Wtwoplain
\fmfipair{wu[]}
\fmfipair{w[]}
\fmfipair{wd[]}
\svertex{w5}{p5}
\svertex{w1}{p1}
\svertex{w8}{p8}
\fmfi{wiggly}{w5..w8}
\wigglywrap{w1}{v5}{v8}{w8}
\end{fmfchar*}}}
\;{\quad}&W_{F15}=&
\raisebox{\eqoff}{%
\fmfframe(3,1)(1,4){%
\begin{fmfchar*}(20,20)
\Wtwoplain
\fmfipair{wu[]}
\fmfipair{w[]}
\fmfipair{wd[]}
\svertex{w5}{p5}
\svertex{w3}{p3}
\dvertex{wu8}{wd8}{p8}
\fmfi{wiggly}{w5..wd8}
\wigglywrap{w3}{v5}{v8}{wu8}
\end{fmfchar*}}}
\;{\quad}&W_{F16}=&
\raisebox{\eqoff}{%
\fmfframe(3,1)(1,4){%
\begin{fmfchar*}(20,20)
\Wtwoplain
\fmfipair{wu[]}
\fmfipair{w[]}
\fmfipair{wd[]}
\svertex{w5}{p5}
\svertex{w1}{p1}
\dvertex{wu8}{wd8}{p8}
\fmfi{wiggly}{w5..wd8}
\wigglywrap{w1}{v5}{v8}{wu8}
\end{fmfchar*}}}
\\
W_{F17}=&
\raisebox{\eqoff}{%
\fmfframe(3,1)(1,4){%
\begin{fmfchar*}(20,20)
\Wtwoplain
\fmfipair{wu[]}
\fmfipair{w[]}
\fmfipair{wd[]}
\svertex{w5}{p5}
\svertex{w4}{p4}
\svertex{w7}{p7}
\svertex{w8}{p8}
\fmfi{wiggly}{w5..w8}
\wigglywrap{w4}{v5}{v8}{w7}
\end{fmfchar*}}}
\;{\quad}&W_{F18}=&
\raisebox{\eqoff}{%
\fmfframe(3,1)(1,4){%
\begin{fmfchar*}(20,20)
\Wtwoplain
\fmfipair{wu[]}
\fmfipair{w[]}
\fmfipair{wd[]}
\svertex{w5}{p5}
\svertex{w3}{p3}
\svertex{w7}{p7}
\svertex{w8}{p8}
\fmfi{wiggly}{w5..w8}
\wigglywrap{w3}{v5}{v8}{w7}
\end{fmfchar*}}}
\;{\quad}&W_{F19}=&
\raisebox{\eqoff}{%
\fmfframe(3,1)(1,4){%
\begin{fmfchar*}(20,20)
\Wtwoplain
\fmfipair{wu[]}
\fmfipair{w[]}
\fmfipair{wd[]}
\svertex{w1}{p1}
\svertex{w5}{p5}
\svertex{w7}{p7}
\svertex{w8}{p8}
\fmfi{wiggly}{w5..w8}
\wigglywrap{w1}{v5}{v8}{w7}
\end{fmfchar*}}}
\;{\quad}&W_{F20}=&
\raisebox{\eqoff}{%
\fmfframe(3,1)(1,4){%
\begin{fmfchar*}(20,20)
\Wtwoplain
\fmfipair{wu[]}
\fmfipair{w[]}
\fmfipair{wd[]}
\svertex{w5}{p5}
\svertex{w6}{p6}
\svertex{w1}{p1}
\svertex{w10}{p10}
\fmfi{wiggly}{w5..w6}
\wigglywrap{w1}{v5}{v8}{w10}
\end{fmfchar*}}}
\\
W_{F21}=&
\raisebox{\eqoff}{%
\fmfframe(3,1)(1,4){%
\begin{fmfchar*}(20,20)
\Wtwoplain
\fmfipair{wu[]}
\fmfipair{w[]}
\fmfipair{wd[]}
\svertex{w5}{p5}
\svertex{w6}{p6}
\svertex{w3}{p3}
\svertex{w8}{p8}
\fmfi{wiggly}{w5..w6}
\wigglywrap{w3}{v5}{v8}{w8}
\end{fmfchar*}}}
\;{\quad}&W_{F22}=&
\raisebox{\eqoff}{%
\fmfframe(3,1)(1,4){%
\begin{fmfchar*}(20,20)
\Wtwoplain
\fmfipair{wu[]}
\fmfipair{w[]}
\fmfipair{wd[]}
\svertex{w5}{p5}
\svertex{w6}{p6}
\svertex{w1}{p1}
\svertex{w8}{p8}
\fmfi{wiggly}{w5..w6}
\wigglywrap{w1}{v5}{v8}{w8}
\end{fmfchar*}}}
\;{\quad}&W_{F23}=&
\raisebox{\eqoff}{%
\fmfframe(3,1)(1,4){%
\begin{fmfchar*}(20,20)
\Wtwoplain
\fmfipair{wu[]}
\fmfipair{w[]}
\fmfipair{wd[]}
\fmfiequ{wu8}{point length(p8)/4 of p8}
\fmfiequ{wd8}{point 3length(p8)/4 of p8}
\vvertex{w9}{wu8}{p3}
\fmfi{wiggly}{w9..wu8}
\wigglywrap{w9}{v5}{v8}{wd8}
\end{fmfchar*}}}
\;{\quad}&W_{F24}=&
\raisebox{\eqoff}{%
\fmfframe(3,1)(1,4){%
\begin{fmfchar*}(20,20)
\Wtwoplain
\fmfipair{wu[]}
\fmfipair{w[]}
\fmfipair{wd[]}
\fmfiequ{wu8}{point length(p8)/4 of p8}
\fmfiequ{wd8}{point 3length(p8)/4 of p8}
\fmfiequ{wu3}{point length(p8)/4 of p3}
\fmfiequ{wd3}{point 3length(p8)/4 of p3}
\fmfi{wiggly}{wd3..wu8}
\wigglywrap{wu3}{v5}{v8}{wd8}
\end{fmfchar*}}}
\\
W_{F25}=&
\raisebox{\eqoff}{%
\fmfframe(3,1)(1,4){%
\begin{fmfchar*}(20,20)
\Wtwoplain
\fmfipair{wu[]}
\fmfipair{w[]}
\fmfipair{wd[]}
\svertex{w3}{p3}
\svertex{w1}{p1}
\fmfiequ{wu8}{point length(p8)/4 of p8}
\fmfiequ{wd8}{point 3length(p8)/4 of p8}
\fmfi{wiggly}{w3..wu8}
\wigglywrap{w1}{v5}{v8}{wd8}
\end{fmfchar*}}}
\;{\quad}&W_{F26}=&
\raisebox{\eqoff}{%
\fmfframe(3,1)(1,4){%
\begin{fmfchar*}(20,20)
\Wtwoplain
\fmfipair{wu[]}
\fmfipair{w[]}
\fmfipair{wd[]}
\svertex{w3}{p3}
\svertex{w8}{p8}
\fmfi{wiggly}{w3..w8}
\wigglywrap{w3}{v5}{v8}{w8}
\end{fmfchar*}}}
\;{\quad}&W_{F27}=&
\raisebox{\eqoff}{%
\fmfframe(3,1)(1,4){%
\begin{fmfchar*}(20,20)
\Wtwoplain
\fmfipair{wu[]}
\fmfipair{w[]}
\fmfipair{wd[]}
\svertex{w1}{p1}
\svertex{w3}{p3}
\svertex{w8}{p8}
\fmfi{wiggly}{w3..w8}
\wigglywrap{w1}{v5}{v8}{w8}
\end{fmfchar*}}}
\;{\quad}&W_{F28}=&
\raisebox{\eqoff}{%
\fmfframe(3,1)(1,4){%
\begin{fmfchar*}(20,20)
\Wtwoplain
\fmfipair{wu[]}
\fmfipair{w[]}
\fmfipair{wd[]}
\fmfiequ{wu8}{point length(p8)/4 of p8}
\fmfiequ{wd8}{point 3length(p8)/4 of p8}
\fmfiequ{wu3}{point length(p8)/4 of p3}
\fmfiequ{wd3}{point 3length(p8)/4 of p3}
\fmfi{wiggly}{wd3..wd8}
\wigglywrap{wu3}{v5}{v8}{wu8}
\end{fmfchar*}}}
\\
W_{F29}=&
\raisebox{\eqoff}{%
\fmfframe(3,1)(1,4){%
\begin{fmfchar*}(20,20)
\Wtwoplain
\fmfipair{wu[]}
\fmfipair{w[]}
\fmfipair{wd[]}
\svertex{w3}{p3}
\svertex{w1}{p1}
\dvertex{wu8}{wd8}{p8}
\fmfi{wiggly}{w3..wd8}
\wigglywrap{w1}{v5}{v8}{wu8}
\end{fmfchar*}}}
\;{\quad}&W_{F30}=&
\raisebox{\eqoff}{%
\fmfframe(3,1)(1,4){%
\begin{fmfchar*}(20,20)
\Wtwoplain
\fmfipair{wu[]}
\fmfipair{w[]}
\fmfipair{wd[]}
\svertex{w3}{p3}
\svertex{w8}{p8}
\svertex{w1}{p1}
\svertex{w7}{p7}
\fmfi{wiggly}{w3..w8}
\wigglywrap{w1}{v5}{v8}{w7}
\end{fmfchar*}}}
\;{\quad}&W_{F31}=&
\raisebox{\eqoff}{%
\fmfframe(3,1)(1,4){%
\begin{fmfchar*}(20,20)
\Wtwoplain
\fmfipair{wu[]}
\fmfipair{w[]}
\fmfipair{wd[]}
\svertex{w1}{p1}
\svertex{w3}{p3}
\svertex{w6}{p6}
\svertex{w8}{p8}
\fmfi{wiggly}{w3..w6}
\wigglywrap{w1}{v5}{v8}{w8}
\end{fmfchar*}}}
\;{\quad}&W_{F32}=&
\raisebox{\eqoff}{%
\fmfframe(3,1)(1,4){%
\begin{fmfchar*}(20,20)
\Wtwoplain
\fmfipair{wu[]}
\fmfipair{w[]}
\fmfipair{wd[]}
\svertex{w3}{p3}
\svertex{w4}{p4}
\svertex{w6}{p6}
\svertex{w7}{p7}
\fmfi{wiggly}{w3..w6}
\wigglywrap{w4}{v5}{v8}{w7}
\end{fmfchar*}}}
\\
W_{F33}=&
\raisebox{\eqoff}{%
\fmfframe(3,1)(1,4){%
\begin{fmfchar*}(20,20)
\Wtwoplain
\fmfipair{wu[]}
\fmfipair{w[]}
\fmfipair{wd[]}
\svertex{w6}{p6}
\svertex{w7}{p7}
\dvertex{wu3}{wd3}{p3}
\fmfi{wiggly}{wu3..w6}
\wigglywrap{wd3}{v5}{v8}{w7}
\end{fmfchar*}}}
\;{\quad}&W_{F34}=&
\raisebox{\eqoff}{%
\fmfframe(3,1)(1,4){%
\begin{fmfchar*}(20,20)
\Wtwoplain
\fmfipair{wu[]}
\fmfipair{w[]}
\fmfipair{wd[]}
\svertex{w3}{p3}
\svertex{w6}{p6}
\svertex{w7}{p7}
\fmfi{wiggly}{w3..w6}
\wigglywrap{w3}{v5}{v8}{w7}
\end{fmfchar*}}}
\;{\quad}&W_{F35}=&
\raisebox{\eqoff}{%
\fmfframe(3,1)(1,4){%
\begin{fmfchar*}(20,20)
\Wtwoplain
\fmfipair{wu[]}
\fmfipair{w[]}
\fmfipair{wd[]}
\svertex{w1}{p1}
\svertex{w3}{p3}
\svertex{w6}{p6}
\svertex{w7}{p7}
\fmfi{wiggly}{w3..w6}
\wigglywrap{w1}{v5}{v8}{w7}
\end{fmfchar*}}}
\;{\quad}&W_{F36}=&
\raisebox{\eqoff}{%
\fmfframe(3,1)(1,4){%
\begin{fmfchar*}(20,20)
\Wtwoplain
\fmfipair{wu[]}
\fmfipair{w[]}
\fmfipair{wd[]}
\svertex{w1}{p1}
\svertex{w2}{p2}
\svertex{w6}{p6}
\svertex{w7}{p7}
\fmfi{wiggly}{w2..w6}
\wigglywrap{w1}{v5}{v8}{w7}
\end{fmfchar*}}}
\\
W_{F37}=&
\raisebox{\eqoff}{%
\fmfframe(3,1)(1,4){%
\begin{fmfchar*}(20,20)
\Wtwoplain
\fmfipair{wu[]}
\fmfipair{w[]}
\fmfipair{wd[]}
\svertex{w4}{p4}
\svertex{w5}{p5}
\svertex{w6}{p6}
\svertex{w7}{p7}
\fmfi{wiggly}{w5..w6}
\wigglywrap{w4}{v5}{v8}{w7}
\end{fmfchar*}}}
\end{aligned}
\end{equation*}
\caption{Wrapping diagrams with chiral structure $\chi(1,3)$}
\label{diagrams-13}
\end{figure}

\clearpage

\begin{table}[p]
\begin{center}
\small
\fbox{
\begin{tabular}{|m{35pt}m{30pt}|m{6pt}|}
\hline
& & \\[-2ex]
$W_{F1\phantom{0}}\rightarrow$ & $\ast\phantom{W_{A}}$ & 1 \\
$W_{F2\phantom{0}}\rightarrow$ & $\ast\phantom{W_{A}}$ & 4 \\
$W_{F3\phantom{0}}\rightarrow$ & $-W_{F5}$ &  \\
$W_{F4\phantom{0}}\rightarrow$ & finite &  \\
$W_{F5\phantom{0}}\rightarrow$ & $-W_{F3}$ &  \\
$W_{F6\phantom{0}}\rightarrow$ & finite &  \\
$W_{F7\phantom{0}}\rightarrow$ & $\ast\phantom{W_{A}}$ & 2 \\
$W_{F8\phantom{0}}\rightarrow$ & $-W_{F11}$ &  \\
$W_{F9\phantom{0}}\rightarrow$ & $\ast\phantom{W_{A}}$ & 4 \\
$W_{F10}\rightarrow$ & $-W_{F21}$ &  \\
\hline
\end{tabular}
\begin{tabular}{|m{35pt}m{30pt}|m{6pt}|}
\hline
& & \\[-2ex]
$W_{F11}\rightarrow$ & $-W_{F8}$ &  \\
$W_{F12}\rightarrow$ & $\ast\phantom{W_{A}}$ & 2 \\
$W_{F13}\rightarrow$ & $-W_{F15}$ &  \\
$W_{F14}\rightarrow$ & $-W_{F16}$ &  \\
$W_{F15}\rightarrow$ & $-W_{F13}$ &  \\
$W_{F16}\rightarrow$ & $-W_{F14}$ &  \\
$W_{F17}\rightarrow$ & $-W_{F18}$ &  \\
$W_{F18}\rightarrow$ & $-W_{F17}$ &  \\
$W_{F19}\rightarrow$ & finite &  \\
$W_{F20}\rightarrow$ & finite &  \\
\hline
\end{tabular}
\begin{tabular}{|m{35pt}m{30pt}|m{6pt}|}
\hline
& & \\[-2ex]
$W_{F21}\rightarrow$ & $-W_{F10}$ &  \\
$W_{F22}\rightarrow$ & finite &  \\
$W_{F23}\rightarrow$ & finite &  \\
$W_{F24}\rightarrow$ & $-W_{F34}$ &  \\
$W_{F25}\rightarrow$ & $-W_{F33}$ &  \\
$W_{F26}\rightarrow$ & finite &  \\
$W_{F27}\rightarrow$ & finite &  \\
$W_{F28}\rightarrow$ & finite &  \\
$W_{F29}\rightarrow$ & finite &  \\
$W_{F30}\rightarrow$ & finite &  \\
\hline
\end{tabular}
\begin{tabular}{|m{35pt}m{30pt}|m{6pt}|}
\hline
& & \\[-2ex]
$W_{F31}\rightarrow$ & finite &  \\
$W_{F32}\rightarrow$ & $\ast\phantom{W_{A}}$ & 4 \\
$W_{F33}\rightarrow$ & $-W_{F25}$ &  \\
$W_{F34}\rightarrow$ & $-W_{F24}$ &  \\
$W_{F35}\rightarrow$ & finite &  \\
$W_{F36}\rightarrow$ & finite &  \\
$W_{F37}\rightarrow$ & $\ast\phantom{W_{A}}$ & 2 \\
& & \\
& & \\
& & \\
\hline
\end{tabular}}
\end{center}
\caption{Results of $D$-algebra for diagrams with structure $\chi(1,3)$}
\label{results-13}
\end{table}
\normalsize

\clearpage

\begin{figure}[p]
\unitlength=0.75mm
\settoheight{\eqoff}{$\times$}%
\setlength{\eqoff}{0.5\eqoff}%
\addtolength{\eqoff}{-12.5\unitlength}%
\settoheight{\eqofftwo}{$\times$}%
\setlength{\eqofftwo}{0.5\eqofftwo}%
\addtolength{\eqofftwo}{-7.5\unitlength}%
\begin{equation*}
\begin{aligned}
W_{G1\phantom{0}}=&
\raisebox{\eqoff}{%
\fmfframe(3,1)(1,4){%
\begin{fmfchar*}(20,20)
\Woneplain
\fmfipair{wu[]}
\fmfipair{w[]}
\fmfipair{wd[]}
\svertex{w5}{p5}
\svertex{w6}{p6}
\svertex{w1}{p1}
\svertex{w7}{p7}
\vvertex{w8}{w5}{p1}
\vvertex{w9}{w5}{p7}
\fmfi{wiggly}{w1..w5}
\fmfi{wiggly}{w7..w6}
\wigglywrap{w8}{v5}{v8}{w9}
\end{fmfchar*}}}
\;{\quad}&W_{G2\phantom{0}}=&
\raisebox{\eqoff}{%
\fmfframe(3,1)(1,4){%
\begin{fmfchar*}(20,20)
\Woneplain
\fmfipair{wu[]}
\fmfipair{w[]}
\fmfipair{wd[]}
\svertex{w5}{p5}
\svertex{w6}{p6}
\svertex{w1}{p1}
\svertex{w7}{p7}
\vvertex{w8}{w5}{p1}
\vvertex{w9}{w5}{p7}
\fmfi{wiggly}{w1..w5}
\fmfi{wiggly}{w9..w6}
\wigglywrap{w8}{v5}{v8}{w9}
\end{fmfchar*}}}
\;{\quad}&W_{G3\phantom{0}}=&
\raisebox{\eqoff}{%
\fmfframe(3,1)(1,4){%
\begin{fmfchar*}(20,20)
\Woneplain
\fmfipair{wu[]}
\fmfipair{w[]}
\fmfipair{wd[]}
\svertex{w5}{p5}
\svertex{w6}{p6}
\svertex{w1}{p1}
\svertex{w7}{p7}
\svertex{w3}{p3}
\vvertex{w8}{w5}{p1}
\vvertex{w9}{w3}{p7}
\fmfi{wiggly}{w1..w5}
\fmfi{wiggly}{w7..w6}
\wigglywrap{w8}{v5}{v8}{w9}
\end{fmfchar*}}}
\;{\quad}&W_{G4\phantom{0}}=&
\raisebox{\eqoff}{%
\fmfframe(3,1)(1,4){%
\begin{fmfchar*}(20,20)
\Woneplain
\fmfipair{wu[]}
\fmfipair{w[]}
\fmfipair{wd[]}
\svertex{w5}{p5}
\svertex{w6}{p6}
\svertex{w1}{p1}
\svertex{w7}{p7}
\vvertex{w8}{w5}{p1}
\vvertex{w9}{w5}{p7}
\fmfi{wiggly}{w8..w5}
\fmfi{wiggly}{w6..w9}
\wigglywrap{w8}{v5}{v8}{w9}
\end{fmfchar*}}}
\\
W_{G5\phantom{0}}=&
\raisebox{\eqoff}{%
\fmfframe(3,1)(1,4){%
\begin{fmfchar*}(20,20)
\Woneplain
\fmfipair{wu[]}
\fmfipair{w[]}
\fmfipair{wd[]}
\svertex{w5}{p5}
\svertex{w6}{p6}
\svertex{w1}{p1}
\svertex{w7}{p7}
\vvertex{w8}{w5}{p1}
\vvertex{w9}{w5}{p7}
\fmfi{wiggly}{w8..w5}
\fmfi{wiggly}{w9..w6}
\wigglywrap{w8}{v5}{v8}{w7}
\end{fmfchar*}}}
\;{\quad}&W_{G6\phantom{0}}=&
\raisebox{\eqoff}{%
\fmfframe(3,1)(1,4){%
\begin{fmfchar*}(20,20)
\Woneplain
\fmfipair{wu[]}
\fmfipair{w[]}
\fmfipair{wd[]}
\svertex{w5}{p5}
\svertex{w6}{p6}
\svertex{w1}{p1}
\svertex{w7}{p7}
\vvertex{w8}{w5}{p1}
\vvertex{w9}{w5}{p7}
\fmfi{wiggly}{w8..w5}
\fmfi{wiggly}{w6..w9}
\wigglywrap{w1}{v5}{v8}{w7}
\end{fmfchar*}}}
\;{\quad}&W_{G7\phantom{0}}=&
\raisebox{\eqoff}{%
\fmfframe(3,1)(1,4){%
\begin{fmfchar*}(20,20)
\Woneplain
\fmfipair{wu[]}
\fmfipair{w[]}
\fmfipair{wd[]}
\svertex{w5}{p5}
\svertex{w4}{p4}
\svertex{w1}{p1}
\svertex{w7}{p7}
\svertex{w6}{p6}
\vvertex{w8}{w5}{p1}
\vvertex{w9}{w6}{p7}
\fmfi{wiggly}{w1..w5}
\fmfi{wiggly}{w7..w4}
\wigglywrap{w8}{v5}{v8}{w9}
\end{fmfchar*}}}
\;{\quad}&W_{G8\phantom{0}}=&
\raisebox{\eqoff}{%
\fmfframe(3,1)(1,4){%
\begin{fmfchar*}(20,20)
\Woneplain
\fmfipair{wu[]}
\fmfipair{w[]}
\fmfipair{wd[]}
\svertex{w5}{p5}
\svertex{w4}{p4}
\svertex{w1}{p1}
\svertex{w7}{p7}
\vvertex{w8}{w5}{p1}
\vvertex{w9}{w5}{p7}
\fmfi{wiggly}{w1..w5}
\fmfi{wiggly}{w7..w4}
\wigglywrap{w8}{v5}{v8}{w7}
\end{fmfchar*}}}
\\
W_{G9\phantom{0}}=&
\raisebox{\eqoff}{%
\fmfframe(3,1)(1,4){%
\begin{fmfchar*}(20,20)
\Woneplain
\fmfipair{wu[]}
\fmfipair{w[]}
\fmfipair{wd[]}
\svertex{w5}{p5}
\svertex{w3}{p3}
\svertex{w4}{p4}
\svertex{w1}{p1}
\svertex{w7}{p7}
\vvertex{w8}{w5}{p1}
\vvertex{w9}{w3}{p7}
\fmfi{wiggly}{w1..w5}
\fmfi{wiggly}{w4..w7}
\wigglywrap{w8}{v5}{v8}{w9}
\end{fmfchar*}}}
\;{\quad}&W_{G10}=&
\raisebox{\eqoff}{%
\fmfframe(3,1)(1,4){%
\begin{fmfchar*}(20,20)
\Woneplain
\fmfipair{wu[]}
\fmfipair{w[]}
\fmfipair{wd[]}
\svertex{w5}{p5}
\svertex{w6}{p6}
\svertex{w4}{p4}
\svertex{w7}{p7}
\vvertex{w8}{w5}{p1}
\vvertex{w9}{w6}{p7}
\fmfi{wiggly}{w8..w5}
\fmfi{wiggly}{w7..w4}
\wigglywrap{w8}{v5}{v8}{w9}
\end{fmfchar*}}}
\;{\quad}&W_{G11}=&
\raisebox{\eqoff}{%
\fmfframe(3,1)(1,4){%
\begin{fmfchar*}(20,20)
\Woneplain
\fmfipair{wu[]}
\fmfipair{w[]}
\fmfipair{wd[]}
\svertex{w5}{p5}
\svertex{w4}{p4}
\svertex{w3}{p3}
\svertex{w7}{p7}
\vvertex{w8}{w5}{p1}
\vvertex{w9}{w3}{p7}
\fmfi{wiggly}{w8..w5}
\fmfi{wiggly}{w7..w4}
\wigglywrap{w8}{v5}{v8}{w7}
\end{fmfchar*}}}
\;{\quad}&W_{G12}=&
\raisebox{\eqoff}{%
\fmfframe(3,1)(1,4){%
\begin{fmfchar*}(20,20)
\Woneplain
\fmfipair{wu[]}
\fmfipair{w[]}
\fmfipair{wd[]}
\svertex{w5}{p5}
\svertex{w4}{p4}
\svertex{w3}{p3}
\svertex{w7}{p7}
\vvertex{w8}{w5}{p1}
\vvertex{w9}{w3}{p7}
\fmfi{wiggly}{w8..w5}
\fmfi{wiggly}{w7..w4}
\wigglywrap{w8}{v5}{v8}{w9}
\end{fmfchar*}}}
\\
W_{G13}=&
\raisebox{\eqoff}{%
\fmfframe(3,1)(1,4){%
\begin{fmfchar*}(20,20)
\Woneplain
\fmfipair{wu[]}
\fmfipair{w[]}
\fmfipair{wd[]}
\svertex{w5}{p5}
\svertex{w3}{p3}
\svertex{w4}{p4}
\svertex{w1}{p1}
\svertex{w7}{p7}
\svertex{w6}{p6}
\vvertex{w8}{w5}{p1}
\vvertex{w9}{w6}{p7}
\fmfi{wiggly}{w8..w5}
\fmfi{wiggly}{w4..w7}
\wigglywrap{w1}{v5}{v8}{w9}
\end{fmfchar*}}}
\;{\quad}&W_{G14}=&
\raisebox{\eqoff}{%
\fmfframe(3,1)(1,4){%
\begin{fmfchar*}(20,20)
\Woneplain
\fmfipair{wu[]}
\fmfipair{w[]}
\fmfipair{wd[]}
\svertex{w5}{p5}
\svertex{w1}{p1}
\svertex{w4}{p4}
\svertex{w7}{p7}
\vvertex{w8}{w5}{p1}
\vvertex{w9}{w1}{p7}
\fmfi{wiggly}{w8..w5}
\fmfi{wiggly}{w7..w4}
\wigglywrap{w1}{v5}{v8}{w7}
\end{fmfchar*}}}
\;{\quad}&W_{G15}=&
\raisebox{\eqoff}{%
\fmfframe(3,1)(1,4){%
\begin{fmfchar*}(20,20)
\Woneplain
\fmfipair{wu[]}
\fmfipair{w[]}
\fmfipair{wd[]}
\svertex{w5}{p5}
\svertex{w4}{p4}
\svertex{w3}{p3}
\svertex{w7}{p7}
\svertex{w1}{p1}
\vvertex{w8}{w5}{p1}
\vvertex{w9}{w3}{p7}
\fmfi{wiggly}{w8..w5}
\fmfi{wiggly}{w7..w4}
\wigglywrap{w1}{v5}{v8}{w9}
\end{fmfchar*}}}
\;{\quad}&W_{G16}=&
\raisebox{\eqoff}{%
\fmfframe(3,1)(1,4){%
\begin{fmfchar*}(20,20)
\Woneplain
\fmfipair{wu[]}
\fmfipair{w[]}
\fmfipair{wd[]}
\svertex{w5}{p5}
\svertex{w4}{p4}
\svertex{w3}{p3}
\svertex{w1}{p1}
\svertex{w7}{p7}
\svertex{w6}{p6}
\vvertex{w8}{w5}{p1}
\vvertex{w9}{w6}{p7}
\vvertex{w10}{w3}{p7}
\fmfi{wiggly}{w1..w5}
\fmfi{wiggly}{w3..w10}
\wigglywrap{w8}{v5}{v8}{w9}
\end{fmfchar*}}}
\\
W_{G17}=&
\raisebox{\eqoff}{%
\fmfframe(3,1)(1,4){%
\begin{fmfchar*}(20,20)
\Woneplain
\fmfipair{wu[]}
\fmfipair{w[]}
\fmfipair{wd[]}
\svertex{w5}{p5}
\svertex{w3}{p3}
\svertex{w4}{p4}
\svertex{w1}{p1}
\svertex{w7}{p7}
\svertex{w6}{p6}
\vvertex{w8}{w5}{p1}
\vvertex{w9}{w3}{p7}
\fmfi{wiggly}{w1..w5}
\fmfi{wiggly}{w3..w9}
\wigglywrap{w8}{v5}{v8}{w9}
\end{fmfchar*}}}
\;{\quad}&W_{G18}=&
\raisebox{\eqoff}{%
\fmfframe(3,1)(1,4){%
\begin{fmfchar*}(20,20)
\Woneplain
\fmfipair{wu[]}
\fmfipair{w[]}
\fmfipair{wd[]}
\svertex{w5}{p5}
\svertex{w3}{p3}
\svertex{w1}{p1}
\svertex{w4}{p4}
\svertex{w7}{p7}
\vvertex{w8}{w5}{p1}
\vvertex{w9}{w3}{p7}
\fmfi{wiggly}{w1..w5}
\fmfi{wiggly}{w7..w3}
\wigglywrap{w8}{v5}{v8}{w9}
\end{fmfchar*}}}
\;{\quad}&W_{G19}=&
\raisebox{\eqoff}{%
\fmfframe(3,1)(1,4){%
\begin{fmfchar*}(20,20)
\Woneplain
\fmfipair{wu[]}
\fmfipair{w[]}
\fmfipair{wd[]}
\svertex{w5}{p5}
\svertex{w4}{p4}
\svertex{w3}{p3}
\svertex{w7}{p7}
\svertex{w1}{p1}
\svertex{w6}{p6}
\vvertex{w8}{w5}{p1}
\vvertex{w9}{w3}{p7}
\vvertex{w10}{w6}{p7}
\fmfi{wiggly}{w8..w5}
\fmfi{wiggly}{w3..w9}
\wigglywrap{w8}{v5}{v8}{w10}
\end{fmfchar*}}}
\;{\quad}&W_{G20}=&
\raisebox{\eqoff}{%
\fmfframe(3,1)(1,4){%
\begin{fmfchar*}(20,20)
\Woneplain
\fmfipair{wu[]}
\fmfipair{w[]}
\fmfipair{wd[]}
\svertex{w5}{p5}
\svertex{w4}{p4}
\svertex{w3}{p3}
\svertex{w1}{p1}
\svertex{w7}{p7}
\svertex{w6}{p6}
\vvertex{w8}{w5}{p1}
\vvertex{w9}{w6}{p7}
\vvertex{w10}{w3}{p7}
\fmfi{wiggly}{w8..w5}
\fmfi{wiggly}{w3..w7}
\wigglywrap{w8}{v5}{v8}{w7}
\end{fmfchar*}}}
\\
W_{G21}=&
\raisebox{\eqoff}{%
\fmfframe(3,1)(1,4){%
\begin{fmfchar*}(20,20)
\Woneplain
\fmfipair{wu[]}
\fmfipair{w[]}
\fmfipair{wd[]}
\svertex{w5}{p5}
\svertex{w3}{p3}
\svertex{w4}{p4}
\svertex{w1}{p1}
\svertex{w7}{p7}
\svertex{w6}{p6}
\vvertex{w8}{w5}{p1}
\vvertex{w9}{w3}{p7}
\fmfi{wiggly}{w8..w5}
\fmfi{wiggly}{w3..w7}
\wigglywrap{w8}{v5}{v8}{w9}
\end{fmfchar*}}}
\;{\quad}&W_{G22}=&
\raisebox{\eqoff}{%
\fmfframe(3,1)(1,4){%
\begin{fmfchar*}(20,20)
\Woneplain
\fmfipair{wu[]}
\fmfipair{w[]}
\fmfipair{wd[]}
\svertex{w5}{p5}
\svertex{w3}{p3}
\svertex{w1}{p1}
\svertex{w4}{p4}
\svertex{w7}{p7}
\vvertex{w8}{w5}{p1}
\vvertex{w9}{w3}{p7}
\fmfi{wiggly}{w8..w5}
\fmfi{wiggly}{w3..w9}
\wigglywrap{w1}{v5}{v8}{w7}
\end{fmfchar*}}}
\;{\quad}&W_{G23}=&
\raisebox{\eqoff}{%
\fmfframe(3,1)(1,4){%
\begin{fmfchar*}(20,20)
\Woneplain
\fmfipair{wu[]}
\fmfipair{w[]}
\fmfipair{wd[]}
\svertex{w5}{p5}
\svertex{w4}{p4}
\svertex{w3}{p3}
\svertex{w7}{p7}
\svertex{w1}{p1}
\svertex{w6}{p6}
\vvertex{w8}{w5}{p1}
\vvertex{w9}{w3}{p7}
\vvertex{w10}{w6}{p7}
\fmfi{wiggly}{w8..w5}
\fmfi{wiggly}{w3..w7}
\wigglywrap{w1}{v5}{v8}{w7}
\end{fmfchar*}}}
\;{\quad}&W_{G24}=&
\raisebox{\eqoff}{%
\fmfframe(3,1)(1,4){%
\begin{fmfchar*}(20,20)
\Woneplain
\fmfipair{wu[]}
\fmfipair{w[]}
\fmfipair{wd[]}
\svertex{w5}{p5}
\svertex{w4}{p4}
\svertex{w3}{p3}
\svertex{w1}{p1}
\svertex{w7}{p7}
\svertex{w6}{p6}
\vvertex{w8}{w5}{p1}
\vvertex{w9}{w3}{p7}
\vvertex{w10}{w3}{p7}
\fmfi{wiggly}{w8..w5}
\fmfi{wiggly}{w3..w7}
\wigglywrap{w1}{v5}{v8}{w9}
\end{fmfchar*}}}
\\
W_{G25}=&
\raisebox{\eqoff}{%
\fmfframe(3,1)(1,4){%
\begin{fmfchar*}(20,20)
\Woneplain
\fmfipair{wu[]}
\fmfipair{w[]}
\fmfipair{wd[]}
\svertex{w5}{p5}
\svertex{w3}{p3}
\svertex{w4}{p4}
\svertex{w6}{p6}
\svertex{w7}{p7}
\svertex{w6}{p6}
\fmfiequ{wu4}{point length(p4)/4 of p4}
\fmfiequ{wd4}{point 3length(p4)/4 of p4}
\vvertex{w8}{w5}{p1}
\vvertex{w9}{w6}{p7}
\vvertex{w10}{wu4}{p1}
\vvertex{w11}{wd4}{p7}
\fmfi{wiggly}{w10..wu4}
\fmfi{wiggly}{w11..wd4}
\wigglywrap{w8}{v5}{v8}{w9}
\end{fmfchar*}}}
\;{\quad}&W_{G26}=&
\raisebox{\eqoff}{%
\fmfframe(3,1)(1,4){%
\begin{fmfchar*}(20,20)
\Woneplain
\fmfipair{wu[]}
\fmfipair{w[]}
\fmfipair{wd[]}
\svertex{w5}{p5}
\svertex{w3}{p3}
\svertex{w4}{p4}
\svertex{w6}{p6}
\svertex{w7}{p7}
\svertex{w6}{p6}
\fmfiequ{wu4}{point length(p4)/4 of p4}
\fmfiequ{wd4}{point 3length(p4)/4 of p4}
\vvertex{w8}{w5}{p1}
\vvertex{w9}{w6}{p7}
\vvertex{w10}{wu4}{p1}
\vvertex{w11}{wd4}{p7}
\fmfi{wiggly}{w10..wu4}
\fmfi{wiggly}{w11..wd4}
\wigglywrap{w8}{v5}{v8}{w11}
\end{fmfchar*}}}
\;{\quad}&W_{G27}=&
\raisebox{\eqoff}{%
\fmfframe(3,1)(1,4){%
\begin{fmfchar*}(20,20)
\Woneplain
\fmfipair{wu[]}
\fmfipair{w[]}
\fmfipair{wd[]}
\svertex{w5}{p5}
\svertex{w3}{p3}
\svertex{w4}{p4}
\svertex{w6}{p6}
\svertex{w7}{p7}
\svertex{w6}{p6}
\fmfiequ{wu4}{point length(p4)/4 of p4}
\fmfiequ{wd4}{point 3length(p4)/4 of p4}
\vvertex{w8}{w5}{p1}
\vvertex{w9}{w6}{p7}
\vvertex{w10}{wu4}{p1}
\vvertex{w11}{wd4}{p7}
\vvertex{w12}{wu4}{p7}
\fmfi{wiggly}{w10..wu4}
\fmfi{wiggly}{w11..wd4}
\wigglywrap{w8}{v5}{v8}{w12}
\end{fmfchar*}}}
\;{\quad}&W_{G28}=&
\raisebox{\eqoff}{%
\fmfframe(3,1)(1,4){%
\begin{fmfchar*}(20,20)
\Woneplain
\fmfipair{wu[]}
\fmfipair{w[]}
\fmfipair{wd[]}
\svertex{w5}{p5}
\svertex{w3}{p3}
\svertex{w4}{p4}
\svertex{w6}{p6}
\svertex{w7}{p7}
\svertex{w6}{p6}
\fmfiequ{wu4}{point length(p4)/4 of p4}
\fmfiequ{wd4}{point 3length(p4)/4 of p4}
\vvertex{w8}{w5}{p1}
\vvertex{w9}{w6}{p7}
\vvertex{w10}{wu4}{p1}
\vvertex{w11}{wd4}{p7}
\vvertex{w12}{wu4}{p7}
\fmfi{wiggly}{w10..wu4}
\fmfi{wiggly}{w11..wd4}
\wigglywrap{w10}{v5}{v8}{w9}
\end{fmfchar*}}}
\\
W_{G29}=&
\raisebox{\eqoff}{%
\fmfframe(3,1)(1,4){%
\begin{fmfchar*}(20,20)
\Woneplain
\fmfipair{wu[]}
\fmfipair{w[]}
\fmfipair{wd[]}
\svertex{w5}{p5}
\svertex{w3}{p3}
\svertex{w4}{p4}
\svertex{w6}{p6}
\svertex{w7}{p7}
\svertex{w6}{p6}
\fmfiequ{wu4}{point length(p4)/4 of p4}
\fmfiequ{wd4}{point 3length(p4)/4 of p4}
\vvertex{w8}{w5}{p1}
\vvertex{w9}{w6}{p7}
\vvertex{w10}{wu4}{p1}
\vvertex{w11}{wd4}{p7}
\fmfi{wiggly}{w10..wu4}
\fmfi{wiggly}{w11..wd4}
\wigglywrap{w10}{v5}{v8}{w11}
\end{fmfchar*}}}
\;{\quad}&W_{G30}=&
\raisebox{\eqoff}{%
\fmfframe(3,1)(1,4){%
\begin{fmfchar*}(20,20)
\Woneplain
\fmfipair{wu[]}
\fmfipair{w[]}
\fmfipair{wd[]}
\svertex{w5}{p5}
\svertex{w3}{p3}
\svertex{w4}{p4}
\svertex{w6}{p6}
\svertex{w7}{p7}
\svertex{w6}{p6}
\fmfiequ{wu4}{point length(p4)/4 of p4}
\fmfiequ{wd4}{point 3length(p4)/4 of p4}
\vvertex{w8}{w5}{p1}
\vvertex{w9}{w6}{p7}
\vvertex{w10}{wu4}{p1}
\vvertex{w11}{wd4}{p7}
\vvertex{w12}{wu4}{p7}
\fmfi{wiggly}{w10..wu4}
\fmfi{wiggly}{w11..wd4}
\wigglywrap{w10}{v5}{v8}{w12}
\end{fmfchar*}}}
\;{\quad}&W_{G31}=&
\raisebox{\eqoff}{%
\fmfframe(3,1)(1,4){%
\begin{fmfchar*}(20,20)
\Woneplain
\fmfipair{wu[]}
\fmfipair{w[]}
\fmfipair{wd[]}
\svertex{w5}{p5}
\svertex{w3}{p3}
\svertex{w4}{p4}
\svertex{w6}{p6}
\svertex{w7}{p7}
\svertex{w6}{p6}
\fmfiequ{wu4}{point length(p4)/4 of p4}
\fmfiequ{wd4}{point 3length(p4)/4 of p4}
\vvertex{w8}{w3}{p1}
\vvertex{w9}{w6}{p7}
\vvertex{w10}{wu4}{p1}
\vvertex{w11}{wd4}{p7}
\vvertex{w12}{wu4}{p7}
\fmfi{wiggly}{w10..wu4}
\fmfi{wiggly}{w11..wd4}
\wigglywrap{w8}{v5}{v8}{w9}
\end{fmfchar*}}}
\;{\quad}&W_{G32}=&
\raisebox{\eqoff}{%
\fmfframe(3,1)(1,4){%
\begin{fmfchar*}(20,20)
\Woneplain
\fmfipair{wu[]}
\fmfipair{w[]}
\fmfipair{wd[]}
\svertex{w5}{p5}
\svertex{w3}{p3}
\svertex{w4}{p4}
\svertex{w6}{p6}
\svertex{w7}{p7}
\svertex{w6}{p6}
\fmfiequ{wu4}{point length(p4)/4 of p4}
\fmfiequ{wd4}{point 3length(p4)/4 of p4}
\vvertex{w8}{w3}{p1}
\vvertex{w9}{w6}{p7}
\vvertex{w10}{wu4}{p1}
\vvertex{w11}{wd4}{p7}
\vvertex{w12}{wu4}{p7}
\fmfi{wiggly}{w10..wu4}
\fmfi{wiggly}{w11..wd4}
\wigglywrap{w8}{v5}{v8}{w11}
\end{fmfchar*}}}
\\
W_{G33}=&
\raisebox{\eqoff}{%
\fmfframe(3,1)(1,4){%
\begin{fmfchar*}(20,20)
\Woneplain
\fmfipair{wu[]}
\fmfipair{w[]}
\fmfipair{wd[]}
\svertex{w2}{p2}
\svertex{w3}{p3}
\svertex{w4}{p4}
\svertex{w6}{p6}
\svertex{w7}{p7}
\svertex{w6}{p6}
\fmfiequ{wu4}{point length(p4)/4 of p4}
\fmfiequ{wd4}{point 3length(p4)/4 of p4}
\vvertex{w8}{w2}{p1}
\vvertex{w9}{w3}{p7}
\vvertex{w10}{wu4}{p1}
\vvertex{w11}{wd4}{p7}
\fmfi{wiggly}{w10..wu4}
\fmfi{wiggly}{w11..wd4}
\wigglywrap{w8}{v5}{v8}{w9}
\end{fmfchar*}}}
\;{\quad}&W_{G34}=&
\raisebox{\eqoff}{%
\fmfframe(3,1)(1,4){%
\begin{fmfchar*}(20,20)
\Woneplain
\fmfipair{wu[]}
\fmfipair{w[]}
\fmfipair{wd[]}
\svertex{w5}{p5}
\svertex{w1}{p1}
\svertex{w4}{p4}
\svertex{w6}{p6}
\svertex{w7}{p7}
\svertex{w6}{p6}
\dvertex{wu4}{wd4}{p4}
\vvertex{w8}{w5}{p1}
\vvertex{w9}{w6}{p7}
\vvertex{w10}{wu4}{p1}
\vvertex{w11}{wd4}{p7}
\vvertex{w12}{wu4}{p7}
\fmfi{wiggly}{w1..w4}
\fmfi{wiggly}{w7..w4}
\wigglywrap{w8}{v5}{v8}{w9}
\end{fmfchar*}}}
\;{\quad}&W_{G35}=&
\raisebox{\eqoff}{%
\fmfframe(3,1)(1,4){%
\begin{fmfchar*}(20,20)
\Woneplain
\fmfipair{wu[]}
\fmfipair{w[]}
\fmfipair{wd[]}
\svertex{w5}{p5}
\svertex{w1}{p1}
\svertex{w4}{p4}
\svertex{w6}{p6}
\svertex{w7}{p7}
\svertex{w6}{p6}
\dvertex{wu4}{wd4}{p4}
\vvertex{w8}{w5}{p1}
\vvertex{w9}{w6}{p7}
\vvertex{w10}{wu4}{p1}
\vvertex{w11}{wd4}{p7}
\vvertex{w12}{wu4}{p7}
\fmfi{wiggly}{w1..w4}
\fmfi{wiggly}{w7..w4}
\wigglywrap{w8}{v5}{v8}{w7}
\end{fmfchar*}}}
\;{\quad}&W_{G36}=&
\raisebox{\eqoff}{%
\fmfframe(3,1)(1,4){%
\begin{fmfchar*}(20,20)
\Woneplain
\fmfipair{wu[]}
\fmfipair{w[]}
\fmfipair{wd[]}
\svertex{w5}{p5}
\svertex{w1}{p1}
\svertex{w4}{p4}
\svertex{w3}{p3}
\svertex{w7}{p7}
\svertex{w6}{p6}
\dvertex{wu4}{wd4}{p4}
\vvertex{w8}{w5}{p1}
\vvertex{w9}{w3}{p7}
\vvertex{w10}{wu4}{p1}
\vvertex{w11}{wd4}{p7}
\vvertex{w12}{wu4}{p7}
\fmfi{wiggly}{w1..w4}
\fmfi{wiggly}{w7..w4}
\wigglywrap{w8}{v5}{v8}{w9}
\end{fmfchar*}}}
\\
W_{G37}=&
\raisebox{\eqoff}{%
\fmfframe(3,1)(1,4){%
\begin{fmfchar*}(20,20)
\Woneplain
\fmfipair{wu[]}
\fmfipair{w[]}
\fmfipair{wd[]}
\svertex{w2}{p2}
\svertex{w1}{p1}
\svertex{w3}{p3}
\svertex{w4}{p4}
\svertex{w6}{p6}
\svertex{w7}{p7}
\svertex{w6}{p6}
\dvertex{wu4}{wd4}{p4}
\vvertex{w8}{w2}{p1}
\vvertex{w9}{w3}{p7}
\vvertex{w10}{wu4}{p1}
\vvertex{w11}{wd4}{p7}
\fmfi{wiggly}{w4..w7}
\fmfi{wiggly}{w1..w4}
\wigglywrap{w1}{v5}{v8}{w7}
\end{fmfchar*}}}
\;{\quad}&W_{G38}=&
\raisebox{\eqoff}{%
\fmfframe(3,1)(1,4){%
\begin{fmfchar*}(20,20)
\Woneplain
\fmfipair{wu[]}
\fmfipair{w[]}
\fmfipair{wd[]}
\svertex{w5}{p5}
\svertex{w1}{p1}
\svertex{w4}{p4}
\svertex{w3}{p3}
\svertex{w7}{p7}
\svertex{w6}{p6}
\dvertex{wu4}{wd4}{p4}
\vvertex{w8}{w5}{p1}
\vvertex{w9}{w3}{p7}
\vvertex{w10}{wu4}{p1}
\vvertex{w11}{wd4}{p7}
\vvertex{w12}{wu4}{p7}
\fmfi{wiggly}{w1..w4}
\fmfi{wiggly}{w7..w4}
\wigglywrap{w1}{v5}{v8}{w9}
\end{fmfchar*}}}
\;{\quad}&W_{G39}=&
\raisebox{\eqoff}{%
\fmfframe(3,1)(1,4){%
\begin{fmfchar*}(20,20)
\Woneplain
\fmfipair{wu[]}
\fmfipair{w[]}
\fmfipair{wd[]}
\svertex{w5}{p5}
\svertex{w1}{p1}
\svertex{w4}{p4}
\svertex{w3}{p3}
\svertex{w7}{p7}
\svertex{w6}{p6}
\dvertex{wu4}{wd4}{p4}
\vvertex{w8}{w3}{p1}
\vvertex{w9}{w3}{p7}
\vvertex{w10}{wu4}{p1}
\vvertex{w11}{wd4}{p7}
\vvertex{w12}{wu4}{p7}
\fmfi{wiggly}{w1..w4}
\fmfi{wiggly}{w7..w4}
\wigglywrap{w8}{v5}{v8}{w9}
\end{fmfchar*}}}
\;{\quad}&W_{G40}=&
\raisebox{\eqoff}{%
\fmfframe(3,1)(1,4){%
\begin{fmfchar*}(20,20)
\Woneplain
\fmfipair{wu[]}
\fmfipair{w[]}
\fmfipair{wd[]}
\svertex{w5}{p5}
\svertex{w1}{p1}
\svertex{w4}{p4}
\svertex{w3}{p3}
\svertex{w7}{p7}
\svertex{w6}{p6}
\dvertex{wu4}{wd4}{p4}
\vvertex{w8}{w5}{p1}
\vvertex{w9}{w3}{p7}
\vvertex{w10}{w6}{p7}
\vvertex{w11}{wd4}{p7}
\vvertex{w12}{wu4}{p7}
\fmfi{wiggly}{w1..w4}
\fmfi{wiggly}{w3..w9}
\wigglywrap{w8}{v5}{v8}{w10}
\end{fmfchar*}}}
\\
\end{aligned}
\end{equation*}
\caption{Wrapping diagrams with chiral structure $\chi(1)$}
\label{diagrams-1}
\end{figure}

\clearpage
\newpage
\thispagestyle{empty}

\begin{figure}[!h]
\unitlength=0.75mm
\settoheight{\eqoff}{$\times$}%
\setlength{\eqoff}{0.5\eqoff}%
\addtolength{\eqoff}{-12.5\unitlength}%
\settoheight{\eqofftwo}{$\times$}%
\setlength{\eqofftwo}{0.5\eqofftwo}%
\addtolength{\eqofftwo}{-7.5\unitlength}%
\begin{equation*}
\begin{aligned}
W_{G41}=&
\raisebox{\eqoff}{%
\fmfframe(3,1)(1,4){%
\begin{fmfchar*}(20,20)
\Woneplain
\fmfipair{wu[]}
\fmfipair{w[]}
\fmfipair{wd[]}
\svertex{w2}{p2}
\svertex{w1}{p1}
\svertex{w3}{p3}
\svertex{w4}{p4}
\svertex{w6}{p6}
\svertex{w7}{p7}
\svertex{w5}{p5}
\dvertex{wu4}{wd4}{p4}
\vvertex{w8}{w5}{p1}
\vvertex{w9}{w3}{p7}
\vvertex{w10}{wu4}{p1}
\vvertex{w11}{wd4}{p7}
\fmfi{wiggly}{w3..w9}
\fmfi{wiggly}{w1..w4}
\wigglywrap{w8}{v5}{v8}{w9}
\end{fmfchar*}}}
\;{\quad}&W_{G42}=&
\raisebox{\eqoff}{%
\fmfframe(3,1)(1,4){%
\begin{fmfchar*}(20,20)
\Woneplain
\fmfipair{wu[]}
\fmfipair{w[]}
\fmfipair{wd[]}
\svertex{w5}{p5}
\svertex{w1}{p1}
\svertex{w4}{p4}
\svertex{w3}{p3}
\svertex{w7}{p7}
\svertex{w6}{p6}
\dvertex{wu4}{wd4}{p4}
\vvertex{w8}{w5}{p1}
\vvertex{w9}{w3}{p7}
\vvertex{w10}{wu4}{p1}
\vvertex{w11}{wd4}{p7}
\vvertex{w12}{wu4}{p7}
\fmfi{wiggly}{w1..w4}
\fmfi{wiggly}{w7..w3}
\wigglywrap{w8}{v5}{v8}{w9}
\end{fmfchar*}}}
\;{\quad}&W_{G43}=&
\raisebox{\eqoff}{%
\fmfframe(3,1)(1,4){%
\begin{fmfchar*}(20,20)
\Woneplain
\fmfipair{wu[]}
\fmfipair{w[]}
\fmfipair{wd[]}
\svertex{w5}{p5}
\svertex{w1}{p1}
\svertex{w4}{p4}
\svertex{w3}{p3}
\svertex{w7}{p7}
\svertex{w6}{p6}
\dvertex{wu4}{wd4}{p4}
\vvertex{w8}{w3}{p1}
\vvertex{w9}{w3}{p7}
\vvertex{w10}{wu4}{p1}
\vvertex{w11}{wd4}{p7}
\vvertex{w12}{wu4}{p7}
\fmfi{wiggly}{w1..w4}
\fmfi{wiggly}{w3..w9}
\wigglywrap{w1}{v5}{v8}{w7}
\end{fmfchar*}}}
\;{\quad}&W_{G44}=&
\raisebox{\eqoff}{%
\fmfframe(3,1)(1,4){%
\begin{fmfchar*}(20,20)
\Woneplain
\fmfipair{wu[]}
\fmfipair{w[]}
\fmfipair{wd[]}
\svertex{w5}{p5}
\svertex{w1}{p1}
\svertex{w4}{p4}
\svertex{w3}{p3}
\svertex{w7}{p7}
\svertex{w6}{p6}
\dvertex{wu4}{wd4}{p4}
\vvertex{w8}{w5}{p1}
\vvertex{w9}{w3}{p7}
\vvertex{w10}{w6}{p7}
\vvertex{w11}{wd4}{p7}
\vvertex{w12}{wu4}{p7}
\fmfi{wiggly}{w1..w4}
\fmfi{wiggly}{w3..w7}
\wigglywrap{w1}{v5}{v8}{w7}
\end{fmfchar*}}}
\\
W_{G45}=&
\raisebox{\eqoff}{%
\fmfframe(3,1)(1,4){%
\begin{fmfchar*}(20,20)
\Woneplain
\fmfipair{wu[]}
\fmfipair{w[]}
\fmfipair{wd[]}
\svertex{w4}{p4}
\svertex{w6}{p6}
\svertex{w1}{p1}
\svertex{w7}{p7}
\svertex{w3}{p3}
\vvertex{w8}{w7}{p1}
\vvertex{w9}{w3}{p7}
\fmfi{wiggly}{w1..w4}
\fmfi{wiggly}{w7..w3}
\wigglywrap{w1}{v5}{v8}{w9}
\end{fmfchar*}}}
\;{\quad}&W_{G46}=&
\raisebox{\eqoff}{%
\fmfframe(3,1)(1,4){%
\begin{fmfchar*}(20,20)
\Woneplain
\fmfipair{wu[]}
\fmfipair{w[]}
\fmfipair{wd[]}
\svertex{w4}{p4}
\svertex{w6}{p6}
\svertex{w1}{p1}
\svertex{w7}{p7}
\svertex{w3}{p3}
\vvertex{w8}{w3}{p1}
\vvertex{w9}{w3}{p7}
\fmfi{wiggly}{w1..w4}
\fmfi{wiggly}{w9..w3}
\wigglywrap{w8}{v5}{v8}{w7}
\end{fmfchar*}}}
\;{\quad}&W_{G47}=&
\raisebox{\eqoff}{%
\fmfframe(3,1)(1,4){%
\begin{fmfchar*}(20,20)
\Woneplain
\fmfipair{wu[]}
\fmfipair{w[]}
\fmfipair{wd[]}
\svertex{w4}{p4}
\svertex{w6}{p6}
\svertex{w1}{p1}
\svertex{w7}{p7}
\svertex{w3}{p3}
\vvertex{w8}{w3}{p1}
\vvertex{w9}{w3}{p7}
\fmfi{wiggly}{w1..w4}
\fmfi{wiggly}{w7..w3}
\wigglywrap{w8}{v5}{v8}{w7}
\end{fmfchar*}}}
\;{\quad}&W_{G48}=&
\raisebox{\eqoff}{%
\fmfframe(3,1)(1,4){%
\begin{fmfchar*}(20,20)
\Woneplain
\fmfipair{wu[]}
\fmfipair{w[]}
\fmfipair{wd[]}
\svertex{w4}{p4}
\svertex{w6}{p6}
\svertex{w1}{p1}
\svertex{w7}{p7}
\svertex{w3}{p3}
\vvertex{w8}{w3}{p1}
\vvertex{w9}{w3}{p7}
\fmfi{wiggly}{w1..w4}
\fmfi{wiggly}{w7..w3}
\wigglywrap{w8}{v5}{v8}{w9}
\end{fmfchar*}}}
\\
W_{G49}=&
\raisebox{\eqoff}{%
\fmfframe(3,1)(1,4){%
\begin{fmfchar*}(20,20)
\Woneplain
\fmfipair{wu[]}
\fmfipair{w[]}
\fmfipair{wd[]}
\svertex{w4}{p4}
\svertex{w2}{p2}
\svertex{w1}{p1}
\svertex{w7}{p7}
\svertex{w3}{p3}
\vvertex{w8}{w2}{p1}
\vvertex{w9}{w3}{p7}
\fmfi{wiggly}{w8..w2}
\fmfi{wiggly}{w9..w3}
\wigglywrap{w1}{v5}{v8}{w7}
\end{fmfchar*}}}
\;{\quad}&W_{G50}=&
\raisebox{\eqoff}{%
\fmfframe(3,1)(1,4){%
\begin{fmfchar*}(20,20)
\Woneplain
\fmfipair{wu[]}
\fmfipair{w[]}
\fmfipair{wd[]}
\svertex{w4}{p4}
\svertex{w2}{p2}
\svertex{w1}{p1}
\svertex{w7}{p7}
\svertex{w3}{p3}
\vvertex{w8}{w2}{p1}
\vvertex{w9}{w3}{p7}
\fmfi{wiggly}{w8..w2}
\fmfi{wiggly}{w3..w7}
\wigglywrap{w1}{v5}{v8}{w7}
\end{fmfchar*}}}
\;{\quad}&W_{G51}=&
\raisebox{\eqoff}{%
\fmfframe(3,1)(1,4){%
\begin{fmfchar*}(20,20)
\Woneplain
\fmfipair{wu[]}
\fmfipair{w[]}
\fmfipair{wd[]}
\svertex{w4}{p4}
\svertex{w2}{p2}
\svertex{w1}{p1}
\svertex{w7}{p7}
\svertex{w3}{p3}
\vvertex{w8}{w2}{p1}
\vvertex{w9}{w3}{p7}
\fmfi{wiggly}{w8..w2}
\fmfi{wiggly}{w7..w3}
\wigglywrap{w1}{v5}{v8}{w9}
\end{fmfchar*}}}
\;{\quad}&W_{G52}=&
\raisebox{\eqoff}{%
\fmfframe(3,1)(1,4){%
\begin{fmfchar*}(20,20)
\Woneplain
\fmfipair{wu[]}
\fmfipair{w[]}
\fmfipair{wd[]}
\svertex{w4}{p4}
\svertex{w2}{p2}
\svertex{w1}{p1}
\svertex{w7}{p7}
\svertex{w3}{p3}
\vvertex{w8}{w2}{p1}
\vvertex{w9}{w3}{p7}
\fmfi{wiggly}{w1..w2}
\fmfi{wiggly}{w7..w3}
\wigglywrap{w1}{v5}{v8}{w7}
\end{fmfchar*}}}
\\
W_{G53}=&
\raisebox{\eqoff}{%
\fmfframe(3,1)(1,4){%
\begin{fmfchar*}(20,20)
\Woneplain
\fmfipair{wu[]}
\fmfipair{w[]}
\fmfipair{wd[]}
\svertex{w4}{p4}
\svertex{w2}{p2}
\svertex{w1}{p1}
\svertex{w7}{p7}
\svertex{w3}{p3}
\vvertex{w8}{w2}{p1}
\vvertex{w9}{w3}{p7}
\fmfi{wiggly}{w1..w2}
\fmfi{wiggly}{w7..w3}
\wigglywrap{w1}{v5}{v8}{w9}
\end{fmfchar*}}}
\;{\quad}&W_{G54}=&
\raisebox{\eqoff}{%
\fmfframe(3,1)(1,4){%
\begin{fmfchar*}(20,20)
\Woneplain
\fmfipair{wu[]}
\fmfipair{w[]}
\fmfipair{wd[]}
\svertex{w4}{p4}
\svertex{w2}{p2}
\svertex{w1}{p1}
\svertex{w7}{p7}
\svertex{w3}{p3}
\vvertex{w8}{w2}{p1}
\vvertex{w9}{w3}{p7}
\fmfi{wiggly}{w1..w2}
\fmfi{wiggly}{w3..w7}
\wigglywrap{w8}{v5}{v8}{w9}
\end{fmfchar*}}}
\end{aligned}
\end{equation*}
\begin{center}
Figure \ref{diagrams-1}: Wrapping diagrams with chiral structure $\chi(1)$ (continued)
\end{center}
\vspace{1cm}
\begin{center}
\small
\fbox{
\begin{tabular}{|m{35pt}m{30pt}|m{6pt}|}
\hline
& & \\[-2ex]
$W_{G1\phantom{0}}\rightarrow$ & $-W_{G34}$ &  \\
$W_{G2\phantom{0}}\rightarrow$ & $-W_{G10}$ &  \\
$W_{G3\phantom{0}}\rightarrow$ & $-W_{G5}$ &  \\
$W_{G4\phantom{0}}\rightarrow$ & finite &  \\
$W_{G5\phantom{0}}\rightarrow$ & $-W_{G3}$ &  \\
$W_{G6\phantom{0}}\rightarrow$ & finite &  \\
$W_{G7\phantom{0}}\rightarrow$ & $-W_{G25}$ &  \\
$W_{G8\phantom{0}}\rightarrow$ & $-W_{G28}$ &  \\
$W_{G9\phantom{0}}\rightarrow$ & $-W_{G12}$ &  \\
$W_{G10}\rightarrow$ & $-W_{G2}$ &  \\
$W_{G11}\rightarrow$ & $\ast\phantom{W_{A}}$ & 2 \\
$W_{G12}\rightarrow$ & $-W_{G9}$ &  \\
$W_{G13}\rightarrow$ & $-W_{G14}$ &  \\
$W_{G14}\rightarrow$ & $-W_{G13}$ &  \\
\hline
\end{tabular}
\begin{tabular}{|m{35pt}m{30pt}|m{6pt}|}
\hline
& & \\[-2ex]
$W_{G15}\rightarrow$ & finite &  \\
$W_{G16}\rightarrow$ & finite &  \\
$W_{G17}\rightarrow$ & finite &  \\
$W_{G18}\rightarrow$ & finite &  \\
$W_{G19}\rightarrow$ & finite &  \\
$W_{G20}\rightarrow$ & finite &  \\
$W_{G21}\rightarrow$ & finite &  \\
$W_{G22}\rightarrow$ & finite &  \\
$W_{G23}\rightarrow$ & finite &  \\
$W_{G24}\rightarrow$ & finite &  \\
$W_{G25}\rightarrow$ & $-W_{G7}$ &  \\
$W_{G26}\rightarrow$ & $-W_{G35}$ &  \\
$W_{G27}\rightarrow$ & $-W_{G30}$ &  \\
$W_{G28}\rightarrow$ & $-W_{G8}$ &  \\
\hline
\end{tabular}
\begin{tabular}{|m{35pt}m{30pt}|m{6pt}|}
\hline
& & \\[-2ex]
$W_{G29}\rightarrow$ & $\ast\phantom{W_{A}}$ & 2 \\
$W_{G30}\rightarrow$ & $-W_{G27}$ &  \\
$W_{G31}\rightarrow$ & $-W_{G32}$ &  \\
$W_{G32}\rightarrow$ & $-W_{G31}$ &  \\
$W_{G33}\rightarrow$ & finite &  \\
$W_{G34}\rightarrow$ & $-W_{G1}$ &  \\
$W_{G35}\rightarrow$ & $-W_{G26}$ &  \\
$W_{G36}\rightarrow$ & $-W_{G38}$ &  \\
$W_{G37}\rightarrow$ & finite &  \\
$W_{G38}\rightarrow$ & $-W_{G36}$ &  \\
$W_{G39}\rightarrow$ & finite &  \\
$W_{G40}\rightarrow$ & finite &  \\
$W_{G41}\rightarrow$ & finite &  \\
$W_{G42}\rightarrow$ & finite &  \\
\hline
\end{tabular}
\begin{tabular}{|m{35pt}m{30pt}|m{6pt}|}
\hline
& & \\[-2ex]
$W_{G43}\rightarrow$ & finite &  \\
$W_{G44}\rightarrow$ & finite &  \\
$W_{G45}\rightarrow$ & finite &  \\
$W_{G46}\rightarrow$ & finite &  \\
$W_{G47}\rightarrow$ & finite &  \\
$W_{G48}\rightarrow$ & finite &  \\
$W_{G49}\rightarrow$ & finite &  \\
$W_{G50}\rightarrow$ & finite &  \\
$W_{G51}\rightarrow$ & finite &  \\
$W_{G52}\rightarrow$ & finite &  \\
$W_{G53}\rightarrow$ & finite &  \\
$W_{G54}\rightarrow$ & finite &  \\
& & \\
& & \\
\hline
\end{tabular}}
\end{center}
\begin{mytable}
Results of $D$-algebra for diagrams with structure $\chi(1)$
\label{results-1}
\end{mytable}
\end{figure}
\normalsize

\clearpage

\begin{table}[p]
\begin{center}
\small
\begin{tabular}{|ll|}
\hline
& \\[-2ex]
$W_{A1\phantom{0}}\rightarrow$ & $(g^2 N)^4 (I_4/2)\,\chi(2,4,1,3)\rightarrow-2(g^2 N)^4 I_4 M$  \\
$W_{A2\phantom{0}}\rightarrow$ & $(g^2 N)^4 I_2\,[\,\chi(1,4,3,2)+\chi(4,1,2,3)\,]\rightarrow-2(g^2 N)^4 I_2 M$  \\
$W_{A3\phantom{0}}\rightarrow$ & $(g^2 N)^4 I_3\,\chi(4,3,1,2)\rightarrow-2(g^2 N)^4 I_3 M$  \\
[1ex]
\hline
\multicolumn{2}{|c|}{}\\[-1ex]
\multicolumn{2}{|c|}{
$\sum W_{A**}
\rightarrow-2(g^2 N)^4 (I_4+I_2+I_3)M
$
} \\[1ex]
\hline
\hline
& \\[-2ex]
$W_{B1\phantom{0}}\rightarrow$ & $-(g^2 N)^4 (I_3+I_5+2I_7) [\chi(1,2,3)+\chi(3,2,1)]\rightarrow-2(g^2 N)^4 (I_3+I_5+2I_7)M$  \\
$W_{B2\phantom{0}}\rightarrow$ & $(g^2 N)^4 I_3 [\chi(1,2,3)+\chi(3,2,1)]\rightarrow2(g^2 N)^4 I_3 M$  \\
$W_{B3\phantom{0}}\rightarrow$ & $-(g^2 N)^4 I_2 [\chi(1,2,3)+\chi(3,2,1)]\rightarrow-2(g^2 N)^4 I_2 M$  \\
$W_{B4\phantom{0}}\rightarrow$ & $(g^2 N)^4 (I_2+I_5+2I_9) [\chi(1,2,3)+\chi(3,2,1)]\rightarrow2(g^2 N)^4 (I_2+I_5+2I_9)M$  \\
[1ex]
\hline
\multicolumn{2}{|c|}{}\\[-1ex]
\multicolumn{2}{|c|}{
$\sum W_{B**}\rightarrow-2(g^2 N)^4 (I_7-I_9)[\chi(1,2,3)+\chi(3,2,1)]\rightarrow-4(g^2 N)^4 (I_7-I_9)M$
} \\[1ex]
\hline
\hline
& \\[-2ex]
$W_{C1\phantom{0}}\rightarrow$ & $-(g^2 N)^4 (2I_2+2I_6) \chi(1,3,2)\rightarrow-2(g^2 N)^4 (2I_2+2I_6)M$  \\
$W_{C2\phantom{0}}\rightarrow$ & $2(g^2 N)^4 I_2 \chi(1,3,2)\rightarrow4(g^2 N)^4 I_2 M$  \\
[1ex]
\hline
\multicolumn{2}{|c|}{}\\[-1ex]
\multicolumn{2}{|c|}{
$\sum W_{C**}\rightarrow-2(g^2 N)^4 I_6\chi(1,3,2)\rightarrow-4(g^2 N)^4 I_6 M$
} \\[1ex]
\hline
\hline
& \\[-2ex]
$W_{D1\phantom{0}}\rightarrow$ & $-2(g^2 N)^4 (I_2+I_8) \chi(2,1,3)\rightarrow-4(g^2 N)^4 (I_2+I_8)M$  \\
$W_{D2\phantom{0}}\rightarrow$ & $2(g^2 N)^4 I_2 \chi(2,1,3)\rightarrow4(g^2 N)^4 I_2 M$  \\
[1ex]
\hline
\multicolumn{2}{|c|}{}\\[-1ex]
\multicolumn{2}{|c|}{
$\sum W_{D**}\rightarrow-2I_8(g^2 N)^4 \chi(2,1,3)\rightarrow-4(g^2 N)^4 I_8 M$
} \\[1ex]
\hline
\hline
& \\[-2ex]
$W_{E2\phantom{0}}\rightarrow$ & $-(g^2 N)^4 (I_2+I_5+2I_9) [\chi(1,2)+\chi(2,1)]\rightarrow2(g^2 N)^4 (I_2+I_5+2I_9)M$  \\
$W_{E5\phantom{0}}\rightarrow$ & $(g^2 N)^4 I_2 [\chi(1,2)+\chi(2,1)]\rightarrow-2(g^2 N)^4 I_2 M$  \\
$W_{E11}\rightarrow$ & $(g^2 N)^4 (I_3+I_5+2I_7) [\chi(1,2)+\chi(2,1)]\rightarrow-2(g^2 N)^4 (I_3+I_5+2I_7)M$  \\
$W_{E14}\rightarrow$ & $-(g^2 N)^4 I_3 [\chi(1,2)+\chi(2,1)]\rightarrow2(g^2 N)^4 I_3 M$  \\
[1ex]
\hline
\multicolumn{2}{|c|}{}\\[-1ex]
\multicolumn{2}{|c|}{
$\sum W_{E**}\rightarrow2(g^2 N)^4 (I_7-I_9)[\chi(1,2)+\chi(2,1)]\rightarrow-4(g^2 N)^4 (I_7-I_9)M$
} \\[1ex]
\hline
\hline
& \\[-2ex]
$W_{F1\phantom{0}}\rightarrow$ & $ (g^2 N)^4 (I_3+I_5+4I_7+2I_{10})\chi(1,3)\rightarrow-2(g^2 N)^4 (I_3+I_5+4I_7+2I_{10})M$  \\
$W_{F2\phantom{0}}\rightarrow$ & $ -2(g^2 N)^4 (I_3+I_5+2I_7)\chi(1,3)\rightarrow4(g^2 N)^4 (I_3+I_5+2I_7)M$  \\
$W_{F7\phantom{0}}\rightarrow$ & $ (g^2 N)^4 (I_4+I_5+2I_{11}-4I_{12})\chi(1,3)\rightarrow-2(g^2 N)^4 (I_4+I_5+2I_{11}-4I_{12})M$  \\
$W_{F9\phantom{0}}\rightarrow$ & $ 2(g^2 N)^4 I_3\chi(1,3)\rightarrow-4(g^2 N)^4 I_3 M$  \\
$W_{F12}\rightarrow$ & $ -(g^2 N)^4 I_3\chi(1,3)\rightarrow2(g^2 N)^4 I_3 M$  \\
$W_{F32}\rightarrow$ & $ -2(g^2 N)^4 I_4\chi(1,3)\rightarrow4(g^2 N)^4 I_4 M$  \\
$W_{F37}\rightarrow$ & $ (g^2 N)^4 I_4\chi(1,3)\rightarrow-2(g^2 N)^4 I_4 M$  \\
[1ex]
\hline
\multicolumn{2}{|c|}{}\\[-1ex]
\multicolumn{2}{|c|}{
$\sum W_{F**}\rightarrow2(g^2 N)^4 (I_{10}+I_{11}-2I_{12})\chi(1,3)\rightarrow-4(g^2 N)^4 (I_{10}+I_{11}-2I_{12})M$
} \\[1ex]
\hline
\hline
& \\[-2ex]
$W_{G11}\rightarrow$ & $ 2(g^2 N)^4 I_1\chi(1)\rightarrow2(g^2 N)^4 I_1M$  \\
$W_{G29}\rightarrow$ & $ -2(g^2 N)^4 I_{2}\chi(1)\rightarrow-2(g^2 N)^4 I_2 M$  \\
[1ex]
\hline
\multicolumn{2}{|c|}{}\\[-1ex]
\multicolumn{2}{|c|}{
$\sum W_{G**}\rightarrow2(g^2 N)^4 (I_1-I_2)\chi(1)\rightarrow2(g^2 N)^4 (I_1-I_2)M$
} \\[1ex]
\hline
\end{tabular}
\end{center}
\caption{Wrapping contributions at four loops}
\label{wrapping-results}
\end{table}

\clearpage

\begin{table}[p]
\settoheight{\eqoff}{$\times$}%
\setlength{\eqoff}{0.5\eqoff}%
\addtolength{\eqoff}{-7.5\unitlength}
\begin{equation*}
\begin{aligned}\label{scalarintegrals}
I_1=J_1=
\raisebox{\eqoff}{%
\begin{fmfchar*}(20,15)
\fmfleft{in}
\fmfright{out}
\fmf{plain}{in,v1}
\fmf{plain,left=0.25}{v1,v2}
\fmf{plain,left=0}{v2,v4}
\fmf{plain,left=0.25}{v4,v3}
\fmf{plain,tension=0.5,right=0.25}{v1,v0,v1}
\fmf{plain,right=0.25}{v0,v3}
\fmf{plain}{v0,v2}
\fmf{plain}{v0,v4}
\fmf{plain}{v3,out}
\fmffixed{(0.9w,0)}{v1,v3}
\fmffixed{(0.4w,0)}{v2,v4}
\fmfpoly{phantom}{v4,v2,v0}
\fmffreeze
\end{fmfchar*}}
&=\frac{1}{(4\pi)^8}\Big(
-\frac{1}{24\varepsilon^4}+\frac{1}{4\varepsilon^3}
-\frac{19}{24\varepsilon^2}
+\frac{5}{4\varepsilon}
\Big)
\\
I_2=\raisebox{\eqoff}{%
\begin{fmfchar*}(20,15)
\fmfleft{in}
\fmfright{out}
\fmf{plain}{in,v1}
\fmf{plain,left=0.25}{v1,v2}
\fmf{plain,left=0.25}{v2,v3}
\fmf{plain,left=0.25}{v3,v4}
\fmf{plain,left=0.25}{v4,v1}
\fmf{plain,tension=0.5,right=0.5}{v2,v0,v2}
\fmf{phantom}{v0,v3}
\fmf{plain}{v1,v0}
\fmf{plain}{v0,v4}
\fmf{plain}{v3,out}
\fmffixed{(0.9w,0)}{v1,v3}
\fmffixed{(0,0.45w)}{v4,v2}
\fmffreeze
\end{fmfchar*}}
&=\frac{1}{(4\pi)^8}\Big(
-\frac{1}{24\varepsilon^4}+\frac{1}{4\varepsilon^3}
-\frac{19}{24\varepsilon^2}
+\frac{1}{\varepsilon}\Big(\frac{5}{4}-\zeta(3)\Big)
\Big)
\\
I_3=J_5=\raisebox{\eqoff}{%
\begin{fmfchar*}(20,15)
\fmfleft{in}
\fmfright{out}
\fmf{plain}{in,v1}
\fmf{plain,left=0.25}{v1,v2}
\fmf{plain,left=0.25}{v2,v3}
\fmf{plain,left=0.25}{v3,v4}
\fmf{plain,left=0.25}{v4,v1}
\fmf{plain,tension=0.5,right=0.25}{v1,v0,v1}
\fmf{phantom}{v0,v3}
\fmf{plain}{v2,v0}
\fmf{plain}{v0,v4}
\fmf{plain}{v3,out}
\fmffixed{(0.9w,0)}{v1,v3}
\fmffixed{(0,0.45w)}{v4,v2}
\fmffreeze
\end{fmfchar*}}
&=\frac{1}{(4\pi)^8}\Big(
-\frac{1}{12\varepsilon^4}+\frac{1}{3\varepsilon^3}
-\frac{5}{12\varepsilon^2}
-\frac{1}{\varepsilon}\Big(\frac{1}{2}-\zeta(3)\Big)\Big)
\\
I_4=\raisebox{\eqoff}{%
\begin{fmfchar*}(20,15)
\fmfleft{in}
\fmfright{out}
\fmf{plain}{in,v1}
\fmf{plain,left=0.25}{v1,v2}
\fmf{plain,left=0.25}{v2,v3}
\fmf{plain,left=0.25}{v3,v4}
\fmf{plain,left=0.25}{v4,v1}
\fmf{plain,tension=0.5,right=0.5}{v2,v0,v2}
\fmf{plain,tension=0.5,right=0.5}{v0,v4,v0}
\fmf{plain}{v3,out}
\fmffixed{(0.9w,0)}{v1,v3}
\fmffixed{(0,0.45w)}{v4,v2}
\fmffreeze
\end{fmfchar*}}
&=\frac{1}{(4\pi)^8}\Big(
-\frac{1}{6\varepsilon^4}+\frac{1}{3\varepsilon^3}
+\frac{1}{3\varepsilon^2}
-\frac{1}{\varepsilon}(1-\zeta(3))
\Big)
\\
I_5=\raisebox{\eqoff}{%
\begin{fmfchar*}(20,15)
\fmfleft{in}
\fmfright{out}
\fmf{plain}{in,v1}
\fmf{plain,left=0.25}{v1,v2}
\fmf{plain,left=0.25}{v2,v3}
\fmf{plain,left=0.25}{v3,v4}
\fmf{plain,left=0.25}{v4,v1}
\fmf{plain}{v2,v4}
\fmf{plain}{v3,v1}
\fmf{plain}{v3,out}
\fmffixed{(0.9w,0)}{v1,v3}
\fmffixed{(0,0.45w)}{v4,v2}
\fmffreeze
\end{fmfchar*}}
&=\frac{1}{(4\pi)^8}
\frac{1}{\varepsilon}5\zeta(5)
\end{aligned}
\end{equation*}
\begin{equation*}\label{derivativeintegrals}
\begin{gathered}
\begin{aligned}
I_6=\raisebox{\eqoff}{%
\begin{fmfchar*}(20,15)
\fmfleft{in}
\fmfright{out}
\fmf{plain}{in,v1}
\fmf{plain,left=0.25}{v1,v2}
\fmf{plain,left=0.25}{v2,v3}
\fmf{derplain,left=0.25}{v4,v1}
\fmf{plain,right=0.25}{v4,v0}
\fmf{plain,right=0.25}{v0,v5}
\fmf{plain,right=0.75}{v4,v5}
\fmf{derplain,right=0.25}{v5,v3}
\fmf{plain}{v3,out}
\fmffixed{(0.9w,0)}{v1,v3}
\fmfpoly{phantom}{v2,v4,v5}
\fmffixed{(0.5w,0)}{v4,v5}
\fmffixed{(0.5w,0)}{v4,v5}
\fmf{plain,tension=0.25,right=0.25}{v2,v0,v2}
\fmffreeze
\fmfshift{(0,0.1w)}{in,out,v1,v2,v3,v4,v5,v0}
\end{fmfchar*}}
&=\frac{1}{(4\pi)^8}\Big(\frac{1}{12\varepsilon^2}
-\frac{7}{12\varepsilon}\Big)
\\
I_8=\raisebox{\eqoff}{%
\begin{fmfchar*}(20,15)
\fmfleft{in}
\fmfright{out}
\fmf{plain}{in,v1}
\fmf{plain,tension=2,left=0.125}{v1,v2c}
\fmf{plain,tension=2,left=0.125}{v2c,v3}
\fmf{plain,tension=1}{v2c,v2}
\fmf{derplain,left=0.25}{v4,v1}
\fmf{plain,right=0.25}{v4,v0}
\fmf{plain,right=0}{v0,v1}
\fmf{plain,right=0.25}{v0,v5}
\fmf{plain,right=0.75}{v4,v5}
\fmf{plain,right=0}{v3,v0}
\fmf{derplain,right=0.25}{v5,v3}
\fmf{phantom}{v3,out}
\fmffixed{(0,0.05w)}{v2c,v2}
\fmffixed{(0.9w,0)}{v1,v3}
\fmfpoly{phantom}{v2c,v4,v5}
\fmffixed{(0.5w,0)}{v4,v5}
\fmffreeze
\fmfshift{(0,0.15w)}{in,out,v1,v2,v2c,v3,v4,v5,v0}
\end{fmfchar*}}
&=
\frac{1}{(4\pi)^8}
\Big(\frac{1}{4\varepsilon^2}
-\frac{11}{12\varepsilon}\Big)
\end{aligned}
\qquad
\begin{aligned}
I_7=\raisebox{\eqoff}{%
\begin{fmfchar*}(20,15)
\fmfleft{in}
\fmfright{out}
\fmf{plain}{in,v1}
\fmf{plain,tension=2,left=0.25}{v1,v2}
\fmf{plain,tension=2,left=0.25}{v2,v3}
\fmf{derplain,left=0.25}{v4,v1}
\fmf{plain,right=0.25}{v4,v0}
\fmf{plain,right=0}{v0,v1}
\fmf{plain,right=0.25}{v0,v5}
\fmf{plain,right=0.75}{v4,v5}
\fmf{phantom,right=0}{v3,v0}
\fmf{derplain,right=0.25}{v5,v3}
\fmf{plain}{v3,out}
\fmffixed{(0.9w,0)}{v1,v3}
\fmfpoly{phantom}{v2,v4,v5}
\fmffixed{(0.5w,0)}{v4,v5}
\fmf{plain,tension=0.5}{v2,v0}
\fmffreeze
\fmfshift{(0,0.15w)}{in,out,v1,v2,v3,v4,v5,v0}
\end{fmfchar*}}
&=
\frac{1}{(4\pi)^8}
\frac{1}{\varepsilon}(-\zeta(3))
\\
I_9=\raisebox{\eqoff}{%
\begin{fmfchar*}(20,15)
\fmfleft{in}
\fmfright{out}
\fmf{plain}{in,v1}
\fmf{derplain,tension=2,right=0.25}{v2,v1}
\fmf{derplain,tension=2,left=0.25}{v2,v3}
\fmf{plain,left=0.25}{v4,v1}
\fmf{plain,right=0.25}{v4,v0}
\fmf{plain,right=0}{v0,v1}
\fmf{plain,right=0.25}{v0,v5}
\fmf{plain,right=0.75}{v4,v5}
\fmf{phantom,right=0}{v3,v0}
\fmf{plain,right=0.25}{v5,v3}
\fmf{plain}{v3,out}
\fmffixed{(0.9w,0)}{v1,v3}
\fmfpoly{phantom}{v2,v4,v5}
\fmffixed{(0.5w,0)}{v4,v5}
\fmf{plain,tension=0.5}{v2,v0}
\fmffreeze
\fmfshift{(0,0.15w)}{in,out,v1,v2,v3,v4,v5,v0}
\end{fmfchar*}}
&=\frac{1}{(4\pi)^8}\frac{1}{\varepsilon}\Big(\frac{1}{2}\zeta(3)
-\frac{5}{2}\zeta(5)\Big)
\end{aligned}
\\
\begin{aligned}
I_{10}=\raisebox{\eqoff}{%
\begin{fmfchar*}(20,15)
\fmfleft{in}
\fmfright{out}
\fmf{plain}{in,v1}
\fmf{derplain,tension=2,right=0.25}{v2,v1}
\fmf{derplain,tension=2,left=0.25}{v3,v4}
\fmf{derplainpt,tension=2,left=0.25}{v5,v1}
\fmf{derplainpt,tension=2,right=0.25}{v6,v4}
\fmf{plain}{v2,v0}
\fmf{plain}{v3,v0}
\fmf{plain}{v5,v0}
\fmf{plain}{v6,v0}
\fmf{plain}{v4,out}
\fmffixed{(0.9w,0)}{v1,v4}
\fmfpoly{phantom}{v3,v2,v5,v6}
\fmf{plain}{v2,v3}
\fmf{plain}{v5,v6}
\fmffixed{(0.4w,0)}{v5,v6}
\fmffreeze
\end{fmfchar*}}
&=
\frac{1}{(4\pi)^8}\frac{1}{\varepsilon}
\Big(-\frac{1}{2}-\frac{1}{2}\zeta(3)+\frac{5}{2}\zeta(5)\Big)
\\
I_{11}=\raisebox{\eqoff}{%
\begin{fmfchar*}(20,15)
\fmfleft{in}
\fmfright{out}
\fmf{plain}{in,v1}
\fmf{derplain,tension=2,right=0.25}{v2,v1}
\fmf{derplain,tension=2,left=0.25}{v3,v4}
\fmf{plain,tension=2,right=0.25}{v1,v5}
\fmf{plain,tension=2,right=0.25}{v6,v4}
\fmf{plain}{v2,v0}
\fmf{plain}{v3,v0}
\fmf{plain}{v5,v0}
\fmf{plain}{v0,v6}
\fmf{plain}{v4,out}
\fmffixed{(0.9w,0)}{v1,v4}
\fmfpoly{phantom}{v3,v2,v5,v6}
\fmf{derplainpt}{v2,v3}
\fmf{derplainpt}{v5,v6}
\fmffixed{(0.4w,0)}{v5,v6}
\fmffreeze
\end{fmfchar*}}
&=
\frac{1}{(4\pi)^8}\frac{1}{\varepsilon}
\Big(-\frac{1}{4}-\frac{3}{2}\zeta(3)
+\frac{5}{2}\zeta(5)\Big)
\\
I_{12}=\raisebox{\eqoff}{%
\begin{fmfchar*}(20,15)
\fmfleft{in}
\fmfright{out}
\fmf{plain}{in,v1}
\fmf{derplain,tension=2,right=0.25}{v2,v1}
\fmf{plain,tension=2,left=0.25}{v3,v4}
\fmf{plain,tension=2,right=0.25}{v1,v5}
\fmf{derplainpt,tension=2,right=0.25}{v6,v4}
\fmf{plain}{v2,v0}
\fmf{plain}{v3,v0}
\fmf{plain}{v5,v0}
\fmf{plain}{v0,v6}
\fmf{plain}{v4,out}
\fmffixed{(0.9w,0)}{v1,v4}
\fmfpoly{phantom}{v3,v2,v5,v6}
\fmf{derplain}{v3,v2}
\fmf{derplainpt}{v5,v6}
\fmffixed{(0.4w,0)}{v5,v6}
\fmffreeze
\end{fmfchar*}}
&=\frac{1}{(4\pi)^8}\frac{1}{\varepsilon}
\Big(-\frac{1}{8}-\frac{1}{4}\zeta(3)
+\frac{5}{4}\zeta(5)\Big)
\\
\end{aligned}
\end{gathered}
\end{equation*}
\caption{Loop integrals for 4-loop wrapping diagrams. 
The arrows of the same type indicate contracted spacetime derivatives}
\label{integrals}
\end{table}

\clearpage

\renewcommand{\thefigure}{D.\arabic{figure}}
\setcounter{figure}{0}
\renewcommand{\thetable}{D.\arabic{table}}
\setcounter{table}{0}

\newpage

\section{Integral calculation via GPXT}

\label{app:GPXT}
We use the Gegenbauer polynomial $x$-space technique (GPXT) to compute
the four-loop propagator-type logarithmic divergent integrals. 
The integrals are dimensionally regularized in $D=4-2\varepsilon=2(\lambda+1)$
spacetime dimensions, i.e.\ the parameter $\lambda$ assumes the value
\begin{equation}
\lambda=1-\varepsilon\pnt
\end{equation}

We briefly describe our notation.
Parentheses around spacetime indices
denote a traceless symmetric 
product. For two vectors it is defined as
\begin{equation}
x_1^{(\mu}x_2^{\nu)}=x_1^\mu x_2^{\nu}-\frac{x_1\cdot x_2}{D}g^{\mu\nu}
\pnt
\end{equation}
In $x$-space the integrals can be solved by an expansion of 
the propagators and traceless symmetric products in the numerator 
in terms of Gegenbauer polynomials $C_n^\lambda$ as follows
\begin{equation}\label{GPexp}
\begin{aligned}
\frac{1}{(x_1-x_2)^{2\lambda}}
=\frac{1}{\max_{12}^\lambda}\sum_{n=0}^\infty 
C_n^\lambda(\hat x_1\cdot\hat x_2)
\Big(\frac{\min_{12}}{\max_{12}}\Big)^{\frac{n}{2}}
\col\\
x_1^{(\mu_1\dots\mu_n)}x_2^{(\mu_1\dots\mu_n)}
=\frac{n!\Gamma(\lambda)}{2^n\Gamma(n+\lambda)}
C_n^\lambda(\hat x_1\cdot\hat x_2)(r_1r_2)^{\frac{n}{2}}
\pnt
\end{aligned}
\end{equation}
where $r_i=x_i^2$ denote the radial coordinates in Euclidean space, and 
$\hat x_i$ is the corresponding unit vector in the direction of $x_i$. 
We have also abbreviated
\begin{equation}
\textstyle
\min_{ij}=\min(r_i,r_j)\col\qquad\max_{ij}=\max(r_i,r_j)\pnt
\end{equation}
The Gegenbauer polynomials fulfill
\begin{equation}
C_n^\lambda(1)=\frac{\Gamma(n+2\lambda)}{n!\Gamma(2\lambda)}
\pnt
\end{equation}
Appearing products of two Gegenbauer polynomials with the 
same argument have to be expanded according to the Clebsch-Gordan series
\begin{equation}\label{CGseries}
\begin{aligned}
C_m^\lambda(x)C_n^\lambda(x)
&=\sum_{\substack{i=|m-n| \\ \frac{i+m+n}{2}\in\mathds{N}}}^{m+n}
D_\lambda(m,n,i)C_i^\lambda(x)\col\\
D_\lambda(m,n,i)
&=
{\textstyle\frac{i!(i+\lambda)\Gamma(\frac{m+n+i}{2}+2\lambda)}
{\Gamma(\lambda)^2\Gamma(\frac{m+n+i}{2}+\lambda+1)\Gamma(i+2\lambda)}
\frac{\Gamma(\frac{-m+n+i}{2}+\lambda)\Gamma(\frac{m-n+i}{2}+\lambda)
\Gamma(\frac{m+n-i}{2}+\lambda)}
{\Gamma(\frac{-m+n+i}{2}+1)\Gamma(\frac{m-n+i}{2}+1)
\Gamma(\frac{m+n-i}{2}+1)}}
\col
\end{aligned}
\end{equation}
where it is important to notice that for positive $\lambda$ the coefficients
$D_\lambda(m,n,i)$ never become singular.

For compact notation, we introduce the normalization factor
\begin{equation}
N_\lambda(P,L)=\frac{\Gamma(\lambda)^P}{(2^{2L}\pi^P)^{1+\lambda}}
\frac{1}{2^{P-L}}\Omega_{D-1}^{P-L}\col
\end{equation}
which depends on the number of propagators $P$ and number of loops $L$.
It arises when the momentum integral is transformed via a Fourier 
transformation into $x$-space, where the integral measure is then rewritten 
in polar coordinates according to
\begin{equation}
\de^Dx=\frac{1}{2}\Omega_{D-1}r^\lambda\de r\de\hat x
\pnt
\end{equation}
Here $\hat x$ is the unit angular vector, and
the volume of the $D-1$-dimensional unit sphere is given by
\begin{equation}
\Omega_{D-1}=\frac{2\pi^{\frac{D}{2}}}{\Gamma(\frac{D}{2})}\pnt
\end{equation}

\subsection{Sample calculation of $I_6$ with two derivatives}

As an example for the computation of the pole part of integrals with two 
derivatives
we show how to obtain $I_6$ of Table \ref{integrals}. 
This integral contains five propagators which do not end at the root vertex.

Shifting the derivatives to the root vertex, the required integral 
can be rewritten as
\begin{equation}\label{I1twodertrafo}
\settoheight{\eqoff}{$\times$}%
\setlength{\eqoff}{0.5\eqoff}%
\addtolength{\eqoff}{-7.5\unitlength}%
\raisebox{\eqoff}{%
\begin{fmfchar*}(20,15)
  \fmfleft{in}
  \fmfright{out}
  \fmf{plain}{in,v1}
  \fmf{plain,left=0.25}{v1,v2}
  \fmf{plain,left=0.25}{v2,v3}
  \fmf{derplain,left=0.25}{v4,v1}
  \fmf{plain,left=0.25}{v0,v4}
  \fmf{plain,right=0.25}{v0,v5}
  \fmf{plain,right=0.75}{v4,v5}
  \fmf{derplain,right=0.25}{v5,v3}
  \fmf{plain}{v3,out}
\fmffixed{(0.9w,0)}{v1,v3}
\fmfpoly{phantom}{v2,v4,v5}
\fmffixed{(0.5w,0)}{v4,v5}
\fmffixed{(0.5w,0)}{v4,v5}
\fmf{plain,tension=0.25,right=0.25}{v2,v0,v2}
\fmffreeze
\fmfshift{(0,0.1w)}{in,out,v1,v2,v3,v4,v5,v0}
\end{fmfchar*}}
{}={}
\raisebox{\eqoff}{%
\begin{fmfchar*}(20,15)
  \fmfleft{in}
  \fmfright{out}
  \fmf{plain}{in,v1}
  \fmf{plain,left=0.25}{v1,v2}
  \fmf{plain,left=0.25}{v2,v3}
  \fmf{plain,right=0.25}{v1,v4}
  \fmf{derplain,left=0.25}{v0,v4}
  \fmf{derplain,right=0.25}{v0,v5}
  \fmf{plain,right=0.75}{v4,v5}
  \fmf{plain,right=0.25}{v5,v3}
  \fmf{plain}{v3,out}
\fmffixed{(0.9w,0)}{v1,v3}
\fmfpoly{phantom}{v2,v4,v5}
\fmffixed{(0.5w,0)}{v4,v5}
\fmffixed{(0.5w,0)}{v4,v5}
\fmf{plain,tension=0.25,right=0.25}{v2,v0,v2}
\fmffreeze
\fmfshift{(0,0.1w)}{in,out,v1,v2,v3,v4,v5,v0}
\end{fmfchar*}}
{}+{}
\raisebox{\eqoff}{%
\begin{fmfchar*}(20,15)
  \fmfleft{in}
  \fmfright{out}
  \fmf{plain}{in,v1}
  \fmf{plain,left=0.25}{v1,v2}
  \fmf{plain,left=0.25}{v2,v3}
  \fmf{plain,right=0.25}{v1,v4}
  \fmf{plain,left=0.25}{v0,v4}
  \fmf{phantom,right=0.25}{v0,v5}
  \fmf{phantom,right=0.75}{v4,v5}
  \fmf{phantom,right=0.25}{v5,v3}
  \fmf{plain}{v3,out}
\fmffixed{(0.9w,0)}{v1,v3}
\fmfpoly{phantom}{v2,v4,v5}
\fmffixed{(0.5w,0)}{v4,v5}
\fmffixed{(0.5w,0)}{v4,v5}
\fmf{plain,tension=0.25,right=0.25}{v2,v0,v2}
\fmffreeze
  \fmf{plain,right=0.25}{v0,v4}
  \fmf{plain,right=0.25}{v0,v3}
\fmfshift{(0,0.1w)}{in,out,v1,v2,v3,v4,v5,v0}
\end{fmfchar*}}
{}-{}
\raisebox{\eqoff}{%
\fmfframe(0,0)(-4,0){
\begin{fmfchar*}(20,15)
  \fmfleft{in}
  \fmfright{out}
  \fmf{plain}{in,v1}
  \fmf{plain,left=0.25}{v1,v2}
  \fmf{phantom,left=0.25}{v2,v3}
  \fmf{plain,right=0.25}{v1,v4}
  \fmf{plain,left=0.25}{v0,v4}
  \fmf{plain,right=0.25}{v0,v5}
  \fmf{plain,right=0.75}{v4,v5}
  \fmf{phantom,right=0.25}{v5,v3}
  \fmf{phantom}{v3,out}
\fmffixed{(0.9w,0)}{v1,v3}
\fmfpoly{phantom}{v2,v4,v5}
\fmffixed{(0.5w,0)}{v4,v5}
\fmffixed{(0.5w,0)}{v4,v5}
\fmf{plain,tension=0.25,right=0.25}{v2,v0,v2}
\fmffreeze
  \fmf{plain,right=0.5}{v5,v2}
\fmfshift{(0,0.1w)}{in,out,v1,v2,v3,v4,v5,v0}
\end{fmfchar*}}}
\pnt
\end{equation}
We hence have to compute the pole part of the integral
\begin{equation}\label{I1twoder}
\settoheight{\eqoff}{$\times$}%
\setlength{\eqoff}{0.5\eqoff}%
\addtolength{\eqoff}{-7.5\unitlength}%
\raisebox{\eqoff}{%
\begin{fmfchar*}(20,15)
  \fmfleft{in}
  \fmfright{out}
  \fmf{plain}{in,v1}
  \fmf{plain,left=0.25}{v1,v2}
  \fmf{plain,left=0.25}{v2,v3}
  \fmf{plain,right=0.25}{v1,v4}
  \fmf{derplain,left=0.25}{v0,v4}
  \fmf{derplain,right=0.25}{v0,v5}
  \fmf{plain,right=0.75}{v4,v5}
  \fmf{plain,right=0.25}{v5,v3}
  \fmf{plain}{v3,out}
\fmffixed{(0.9w,0)}{v1,v3}
\fmfpoly{phantom}{v2,v4,v5}
\fmffixed{(0.5w,0)}{v4,v5}
\fmffixed{(0.5w,0)}{v4,v5}
\fmf{plain,tension=0.25,right=0.25}{v2,v0,v2}
\fmffreeze
\fmfshift{(0,0.1w)}{in,out,v1,v2,v3,v4,v5,v0}
\end{fmfchar*}}
=-\lambda^2\frac{\Gamma(\lambda)^9}{(2^8\pi^9)^{1+\lambda}}
\int\frac{\de^Dx_1\de^Dx_2\de^Dx_3\de^Dx_4\de^Dx_5\,
x_4\cdot x_5\e^{2ip\cdot(x_3-x_1)}}
{x_2^{4\lambda}(x_4^2x_5^2)^{1+\lambda}
(\Delta_{12}^2\Delta_{23}^2\Delta_{34}^2\Delta_{45}^2\Delta_{51}^2)^\lambda}
\pnt
\end{equation}

After neglecting the exponential function, and introducing an IR cutoff $R$, the expansion of this integral in Gegenbauer polynomials assumes the form
\begin{equation}\label{Itwoderexpand}
\begin{aligned}
I''&=-\frac{\lambda}{2}
N_\lambda(9,4)
\sum_{i,j,k,l,r=0}^\infty R_\lambda(i,j,k,l,r)A_\lambda(i,j,k,l,r)
\col
\end{aligned}
\end{equation}
where $R_\lambda(i,j,k,l,r)$ and $A_\lambda(i,j,k,l,r)$ denote the corresponding
contributions from the radial and angular integrals. They are given by
\begin{equation}
\begin{aligned}\label{I1RA}
R_\lambda(i,j,k,l,r)
&=\int_0^R\frac{\de r_1\de r_2\de r_3\de r_4\de r_5\,
r_1^{\lambda}r_2^{-\lambda} r_3^{\lambda}
r_4^{-\frac{1}{2}} r_5^{-\frac{1}{2}}}
{(\max_{12}\max_{23}\max_{34}\max_{45}\max_{51})^\lambda}\\
&\phantom{{}={}\int_0^R}
\Big(\frac{\min_{12}}{\max_{12}}\Big)^{\frac{i}{2}}
\Big(\frac{\min_{23}}{\max_{23}}\Big)^{\frac{j}{2}}
\Big(\frac{\min_{34}}{\max_{34}}\Big)^{\frac{k}{2}}
\Big(\frac{\min_{45}}{\max_{45}}\Big)^{\frac{l}{2}}
\Big(\frac{\min_{51}}{\max_{51}}\Big)^{\frac{r}{2}}
\col\\
A_\lambda(i,j,k,l,r)
&=\sum_{\substack{m=|l-1| \\ m\neq l}}^{l+1}D_\lambda(1,l,m)
\int\de\hat x_1\dots
\de\hat x_5\,
C_i^\lambda(\hat x_1\cdot\hat x_2)C_j^\lambda(\hat x_2\cdot\hat x_3)
C_k^\lambda(\hat x_3\cdot\hat x_4)\\
&\hphantom{{}={}\sum_{\substack{m=|l-1| \\ m\neq l}}^{l+1}D_\lambda(1,l,m)
\int\de\hat x_1\dots
\de\hat x_5\,}
C_m^\lambda(\hat x_4\cdot\hat x_5)C_r^\lambda(\hat x_5\cdot\hat x_1)\\
&=\delta_{ij}\delta_{ik}\delta_{im}\delta_{ir}
\sum_{\substack{i=|l-1| \\ i\neq l}}^{l+1}\frac{\lambda^4}{(i+\lambda)^4}
D_\lambda(1,l,i)C_i^\lambda(1)
\pnt
\end{aligned}
\end{equation}
Evaluating these expressions, substituting them into \eqref{Itwoderexpand}, and expanding the result into a power series with negative powers in $\varepsilon$, we obtain for the pole part after subtraction of the subdivergences
\begin{equation}
\begin{aligned}
\settoheight{\eqoff}{$\times$}%
\setlength{\eqoff}{0.5\eqoff}%
\addtolength{\eqoff}{-7.5\unitlength}%
\raisebox{\eqoff}{%
\begin{fmfchar*}(20,15)
  \fmfleft{in}
  \fmfright{out}
  \fmf{plain}{in,v1}
  \fmf{plain,left=0.25}{v1,v2}
  \fmf{plain,left=0.25}{v2,v3}
  \fmf{plain,right=0.25}{v1,v4}
  \fmf{derplain,left=0.25}{v0,v4}
  \fmf{derplain,right=0.25}{v0,v5}
  \fmf{plain,right=0.75}{v4,v5}
  \fmf{plain,right=0.25}{v5,v3}
  \fmf{plain}{v3,out}
\fmffixed{(0.9w,0)}{v1,v3}
\fmfpoly{phantom}{v2,v4,v5}
\fmffixed{(0.5w,0)}{v4,v5}
\fmffixed{(0.5w,0)}{v4,v5}
\fmf{plain,tension=0.25,right=0.25}{v2,v0,v2}
\fmffreeze
\fmfshift{(0,0.1w)}{in,out,v1,v2,v3,v4,v5,v0}
\end{fmfchar*}}
&=\frac{1}{(4\pi)^8}\Big(\frac{1}{6\varepsilon^4}
-\frac{1}{6\varepsilon^3}-\frac{3}{4\varepsilon^2}
+\frac{1}{\varepsilon}\Big(\frac{11}{12}-\zeta(3)\Big)\Big)
\col
\end{aligned}
\end{equation}
where the higher negative powers in $\varepsilon$ indicate that the integral
\eqref{I1twoder} contained subdivergences.
Substituting the above result into \eqref{I1twodertrafo}, we find for the 
required integral
\begin{equation}
\settoheight{\eqoff}{$\times$}%
\setlength{\eqoff}{0.5\eqoff}%
\addtolength{\eqoff}{-7.5\unitlength}%
\raisebox{\eqoff}{%
\begin{fmfchar*}(20,15)
  \fmfleft{in}
  \fmfright{out}
  \fmf{plain}{in,v1}
  \fmf{plain,left=0.25}{v1,v2}
  \fmf{plain,left=0.25}{v2,v3}
  \fmf{derplain,left=0.25}{v4,v1}
  \fmf{plain,right=0.25}{v4,v0}
  \fmf{plain,right=0.25}{v0,v5}
  \fmf{plain,right=0.75}{v4,v5}
  \fmf{derplain,right=0.25}{v5,v3}
  \fmf{plain}{v3,out}
\fmffixed{(0.9w,0)}{v1,v3}
\fmfpoly{phantom}{v2,v4,v5}
\fmffixed{(0.5w,0)}{v4,v5}
\fmffixed{(0.5w,0)}{v4,v5}
\fmf{plain,tension=0.25,right=0.25}{v2,v0,v2}
\fmffreeze
\fmfshift{(0,0.1w)}{in,out,v1,v2,v3,v4,v5,v0}
\end{fmfchar*}}
=\frac{1}{(4\pi)^8}\Big(\frac{1}{12\varepsilon^2}
-\frac{7}{12\varepsilon}\Big)\pnt
\end{equation}

\subsection{Sample calculation of $I_{11}$ with four derivatives}

As an example for the computation of the pole part of integrals with four 
derivatives we show how to obtain $I_{11}$ of Table \ref{integrals}. 
This integral contains six propagators which do not end at the 
root vertex.
It is the most complicated one, which we have to compute at four loops. 

Shifting the derivatives to the root vertex, the required integral 
can be rewritten as
\begin{equation}
\begin{aligned}\label{I2fourdertrafo}
{}&
\settoheight{\eqoff}{$\times$}%
\setlength{\eqoff}{0.5\eqoff}%
\addtolength{\eqoff}{-7.5\unitlength}%
\raisebox{\eqoff}{%
\begin{fmfchar*}(20,15)
  \fmfleft{in}
  \fmfright{out}
  \fmf{plain}{in,v2}
  \fmf{derplainpt,tension=2,right=0.25}{v1,v2}
 \fmf{plain,tension=2,right=0.25}{v2,v3}
  \fmf{derplainpt,tension=2,right=0.25}{v4,v5}
  \fmf{plain,tension=2,left=0.25}{v6,v5}
  \fmf{plain}{v0,v1}
  \fmf{plain}{v0,v3}
  \fmf{plain}{v0,v4}
  \fmf{plain}{v0,v6}
  \fmf{plain}{v5,out}
\fmffixed{(0.9w,0)}{v2,v5}
\fmfpoly{phantom}{v6,v1,v3,v4}
  \fmf{derplain}{v3,v4}
  \fmf{derplain}{v6,v1}
\fmffixed{(0.4w,0)}{v3,v4}
\fmffreeze
\end{fmfchar*}}\\
&=
\settoheight{\eqoff}{$\times$}%
\setlength{\eqoff}{0.5\eqoff}%
\addtolength{\eqoff}{-7.5\unitlength}%
\raisebox{\eqoff}{%
\begin{fmfchar*}(20,15)
  \fmfleft{in}
  \fmfright{out}
  \fmf{plain}{in,v2}
  \fmf{plain,tension=2,left=0.25}{v2,v1}
 \fmf{plain,tension=2,right=0.25}{v2,v3}
  \fmf{plain,tension=2,right=0.25}{v4,v5}
  \fmf{plain,tension=2,left=0.25}{v6,v5}
  \fmf{dblderplain}{v0,v1}
  \fmf{derplainpt}{v0,v3}
  \fmf{plain}{v0,v4}
  \fmf{derplain}{v0,v6}
  \fmf{plain}{v5,out}
\fmffixed{(0.9w,0)}{v2,v5}
\fmfpoly{phantom}{v6,v1,v3,v4}
  \fmf{plain}{v3,v4}
  \fmf{plain}{v1,v6}
\fmffixed{(0.4w,0)}{v3,v4}
\fmffreeze
\end{fmfchar*}}
+\frac{1}{2}
\raisebox{\eqoff}{%
\begin{fmfchar*}(20,15)
  \fmfleft{in}
  \fmfright{out}
  \fmf{plain}{in,v2}
  \fmf{plain,tension=2,left=0.25}{v2,v1}
 \fmf{plain,tension=2,right=0.25}{v2,v3}
  \fmf{plain,tension=2,right=0.25}{v4,v5}
  \fmf{phantom,tension=2,left=0.25}{v6,v5}
  \fmf{phantom}{v0,v1}
  \fmf{plain}{v0,v3}
  \fmf{plain}{v0,v4}
  \fmf{phantom}{v0,v6}
  \fmf{plain}{v5,out}
\fmffixed{(0.9w,0)}{v2,v5}
\fmfpoly{phantom}{v6,v1,v3,v4}
  \fmf{plain}{v3,v4}
  \fmf{phantom}{v1,v6}
\fmffixed{(0.4w,0)}{v3,v4}
\fmffreeze
  \fmf{derplain}{v0,v5}
  \fmf{plain,left=0.25}{v0,v1}
  \fmf{derplain,right=0.25}{v0,v1}
\fmffreeze
\end{fmfchar*}}
-
\raisebox{\eqoff}{%
\begin{fmfchar*}(20,15)
  \fmfleft{in}
  \fmfright{out}
  \fmf{plain}{in,v2}
  \fmf{plain,tension=2,left=0.25}{v2,v1}
 \fmf{plain,tension=2,right=0.25}{v2,v3}
  \fmf{plain,tension=2,right=0.25}{v4,v5}
  \fmf{plain,tension=2,left=0.25}{v1,v5}
  \fmf{plain,left=0.5}{v0,v1}
  \fmf{plain,right=0.5}{v0,v1}
  \fmf{derplain}{v0,v3}
  \fmf{derplain}{v0,v4}
  \fmf{plain}{v5,out}
\fmffixed{(0.9w,0)}{v2,v5}
\fmfpoly{phantom}{v1,v3,v4}
  \fmf{plain}{v3,v4}
\fmffixed{(0.4w,0)}{v3,v4}
\fmffreeze
\end{fmfchar*}}
-
\raisebox{\eqoff}{%
\begin{fmfchar*}(20,15)
  \fmfleft{in}
  \fmfright{out}
  \fmf{plain}{in,v2}
  \fmf{plain,tension=2,left=0.25}{v2,v1}
 \fmf{plain,tension=2,right=0.25}{v2,v3}
  \fmf{plain,tension=2,right=0.25}{v4,v5}
  \fmf{phantom,tension=2,left=0.25}{v6,v5}
  \fmf{phantom}{v0,v1}
  \fmf{plain}{v0,v3}
  \fmf{derplain}{v0,v4}
  \fmf{phantom}{v0,v6}
  \fmf{plain}{v5,out}
\fmffixed{(0.9w,0)}{v2,v5}
\fmfpoly{phantom}{v6,v1,v3,v4}
  \fmf{plain}{v3,v4}
  \fmf{phantom}{v1,v6}
\fmffixed{(0.4w,0)}{v3,v4}
\fmffreeze
  \fmf{plain}{v0,v5}
  \fmf{plain,left=0.25}{v0,v1}
  \fmf{derplain,right=0.25}{v0,v1}
\fmffreeze
\end{fmfchar*}}
-\negthickspace\negthickspace
\raisebox{\eqoff}{%
\begin{fmfchar*}(20,15)
  \fmfleft{in}
  \fmfright{out}
  \fmf{phantom}{in,v2}
  \fmf{phantom,tension=2,left=0.25}{v2,v1}
 \fmf{phantom,tension=2,right=0.25}{v2,v3}
  \fmf{plain,tension=2,right=0.25}{v4,v5}
  \fmf{plain,tension=2,left=0.25}{v6,v5}
  \fmf{plain}{v0,v1}
  \fmf{plain}{v0,v3}
  \fmf{derplain}{v0,v4}
  \fmf{derplain}{v0,v6}
  \fmf{plain}{v5,out}
\fmffixed{(0.9w,0)}{v2,v5}
\fmfpoly{phantom}{v6,v1,v3,v4}
  \fmf{plain}{v3,v4}
  \fmf{plain}{v1,v6}
\fmffixed{(0.4w,0)}{v3,v4}
\fmffreeze
\fmf{plain,right=0.5}{v1,v3}
\end{fmfchar*}}\\
&\phantom{{}={}}
\settoheight{\eqoff}{$\times$}%
\setlength{\eqoff}{0.5\eqoff}%
\addtolength{\eqoff}{-7.5\unitlength}%
-\negthickspace\negthickspace
\raisebox{\eqoff}{%
\begin{fmfchar*}(20,15)
  \fmfleft{in}
  \fmfright{out}
  \fmf{phantom}{in,v2}
  \fmf{phantom,tension=2,left=0.25}{v2,v1}
 \fmf{phantom,tension=2,right=0.25}{v2,v3}
  \fmf{plain,tension=2,right=0.25}{v4,v5}
  \fmf{phantom,tension=2,left=0.25}{v6,v5}
  \fmf{phantom}{v0,v1}
  \fmf{plain}{v0,v3}
  \fmf{plain}{v0,v4}
  \fmf{phantom}{v0,v6}
  \fmf{plain}{v5,out}
\fmffixed{(0.9w,0)}{v2,v5}
\fmfpoly{phantom}{v6,v1,v3,v4}
  \fmf{plain}{v3,v4}
  \fmf{phantom}{v1,v6}
\fmffixed{(0.4w,0)}{v3,v4}
\fmffreeze
  \fmf{plain,right=0.5}{v1,v3}
  \fmf{plain}{v0,v5}
  \fmf{plain,left=0.25}{v0,v1}
  \fmf{plain,right=0.25}{v0,v1}
\fmffreeze
\end{fmfchar*}}
+\negthickspace\negthickspace\negthickspace
\raisebox{\eqoff}{%
\begin{fmfchar*}(20,15)
  \fmfleft{in}
  \fmfright{out}
  \fmf{phantom}{in,v2}
  \fmf{phantom,tension=2,left=0.25}{v2,v1}
 \fmf{phantom,tension=2,right=0.25}{v2,v3}
  \fmf{plain,tension=2,right=0.25}{v4,v5}
  \fmf{plain,tension=2,left=0.25}{v1,v5}
  \fmf{plain,left=0.5}{v0,v1}
  \fmf{plain,right=0.5}{v0,v1}
  \fmf{plain}{v0,v3}
  \fmf{plain}{v0,v4}
  \fmf{plain}{v5,out}
\fmffixed{(0.9w,0)}{v2,v5}
\fmfpoly{phantom}{v1,v3,v4}
  \fmf{plain}{v3,v4}
\fmffixed{(0.4w,0)}{v3,v4}
\fmffreeze
\fmf{plain,right=0.5}{v1,v3}
\end{fmfchar*}}
-
\raisebox{\eqoff}{%
\begin{fmfchar*}(20,15)
  \fmfleft{in}
  \fmfright{out}
  \fmf{plain}{in,v2}
  \fmf{plain,tension=2,left=0.25}{v2,v1}
 \fmf{plain,tension=2,right=0.25}{v2,v3}
  \fmf{phantom,tension=2,right=0.25}{v4,v5}
  \fmf{plain,tension=2,left=0.25}{v1,v5}
  \fmf{plain,left=0.5}{v0,v1}
  \fmf{plain,right=0.5}{v0,v1}
  \fmf{phantom}{v0,v3}
  \fmf{phantom}{v0,v4}
  \fmf{plain}{v5,out}
\fmffixed{(0.9w,0)}{v2,v5}
\fmfpoly{phantom}{v1,v3,v4}
  \fmf{phantom}{v3,v4}
\fmffixed{(0.4w,0)}{v3,v4}
\fmffreeze
\fmf{plain}{v0,v5}
  \fmf{plain,left=0.25}{v0,v3}
  \fmf{plain,right=0.25}{v0,v3}
\end{fmfchar*}}
-\frac{1}{4}
\raisebox{\eqoff}{%
\begin{fmfchar*}(20,15)
  \fmfleft{in}
  \fmfright{out}
  \fmf{plain}{in,v2}
  \fmf{plain,tension=2,left=0.25}{v2,v1}
 \fmf{plain,tension=2,right=0.25}{v2,v3}
  \fmf{phantom,tension=2,right=0.25}{v4,v5}
  \fmf{phantom,tension=2,left=0.25}{v6,v5}
  \fmf{phantom}{v0,v1}
  \fmf{phantom}{v0,v3}
  \fmf{phantom}{v0,v4}
  \fmf{phantom}{v0,v6}
  \fmf{plain}{v5,out}
\fmffixed{(0.9w,0)}{v2,v5}
\fmfpoly{phantom}{v6,v1,v3,v4}
  \fmf{phantom}{v3,v4}
  \fmf{phantom}{v1,v6}
\fmffixed{(0.4w,0)}{v3,v4}
\fmffreeze
\fmf{plain,right=0.25}{v0,v5}
\fmf{plain,left=0.25}{v0,v5}
  \fmf{plain,left=0.25}{v0,v1}
  \fmf{plain,right=0.25}{v0,v1}
  \fmf{plain,left=0.25}{v0,v3}
  \fmf{plain,right=0.25}{v0,v3}
\end{fmfchar*}}
+\frac{1}{4}
\raisebox{\eqoff}{%
\begin{fmfchar*}(20,15)
  \fmfleft{in}
  \fmfright{out}
  \fmf{plain}{in,v2}
  \fmf{plain,tension=2,left=0.25}{v2,v1}
 \fmf{plain,tension=2,right=0.25}{v2,v3}
  \fmf{phantom,tension=2,right=0.25}{v4,v5}
  \fmf{phantom,tension=2,left=0.25}{v6,v5}
  \fmf{plain}{v0,v1}
  \fmf{phantom}{v0,v3}
  \fmf{phantom}{v0,v4}
  \fmf{phantom}{v0,v6}
  \fmf{phantom}{v5,out}
\fmffixed{(0.9w,0)}{v2,v5}
\fmfpoly{phantom}{v6,v1,v3,v4}
  \fmf{phantom}{v3,v4}
  \fmf{plain}{v1,v6}
\fmffixed{(0.4w,0)}{v3,v4}
\fmffreeze
\fmf{plain,right=0.25}{v0,v6}
\fmf{plain,left=0.25}{v0,v6}
  \fmf{plain,left=0.25}{v0,v3}
  \fmf{plain,right=0.25}{v0,v3}
\end{fmfchar*}}
\negthickspace\negthickspace\negthickspace\negthickspace
+\frac{1}{4}
\raisebox{\eqoff}{%
\begin{fmfchar*}(20,15)
  \fmfleft{in}
  \fmfright{out}
  \fmf{plain}{in,v2}
  \fmf{plain,tension=2,left=0.25}{v2,v1}
 \fmf{phantom,tension=2,right=0.25}{v2,v3}
  \fmf{plain,tension=2,right=0.25}{v4,v5}
  \fmf{phantom,tension=2,left=0.25}{v6,v5}
  \fmf{phantom}{v0,v1}
  \fmf{phantom}{v0,v3}
  \fmf{phantom}{v0,v4}
  \fmf{phantom}{v0,v6}
  \fmf{plain}{v5,out}
\fmffixed{(0.9w,0)}{v2,v5}
\fmfpoly{phantom}{v6,v1,v3,v4}
  \fmf{phantom}{v3,v4}
  \fmf{phantom}{v1,v6}
\fmffixed{(0.4w,0)}{v3,v4}
\fmffreeze
\fmf{plain}{v0,v2}
\fmf{plain}{v0,v5}
\fmf{plain,right=0.25}{v0,v1}
\fmf{plain,left=0.25}{v0,v1}
  \fmf{plain,left=0.25}{v0,v4}
  \fmf{plain,right=0.25}{v0,v4}
\end{fmfchar*}}
\pnt
\end{aligned}
\end{equation}
We hence have to compute the pole part of the integral
\begin{equation}\label{I2fourder}
\begin{aligned}
\settoheight{\eqoff}{$\times$}%
\setlength{\eqoff}{0.5\eqoff}%
\addtolength{\eqoff}{-7.5\unitlength}%
\raisebox{\eqoff}{%
\fmfframe(0,0)(0,0){%
\begin{fmfchar*}(20,15)
  \fmfleft{in}
  \fmfright{out}
  \fmf{plain}{in,v2}
  \fmf{plain,tension=2,left=0.25}{v2,v1}
 \fmf{plain,tension=2,right=0.25}{v2,v3}
  \fmf{plain,tension=2,right=0.25}{v4,v5}
  \fmf{plain,tension=2,left=0.25}{v6,v5}
  \fmf{dblderplain}{v0,v1}
  \fmf{derplainpt}{v0,v3}
  \fmf{plain}{v0,v4}
  \fmf{derplain}{v0,v6}
  \fmf{plain}{v5,out}
\fmffixed{(0.9w,0)}{v2,v5}
\fmfpoly{phantom}{v6,v1,v3,v4}
  \fmf{plain}{v3,v4}
  \fmf{plain}{v1,v6}
\fmffixed{(0.4w,0)}{v3,v4}
\fmffreeze
\end{fmfchar*}}}
=\lambda^3(1+\lambda)
\frac{\Gamma(\lambda)^{10}}{(2^8\pi^{10})^{1+\lambda}}
\int\frac{\de^Dx_1\dots\de^Dx_6\,x_3\cdot x_{(2}\,
x_{2)}\cdot x_6
\e^{2ip\cdot(x_4-x_1)}}
{x_2^{2(\lambda-\alpha+2)}(x_3^2x_6^2)^{1+\lambda}
(x_5^2\Delta_{12}^2\Delta_{23}^2\Delta_{34}^2\Delta_{45}^2\Delta_{56}^2\Delta_{61}^2)^\lambda}
\col
\end{aligned}
\end{equation}
where the parameter $\alpha$ has to be introduced to regularize additional 
divergences which arise when we neglect the exponential function and introduce the IR cutoff $R$. It should assume the value $\alpha=0$. 
The parameter $\alpha$ also deforms the (traceless and symmetric) products 
in the numerator as
\begin{equation}
x_3\cdot x_{(2}x_{2)}\cdot x_6
=x_2\cdot x_3\,x_2\cdot x_6-\frac{1}{2(1-\alpha+\lambda)}x_2^2\,x_3\cdot x_6
\pnt
\end{equation}
The expansion of the integral is then given by
\begin{equation}\label{I2fourderexpand}
\begin{aligned}
I''''&=\frac{\lambda(1+\lambda)}{4}
N_\lambda(10,4)
\sum_{i,j,k,l,r,s=0}^\infty
\Big(A_\lambda(i,j,k,l,r,s)\\
&\hphantom{{}={}\frac{\lambda(1+\lambda)}{4}
N_\lambda(10,4)\sum_{i,j,k,l,r,s=0}^\infty\Big(}
-\frac{\lambda}{1-\alpha+\lambda}
A^{\tr}_\lambda(i,j,k,l,r,s)\Big)R_\lambda(i,j,k,l,r,s)
\col
\end{aligned}
\end{equation}
where the radial and angular integrals are respectively given by 
\begin{equation}
\begin{aligned}
R_\lambda(i,j,k,l,r,s)
&=\int_0^R\frac{\de r_1\de r_2\de r_3\de r_4\de r_5\de r_6\,
r_1^\lambda r_4^\lambda r_2^{\alpha-1}r_3^{-\frac{1}{2}}r_6^{-\frac{1}{2}}}
{(\max_{12}\max_{23}\max_{34}\max_{45}\max_{56}
\max_{61})^\lambda}\\
&\phantom{{}={}\int_0^R}
\Big(\frac{\min_{12}}{\max_{12}}\Big)^{\frac{i}{2}}
\Big(\frac{\min_{23}}{\max_{23}}\Big)^{\frac{j}{2}}
\Big(\frac{\min_{34}}{\max_{34}}\Big)^{\frac{k}{2}}
\Big(\frac{\min_{45}}{\max_{45}}\Big)^{\frac{l}{2}}
\Big(\frac{\min_{56}}{\max_{56}}\Big)^{\frac{r}{2}}
\Big(\frac{\min_{61}}{\max_{61}}\Big)^{\frac{s}{2}}
\col\\
A_\lambda(i,j,k,l,r,s)
&=\delta_{is}\delta_{mk}\delta_{ml}\delta_{mr}
\sum_{\substack{m=|i-1| \\ m\neq i}}^{i+1}
\sum_{\substack{m=|j-1| \\ m\neq j}}^{j+1}
\frac{\lambda^5}{(i+\lambda)(m+\lambda)^4}
D_\lambda(1,i,m)
D_\lambda(1,j,m)C_m^\lambda(1)
\col\\
A^{\tr}_\lambda(i,j,k,l,r,s)
&=\delta_{ij}\delta_{is}\delta_{mk}\delta_{ml}\delta_{mr}
\frac{\lambda^5}{(i+\lambda)^2(m+\lambda)^3}
\sum_{\substack{m=|i-1| \\ m\neq i}}^{i+1}D_\lambda(1,i,m)C_m^\lambda(1)
\pnt
\end{aligned}
\end{equation}
Using the identity
\begin{equation}\label{doublesumid}
\begin{aligned}
{}&
\sum_{m,i,j=0}^n 
\sum_{\substack{m=|i-1|\\ m\neq i}}^{i+1}\sum_{\substack{m=|j-1| \\ m\neq j}}^{j+1}
f_\lambda(i,j,m)\\
&\qquad\qquad=
f_\lambda(1,1,0)
+\sum_{m=1}^{n-1}\sum_{a,b=\pm1}f_\lambda(m+a,m+b,m)
+\sum_{m=n}^{n+1}f_\lambda(m-1,m-1,m)
%
\col
\end{aligned}
\end{equation}
the integral can be cast into the following form
\begin{equation}
\begin{aligned}
I''''
&=\frac{\lambda(1+\lambda)}{4}
N_\lambda(10,4)\Big(
\Delta f_\lambda(1,0)+\sum_{m=1}^\infty(\Delta f_\lambda(m+1,m)+\Delta f_\lambda(m-1,m))\\
&\phantom{{}={}\frac{\lambda(1+\lambda)}{4}
N_\lambda(10,4)\Big(}
+\sum_{m=1}^\infty(f_\lambda(m+1,m-1,m)+f_\lambda(m-1,m+1,m))
\Big)
\col
\end{aligned}
\end{equation}
where we have abbreviated
\begin{equation}
\begin{aligned}
f_\lambda(i,j,m)
&=
\frac{\lambda^5}{(i+\lambda)(m+\lambda)^4}D_\lambda(1,i,m)D_\lambda(1,j,m)C_m^\lambda(1)
R_\lambda(i,j,m,m,m,i)\col\\
g_\lambda(i,m)
&=
\frac{\lambda^6}{(1-\alpha+\lambda)(i+\lambda)^2(m+\lambda)^3}D_\lambda(1,i,m)C_m^\lambda(1)R_\lambda(i,i,m,m,m,i)\col\\
\Delta f_\lambda(i,m)&=f_\lambda(i,i,m)-g_\lambda(i,m)
\pnt
\end{aligned}
\end{equation}
The explicit forms of the functions which appear in the above expressions 
read
\begin{equation}
\begin{aligned}
\Delta f_\lambda(m-1,m)
&=\frac{m((m-1)\lambda-m\alpha)\lambda^7}{(1-\alpha+\lambda)(m-1+\lambda)^3(m+\lambda)^4}
C_m^\lambda(1)\\
&\phantom{{}={}}
R_\lambda(m-1,m-1,m,m,m,m-1)\col\\
\Delta f_\lambda(m+1,m)
&=\frac{(m+2\lambda)(1+m+2\lambda)\lambda^8}{(1+\lambda)(m+1+\lambda)^3(m+\lambda)^4}C_m^\lambda(1)\\
&\phantom{{}={}}
R_\lambda(m+1,m+1,m,m,m,m+1)\col\\
f_\lambda(m-1,m+1,m)
&=\frac{\lambda^5}{(m-1+\lambda)(m+\lambda)^4}D_\lambda(1,m-1,m)D_\lambda(1,m+1,m)C_m^\lambda(1)\\
&\phantom{{}={}}
R_\lambda(m-1,m+1,m,m,m,m-1)\col\\
f_\lambda(m+1,m-1,m)
&=\frac{\lambda^5}{(m+1+\lambda)(m+\lambda)^4}D_\lambda(1,m+1,m)D_\lambda(1,m-1,m)C_m^\lambda(1)\\
&\phantom{{}={}}
R_\lambda(m+1,m-1,m,m,m,m+1)
\pnt
\end{aligned}
\end{equation}
It is important to remark that one can set $\alpha$ to its correct value 
$\alpha=0$ in nearly all contributions, except in $\Delta f_\lambda(0,1)$, 
where 
$m=1$. In this case, some integration domains of the radial integral 
$R_\lambda(0,0,1,1,1,0)$ become divergent as $\frac{1}{\alpha}$ for 
$\alpha\to 0$. This divergence is precisely canceled by the prefactor which
becomes proportional to $\alpha$ in this case. A finite contribution hence
remains, and it has to be considered to obtain the correct answer for the 
integral. We will comment further on this in a forthcoming publication 
\cite{usGPXT}.

Expanding the result into a power series with negative powers in $\varepsilon$, we obtain for the pole part after subtraction of the subdivergences
\begin{equation}
\begin{aligned}
\settoheight{\eqoff}{$\times$}%
\setlength{\eqoff}{0.5\eqoff}%
\addtolength{\eqoff}{-7.5\unitlength}%
\raisebox{\eqoff}{%
\begin{fmfchar*}(20,15)
  \fmfleft{in}
  \fmfright{out}
  \fmf{plain}{in,v2}
  \fmf{plain,tension=2,left=0.25}{v2,v1}
 \fmf{plain,tension=2,right=0.25}{v2,v3}
  \fmf{plain,tension=2,right=0.25}{v4,v5}
  \fmf{plain,tension=2,left=0.25}{v6,v5}
  \fmf{dblderplain}{v0,v1}
  \fmf{derplainpt}{v0,v3}
  \fmf{plain}{v0,v4}
  \fmf{derplain}{v0,v6}
  \fmf{plain}{v5,out}
\fmffixed{(0.9w,0)}{v2,v5}
\fmfpoly{phantom}{v6,v1,v3,v4}
  \fmf{plain}{v3,v4}
  \fmf{plain}{v1,v6}
\fmffixed{(0.4w,0)}{v3,v4}
\fmffreeze
\end{fmfchar*}}
&=\frac{1}{(4\pi)^8}
\Big(-\frac{1}{32\varepsilon^4}+\frac{1}{12\varepsilon^3}
-\frac{19}{96\varepsilon^2}
-\frac{1}{\varepsilon}\Big(\frac{29}{48}+\frac{5}{2}\zeta(3)
-\frac{5}{2}\zeta(5)\Big)
\Big)
\pnt
\end{aligned}
\end{equation}
Inserting this result into \eqref{I2fourdertrafo}, the
result for the required integral is hence found to be given by
\begin{equation}
\settoheight{\eqoff}{$\times$}%
\setlength{\eqoff}{0.5\eqoff}%
\addtolength{\eqoff}{-7.5\unitlength}%
\raisebox{\eqoff}{%
\begin{fmfchar*}(20,15)
  \fmfleft{in}
  \fmfright{out}
  \fmf{plain}{in,v2}
  \fmf{derplainpt,tension=2,right=0.25}{v1,v2}
 \fmf{plain,tension=2,right=0.25}{v2,v3}
  \fmf{derplainpt,tension=2,right=0.25}{v4,v5}
  \fmf{plain,tension=2,left=0.25}{v6,v5}
  \fmf{plain}{v0,v1}
  \fmf{plain}{v0,v3}
  \fmf{plain}{v0,v4}
  \fmf{plain}{v0,v6}
  \fmf{plain}{v5,out}
\fmffixed{(0.9w,0)}{v2,v5}
\fmfpoly{phantom}{v6,v1,v3,v4}
  \fmf{derplain}{v3,v4}
  \fmf{derplain}{v6,v1}
\fmffixed{(0.4w,0)}{v3,v4}
\fmffreeze
\end{fmfchar*}}
=\frac{1}{(4\pi)^8\varepsilon}\Big(-\frac{1}{4}
-\frac{3}{2}\zeta(3)+\frac{5}{2}\zeta(5)\Big)
\pnt
\end{equation}

\end{fmffile}

\footnotesize
\bibliographystyle{JHEP}
\bibliography{references}

\end{document}